\documentclass[timesnewroman,12pt, twoside]{report}
\usepackage[titletoc]{appendix}
\usepackage[a4paper]{geometry}
\usepackage{authblk,lineno,color,lscape,url,sectsty}
\usepackage[autostyle]{csquotes}
\usepackage[dvipsnames]{xcolor}
\usepackage[labelfont=bf]{caption}
\usepackage{emptypage}
\usepackage{array,float,multirow,graphicx,setspace,cite,url,titlesec,tocloft}
\usepackage{amsmath,amsfonts,mathtools,notoccite}
\usepackage{chngcntr,rotating,pdflscape,enumitem,algpseudocode,algorithm,subcaption}
\usepackage[square,sort,comma,numbers,sort&compress]{natbib}
\usepackage[colorlinks=True,citecolor=blue,linkcolor=blue,anchorcolor=blue,filecolor=blue,urlcolor=blue]{hyperref}
\usepackage{tensor}
\usepackage{braket}
\usepackage{booktabs}
\usepackage{lipsum}
\usepackage[Sonny]{fncychap}
\newcommand{\ym}{\widetilde{y}}

\newcommand{\vect}[1]{\boldsymbol{\mathrm{#1}}}
\begin{document}
%%%%%%%%%%%%%%%%%%%  Title page %%%%%%%%%%%%%%%%%%%%%%%%%%%%%%%%%%%
\pagenumbering{roman}
\thispagestyle{empty}
%%%%%%%%%%%%%%%%%%%%%%%%%%%%%%%%%%%%%%%
\thispagestyle{empty}
\graphicspath{{Figures/PNG/}{Figures/}}
%%%%%%%%%%%%%%%%%%%%%%%%%%%%%%%%%%%%%%%
\begin{center}
{\bf {\Large Impacts of dark matter interaction on nuclear and neutron star matter within the relativistic mean-field model} }
\end{center}
%%%%%%%%%%%%
\vspace{0.3 cm}
\begin{center}
    {\it \large By}
\end{center}
%%%%%%%%%%%%%%
\begin{center}
    {\bf {\Large Harish Chandra Das } \\ PHYS07201804002}
\end{center}
%%%%%%%%%%%%%%%
\begin{center}
\bf {{\large Institute of Physics, Bhubaneswar, India }}
\end{center}
%%%%%%%%%%%%%%
\vskip 2.0 cm
\begin{center}
%%%%%%%%%%%%%%
\large{
{ A thesis submitted to the }  \\
 {Board of studies in Physical Sciences }\\

In partial fulfillment of requirements \\
For the Degree of } \\
{\bf  DOCTOR OF PHILOSOPHY} \\
\emph{of} \\
{\bf HOMI BHABHA NATIONAL INSTITUTE}
%%%%%%%%%%%%%%%%%
\vskip 1.0 cm
%%%%%%%%%%%%%%%%%
\begin{figure}[H]
	\begin{center}
    \includegraphics[width=4.0cm, height= 4.0cm]{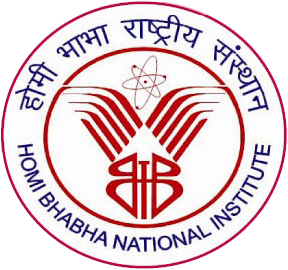}
	\end{center}
\end{figure}
%%%%%%%%%%%%%%%%%
\vskip 0.2 cm
{\bf \today}
\end{center}
\thispagestyle{empty}
%%%%%%%%%%%%%%%%%%%%%  
%\blankpage     
% %%%%%%%%%%%%%%%%%%%% Recommendation page %%%%%%%%%%%%%%%%%%%%%%%%%%%
% \input{recomendation.tex}
% \blankpage
% %%%%%%%%%%%%%%%%%%% Statement page %%%%%%%%%%%%%%%%%%%%%%%%%%%%%%%%
% \input{SBA.tex} 
% \blankpage   
% %%%%%%%%%%%%%%%%%%% Declaration page %%%%%%%%%%%%%%%%%%%%%%%%%%%%%%
% \input{declaration.tex}     
% \blankpage
% %%%%%%%%%%%%%%%%%%% List of publications %%%%%%%%%%%%%%%%%%%%%%%%%%
% \input{listofpublications.tex}
% %\blankpage
% %%%%%%%%%%%%%%%%%%% Dedication page %%%%%%%%%%%%%%%%%%%%%%%%%%%%%%%
% \input{dedication.tex}    
% \blankpage  
% %%%%%%%%%%%%%%%%%%% Acknowledgement page %%%%%%%%%%%%%%%%%%%%%%%%%%
% \input{acknowledgment.tex}
% \blankpage
%%%%%%%%%%%%%%%%% Table of contents %%%%%%%%%%%%%%%%%%%%%%%%%%%%%
\hypersetup{linkcolor=blue}
\tableofcontents
%\blankpage 
%%%%%%%%%%%%%%%%%%%%%%%%%%%%%%%%
\numberwithin{equation}{chapter}
\numberwithin{figure}{chapter}
\numberwithin{table}{chapter}
%%%%%%%%%%%%%%%%%%%%%%%%%%%%%%
% %%%%%%%%%%%%%%%%%%%% Synopsis %%%%%%%%%%%%%%%%%%%%%%%%%%%%%%%%%%%%%%
% \input{summary.tex}
% \blankpage 
% %%%%%%%%%%%%%%%%%%% Table of figures %%%%%%%%%%%%%%%%%%%%%%%%%%%%%%
% \addcontentsline{toc}{chapter}{List of Figures}
% \listoffigures
% \blankpage 
% %%%%%%%%%%%%%%%%%%% Table of tables %%%%%%%%%%%%%%%%%%%%%%%%%%%%%%%
% \addcontentsline{toc}{chapter}{List of Tables}
% \listoftables
%\blankpage 
%%%%%%%%%%%%%%%%%%%%%%%%%%%%%%%%%%%%%%%%%%%%%%%%%%%%%%%%%%%%%%%%%%%
\newpage
\setcounter{page}{1}
\pagenumbering{arabic}
%%%%%%%%%%%%%%%%%%%%%%%%% Chapter-1 %%%%%%%%%%%%%%%%%%%%%%%%%%%%%%%
%\blankpage
%%%%%%%%%%%%%%%%%%%%%%% CHAPTER - 1 %%%%%%%%%%%%%%%%%%%%\\
\chapter{Introduction}
\label{C1} 
%%%%%%%%%%%%%%%%%%%%%%%%%%%%
\graphicspath{{Figures/Chapter-1figs/PDF/}{Figures/Chapter-1figs/}}
%%%%%%%%%%%%%%%%%%%%%%%%%%%%%%%%%%%%%%%%%%%%%%%%%%%%%%%%%%%%%%%%%%%
%%%%%%%%%%%%%%%%%%%%
\section{Background}
\begin{figure}
	\centering
	\includegraphics[width=0.5\columnwidth]{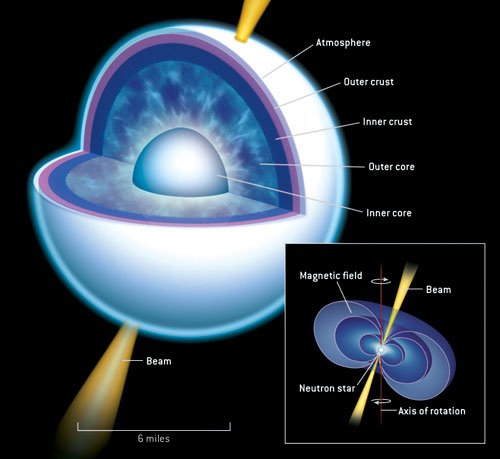}
	\caption{Schematic structure of the NS. This figure is taken from \cite{sigg_francisco_2005}}
	\label{fig:NS_structure}
\end{figure}

%%%%%%%%%%%%%%%%%%%%
Neutron star (NS) is one of the most enigmatic stellar compact object forms after the collapse of the star having the masses $> 8-20 \ M_\odot$ during the type II supernovae ~\cite{Burrows_1986}. It is mainly composed of neutrons and a few amounts of protons and leptons to maintain the charge neutrality. The density inside the NS is mainly $5-10$ times the nuclear matter (NM) saturation density \cite{Lattimer_2004}. It can be observed as a hot object having its surface temperature $\sim6\times10^5$ K. The magnetic field at the surface of the NS is $\sim 10^{15}$ G, while it is roughly $10^{18}$ G in the core. It is approximately $10^8-10^{15}$ times stronger than the Earth's magnetic field. The gravitational field at the surface of the NS is about $\sim200$ billion times larger than the field at the surface of the Earth. The internal structure is divided mainly into four parts such as (i) outer crust, (ii) inner crust, (iii) outer core, and (iv) inner core, as shown in Fig. \ref{fig:NS_structure}. Because of the density variation, different shapes and non-spherical configurations, known as pasta phases, are formed inside the crust. A large number of exotic particles, such as hyperons and kaon condensations, appear, mainly in the core. The hadrons are converted to quarks due to very high baryon density. Therefore, different areas of physics require exploration of the various properties of NS.
%%%%%%%%%%%%%%
%%%%%%%%%%%%%%
%%%%%%%%%%%%%%%%%%%%%%%%%%%%%%%%%%%%%%%%%
\subsection{History of the Neutron Stars}
%%%%%%%%%%%%%%%%%%%%%%%%%%%%%%%%%%%%%%%%%
For the first time, in February 1931, Landau theoretically postulated the compact star as a gigantic nucleus. He hypothesized that the stars composed of NM might exist, which are more compact than white dwarfs. Accordingly, he asserted that {\it``all stars heavier than 1.5 times the sun's mass undoubtedly have zones where quantum statistics principles are broken"}. However, in the last part of the paper, he concluded that {\it ``the density of matter becomes so great that atomic nuclei come in close contact, forming one gigantic nucleus"} \cite{Haensel_2007}. This work was coincidentally published on February 1932, just a few days after the publication of the discovery of neutron \cite{Chadwick_1932}.

After two years, in 1934, Baade and Zwicky analyzed the emission of tremendous amounts of energy in supernovae explosions \cite{Baade_1934_1, Baade_1934_2}. They explained the supernovae explosions as the transitioning from ordinary stars to objects with tightly packed neutrons in their final stages, hence the name NS. Subsequently, it is proposed that such stars could have very small radii and be highly dense. Furthermore, it is also suggested that the gravitational packing energy could be immense and exceed the ordinary  packing fractions because neutrons can be packed more efficiently than nuclei and electrons. As a result, the NS may be represented as the configuration of the most stable matter. Finally, on November 28, 1967, Jocelyn Bell discovered the NSs as radio pulsars. 
%%%%%%%%%%%%%%%%%%%%%%%%%%%%%%%%%%%
\subsection{Birth of Neutron Stars}
\begin{figure}
	\centering
	\includegraphics[width=0.5\columnwidth]{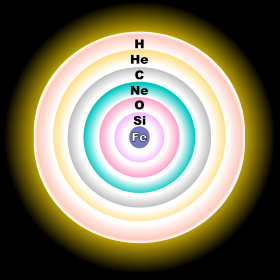}
	\caption{Onion-like structures for the nuclear fusion inside the star. The image is taken from \cite{Sutter_2020}.}
	\label{fig:fusion_shell}
\end{figure}

%%%%%%%%%%%%%%%%%%%%%%%%%%%%%%%%%%%
The death of a massive star with a mass greater than $8-20 \ M_\odot$ produces a NS in the supernovae ~\cite{Burrows_1986}. The evolution of a massive star is a steady process which accelerates towards higher temperature and density, mainly in the core part. When the core temperature of the star reaches around $10^7$ K, four hydrogen nuclei fuse into a helium nucleus, releasing thermal energy that heats the star's core and provides outward thermal pressure that protects the star from gravitational collapse. The star spends nearly $10^7$ years of its life in the hydrogen-burning phase. Because the temperature in the core is still insufficient for the helium to fuse, the generated helium accumulates there. When the core of the hydrogen runs out, it compresses and heats up to a temperature where helium may fuse with carbon. The star has been in the helium-burning phase for over $10^6$ years. This cycle is repeated at a progressively faster rate through the stages of burning carbon, oxygen, and silicon. The final stage of silicon burning produces a core of iron from which no further energy can be extracted through nuclear burning, and the fusion stops in the core. The iron core of the star is currently encased in a mantle made of silicon, sulfur, oxygen, neon, carbon, helium, and an attenuated envelope of hydrogen. The star has an onion-like structure, as shown in Fig. \ref{fig:fusion_shell}. The star is compressed by gravitational pressure to an extremely high density, which causes the electrons to adopt a relativistic state. The protons capture the relativistic electrons by the inverse beta decay process, which reduces the electron pressure and falls to the point at which any further increase in the mass of the core is supported against gravity. At this stage, the maximum mass of the core is between $1.2-1.5 \ M_\odot$. Finally, the core rapidly undergoes an implosion in less than a second. During the collapse, many energetic neutrinos escape from the core due to the neutronization in inverse beta decay.
%%%%%%%%%%%%%%
%%%%%%%%%%%%%%

The amount of energy released in these supernovae is $~10^{53}$ ergs mainly via neutrino emission ~\cite{Janka_2012}. At the initial stage after the SN, the environment contains almost protons, leptons, and few neutrons, termed proto-NS (PNS). The newly born PNS is very hot (temperature is $\sim 5-40$ MeV), having a density in the range ($2-6\times 10^{14}$ gm/cm$^3$). The high-energy neutrinos escape due to their non-interacting character, which carries enormous energy which cools the PNS. It takes only a few milliseconds to tens of seconds for the lepton-rich cores to the hydrostatic equilibrium. The cooling of NSs depends on the state of super dense matter in the interior, which mainly controls the neutrino emission, and on the structure of the outer layers where photon emission is controlled. The cooling simulations can provide important information about the various physical processes in the interior of NSs. It is confronted with observations in different regions of the electromagnetic spectrum like soft X-ray, UV, extreme UV, and optical observations of the thermal photon flux emitted from the surface. For newly born hot NSs, neutrino emission is the predominant cooling mechanism with an initial time scale of a few seconds. The neutrino cooling continues to dominate for at least initial thousand years or even longer for the slow cooling scenarios. The photon emission overtakes neutrino emission after the internal temperature has dropped sufficiently.

The equation of state (EOS) of dense matter and its associated neutrino opacity are the essential micro-physical ingredients that govern the evolution of the PNS during the so-called Kelvin-Helmholtz epoch, which changes the remnant from a hot and bloated lepton-rich PNS into a cold and deleptonized NS ~\cite{Pons_1999, Burrows_1986}. The lifetime of the NS is around a billion years. In these years, it cools with the emission of neutrinos, and photons ~\cite{Page_2004, Yakovlev_2004, Yakovlev_2005}. Some processes are direct URCA, modified URCA, x-ray emission, which cools to shallow temperatures ~\cite{Yakovlev_2004}, and finally, collapse to a BH. 
%%%%%%%%%%%%%%%%%%%%%%%%%%%%%%%%%%%%%%%%%%
\subsection{Detection of the Neutron Stars}
%%%%%%%%%%%%%%%%%%%%%%%%%%%%%%%%%%%%%%%%%%
A typical NS has a mass $\sim 2 M_\odot$, and a radius is $\sim10$ km. It has a central density of $5-10$ times the nuclear saturation density. Therefore, it is also called the most known densest object in the Universe. It was first detected as a pulsar (PSR). The pulsar is a rotating NS that emits light from the pole due to its vast magnetic field ($10^8-10^{15}$ times more potent than Earth's magnetic field), and the rotational speed of the pulsar is very fast. The fastest pulsar we detected is PSR J1748-2446ad, with a frequency of 716 Hz or 43000 revolutions per minute.

Different ways to detect the NS, such as hot-spot measurement, pulsars timing measurement, x-ray wavelength, and gravitational waves measurement, are useful techniques used by different observatories and telescopes. Using the Shapiro delay techniques, the mass of PSR J1614–2230 is estimated by Demorest {\it et al.} $M=1.97\pm0.04 M_\odot$ ~\cite{Demorest_2010}. Further, the mass range of PSR J1614–2230 is modified using the high-precision timing of the pulsar techniques, and it is found to be $M = 1.908\pm0.016 \ M_\odot$ in 2018~\cite{Arzoumanian_2018}. In 2013, another pulsar (PSR J0348+0432) having mass $M = 2.01\pm 0.04 \ M_\odot$, which is slightly higher was observed by Antoniadis {\it et al.} ~\cite{Antoniadis_2013}. Using the same Shapiro delay measurement, Cromartie {\it et al.} \cite{Cromartie_2020} deduced the mass of the millisecond pulsar (MSP) J0740+6620 to be $M=2.14_{-0.09}^{+0.10} \ M_\odot$ with 68.3\% credibility interval (CI), which is an important constraint on the equation of state of ultra-dense matter and has implications for our understanding of NSs. Recently, the limit has been refined with mass $M=2.08_{-0.07}^{+0.07} \ M_\odot$ with 68.3\% CI \cite{Fonseca_2021}. The heaviest and fastest pulsar is  PSR J0952–0607, with a reported mass of $2.35\pm0.17 \ M_\odot$ has been detected in the Milk Way Galaxy disc, which comes in the categories of ``black widow" pulsars \cite{Romani_2022}. The simultaneous measurement of mass and radius of the PSR J0030+0451 by the NASA Interior Composition Explorer (NICER), a telescope fixed in the international space station, has provided valuable insights into the properties of NSs and the nature of matter at extreme densities. The telescope measures the hot spot in the NS, which has been fitted for soft X-ray waveforms of rotation-powered pulsars observed and estimated the mass and its corresponding radius is given by Miller {\it et al.} are $M=1.44_{-0.14}^{+0.15} \ M_\odot$, and $R=13.02_{-1.06}^{+1.24}$ km ~\cite{Miller_2019} respectively. Another similar estimation has been provided by Riley {\it et al.} is $M=1.34_{-0.16}^{+0.15} \ M_\odot$, and $R=12.71_{-1.19}^{+1.14}$ km respectively ~\cite{Riley_2019}. Recently, Miller {\it et al.} put another radius constraint on both for the canonical ($R_{1.4} = 12.45 \pm 0.65$ \ km) and maximum NS ($R_{2.08} = 12.35 \pm 0.75$ \ km) from the NICER and X-ray Multi-Mirror (XMM) Newton data ~\cite{Miller_2021}.

In 2017, the two terrestrial detectors, such as advanced LIGO and Virgo, detected the gravitational waves (GWs) from the binary NS (BNS) merger event, opening a new era in NS physics. The best-measured quantity is the chirp mass, ${\cal M}=1.188_{-0.002}^{+0.004} \ M_\odot$ in the inspiral phase since they chirp like birds. Another quantity measured is the combined dimensionless tidal deformability ($\Tilde{\Lambda}\leq800$) for the low spin prior case with 90\% confidence limit ~\cite{Abbott_2017}. The translated value of canonical tidal deformability is found to be ($\Lambda_{1.4}=190_{-120}^{+390}$) ~\cite{Abbott_2018}. The GW170817 has put a limit on the tidal deformability of the canonical NS, which is used to constrain the EOS of neutron-rich matter at 2-3 times the nuclear saturation densities ~\cite{Abbott_2017}. Due to a strong dependence on tidal deformability with radius ($\Lambda \sim R^5$), it can put stringent constraints on the EOS. Several approaches ~\cite{Bauswein_2017, Annala_2018, Fattoyev_2018, Radice_2018, Mallik_2018, Most_2018, Tews_2018, Nandi_2019, Capano_2020} have been tried to constraint the EOS on the tidal deformability bound by the GW170817 and it discarded tons of EOSs.

The LIGO and Virgo collaborations discovered the most mysterious compact binary coalescence event (GW190814), which involved a black hole and a compact object with masses of 22.2--24.3 and 2.50--2.67 $M _\odot$, respectively ~\cite{RAbbott_2020}. Because the electromagnetic counterpart has yet to be discovered and there is no detectable signal of GW tidal deformations, the secondary component might be the lightest BH or the heaviest NS. As a result, the enigma of the secondary component of the GW190814 event has sparked much debate ~\cite{Most_2020, Vattis_2020, Tews_2021, ZhangAAS_2020, Lim_2021, Godzieba_2021, Huang_2020, Tan_2020, Fattoyev_2020, Roupas_2021, Biswas_2021}. We hope the more binary NS merger soon may tightly constrain the properties of the NS. Nevertheless, we have to establish enough theoretical models which can help to understand the observational data more clearly. 
%%%%%%%%%%%%%%%%%%%%%%%%%%%%%%%%%%%%%%%%%%%%%%%%%%%%
\subsection{Internal Structure of the Neutron Stars}
%%%%%%%%%%%%%%%%%%%%%%%%%%%%%%%%%%%%%%%%%%%%%%%%%%%%
The internal structure of NS is divided mainly into four parts, as shown in Fig. \ref{fig:NS_structure}. Many different complex phenomena/processes such as hyperons productions ~\cite{NKGh_1985, Schaffner_1996, Dover_1984, Schaffner_2000, Weissenborn_2012, WeissenbornNPA_2012, Biswal_2016, Biswal_2019}, kaons condensations ~\cite{NKG_kaon_1985, NKGk1_1998, NKGk2_1999, Kaplan_1998, Gupta_1997, Gupta_2012, Pal_2000}, quarks deconfinement ~\cite{Orsaria_2014, Mellinger_2017, Sharma_2007, Collins_1975}, superfluidity \cite{BAYM_1969, Chamel_2017, Haskell_2018}, pasta structures \cite{Ravenhall_1983, Lorenz_1993, Avancini_2008, Pons_2013, Newton_2022, Parmar_pasta_2022}, glitch activity \cite{Fuentes_2017, Mongiovi_2017, Basu_2018}, phase transitions \cite{Dexheimer_2018}, oscillations \cite{Anderson_1996, Anderson_1998}, magnetic field \cite{Reisenegger_2003, Peng_2007, Chatterjee_2015, Rather_2021}, anisotropy \cite{Bowers_1974, Cosenza_1981, Horvat_2010, Silva_2015, Doneva_2012, Biswas_2019, Das_2022}, dark matter (DM) capture \cite{Goldman_1989, Kouvaris_2008, Kouvaris_2010, Kouvaris_2011, Kouvaris_2012} etc. are occurring inside the NS. Therefore, exploring the NS properties requires different areas of physics due to its complex behavior. One can treat it as the best laboratory for studying the above-mentioned processes. 

In this study, we take the DM as an extra candidate inside the NS, which affects some properties of the NS. The interacting Lagrangian density is modeled with the assumptions as discussed in Chapter-\ref{C3}. The interaction between nucleons is also taken care of with the help of the relativistic mean-field model, as discussed in the following section.
%%%%%%%%%%%%%%%%%%%%%%%%%%%%%%%%%%%%%%%%%% 
\section{Relativistic mean-field model}
%%%%%%%%%%%%%%%%%%%%%%%%%%%%%%%%%%%%%%%%%%
In recent years, nuclear astrophysics has been well described by the self-consistent effective mean-field models. Among the effective theories, the relativistic mean-field (RMF) approach is one of the most successful formalisms that is currently attracting attention to theoretical analyses of systems such as finite nuclei, NM, and the NS. However, the results for finite nuclei are often reasonably similar to each other, even though the formulation of the energy density functional for the RMF model differs from those for the non-relativistic models, such as Skyrme \cite{Skyrme_1956, Skyrme_1958} and Gogny \cite{Gogny_1980} interactions. The characteristics of NSs may also be predicted with the same degree of accuracy. The RMF model adequately takes into account relativistic effects at higher densities. The interchange of mesons is the basis for the RMF model's description of nucleon interactions. Collectively, these mesons are referred to as effective fields and are represented by conventional classical numbers. In a nutshell, the one-boson exchange (OBE) theory of nuclear interactions approximates the relativistic Hartree or Hartree-Fock method using the RMF formalism. In OBE theory, the nucleons communicate with one another by exchanging isovector mesons like $\pi$, $\rho$, and $\delta$ as well as isoscalar mesons like $\eta$ and $\omega$. The ground-state parity symmetry is not seen for the $\pi$ and $\eta$ mesons because they are pseudo-scalar in nature. They do not impact the ground-state characteristics of even nuclei at the mean-field level. 

By accounting for the contributions of the $\sigma$, $\omega$, and $\rho$ mesons with excluding any nonlinear term for the Lagrangian density, the first and simplest relativistic Lagrangian is given by Walecka \cite{Walecka_74}. This model predicts excessively high incompressibility ($K$) of $\sim 550$ MeV for the infinite NM at saturation. Boguta and Bodmer \cite{Boguta_1977} include the self-coupling factors in the $\sigma$-mesons, which not only lowers the value of $K$ to an acceptable range but also improves the quality of finite nuclei results significantly. Numerous non-linear (NL) parameter sets have been calibrated based on this Lagrangian density \cite{Reinhard_1986, Reinhard_1988, Reinhard_1989, Lalazissis_1997}.  However, at supra-normal densities, the EOSs are surprisingly stiff. As a result, the Lagrangian density has been modified to include vector meson self-coupling, and several parameter sets are created \cite{Bodmer_1991, Gmuca_1992, Gmuca_2003, Sugahara_1994}. These forces explain finite nuclei and NM features to a large extent, although the presence of the Coester band \cite{Coester_1970} and three-body effects must be addressed. The EOS derived by the realistic models (like Bonn and Paris potential) do not reach the Coester band or empirical saturation threshold of symmetric NM ($E/A \simeq 16.0$ MeV at $\rho_0=0.16$ fm$^{-3}$). The accuracy of these models is called into question because, in the greater density domain, all of them exhibit distinct natures than those utilized for NS matter. By creating the density-dependent coupling constants  \cite{Typel_2005} and the effective-field-theory motivated relativistic mean field (E-RMF) model \cite{Furn_1996, Frun_1997, Vretenar_2000}, nuclear physicists also altered their perspective and used new techniques to enhance the outcome.

Additionally, driven by the effective field theory, Furnstahl {\it et al.} \cite{Furn_1996, Frun_1997} utilized all couplings up to the fourth order of the expansion, taking advantage of the naive dimensional analysis and naturalness and derived the G1 and G2 parameter sets. They only considered the isoscalar-isovector cross-coupling contributions to the Lagrangian density since it significantly impacts the neutron radius and EOS of asymmetric NM \cite{Todd_2005, Horowitz_2001}. Later, it is understood that $\delta$-meson contributions are also required to explain specific characteristics of NM under extreme circumstances \cite{Kubis_1997, Singh_2013, Kumar_2017}. Although the $\delta$ mesons' contributions to the bulk properties of normal NM are negligible, they significantly impact on highly asymmetric cases such as NS. Finally, our group's efforts to find a suitable parameter set which are free from all previously mentioned flaws and being inspired by all the previous parameter sets, the new parameter sets G3 and IOPB-I for finite and infinite NM as well as NS systems within the E-RMF model are created. All essential terms, such as $\sigma-\rho$, $\omega-\rho$ cross-couplings, and $\delta$ meson are present in the G3 model Lagrangian \cite{Kumar_2017}. These cross-couplings altered the symmetry energy and density-dependent symmetry parameters, changed the nature of the neutron skin-thickness for finite nuclei, and placed restrictions on the EOS for pure neutron matter \cite{Kumar_2017, Kumar_2018}. The detailed description is given in Chapter-\ref{C2}. 
%%%%%%%%%%%%%%%%%%%%%%%%%%%%%%%%%%%%%%%%%%%%%%%%%%
\section{Motivation for the present research work}
\label{S1.2} 
%%%%%%%%%%%%%%%%%%%%%%%%%%%%%%%%%%%%%%%%%%%%%%%%%%
It is believed that DM occupies $\sim 27\%$ of the matter density in the Universe others are dark energy and visible matter. The DM particles are found in halo forms in the dense region of the Universe. The typical NS has a lifetime of around one billion years. In its evolving stage, theoretically, it has been hypothesized that some DM particles are captured inside the NS due to its huge baryon density, and immense gravitational potential \cite{Goldman_1989, Kouvaris_2008, Kouvaris_2010, Raj_2018, Bell_2021, Bertone_2008}. The accretion also depends on the amount of DM halo present in that environment and the mechanism of accretion processes, such as thermal and non-thermal. The accreted DM particles are either fermionic \cite{Sandin_2009, Das_2019, Das_2020, Das_2020}, or bosonic \cite{Bertoni_2013}. After accretions, the DM particles interact with nucleons, affecting the NS properties. It is the primary motivation for the present thesis. However, other experimental and observational results also open a window to explore the NS properties differently.

The detection of GWs from the merger of the BNS in the GW170817 event has opened up the possibility that NSs can accumulate some amount of DM. This is because the inspiral and merger of the BNS system can cause the capture of DM particles by the NSs, which would affect their masses and properties. This has important implications for our understanding of the nature of DM and its interactions with baryonic matter, as well as for the astrophysical and cosmological implications of NSs. Another event, GW190814, also opens a debatable issue of whether its secondary component is the lightest black hole or the heaviest NS. There are various theoretical insights provided that the NS is a natural DM detector since it captures a certain amount of DM due to its huge baryon density and gravitational potential \cite{Raj_2018, Bell_2021}. In this work, we use the RMF model to calculate the nuclear interactions inside the NS. However, for the DM case, we choose a simple model, assuming that the DM particles are already accreted inside it. Also, we want to constrain the amount/percentage of DM within the NS with the help of different observational data. It is the second motivation for this thesis. It has been observed that with the addition of DM, the macroscopic properties such as mass ($M$), radius ($R$), tidal deformability ($\Lambda$), the moment of inertia ($I$), surface curvature, oscillations frequency are significantly affected with increasing its percentage. In the subsequent chapters, we will discuss the effects of DM on the various NS properties. However, we want to give brief work details in the following section.
%%%%%%%%%%%%%%%%%%%%%%%%%%%
\section{Problem statement}
\label{S1.3}
%%%%%%%%%%%%%%%%%%%%%%%%%%%
Here, we calculate the various properties of the DM admixed NS (DMANS) by changing the DM fractions. To calculate the NS EOS, we use the RMF model with different forces. The Lagrangian density is constructed with the assumption that the DM particles interact with nucleons by exchanging the standard model Higgs \cite{Panotopoulos_2017, Das_2019, Das_2020}. Hence, the Lagrangian density of the DM admixed NS is the addition of both NS and DM (see subsection-\ref{DM_lagrangian}). In this model, we assume that the DM density is constant throughout the NS having its fraction one-third of the total baryon density \cite{Das_2019}. We divide the effects of DM into the following categories of NS: 
%%%%%%%%%%%%%%%%%
\begin{enumerate}
	\item Isolated static/rotating, 
	\item Oscillating but static, 
	\item Binary NS properties.
\end{enumerate}
%%%%%%%%%%%%%%%%
The properties of static and isolated NS, such as mass, radius, tidal deformability, the moment of inertia, curvature, and oscillations, are explored in detail with different fractions of DM. For that, we use the RMF and E-RMF equation of states. The binary NS properties, such as tidal Love number and its deformability, gravitational waves frequency, phase, and last time in the inspiral orbit, are obtained using the post-Newtonian method. It is noticed that DM has significant effects on the various properties, and the magnitude of macroscopic observables reduces with the percentage of DM \cite{Das_2020, Das_2021, DasPRD_2021}. Also, the higher percentage of DMANS sustains more time in the inspiral phase \cite{DasMNRAS_2021}. Another important finding is that DMANS cools more rapidly due to the acceleration of the URCA process, and the time taken for thermal relaxation between the core and the surface is almost 300 years less than without DM contains \cite{Kumar_2022}.
%%%%%%%%%%%%%%%%%%%%%%%%%%%%%%%%%%%%
\section{Organization of the thesis}
\label{S1.4}
%%%%%%%%%%%%%%%%%%%%%%%%%%%%%%%%%%%%
The research work presented in the thesis is organized as follows. After a brief introduction, we outline the E-RMF models and their applications on finite nuclei to NSs in Chapter-\ref{C2}. In this chapter, we provide the detailed formalism for RMF and E-RMF theory. We test the validity of these models by applying them to different nuclear systems and comparing their numerical values with available experimental data. Many parameter sets, which are currently used in the literature and found to be successful, are implemented for the calculations of finite nuclei, NM, and NSs. These models are further used to explore various quantities presented in the forthcoming chapters. 

In Chapter-\ref{C3}, we describe the effects of DM on the NM and its properties. For DM interactions with nucleons, we choose a simple model. The NM quantities such as EOS, binding energy, effective mass, incompressibility, symmetry energy, and its different coefficients are calculated by changing DM fractions as well as the asymmetry parameter. It is to be noted that the symmetry energy and its slope play a great role in the cooling as well as the radius of the NS. 

In Chapter-\ref{C4}, the structural properties of the isolated, static, and rotating DM admixed NS properties are calculated by solving the differential equations with various EOSs obtained from the RMF and E-RMF models with a variation in the DM fractions. The obtained properties are constrained with the help of observational data. The moment of inertia is calculated with a slow rotation approximation of the NS by changing the DM percentages. The curvature inside and outside the NS are also evaluated by solving Einstein's equations. The binding energy of the DMANS is calculated since it is a crucial quantity for a star. 

NSs oscillate with different modes of frequencies after their formation. In Chapter-\ref{C5},  we explore the oscillation properties of hyperon stars in the presence of DM. With the density gradient, different hyperons appear in the core of the NS. Therefore, the inclusion of hyperons inside the star is needed to analyze its effects on the oscillation frequency with the presence of DM. We use the relativistic Cowling approximations to obtain the $f$-mode oscillation frequency for various EOS by changing the DM fractions and constraining its range with the help of observational data. 

The binary NS emits gravitational waves in the inspiral phases. In Chapter-\ref{C6}, we calculate the binary NS inspiral properties for DMANS with various EOSs. The post-Newtonian method is taken to obtain elapsed time, gravitational wave frequency, phases, etc., in the presence of DM. The gravitational waveforms such as $h_+$, $h_\times$, and $h_{22}$ are analyzed with and without DM. Finally, a summary and future perspective are given in the last chapter (Chapter-\ref{C7}).
%\blankpage
%%%%%%%%%%%%%%%%%%%%%%%%%% Chapter-2 %%%%%%%%%%%%%%%%%%%%%%%%%%%%%%
%%%%%%%%%%%%%%%%%%%%%%% CHAPTER - 2 %%%%%%%%%%%%%%%%%%%%%%%%%%
\chapter{Relativistic mean-field models and its applications: From finite nuclei to neutron star}
\label{C2} 
%%%%%%%%%%%%%%%%%%%%%%%%%%%%%%%%%%%%%%%%%%%%%%%%%%%%%%%%%%%%%%
In this chapter, we outline the formalism used for the study of finite nuclei, infinite nuclear matter, and their applications to NSs. The theory is based on quantum hadrodynamics, where nucleons and mesons are the basic degrees of freedom. Walecka formulated the foundation of the RMF theory in 1974 for the simplest $\sigma-\omega$ model \cite{Walecka_74}. In this model, the nucleons interact by exchanging $\sigma$ and $\omega$ mesons. Hence, the total energy density of the system is the contribution of both mesons and nucleons. By inclusion of non-linearity of the $\sigma$-meson, various models have already been constructed \cite{Boguta_1977, Lalazissis_1997}. However, the renormalization of these models is not possible. This is because significant effects from loop integrals that consider the dynamics of the quantum vacuum have caused problems for renormalizable. Therefore, effective field theory is the alternative option. We use the extended RMF model by including two extra mesons such as $\rho$, and $\delta$ \cite{Frun_1997, Singh_2013, Kumar_2017, Kumar_2018}. The self and cross-couplings between mesons up to $4^{\rm th}$ order are considered in the calculations. From the Lagrangian density, the energy density and pressure of the system are calculated for different conditions. Next, we discuss various properties of the finite nuclei, such as binding energy per particle, neutron skin thickness, single particle energy, two neutron separation energy, and charge radius for nuclei throughout the mass table. We also analyzed the NM properties for extreme conditions and, finally, extended the study to explore NS. 
%%%%%%%%%%%%%%%%%%%%%%%
\section{RMF Formalism}
\label{RMF_formalism}
%%%%%%%%%%%%%%%%%%%%%%%%%%%%%%%%%%%%%%
\subsection{Energy density functional}
\label{rmf_edf}
%%%%%%%%%%%%%%%%%%%%%%%%%%%%%%%%%%%%%%
The RMF model assumes that the nucleons interact by exchanging different mesons such as $\sigma$, $\omega$, $\rho$, and $\delta$. The interacting Lagrangian is formed by including all the interactions between all mesons and nucleons. In this case, we take the self ($\sigma^2$, $\sigma^3$, $\sigma^4$, $\omega^2$, $\omega^4$, and $\rho^2$) and cross ($\sigma^2-\omega^2$, $\omega^2-\rho^2$,  $\sigma-\omega^2$ and $\sigma-\rho^2$) couplings between mesons up to fourth order which gives significant contribution and some of the interactions considered to have marginal effects are ignored in the energy density functional ~\cite{Furnstahl_1996, Singh_2014, Kumar_2017, Kumar_2018, DasBig_2021}:
%%%%%%%%%%%%%
\begin{align}
{\cal E}(r)&=\sum_{j=p,n} \varphi_j^\dagger(r)\Bigg\{-i \vect{\alpha}\cdot \vect{\nabla}
+\beta \big[M_n -\Phi(r)-\tau_3 D(r)\big]+ W(r) +\frac{1}{2}\tau_3 R(r)
\nonumber\\
&
+\bigg(\frac{1+\tau_3}{2}\bigg)A(r)
-\frac{i\beta\vect{\alpha}}{2M_n} \cdot \left(f_\omega \vect{\nabla} W(r)
+\frac{1}{2}f_\rho\tau_3 \vect{\nabla} R(r)
+\lambda\vect{\nabla} A(r)\right)
+\frac{1}{2 M_n^2} \big(\beta_\sigma
\nonumber\\
&
+\beta_\omega \tau_3\big)\Delta A(r)\Bigg\}\varphi_j(r)
-\frac{\zeta_0}{4!}\frac{1}{g_\omega^2}W^4(r)
+\left(\frac{1}{2}+\frac{\kappa_3}{3!}\frac{\Phi(r)}{M_n}
+\frac{\kappa_4}{4!}\frac{\Phi^2(r)}{M_n^2}\right)
\frac{m_\sigma^2}{g_\sigma^2}\Phi^2({r})
\nonumber\\
&
+\frac{1}{2g_\sigma^2}\left(1+\alpha_1\frac{\Phi(r)}{M_n}\right)\big(\vect{\nabla}\Phi(r)\big)^2
-\frac{1}{2g_\omega^2}\left(1+\alpha_2\frac{\Phi(r)}{M_n}\right)\big(\vect{\nabla}W(r)\big)^2
\nonumber\\
&
-\frac{1}{2}\left(1+\eta_1\frac{\Phi(r)}{M_n}
+\frac{\eta_2}{2}\frac{\Phi^2(r)}{M_n^2}\right)
\frac{m_\omega^2}{g_\omega^2} W^2(r)
-\frac{1}{2e^2}\big(\vect{\nabla} A(r)\big)^2
-\frac{1}{2g_\rho^2} \left(\vect{\nabla} R(r)\right)^2
\nonumber\\
& 
-\frac{1}{2} \left(1+\eta_\rho\frac{\Phi(r)}{M_n}\right)
\frac{m_\rho^2}{g_\rho^2}R^2(r)
-\Lambda_{\omega}\big(R^2(r)\times W^{2}(r)\big)
+\frac{1}{2g_\delta^2}\big(\vect{\nabla}D(r)\big)^2
\nonumber \\
&
+\frac{m_\delta^2}{2g_\delta^2}D^2(r)
-\frac{1}{2e^2}\left(\vect{\nabla}A(r)\right)^2
+\frac{1}{3g_\gamma g_\omega} A(r) \Delta W +\frac{1}{g_\gamma g_\rho} A(r) \Delta R(r) \; ,
\label{eq:rmf_edf}
\end{align}
%%%%%%%%%%%
where $\Phi$, $W$, $R$, $D$, and $A$ are the fields with $\Phi=g_{\sigma}\sigma, W=g_{\omega} \omega, R=g_{\rho} \rho, {\rm and}\ D=g_{\delta}\delta$ respectively; $g_\sigma$, $g_\omega$, $g_\rho$, $g_\delta$ are the coupling constants; and $m_\sigma$, $m_\omega$, $m_\rho$, and $m_\delta$ are the masses for $\sigma$, $\omega$, $\rho$, and $\delta$ mesons respectively. $\kappa_3$ (or $\kappa_4$) and $\zeta_0$ are the self-interacting coupling constants of the $\sigma$ and $\omega$ mesons respectively. The parameters, such as $\eta_1, \eta_2, \eta_\rho, \Lambda_\omega, \alpha_1, \alpha_2, f_\omega$, and $f_\rho$ are the different cross-coupling terms. The term $\frac{e^2}{4\pi}$ is the photon coupling constant. $M_n$ represents the mass of the nucleon taken as 939 MeV. The $\tau_3$ are the Pauli matrices and behave as the isospin operator when it operates on neutron, and proton gives, $\tau_3\ket{p}=(+1)\ket{p}$ and $\tau_3\ket{n}=(-1)\ket{n}$. Furthermore, the parameters such as $g_\gamma, \lambda, \beta_\sigma $ and $\beta_\omega$ are responsible for the effects related to the electromagnetic structure of the pion and nucleon~\cite{Furnstahl_1996, Frun_1997}. We need to get the constant $\lambda$ to reproduce the magnetic moments of the nuclei and is defined by
%%%%%%%%%%%%%%%%
\begin{eqnarray}
\lambda=\frac{1}{2}\lambda_p (1+\tau_3)+\frac{1}{2}\lambda_n (1-\tau_3)
\label{eq:norm_const}
\end{eqnarray}
%%%%%%%%%%%%%%%
with $\lambda_p=1.793$ and $\lambda_n=-1.913$ the anomalous magnetic moments for the proton and neutron, respectively~\cite{Furnstahl_1996, Frun_1997}. The coupling constants and different meson masses are given in Table \ref{tab:rmf_mass_coupling}. The parameters are obtained by fitting the data for a few spherically known nuclei ($^{16}$O, $^{40}$Ca, $^{48}$Ca, $^{68}$Ni, $^{90}$Zr, $^{100,132}$Sn and $^{208}$Pb) along with the heavy-ion collision (HIC) data. 
%%%%%%%%%%%%%
\begin{table}[t] 
\centering
\caption{The coupling parameters for some of the well-known RMF and E-RMF parameter sets such as NL3~\cite{Lalazissis_1997}, FSUGarnet~\cite{Chen_2015}, G3~\cite{Kumar_2017}, IOPB-I~\cite{Kumar_2018} and BigApple~\cite{Fattoyev_2020, DasBig_2021} are listed.}
%\scalebox{1.3}{
\begin{tabular}{|c|c|c|c|c|c|c|c|c|c|c|}
\hline
\multicolumn{1}{|c}{Parameter}
&\multicolumn{1}{|c}{NL3}
&\multicolumn{1}{|c}{FSUGarnet}
&\multicolumn{1}{|c}{G3}
&\multicolumn{1}{|c}{IOPB-I}
&\multicolumn{1}{|c|}{BigApple}\\ \hline
$m_\sigma/M_n$  &  0.541  &  0.529&  0.559&0.533& 0.525 \\ 
$m_{\omega}/M_n$& 0.833  & 0.833 &  0.832&0.833& 0.833 \\ 
$m_{\rho}/M_n$&  0.812 & 0.812 &  0.820&0.812& 0.812 \\ 
$m_{\delta}/M_n$ & 0.0  &  0.0&   1.043&0.0&0.0  \\ 
$g_{\sigma}/4 \pi$  &  0.813  &  0.837 &  0.782 &0.827&0.769 \\ 
$g_{\omega}/4\pi$ & 1.024 & 1.091 &  0.923&1.062&0.980 \\ 
$g_{\rho}/4 \pi$&  0.712  & 1.105&  0.962 &0.885&1.126 \\ 
$g_{\delta}/4 \pi$  &  0.0  &  0.0&  0.160& 0.0&0.0 \\ 
$k_{3} $   &  1.465  & 1.368&    2.606 &1.496&1.878 \\ 
$k_{4}$  &  -5.688  &  -1.397& 1.694 &-2.932&-7.382  \\ 
$\zeta_{0}$  &  0.0  &4.410&  1.010  &3.103&0.106 \\
$\eta_{1}$  &  0.0  & 0.0&  0.424 &0.0& 0.0 \\ 
$\eta_{2}$  &  0.0  & 0.0&  0.114 &0.0&  0.0\\ 
$\eta_{\rho}$  &  0.0  & 0.0&  0.645& 0.0 &0.0 \\ 
$\Lambda_{\omega}$& 0.0  &0.043 &  0.038&0.024& 0.047 \\ 
$\alpha_{1}$  &  0.0  & 0.0 &   2.000& 0.0&  0.0    \\ 
$\alpha_{2}$  &  0.0  & 0.0 &   -1.468 &0.0&  0.0   \\
$f_\omega/4$  &  0.0  &  0.0 &   0.220 &0.0& 0.0   \\
$f_\rho/4$  &  0.0  &  0.0 &  1.239& 0.0& 0.0   \\ 
$\beta_\sigma$  &  0.0  &  0.0 & -0.087& 0.0& 0.0   \\ 
$\beta_\omega$  &  0.0  &  0.0 & -0.484 &0.0& 0.0   \\
\hline 
\end{tabular}%}
\label{tab:rmf_mass_coupling}
\end{table}
%%%%%%%%%%%%%%%%%%%%%%

The vacuum polarization effects of nucleons and mesons have been neglected because they are composite particles. Also, the negative energy states do not contribute to the densities and currents~\cite{Reinhard_1989}. During the fitting phase, the coupling constants of the effective Lagrangian are calculated using a collection of experimental data that includes a considerable portion of the vacuum polarisation effects in the no-sea approximation. The no-sea approximation is required to obtain the stationary solutions of the relativistic mean-field equations that characterize the ground-state characteristics of the nucleus. The Dirac sea contains the negative-energy eigenvectors of the Dirac Hamiltonian, which vary depending on the nuclei. As a result, it is determined by the specific solution of the set of nonlinear RMF equations. The Dirac spinors can be expanded in terms of vacuum solutions, forming a complete set of plane wave functions in spinor space. This set will be complete when the states with negative energies are part of the positive energy states and create the Dirac sea of the vacuum. 
%%%%%%%%%%%%%%%%%%%%%%%%%%%%%%%%%%%%%%%%%%%%%%%%%%%%
\subsection{Field equations for nucleons and mesons}
\label{rmf_field_equations}
%%%%%%%%%%%%%%%%%%%%%%%%%%%%%%%%%%%%%%%%%%%%%%%%%%%%
To solve the field equations for nucleons, we use the variational method. Hence, the single-particle energy of the nucleons can be obtained using the Lagrange multiplier $\varepsilon_j$, which is the energy eigenvalue of the Dirac equation constraining the normalization condition.
%%%%%%%%%%%%%
\begin{align}
    \sum_j\varphi_j^\dagger(r)\varphi_j(r)=1.
    \label{eq:norm_condition}
\end{align}
%%%%%%%%%%%
The Dirac equation for the wave function $\varphi_j(r)$ becomes
%%%%%%%%%%%%%%%%
\begin{align}
\frac{\partial}{\partial\varphi_j^\dagger(r)}\Bigg[{\cal E}(r)-\sum_j\varphi_j^\dagger(r) \varphi_j({r})\Bigg]= 0,
\label{eq:nucleon_minima}
\end{align}
%%%%%%%%%%%%%%%%
which implies that 
%%%%%%%%%%%%%
\begin{align}
\Bigg\{&-i \vect{\alpha} \cdot \vect{\nabla} \, 
+\beta [M_n - \Phi(r) - \tau_3 D(r)] + W(r)
+\frac{1}{2} \tau_3 R(r) + \frac{1 +\tau_3}{2}A(r)
\nonumber\\
&-\frac{i\beta\vect{\alpha}}{2M_n}\cdot\left(f_\omega \vect{\nabla} W(r)
+\frac{1}{2}f_{\rho} \tau_3 \vect{\nabla} R(r)
+\lambda\vect{\nabla} A(r)\right)
+\frac{1}{2 M_n^2} \big(\beta_\sigma
+\beta_\omega \tau_3\big)\Delta A(r)\Bigg\} \varphi_j (r)
\nonumber \\
&
= \varepsilon_j \, \varphi_j (r) \,.
\label{eq:nucleon_wavefn}
\end{align}
%%%%%%%%%%%
The mean-field equations for $\Phi(r)$, $W(r)$, $R(r)$, $D(r)$, and $A(r)$ are obtained by solving the $\left(\partial {\cal E}/\partial X_l\right)=0$, where $l$ represent for different mesons~\cite{Kumar_2018}
%%%%%%%%%%%%%
\begin{align}
-\Delta \Phi(r) + m_\sigma^2 \Phi(r) &= g_\sigma^2 \rho_s(r)
-\frac{m_\sigma^2}{M_n}\Phi^2 (r)\left(\frac{\kappa_3}{2}
+\frac{\kappa_4}{3!}\frac{\Phi(r)}{M_n}\right)
\nonumber \\
&
+\frac{g_\sigma^2}{2M_n}\left(\eta_1+\eta_2\frac{\Phi(r)}{ M_n}\right)\frac{m_\omega^2}{g_\omega^2} W^2(r)
+\frac{\eta_\rho}{2M_n}\frac{g_\sigma^2}{g_\rho^2}{m_\rho^2 } R^2(r)
\nonumber\\
&
+\frac{\alpha_1}{2M_n}\left[(\vect{\nabla}\Phi(r))^2
+2\Phi(r)\Delta\Phi(r)\right]
+\frac{\alpha_2}{2M_n}\frac{g_\sigma^2}{g_\omega^2}\left(\vect{\nabla}W(r)\right)^2, 
\label{eq:sigma_field}
\end{align}
%%%%%%%%%%%%%
\begin{align}
-\Delta W(r)+m_\omega^2 W(r)&=g_\omega^2 \left(\rho_b(r)
+\frac{f_\omega}{2} \rho_T(r) \right)
-\left(\eta_1+\frac{\eta_2}{2}\frac{\Phi(r)}{M_n} \right)\frac{\Phi(r)}{M_n} m_\omega^2 W(r)
\nonumber\\
&
-\frac{1}{3!}\zeta_0 W^3(r)
+\frac{\alpha_2}{M_n} \big[\vect{\nabla}\Phi(r) \cdot\vect{\nabla}W(r)
+\Phi(r)\Delta W(r)\big]
\nonumber\\
&
-2\Lambda_\omega g_\omega^2 R^2(r)W(r) \,,
\label{eq:omega_field}
\end{align}
%%%%%%%%%%%%%
\begin{align}
-\Delta R(r)+ m_\rho^2 R(r)&=\frac{1}{2}g_\rho^2 \left (\rho_{b3}(r) + \frac{1}{2}f_{\rho}\rho_{T,3}(r)\right)
-\eta_\rho \frac{\Phi(r)}{M_n}m_\rho^2 R(r)
\nonumber\\
&
-2\;\Lambda_\omega g_\rho^2 R(r) W^2(r) \,,
\label{eq:rho_field}
\end{align}
%%%%%%%%%%%%%
\begin{align}
-\Delta D(r)+m_\delta^2 D(r) = g_\delta^2\rho_{s3} \, ,
\label{eq:delta_field}
\end{align}
%%%%%%%%%%%%%
\begin{align}
-\Delta A(r)=e^2 \rho_p(r) \,, 
\label{eq:photon_field}
\end{align}
%%%%%%%%%%%
%%%%%%%%%%%%%%%%%%%%%%%%%%%%%%%%%%%%%%%%%%%
\subsection{Different densities parameters}
\label{rmf_different_densties}
%%%%%%%%%%%%%%%%%%%%%%%%%%%%%%%%%%%%%%%%%%%
The baryon, scalar, iso-vector, iso-scalar, proton, tensor, and iso-tensor densities are given as~\cite{Singh_2014, Kumar_2018}
%%%%%%%%%%%%
\begin{align}
\rho_b(r) = \sum_{j=p,n} \left\langle \varphi_j^\dagger(r) \varphi_j(r) \right\rangle\,
=\rho_p(r)+\rho_n(r)
=\sum_{j=p,n}\frac{1}{\pi^2}\int_{0}^{k_j}k^2 dk \, ,
\label{eq:baryon_density} 
\end{align}
%%%%%%%%%%%%
\begin{align}
\rho_s(r)=\sum_{j=p,n} \left\langle \varphi_j^\dagger(r) \beta \varphi_j(r)\right\rangle
=\rho_{sp}(r)+\rho_{sn}(r)
=\sum_{j=p,n} \frac{M_j^{\ast}}{\pi^2}\int_{0}^{k_j} dk \, \frac{k^2}{E_j^*} \, ,
\label{eq:scalar_density} 
\end{align}
%%%%%%%%%%%%%
\begin{align}
\rho_{b3} (r)=\sum_{j=p,n} \left\langle\varphi_j^\dagger(r) \tau_3 \varphi_j(r)\right\rangle
=\rho_p(r) - \rho_n(r)\, ,
\label{eq:isovector_density}
\end{align}
%%%%%%%%%%%%
\begin{align}
\rho_{s3} (r)=\sum_{j=p,n}  \left\langle\varphi_j^\dagger(r) \tau_3 \beta \varphi_j(r)\right\rangle
=\rho_{sp}(r)-\rho_{sn}(r) \, ,
\label{eq:isoscalar_density}
\end{align}
%%%%%%%%%%%%
\begin{align}
\rho_p(r)=\sum_j \left\langle\varphi_j^\dagger(r) \left (\frac{1 +\tau_3}{2}
\right) \varphi_j(r)\right\rangle\, ,
\label{eq:proton_density} 
\end{align}
%%%%%%%%%%%%%
\begin{align}
\rho_T(r)=\sum_{j=p,n} \left\langle \frac{i}{M_n} \vect{\nabla} \cdot
\left[\varphi_j^\dagger(r) \beta\vect{\alpha}\varphi_j(r) \right]\right\rangle \, ,
\label{eq:tensor_density} 
\end{align}
%%%%%%%%%%%
and
%%%%%%%%%%%%%
\begin{align}
\rho_{T,3}(r)=\sum_{j=p,n} \left\langle \frac{i}{M_n} \vect{\nabla} \!\cdot\!
\left[\varphi_j^\dagger(r) \beta \vect{\alpha}
\tau_3 \varphi_j(r) \right] \right\rangle \, ,  
\label{eq:isotensor_density} 
\end{align}
%%%%%%%%%%%   
where $E_j^*=\sqrt{k_j^2+M_j^{*2}}$ is the effective energy of the nucleons. $k_j$ is the Fermi momentum of the nucleons, $\sum_j$ is over all occupied states, and $M_j$ is the effective mass defined in the following sub-section. 
%%%%%%%%%%%%%%%%%%%%%%%%%%%
\subsection{Effective mass}
\label{rmf_effective_mass}
%%%%%%%%%%%%%%%%%%%%%%%%%%%
The effective mass of the nucleon is represented as~\cite{Furnstahl_1996,  NKGb_1997, Kumar_2017, Kumar_2018, DasBig_2021}
%%%%%%%%%%%%%
\begin{align}
M_{p,n}^\star=M_{p,n} - \Phi_0 \mp D_0 \, , 
\label{eq:effm_nucleons}
\end{align}
%%%%%%%%%%%
where the $\Phi_0$ and $D_0$ are the field equations of $\sigma$ and $\delta$ mesons, respectively. In the RMF model, it is assumed that the meson fields are replaced by their expectation values. 
%%%%%%%%%%%%%%%%%%%%%%%%%%
\subsection{Finite Nuclei}
\label{rmf_finite_nuclei}
%%%%%%%%%%%%%%%%%%%%%%%%%%
The ground-state properties of the finite nucleus are obtained numerically in a self-consistent iterative method. The total binding energy of the nucleus is written as 
%%%%%%%%%%%%%%%%
\begin{eqnarray}
E_{\rm total} = E_{\rm part}+E_{\sigma}+E_{\omega}
+E_{\rho}+E_{\delta}+E_{c}+E_{\rm pair}+E_{\rm cm},
\end{eqnarray}
%%%%%%%%%%%%%%
where $E_{\rm part}$ is the sum of the single-particle energies of the nucleons and $E_{\sigma}$, $E_{\omega}$, $E_{\rho}$, $E_{\delta}$ are the energies of the respective mesons. $E_c$ is the energy from the Columbic repulsion due to protons. $E_{\rm cm}=41A^{-1/3}$ is the center of mass energy correction evaluated with a non-relativistic approximation~\cite{Negele_1970,MCentelles_2001}. The pairing energy $E_{\rm pair}$ is calculated by assuming a few quasi-particle levels as developed in Refs.~\cite{DelEstal_2001, Kumar_2018}. Here, we take the pairing between proton-proton and neutron-neutron, which are invariant under time-reversal symmetry. The pairing can not be ignored for nuclei near to drip line because they have quasi-particle states near the Fermi surface. The simple BCS approximation is appropriate for nuclei near the stability line ~\cite{Gambhir_1990, Sugahara_1994}. However, it breaks down near the drip line. This is because the Fermi level approaches zero, and the number of available states above the Fermi surface will decrease. In this case, the particle-hole and pair excitations reach the continuum, and their wave functions are not localized in a region, which gives rise to unphysical neutron and proton gas around the nucleus. To overcome this situation, one has to take the BCS calculations with quasi-particle states to take care of the pairing interaction~\cite{Dobaczewski_1984}.  

To deal with pairing contribution, we use a simple approach found successful in Refs. ~\cite{MCentelles_2001, DelEstal_2001} to explain both $\beta$-stable and $\beta$-unstable nuclei. This method is similar to the prescription adopted by Chabanat {\it et al.}~\cite{Chabanta_1998}. A constant pairing matrix element $G_q$ is assumed for each kind of nucleon, which gives the zero range pairing force. On top of it, we include a few quasi-bound levels in the BCS calculations~\cite{MCentelles_2001,DelEstal_2001,Sil_2004}. These quasi-bound states are generated by the centrifugal barrier for neutrons and centrifugal plus Coulomb for protons, which mock up the continuum states in the correlations. These wave functions of the quasi-bound states are generally localized in the classically allowed region and sharply decrease outside it. As a result, the contribution of the unphysical nucleon gas surrounding the nucleus is eliminated from the BCS calculations~\cite{Demorest_2010}. We take one harmonic oscillator shell above and below the Fermi surface in our calculation. A detailed description of the method is given in Refs.~\cite{MCentelles_2001,DelEstal_2001,Sil_2004}. Other properties, such as charge radius, neutron-skin thickness, two-nucleon separation energy, and isotopic shift, are calculated from the binding energy and distributions of the nuclei inside the nucleus.
%%%%%%%%%%%%%%%%%%%%%%%%%%%
\subsection{Nuclear Matter}
\label{rmf_nuclear_matter}
%%%%%%%%%%%%%%%%%%%%%%%%%%%
A hypothetical medium consists of only protons and neutrons with no surface. To study NM properties, one has to switch off all the electromagnetic interactions and neglect the $r$ dependencies in all meson fields in Eq. (\ref{eq:rmf_edf}). From the E-RMF energy density functional (in Eq. \ref{eq:rmf_edf}), one can write the NM Lagrangian density as in Refs.~\cite{Kumar_2020, Das_2021}
%%%%%%%%%%%%%
\begin{align}
{\cal L}_{\rm NM} & = \sum_{j=p,n} \bar\psi_{j}
\Bigg\{\gamma_{\mu}\bigg(i\partial^{\mu}-g_{\omega}\omega^{\mu}-\frac{1}{2}g_{\rho}\vec{\tau}_{j}\!\cdot\!\vec{\rho}^{\,\mu}\bigg)-\bigg(M_n-g_{\sigma}\sigma-g_{\delta}\vec{\tau}_{j}\!\cdot\!\vec{\delta}\bigg)\Bigg\} \psi_{j}
\nonumber \\
&+\frac{1}{2}\partial^{\mu}\sigma\,\partial_{\mu}\sigma-\frac{1}{2}m_{\sigma}^{2}\sigma^2+\frac{\zeta_0}{4!}g_\omega^2(\omega^{\mu}\omega_{\mu})^2-\frac{\kappa_3}{3!}\frac{g_{\sigma}m_{\sigma}^2\sigma^3}{M_n}-\frac{\kappa_4}{4!}\frac{g_{\sigma}^2m_{\sigma}^2\sigma^4}{M_n^2}
\nonumber\\
&+\frac{1}{2}m_{\omega}^{2}\omega^{\mu}\omega_{\mu}-\frac{1}{4}W^{\mu\nu}W_{\mu\nu}+\frac{\eta_1}{2}\frac{g_{\sigma}\sigma}{M_n}m_\omega^2\omega^{\mu}\omega_{\mu}+\frac{\eta_2}{4}\frac{g_{\sigma}^2\sigma^2}{M_n^2}m_\omega^2\omega^{\mu}\omega_{\mu}
\nonumber\\
&+\frac{\eta_{\rho}}{2}\frac{m_{\rho}^2}{M_n}g_{\sigma}\sigma\big(\vec\rho^{\,\mu}\!\cdot\!\vec\rho_{\mu}\big)+\frac{1}{2}m_{\rho}^{2}\big(\vec\rho^{\mu}\!\cdot\!\vec\rho_{\mu}\big)-\frac{1}{4}\vec R^{\mu\nu}\!\cdot\!\vec R_{\mu\nu}
\nonumber\\
&+\Lambda_{\omega}g_{\omega}^2g_{\rho}^2\big(\omega^{\mu}\omega_{\mu}\big)
\big(\vec\rho^{\,\mu}\!\cdot\!\vec\rho_{\mu}\big)+\frac{1}{2}\partial^{\mu}\vec\delta\,\partial_{\mu}\vec\delta-\frac{1}{2}m_{\delta}^{2}\vec\delta^{\,2},
\label{rmflag}
\end{align}
%%%%%%%%%%%%%%
For a static, infinite, uniform, and isotropic NM, one has to vanish all the gradients of the fields in Eqs. (\ref{eq:sigma_field}-\ref{eq:delta_field}). Each meson field can be calculated by solving the mean-field equation of motions in a self-consistent way~\cite{Kumar_2017, Kumar_2018, Das_2019}. To get the energy density (${\cal{E}}_{\rm NM}$) and pressure ($P_{\rm NM}$) for the NM system, one has to solve the energy-momentum stress-tensor equation, which is given by~\cite{Walecka_74, NKGb_1997}
%%%%%%%%%%%%%%%%
\begin{align}
T_{\mu \nu}=\sum_{l}{\partial_{\nu} X_{l}} 
\frac{ {\partial \cal L}_{\rm NM}} {\partial \left(\partial^{\mu} X_l\right)} 
-g_{\mu \nu}{\cal L}_{\rm NM}.
\label{eq:energy_momentum}
\end{align}
%%%%%%%%%%%%%%
The zeroth component of the energy-momentum tensor $\langle T_{00} \rangle$ gives the energy density, and the spatial component $\langle T_{ii} \rangle$ determines the pressure of the system. The energy density (${\cal{E}}_{\rm NM}$) and pressure ($P_{\rm NM}$) within mean-fields approximations are given by~\cite{Kumar_2017, Kumar_2018, Das_2020, Das_2021, Kumar_2020}
%%%%%%%%%%%%
\begin{align}
{\cal{E}}_{\rm NM} & =\sum_{j=p,n} \frac{1}{\pi^2}\int_{0}^{k_{j}} k^2\, dk \, E_j^* +\rho_b W+\frac{1}{2}\rho_{3b} R
-\frac{1}{4!}\frac{\zeta_{0}W^{4}}{g_{\omega}^2}-\Lambda_{\omega}\left(R^{2} W^{2}\right)
\nonumber\\
&
+\frac{m_{\sigma}^2\Phi^{2}}{g_{\sigma}^2}\left(\frac{1}{2}+\frac{\kappa_{3}}{3!}\frac{\Phi }{M_n}
+\frac{\kappa_4}{4!}\frac{\Phi^2}{M_n^2}\right)-\frac{1}{2}m_{\omega}^2\frac{W^{2}}{g_{\omega}^2}\left(1+\eta_{1}\frac{\Phi}{M_n}+\frac{\eta_{2}}{2}\frac{\Phi ^2}{M_n^2}\right)
\nonumber\\
&
-\frac{1}{2}\left(1+\frac{\eta_{\rho}\Phi}{M_n}\right)\frac{m_{\rho}^2}{g_{\rho}^2}R^{2}+\frac{1}{2}\frac{m_{\delta}^2}{g_{\delta}^{2}}D^{2} \, ,
\label{eq:ener_NM}
\end{align}
%%%%%%%%%%%
\begin{align}
P_{{\rm NM}}&=\sum_{j=p,n} \frac{1}{3\pi^2}\int_{0}^{k_{j}} dk\, \frac{k^4}{E_j^\ast} +\frac{1}{4!}\frac{\zeta_{0}W^{4}}{g_{\omega}^2} +\Lambda_{\omega}(R^{2} W^{2})
\nonumber\\
&
-\frac{m_{s}^2\Phi^{2}}{g_{s}^2}\left(\frac{1}{2}+\frac{\kappa_{3}}{3!}
\frac{\Phi }{M_n}+ \frac{\kappa_4}{4!}\frac{\Phi^2}{M_n^2}  \right)+\frac{1}{2}m_{\omega}^2\frac{W^{2}}{g_{\omega}^2}\left(1+\eta_{1}\frac{\Phi}{M_n}+\frac{\eta_{2}}{2}\frac{\Phi ^2}{M_n^2}\right)
\nonumber\\
&
+\frac{1}{2}\left(1+\frac{\eta_{\rho}\Phi}{M_n}\right)\frac{m_{\rho}^2}{g_{\rho}^2}R^{2}-\frac{1}{2}\frac{m_{\delta}^2}{g_{\delta}^{2}}D^{2},
\label{eq:press_NM}
\end{align}
%%%%%%%%%%
%%%%%%%%%%%%%%%%%%%%%%%%%%%%%%%%%%
\section{Results and Discussions}
\label{r&d}
%%%%%%%%%%%%%%%%%%%%%%%%%%%%%%%%%%
This section discusses the properties of finite nuclei, NM, and NS matter. Some finite nuclei properties like binding energy per particle, charge radius, neutron-skin thickness, single-particle energy, and two neutron separation energy for a few spherical nuclei are analyzed. The NM parameters, such as incompressibility, symmetry energy, and its different coefficients, are studied for symmetric NM (SNM) and pure neutron matter (PNM). Finally, we extend our calculations to the NS and find its EOS, mass, radius, tidal deformability, and moment of inertia in detail.
%%%%%%%%%%%%%%%%%%%%%%%%%%%
\subsection{Finite Nuclei}
%%%%%%%%%%%%%%%%%%%%%%%%%%%
%%%%%%%%%%%%%%%%%%%%%%%%%%%%%%%%%%%%%%%%%%%%%%%%%%%%%%%%%%%%%
\subsubsection{a. Binding energies, charge radii, and neutron-skin thickness:-}
%%%%%%%%%%%%%%%%%%%%%%%%%%%%%%%%%%%%%%%%%%%%%%%%%%%%%%%%%%%%%
Here, we calculate binding energy ($B/A$), charge radii ($R_c$), and neutron skin thickness ($\Delta r_{np}$) for eight spherical nuclei and compare with the experimental results as given in Table \ref{tab:finite_nuclei_properties}. The BigApple parameter set satisfies the $B/A$ and $R_c$ of the listed nuclei compared to other parameter sets. 

\begin{figure}
\centering
\includegraphics[width=0.6\textwidth]{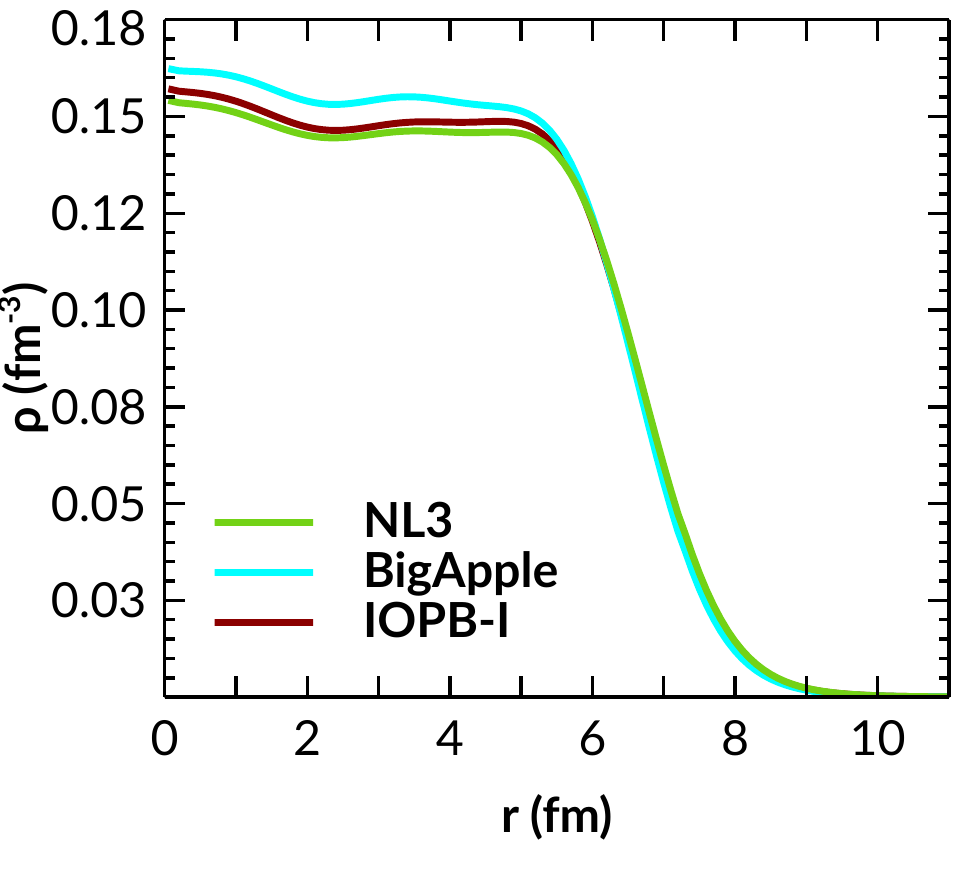}
\caption{The density of the $^{208}$Pb is shown for BigApple, NL3, and IOPB-I parameter sets.}
\label{fig:den}
\end{figure}

The density profile of the $^{208}$Pb nucleus, shown in Fig. \ref{fig:den} for BigApple with NL3 and IOPB-I parameter sets for comparison. The central density of the nucleus is more significant for the BigApple case than NL3 and IOPB-I. That means the nucleus will saturate at a higher density for BigApple than NL3 and IOPB-I. 
%%%%%%%%%%%%%%
\begin{table}[t] 
\centering
\caption{The numerical values of B/A (MeV) and $R_c$ (fm) and $\Delta r_{np}$ (fm) are listed with the available experimental data~\cite{Wang_2012,Angeli_2013}.}
\scalebox{0.9}{
\begin{tabular}{|c|c|c|c|c|c|c|c|c|c|}
\hline
%\hline
\multicolumn{1}{|c|}{Nucleus}& 
\multicolumn{1}{c|}{Obs.}&
\multicolumn{1}{c|}{Expt.}&
\multicolumn{1}{c|}{NL3}&
\multicolumn{1}{c|}{FSUGarnet}&
\multicolumn{1}{c|}{G3}&
\multicolumn{1}{c|}{IOPB-I}&
\multicolumn{1}{c|}{BigApple}\\
\hline
         &B/A & 7.976 &7.917&7.876 & 8.037&7.977& 7.882 \\
$^{16}$O & R$_{c}$& 2.699 & 2.714&2.690   & 2.707&2.705&2.713  \\
         & $\Delta r_{np}$ & ------ & -0.026&-0.028  & -0.028&-0.027&-0.027  \\\hline
         & B/A  & 8.551 & 8.540&8.528  &8.561&8.577 &8.563 \\
$^{40}$Ca& R$_{c}$  & 3.478 &  3.466&3.438  & 3.459&3.458& 3.447 \\
         & $\Delta r_{np}$  & ------ & -0.046&-0.051  & -0.049&-0.049& -0.049 \\\hline
         & B/A  & 8.666 & 8.636&8.609  &8.671& 8.638& 8.547\\
$^{48}$Ca& R$_{c}$  & 3.477 & 3.443&3.426   & 3.466 &3.446& 3.447\\
         & $\Delta r_{np}$  & ------ &  0.229&0.169 & 0.174&0.202& 0.170 \\\hline
         & B/A  & 8.682 & 8.698&8.692  & 8.690&8.707& 8.669 \\
$^{68}$Ni& R$_{c}$  &------  & 3.870 &3.861  &3.892&3.873&3.877  \\
         & $\Delta r_{np}$  &  &0.262 &0.184 &0.190&0.223&0.171  \\\hline
         & B/A  & 8.709 & 8.695&8.693  & 8.699&8.691& 8.691 \\
$^{90}$Zr& R$_{c}$  & 4.269 & 4.253&4.231  & 4.276& 4.253& 4.239\\
         & $\Delta r_{np}$  &------  & 0.115&0.065   & 0.068&0.091& 0.069\\\hline
         & B/A  & 8.258 & 8.301& 8.298   & 8.266&8.284&8.259 \\
$^{100}$Sn& R$_{c}$  & ------ & 4.469&4.426   &4.497&4.464&4.445  \\
         & $\Delta r_{np}$  &------  & -0.073&-0.078  &  -0.079&-0.077& 0.076\\\hline
         & B/A  & 8.355 & 8.371&8.372  & 8.359&8.352&8.320 \\
$^{132}$Sn& R$_{c}$  & 4.709 & 4.697&4.687   & 4.732& 4.706&4.695 \\
         & $\Delta r_{np}$  &------  & 0.349&0.224   & 0.243&0.287& 0.213\\\hline
         & B/A  & 7.867 &7.885& 7.902 & 7.863&7.870& 7.894 \\
$^{208}$Pb& R$_{c}$  &5.501  & 5.509&5.496 & 5.541 &5.52& 5.495\\
         & $\Delta r_{np}$  &------  &   0.283& 0.162 & 0.180 &0.221&0.151\\ \hline
\end{tabular}
\label{tab:finite_nuclei_properties}}
\end{table}
%%%%%%%%%%%%%
%%%%%%%%%%%%%
%%%%%%%%%%%

\begin{figure}
\centering
\includegraphics[width=0.7\textwidth]{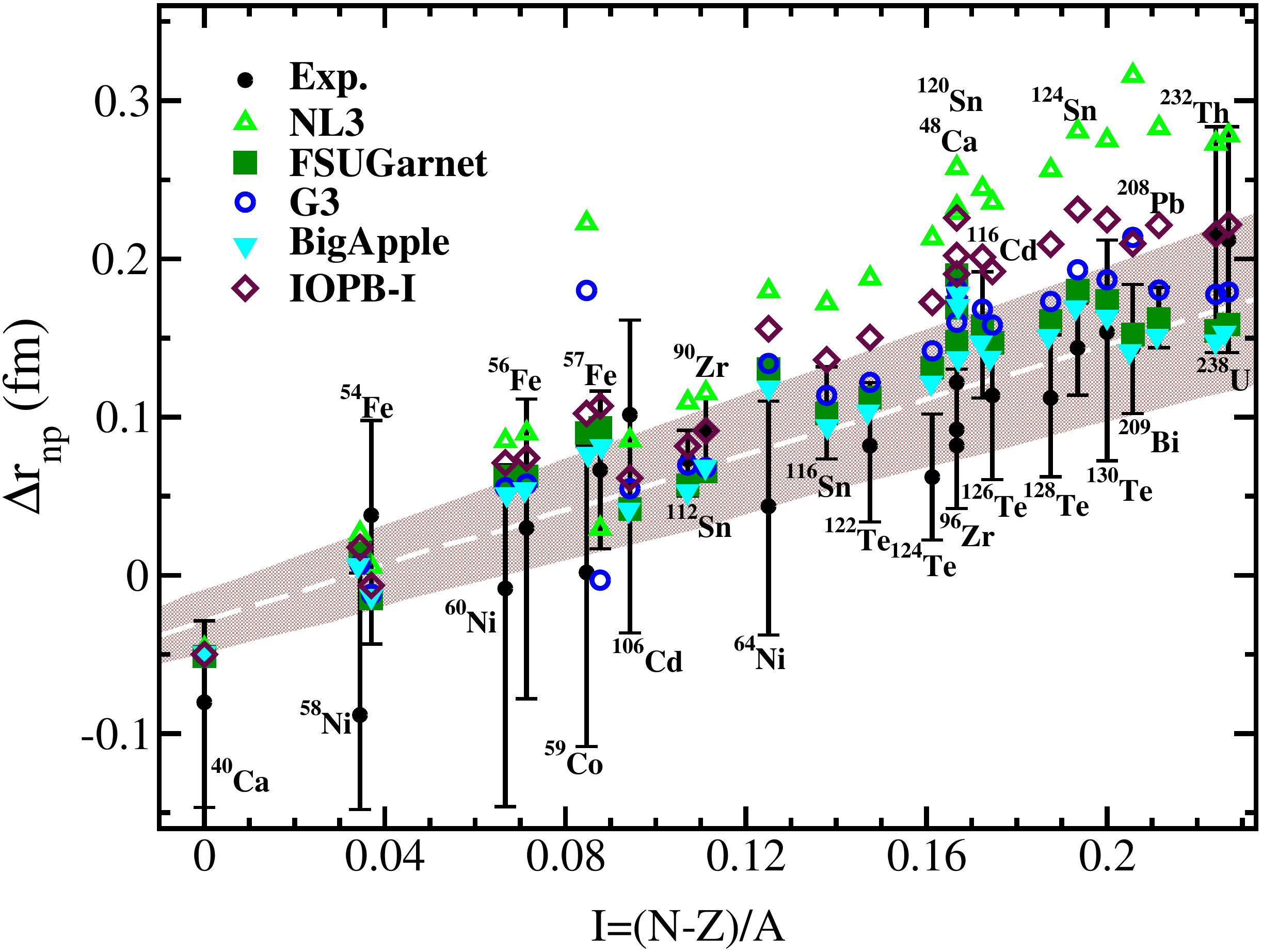}
\caption{The neutron-skin thickness as a function of the $I$. Results obtained with five different parameter sets are compared with experimental data~\cite{Trzci_2001}. The shaded region is depicted using the fitting formula in Eq. (\ref{eq:rnp}).}
\label{fig:skin}
\end{figure}

The neutron skin thickness ($\Delta r_{np}$) is defined as the root mean square radii difference of neutron and proton distribution, i.e., $\Delta r_{np}=R_n-R_p$. Electron scattering experiments can determine the charge distribution of protons in the nucleus. However, it is not straightforward to calculate the neutron distributions in nuclei in a model-independent way. The Lead Radius Experiment (PREX) at JLAB has been designed to measure the neutron distribution radius in $^{208}$Pb from parity violation by the weak interaction. This measurement determined significant uncertainties in the measurement of the neutron radius of $^{208}$Pb~\cite{Abrahamyan_2012}. 
Recently, the PREX-II experiment gave the neutron skin thickness is~\cite{Adhikari_2021}
%%%%%%%%%%%%%%%%
\begin{equation}
    \Delta r_{np}=R_n-R_p=0.283\pm0.071 \, \mathrm{fm},
    \label{eq:Delta_r_np}
\end{equation}
%%%%%%%%%%%%%%%%
with $1\sigma$ uncertainty. Using PREX-II data, researchers have tried to constrain some NM and NS properties, which improve the understanding of the EOS for NM and NS~\cite{Reed_2021, Pattnaik_2022}. The value of $\Delta r_{np}$ for $^{208}$Pb is 0.283, 0.162, 0.180, 0.221, and 0.151 for NL3, FSUGarnet, G3, IOPB-I, and BigApple respectively. Only NL3 and IOPB-I parameter sets well reproduce the neutron skin thickness of the lead nucleus, which is consistent with PREX-II data. On the other hand, the neutron-skin thickness of 26 stable nuclei starting from $^{40}$Ca to $^{238}$U has been deduced by using an anti-protons experiment from the low Energy anti-proton ring at CERN~\cite{Trzci_2001}. The numerically calculated results and experimental data with an error bar are shown in Fig. \ref{fig:skin} with the equation.
%%%%%%%%%%%%%%%%
\begin{equation}
    \Delta r_{np} = (0.90\pm 0.15)I+(-0.03\pm 0.02) \ {\rm fm}.
    \label{eq:rnp}
\end{equation}
%%%%%%%%%%%%%%%
The fitted data for $\Delta r_{np}$ using Eq. (\ref{eq:rnp}) is put as a band in Fig. \ref{fig:skin}. The calculated skin-thickness of the 26 nuclei for the BigApple parameter set matches well with other sets. The skin-thickness for $^{208}$Pb nuclei with BigApple is 0.151 fm, which lies in the range given by the proton elastic scattering experiment~\cite{Zenihiro_2010},  $\Delta r_{np} = 0.148-0.265$ fm.
%%%%%%%%%%%%%
%%%%%%%%%%%

From Fig. \ref{fig:skin}, we observe that the $\Delta r_{np}$ by different parameter sets coincide with each other and also with the experimental data for the nuclei with zero isospins, like $^{40}$Ca. However, as the isospin asymmetry increases, the results from different parameter sets diverge from each other. Some stiff EOS like NL3, which gives a large maximum allowed mass for the NS, shows a severe divergence from the experimental data and lie outside the fitted region. 
%%%%%%%%%%%%%%%%%%%%%%%%%%%%%%%%%%%%%%%%%%%
\subsubsection{b. Single-particle energy:-}
\label{rd:single}
\begin{figure}
    \centering
    \includegraphics[width=0.5\textwidth]{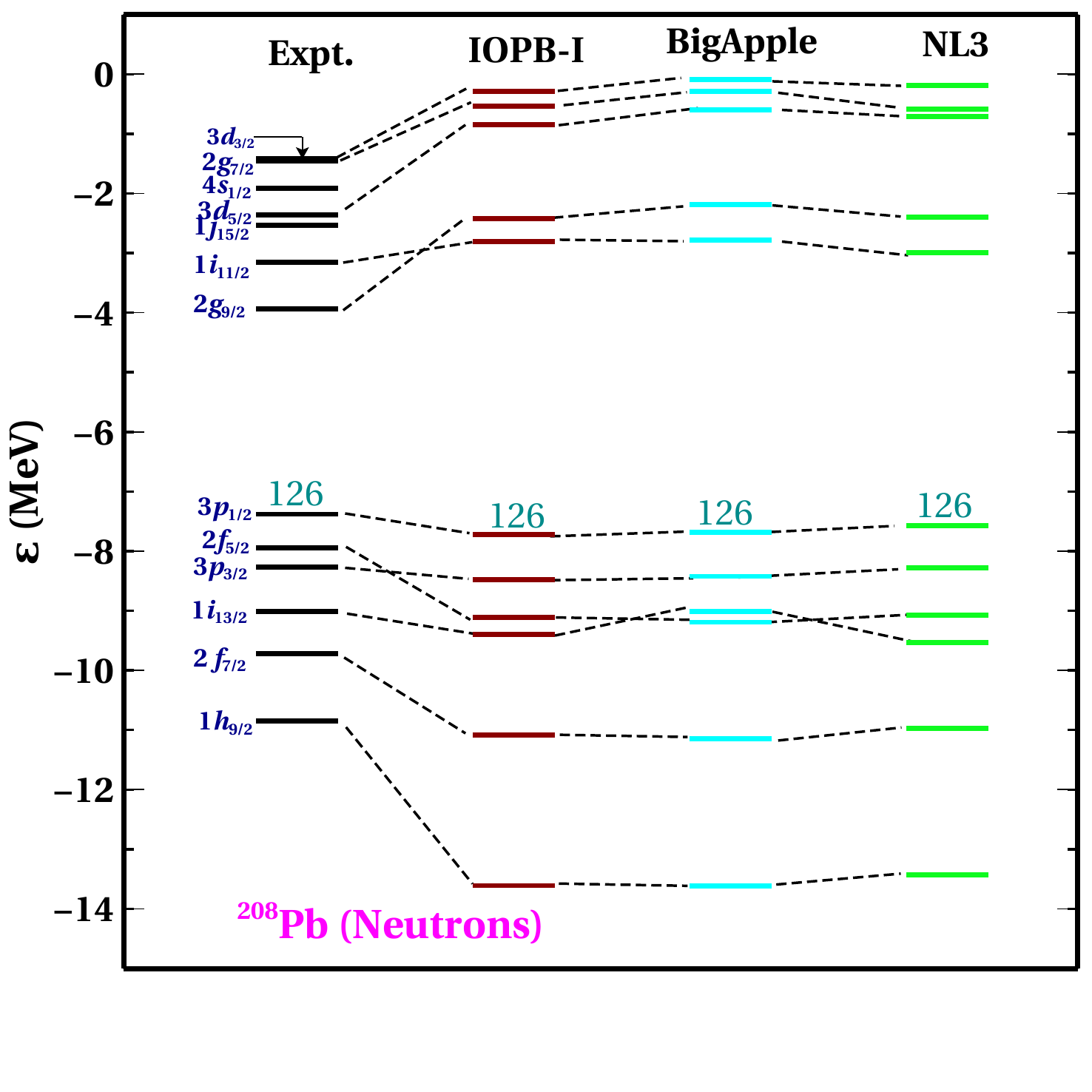}%
    \includegraphics[width=0.5\textwidth]{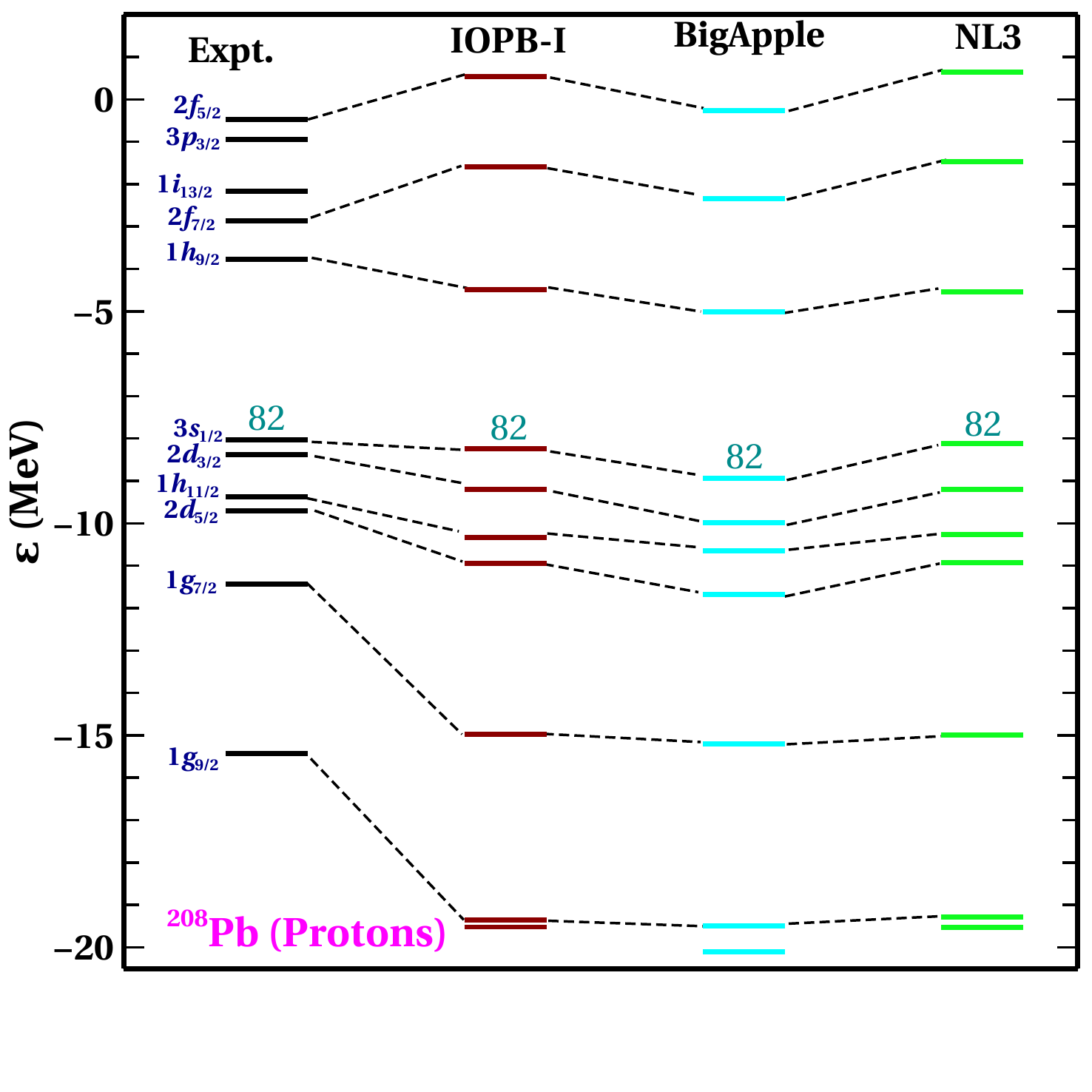}
    \caption{The single-particle energies of $^{208}$Pb for IOPB-I, BigApple, and NL3 are compared with experimental data~\cite{Vautherin_1972}. The last occupied level is also shown with the numbers 126 for neutrons and 82 for protons.}
    \label{fig:pb}
\end{figure}

\begin{figure}
    \centering
    \includegraphics[width=0.5\textwidth]{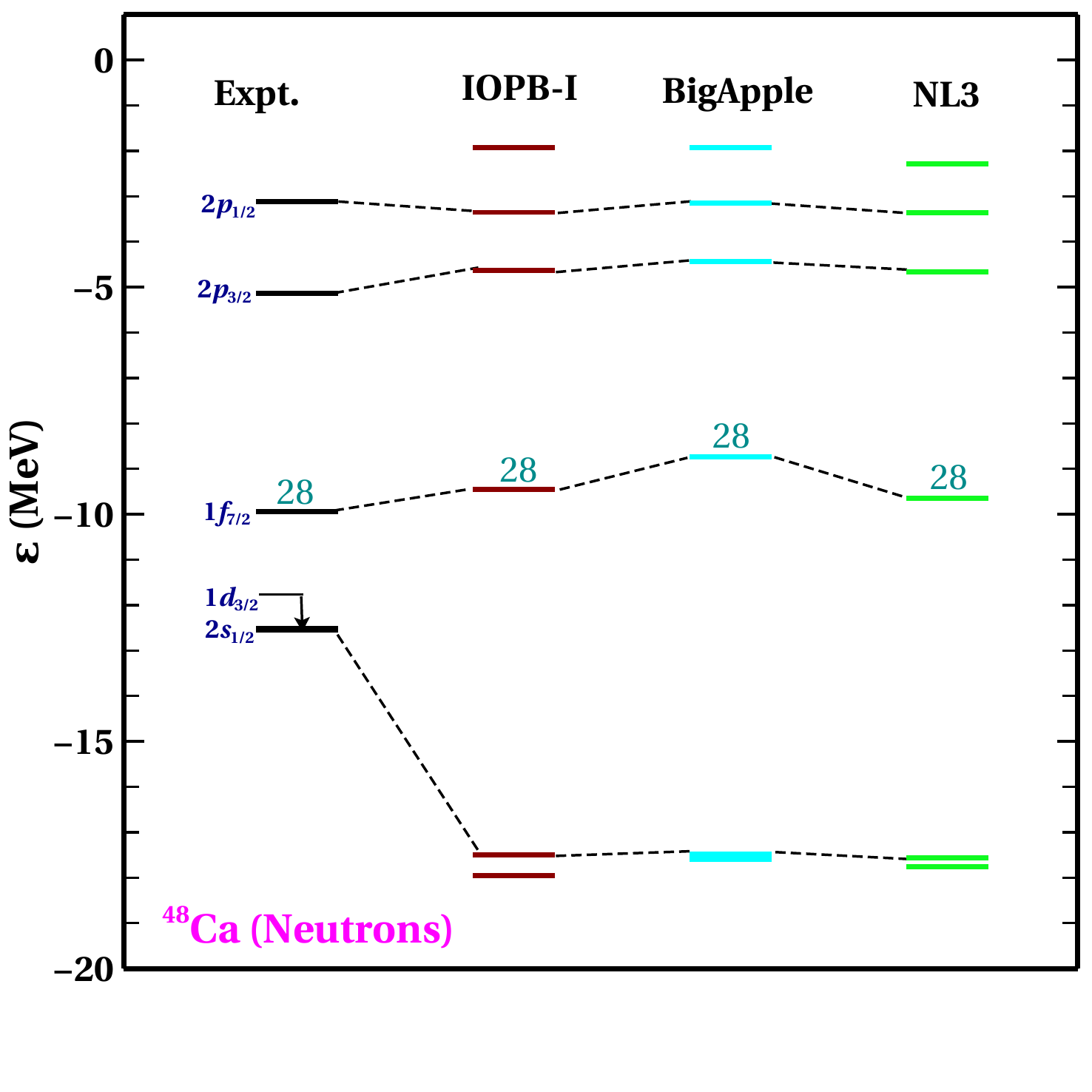}%
    \includegraphics[width=0.5\textwidth]{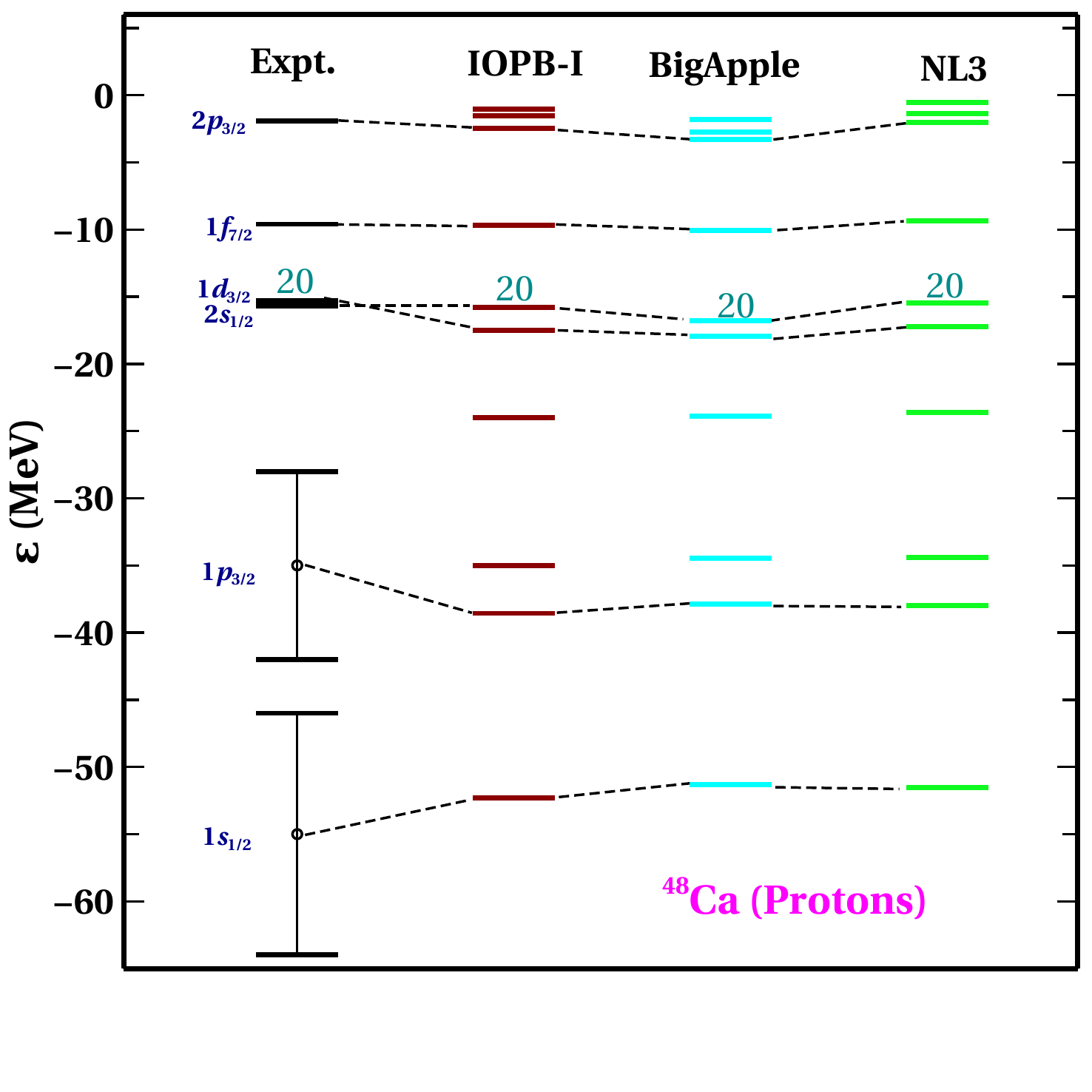}
    \caption{Same as Fig. \ref{fig:pb}, but for $^{48}$Ca.}
    \label{fig:ca}
\end{figure}

%%%%%%%%%%%%%%%%%%%%%%%%%%%%%%%%%%%%%%%%%%%
The study of single-particle energies for nuclei gives us an indication of shell closer. From this, we can identify the large shell gaps and predict the presence of the magic numbers. Here, we calculate the single-particle energies of two doubly magic nuclei as representative cases, e.g., $^{48}$Ca and $^{208}$Pb for IOPB-I, BigApple, and NL3 parameter sets. The predicted single-particle energies for both protons and neutrons of $^{48}$Ca and $^{208}$Pb are compared with the experimental data~\cite{Vautherin_1972} in Figs. \ref{fig:pb} and \ref{fig:ca}. The BigApple set well predicted the magicity compared to the other parameter sets. All three parameter sets reproduce the known magic numbers 20, 28, 82, and 126. The nuclei, $^{48}$Ca and $^{208}$Pb, are doubly closed and considered perfectly spherical.
%%%%%%%%%%%%%%
%%%%%%%%%%%%%
%%%%%%%%%%%%%
%%%%%%%%%%%%%
%%%%%%%%%%%%%%%%%%%%%%%%%%%%%%%%%%%%%%%%%%%%%%%%%%%%%%%%%%%%%%%%%
\subsubsection{c. Two-neutron separation energy $S_{2n}(N, Z)$:-}
\begin{figure} 
\centering
\includegraphics[width=0.7\textwidth]{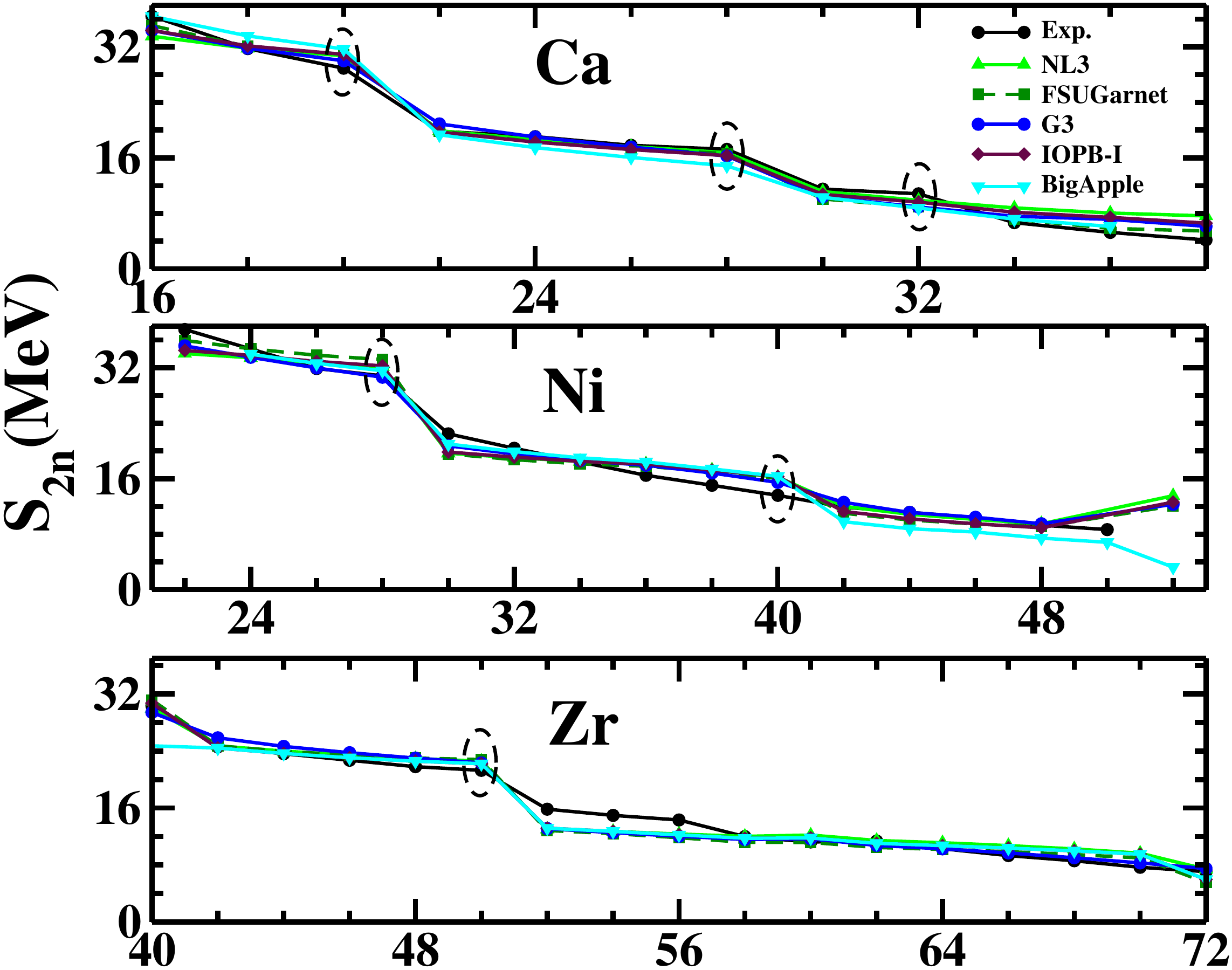}
\includegraphics[width=0.7\textwidth]{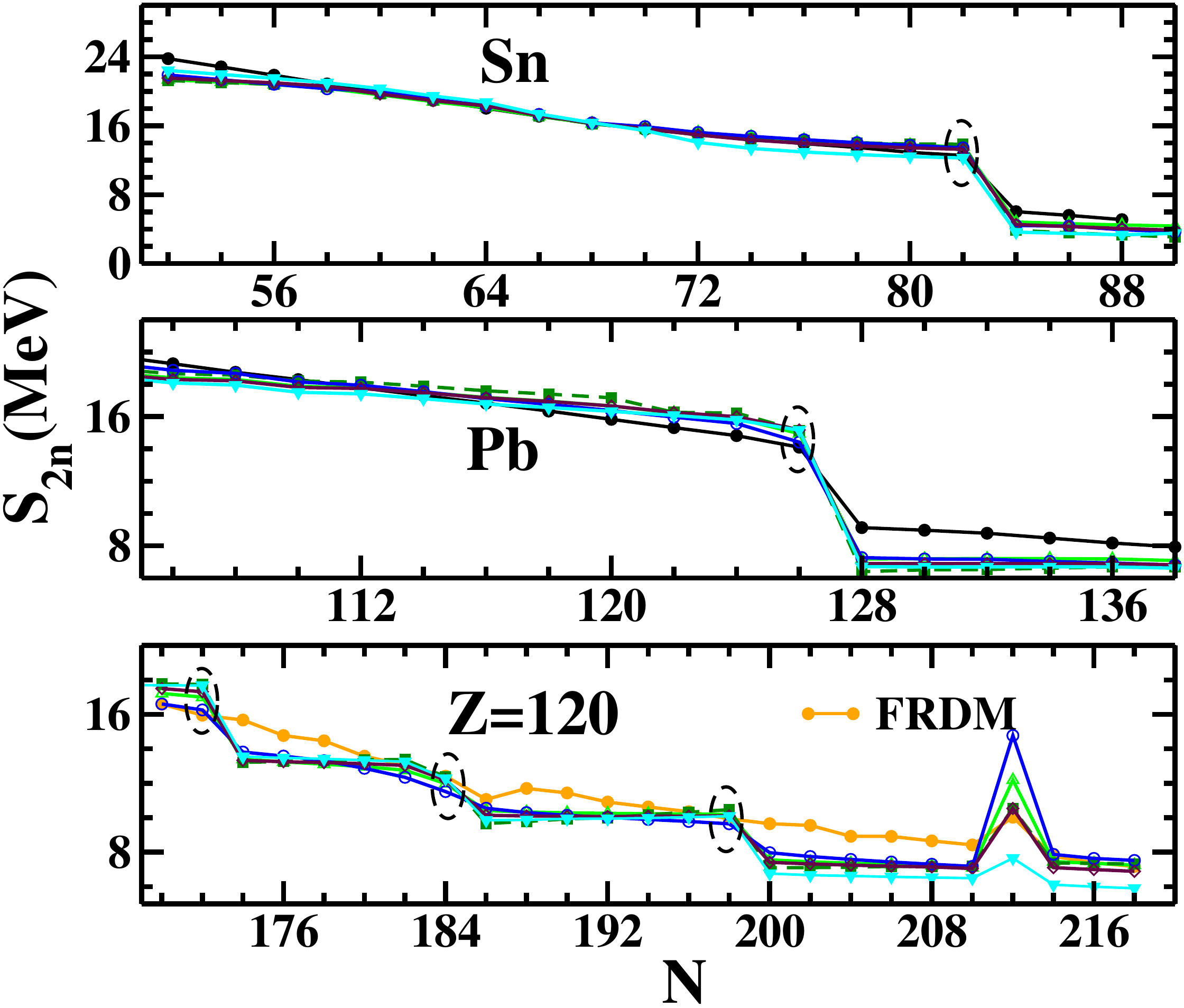}
\caption{The two-neutron separation energy as a function of neutron number for the isotopic nuclei like Ca, Ni, Zr, Sn, and Pb for five different parameter sets. The FRDM~\cite{Moller_2016} data and experimental data~\cite{Wang_2012} are also given for comparison. The circle represents the magicity of the nuclei.}
\label{fig:s2n}
\end{figure}

%%%%%%%%%%%%%%%%%%%%%%%%%%%%%%%%%%%%%%%%%%%%%%%%%%%%%%%%%%%%%%%%%
The two neutron separation energy $S_{2n}(N, Z)$ is the energy required to remove two neutrons from a nucleus with $N$ neutrons and $Z$ protons, i.e. 
%%%%%%%%%%%%%%%%
\begin{equation}
    S_{2n}(N,Z)=B(N,Z)-B(N-2,Z).
\end{equation}
%%%%%%%%%%%%%%%%
The study of neutron separation energy is essential to explore the nuclear structure near the drip line. A sudden drop in $S_{2n}(N, Z)$ represents the beginning of a new shell. The large shell gap in single-particle energy levels indicates the magic number, which is responsible for the extra stability of the magic nuclei. We calculate the $S_{2n}$ for six isotopic chains Ca, Ni, Zr, Sn, Pb, and $Z=120$, which are shown in Fig. \ref{fig:s2n} and compared with experimental data given in Ref.~\cite{Wang_2012}. We also compare the results obtained for the $Z=120$ isotopic chain results with the finite range droplet model (FRDM)~\cite{Moller_2016}. From Fig. \ref{fig:s2n}, it is clear that the value of $S_{2n}$ decreases with the increase of the neutron number, i.e., towards the neutron drip line. All the magic characters appear at neutron number $N=20, 28, 32, 40, 50, 82, 126$. In the last part of Fig. \ref{fig:s2n}, the magicity is found at $N=172, 184, 198$ for $Z=120$ nuclei. In this case, we compare the calculated data with FRDM~\cite{Moller_2016}; no experimental data is available for the Z=120. There is a sharp fall in the $S_{2n}$ for five different parameter sets, which are consistent with the prediction of various models in the superheavy mass region~\cite{Rutz_1997, Gupta_1997, Patra_1999, Mehta_2015}. Bhuyan {\it {\it et al.}.}~\cite{Bhuyan_2012} have predicted that $Z=120$ is next the magic number after $Z=82$, which lies in the superheavy region. Also, Mehta  {\it {\it et al.}.}~\cite{Mehta_2015} have predicted that $Z=120$ nuclei are spherical in their ground state and possible proton magic number at $Z=120$. We hope that future experiments may answer the shell closure at $N=172, 184$, and $198$.
%%%%%%%%%%%%%%%
%%%%%%%%%%%%
%%%%%%%%%%%%%%%%%%%%%%%%%%%%%%%%%%%
\subsubsection{d. Isotopic shift:-}
%%%%%%%%%%%%%%%%%%%%%%%%%%%%%%%%%%%
The Isotopic shift is defined as, $\Delta r_c^2= R_c^2(208)-R_c^2(A)$, where we take $^{208}$Pb as reference nucleus. In Fig \ref{fig:s2n}, we plot the $\Delta r_c^2$ for Pb isotopes for five parameter sets. The experimental data are also given for comparison. The predicted $\Delta r_c^2$ by BigApple set well matches the NL3 set. The iso-spin-dependent term in the nuclear interaction results in the kink in the isotopic shift graph. In the conventional RMF model, the spin-orbital term is included automatically by assuming nucleons as the Dirac spinor. However, the situation is not the same in the Skyrme model, where one must add the spin-orbital part to match the experimental results~\cite{Sharma_1995}.
%%%%%%%%%%%%%%
\begin{figure}
\centering
\includegraphics[width=0.65\textwidth]{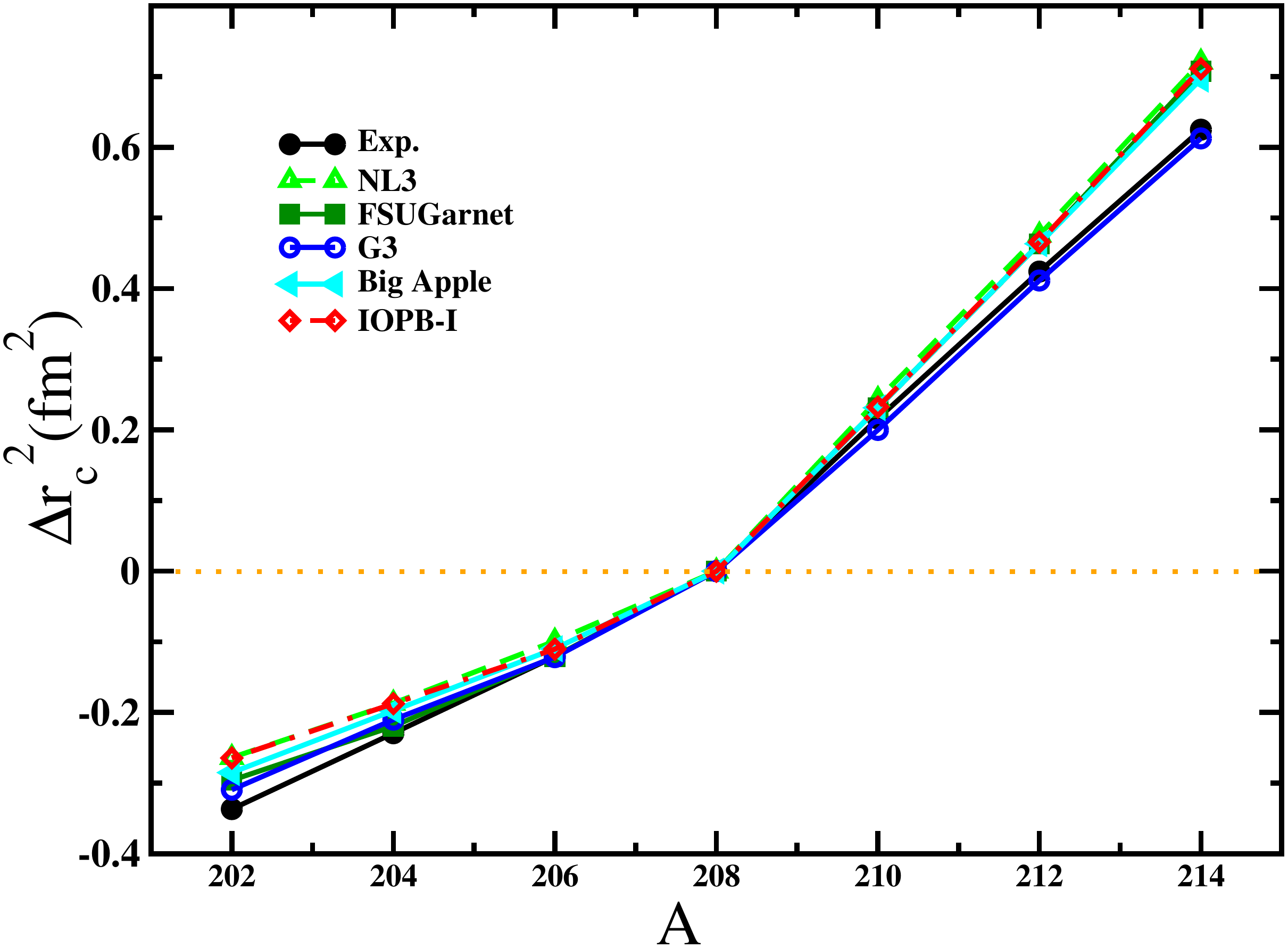}
\caption{Isotopic shift for Pb isotopes for five parameter sets is shown taking $^{208}$Pb as reference.}
\label{fig:isotopic}
\end{figure}
%%%%%%%%%%%%%

In conclusion, we calculate some finite nuclei properties such as $B/A$, $R_c$, $\Delta r_{np}$, single-particle energy, $S_{2n}$, $\Delta r_c^2$ for BigApple parameter set along with other four sets. We find that the $B/A$, $R_c$ for some nuclei are well reproduced by the BigApple like other parameter sets. The skin thickness of lead nuclei is 0.151 fm for BigApple set, which is inconsistent with the PREX-II data~\cite{Adhikari_2021} but satisfies the CERN data~\cite{Trzci_2001}. The single-particle energy for $^{48}$Ca and $^{128}$Pb are well reproduced by the BigApple set. It also predicts the two neutron separation energies for series nuclei, including $Z=120$, compared to other sets. Finally, the isotopic shift is almost consistent with experimental data. From the above studies, we conclude that one can take BigApple set to calculate finite nuclei properties.
%%%%%%%%%%%%%%%%%%%%%%%%%%%%%%%%%%%%%%
\subsection{Nuclear Matter properties}
\label{sub:NM}
\begin{figure}[t]
\centering
\includegraphics[width=0.7\textwidth]{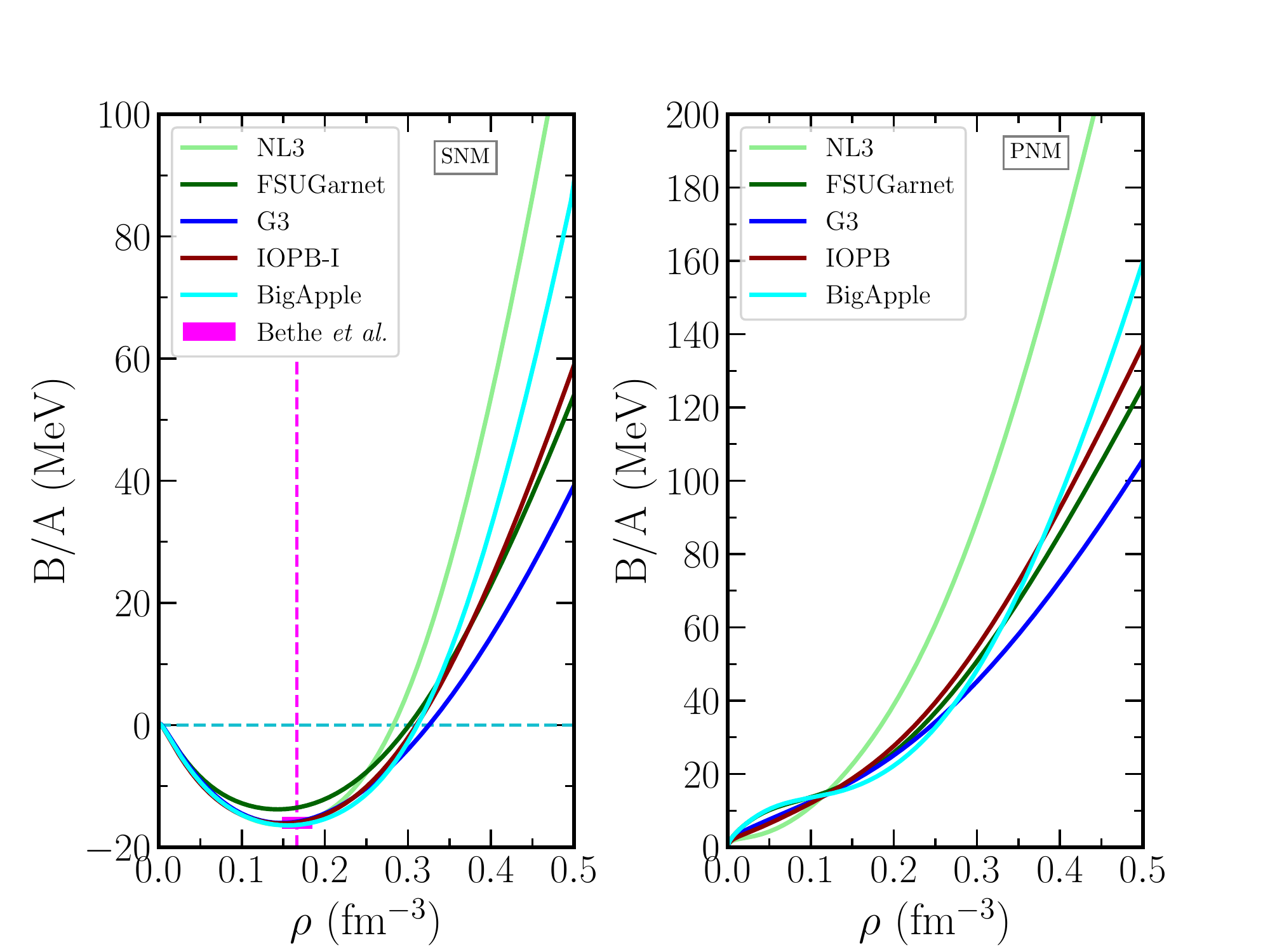}
\caption{The $B/A$ as a function of density for both SNM (left) and PNM (right) for five different parameter sets. The magenta box represents the empirical value given by Bethe {\it et al.}~\cite{Bethe_1971} and its values are given in Table \ref{tab:nuclear_matter_properties}.}
\label{fig:eos_nm}
\end{figure}

%%%%%%%%%%%%%%%%%%%%%%%%%%%%%%%%%%%%%%
In the following sub-section, the formalism used to calculate symmetric and asymmetric NM properties has been discussed. To describe the asymmetric NM, we introduce the asymmetric parameter $\zeta$, which is defined as
%%%%%%%%%%%%%
\begin{align}
\zeta = \frac{\rho_n-\rho_p}{\rho_n+\rho_p}.
\end{align}
%%%%%%%%%%%%
One can change the asymmetry parameter $\zeta$ between the SNM, where $\zeta=0$, to the PNM case ($\zeta=1$). The binding energy per nucleon ($B/A$) is defined  ${\cal E}/\rho-M_n$, which is shown as a function of density in Fig. \ref{fig:eos_nm} for both SNM and PNM. 
%%%%%%%%%%%%%
%%%%%%%%%%%
%%%%%%%%%%%%%
\begin{align}
{\cal{E}}(\rho,\zeta) = {\cal{E}}(\rho,\zeta=0)+S(\rho) \zeta^2+ {\cal{O}}(\zeta^4),
\label{eq:energy_taylor}
\end{align}
%%%%%%%%%%%
where ${\cal{E}}(\rho,\zeta = 0)$ is the energy of SNM, $\rho$ is the baryonic density and $S(\rho)$ is the symmetry energy, which is defined as 
%%%%%%%%%%%%
\begin{align}
S(\rho) = \frac{1}{2}\Bigg(\frac{\partial^2{\cal{E}}}{\partial\zeta^2}\Bigg)_{\zeta=0}
\label{eq:sym_derivative}
\end{align}
%%%%%%%%%%%%
The symmetry energy is  also written as the energy difference between PNM and SNM or vice-versa through parabolic approximation, i.e.
%%%%%%%%%%%%
\begin{align}
S(\rho) = \frac{{\cal{E}}(\rho,\zeta = 1)}{\rho}-\frac{{\cal{E}}(\rho,\zeta = 0)}{\rho}.
\label{eq:sym_energy_diff}
\end{align}
%%%%%%%%%%%%
The isospin asymmetry arises due to the difference in densities and masses of the neutron and proton. The density-type isospin asymmetry is taken care of by $\rho$ meson (isovector-vector meson) and mass asymmetry by the $\delta$ meson (isovector-scalar meson). The general expression for symmetry energy $S(\rho)$ 
is a combined expression of $\rho$ and $\delta$ mesons, which is defined as~\cite{Matsui_1981, Kubis_1997, MCentelles_2001, Chen_2014, Kumar_2018}
%%%%%%%%%%%%
\begin{align}
S(\rho)=S^{\rm kin}(\rho) + S^{\rho}({\rho})+S^{\delta}
(\rho),
\label{eq:sym_divison1}
\end{align}
%%%%%%%%%%%
with
%%%%%%%%%%%
\begin{align}
S^{\rm kin}(\rho)=\frac{k_f^2}{6E_f^*},\;
S^{\rho}({\rho})=\frac{g_{\rho}^2\rho}{8m_{\rho}^{*2}}
\label{eq:sym_divison2}
\end{align}
%%%%%%%%%%%%
and
%%%%%%%%%%%%
\begin{align}
S^{\delta}(\rho)=-\frac{1}{2}\rho \frac{g_\delta^2}{m_\delta^2}\left(
\frac{M_n^*}{E_f}\right)^2 u_\delta \left(\rho,M_n^*\right).
\label{eq:sym_divison3}
\end{align}
%%%%%%%%%%%%
The last function $u_\delta$ is from the discreteness of the Fermi momentum. This large momentum in the NM can be treated as a continuum and continuous system. The function $u_\delta$ is defined as
%%%%%%%%%%%%
\begin{align} 
u_\delta \left(\rho,M_n^*\right)=\frac{1}{ 1+ 3 \frac{g_\delta^2}{m_\delta^2}
\left(\frac{\rho^s}{M_n^*}-\frac{\rho}{E_f}\right)}.
\label{eq:u_delta}
\end{align}
%%%%%%%%%%%%
In the limit of the continuum, the function $u_\delta \approx 1$. 
The whole symmetry energy ($S^{\rm kin}+S^{\rm pot}$) arises from 
$\rho$ and $\delta$ mesons and are given as
%%%%%%%%%%%%
\begin{align}
S(\rho)=\frac{k_f^2}{6E_f^*}+\frac{g_{\rho}^2\rho}{8{m_{\rho}^{*2}}}  
-\frac{1}{2}\rho \frac{g_\delta^2}{m_\delta^2}\left(\frac{M_n^*}{E_f}\right)^2,
\label{eq:sym_addition}
\end{align}
%%%%%%%%%%%%
where $E_f^*$ is the Fermi energy and $k_f$ is the Fermi momentum. The mass of the $\rho$ meson was modified because the cross-coupling of $\rho-\omega$ fields is given by
%%%%%%%%%%%%
\begin{align}
m_{\rho}^{*2}=\left(1+\eta_{\rho}\frac{\Phi}{M_n}\right)m_{\rho}^2
+2g_{\rho}^2(\Lambda_\omega W^2).
\label{eq:m_rho2}
\end{align}
%%%%%%%%%%%%
Although the value of symmetry energy is fairly known at the saturation density ($\rho_0$), its density dependence nature is not well known. The behavior of $S(\rho)$ in high density, both qualitatively and quantitatively, shows a great diversion depending on the model used~\cite{BaoLi_2019}. Similar to the binding energy, the $S(\rho)$ can also be expressed in a leptodermous expansion near the NM saturation density. The analytical expression  of density dependence symmetry energy is written as~\cite{Matsui_1981,Kubis_1997,MCentelles_2001,Chen_2014,Kumar_2018}:
%%%%%%%%%%%%
\begin{align}
S(\rho) = J+L_{\rm sym}\zeta+\frac{1}{2}K_{\rm sym}\zeta^2+\frac{1}{6}Q_{\rm sym}\zeta^3+{\cal{O}}(\zeta^4),
\label{eq:sym_den_dep}
\end{align}
%%%%%%%%%%%%
where $\zeta$=$\frac{\rho-\rho_0}{3\rho_0}$, $J$ = $S(\rho_0$) and the parameters like slope ($L$), curvature ($K_{\rm sym}$) and skewness ($Q_{\rm sym}$) of $S(\rho$) are
%%%%%%%%%%%%
\begin{align}
L=3\rho\frac{\partial S(\rho)}{\partial\rho}, \label{eq:L_sym}\\
K_{\rm sym}=9\rho^2\frac{\partial^2 S(\rho)}{\partial\rho^2}, \label{eq:K_sym}\\
Q_{\rm sym}=27\rho^3\frac{\partial^3 S(\rho)}{\partial\rho^3}. \label{eq:Q_sym}
\end{align}
%%%%%%%%%%%%
The NM incompressibility ($K$) is defined as~\cite{Chen_2014}
%%%%%%%%%%%%
\begin{align}
K=9\rho^2\frac{\partial}{\partial \rho}\Bigg(\frac{P}{\rho^2}\Bigg).
\label{eq:incompressibility}
\end{align}
%%%%%%%%%%%%
Similarly, we can expand the asymmetric NM incompressibility $K(\xi)$ as 
%%%%%%%%%%%%%
\begin{align}
K(\xi)=K+K_\tau \xi^2+{\cal{O}}(\xi^4) \, .
\label{eq:K_zeta}
\end{align}
%%%%%%%%%%%%%
%%%%%%%%%%%%%
\begin{align}
K_\tau= K_{\rm sym.}-6L-\frac{Q_0 L}{K},
\label{eq:k_tau}
\end{align}
%%%%%%%%%%%%%
and $Q_0=27\rho^3\frac{\partial^3 \cal{E}}{\partial {\rho}^3}$ in SNM at saturation density. In this sub-section, we study NM parameters like $B/A$, incompressibility $K$, density-dependent symmetry energy $S(\rho)$ and its different coefficients like slope $L$, curvature $K_{\rm sym}$, skewness $Q_{\rm sym}$ in detail. Here we give a  special emphasis on the newly developed parameter set BigApple~\cite{Fattoyev_2020}. The values of NM quantities are given in Table \ref{tab:nuclear_matter_properties}. First, we discussed the incompressibility of the NM. For BigApple, the value of $K=227$ MeV, which lies in the experimental data range obtained from the excitation energy of the isoscalar giant monopole resonances of $^{208}$Pb and $^{90}$Zr,  and its value is $K=240\pm20$ MeV~\cite{Colo_2014, Stone_2014}. Recently the value of $K_{\rm sym}$ is constrained by Zimmerman {\it et al.}~\cite{Zimmerman_2020} combining the GW170817 and NICER data, and it is found to be $102_{-72}^{+71}$ MeV at $1\sigma$ level. The values of symmetry energy and its slope for BigApple are 33.32 and 39.80 MeV, respectively, which also lie in the range given by Danielewicz and Lee~\cite{Danielewicz_2014} at the saturation density (see Table \ref{tab:nuclear_matter_properties}). 
%%%%%%%%%%%%%%%%      
\begin{table}[t]
\centering
\caption{The NM properties such as binding energy per particle $B/A$, $K$, effective mass ratio $M^*/M$, symmetry energy $J$, and its different coefficients are listed at the saturation density for five different parameter sets. All the parameters have MeV unit except $\rho_0$ (fm$^{-3}$) and $M^{\star}/M$ (dimensionless).}
 \scalebox{0.95}{
\begin{tabular}{|c|c|c|c|c|c|c|c|c|c|c|}
\hline
\multicolumn{1}{|c|}{Parameter}
&\multicolumn{1}{c|}{NL3}
&\multicolumn{1}{c|}{FSUGarnet}
&\multicolumn{1}{c|}{G3}
&\multicolumn{1}{c|}{IOPB-I}
&\multicolumn{1}{c|}{BigApple}
&\multicolumn{1}{c|}{Emp./expt.}\\
\hline
	$\rho_{0}$ (fm$^{-3})$ &  0.148  &  0.153&  0.148&0.149& 0.155&0.148--0.185~\cite{Bethe_1971} \\ 
$B/A$  &  -16.29  & -16.23 &  -16.02&-16.10&-16.34&-15.00-- -17.00~\cite{Bethe_1971}  \\ 
$M^{*}/M$  &  0.595 & 0.578 &  0.699&0.593&0.608&---  \\ 
$J$  & 37.43  &  30.95&  31.84&33.30&31.32& 30.20--33.70~\cite{Danielewicz_2014} \\ 
$L$ &  118.65  &  51.04 &  49.31&63.58 &39.80&35.00--70.00~\cite{Danielewicz_2014}\\ 
$K_{\rm sym}$  &  101.34  & 59.36 & -106.07&-37.09&90.44&-174-- -31~\cite{Zimmerman_2020} \\ 
$Q_{\rm sym}$  &  177.90  & 130.93&  915.47 &862.70&1114. 74&--- \\
$K$ & 271.38  &  229.5&  243.96& 222.65 &227.00&220--260~\cite{Colo_2014}\\ 
$Q_{0}$ &  211.94  & 15.76&   -466.61 &-101.37&-195.67& ---\\
$K_{\tau}$ &  -703.23  &  -250.41&-307.65 &-389.46& -116.34&-840-- -350~\cite{Stone_2014,Pearson_2010}\\
\hline
\end{tabular}}
\label{tab:nuclear_matter_properties}
\end{table}
%%%%%%%%%%%%%%%%%%%%%%
%%%%%%%%%%%%%%%%%%%
\begin{figure}
\centering
\includegraphics[width=0.7\textwidth]{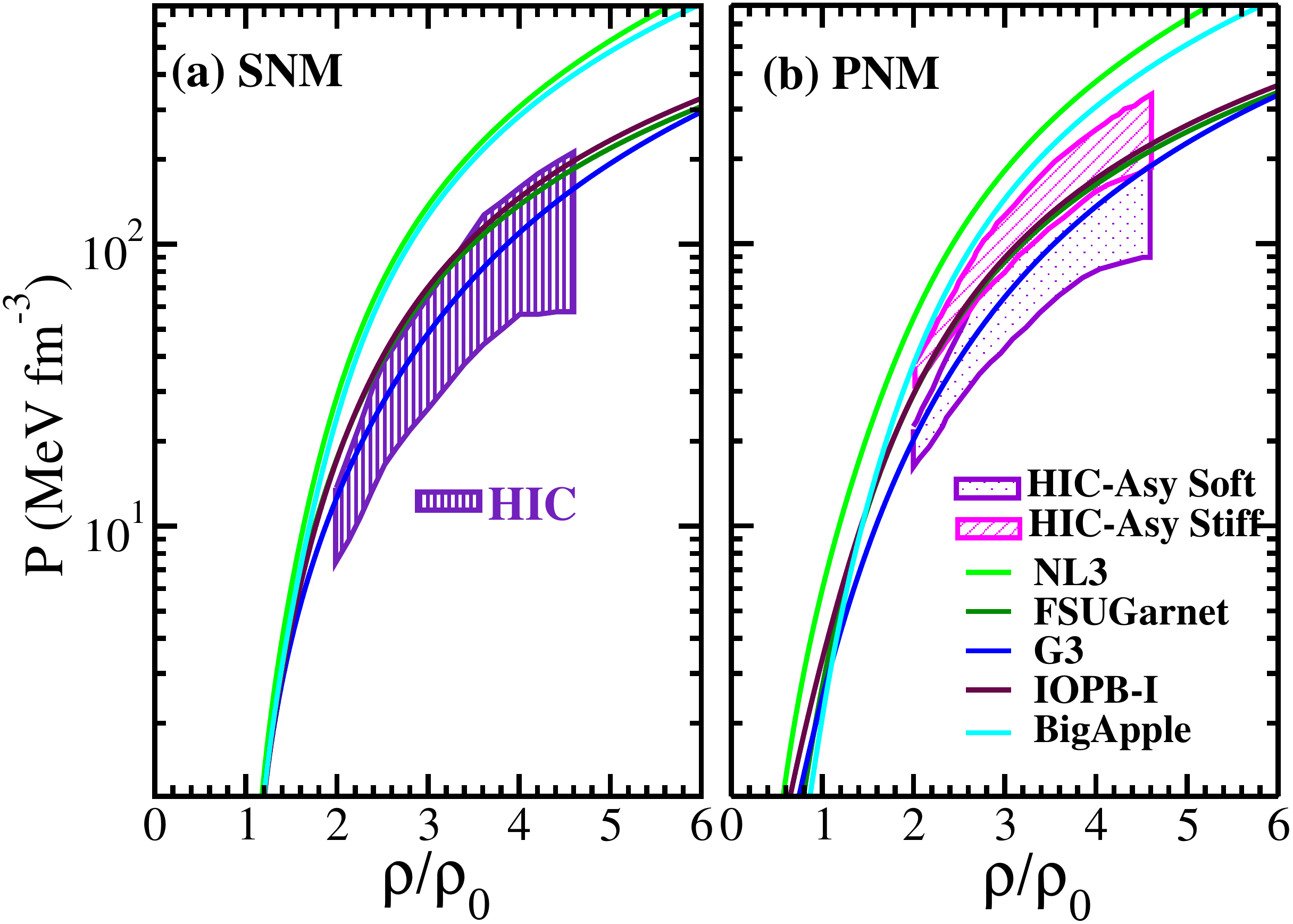}
\caption{Calculated pressure with the variation of baryon density. The results for NL3, BigApple, FSUGarnet, G3, and IOPB-I are compared with HIC data~\cite{Danielewicz_2002} both for SNM (left) and PNM (right). For the PNM case, the same data is divided into (i) HIC-Asy soft and (ii) HIC-Asy stiff, mainly based on the density-dependence symmetry energy.}
\label{fig:eosnm}
\end{figure}
%%%%%%%%%%%%%

In Fig. \ref{fig:eosnm}, we plot the pressure with the variation of the baryon density for SNM and PNM and compare it with the experimental flow data~\cite{Danielewicz_2002}. The calculated pressure by the G3 set is consistent with HIC data for the whole densities range for SNM (shown in Fig. \ref{fig:eosnm}). Although parameter sets like IOPB-I and FSUGarnet reproduce stiffer EOS compared to G3, their calculated pressure still matches HIC data. NL3 and BigApple are the stiffer EOSs as compared to others. So they disagree with the HIC data both for SNM and PNM cases. Although the EOS corresponds to the BigApple parameter set and does not pass through the experimental shaded regions given by HIC; still, it predicts the value of $K$, $J$, and $L$, which is reasonably within the empirical/experimental limit as given in Table \ref{tab:nuclear_matter_properties}.

\begin{figure}
\centering
\includegraphics[width=0.5\textwidth]{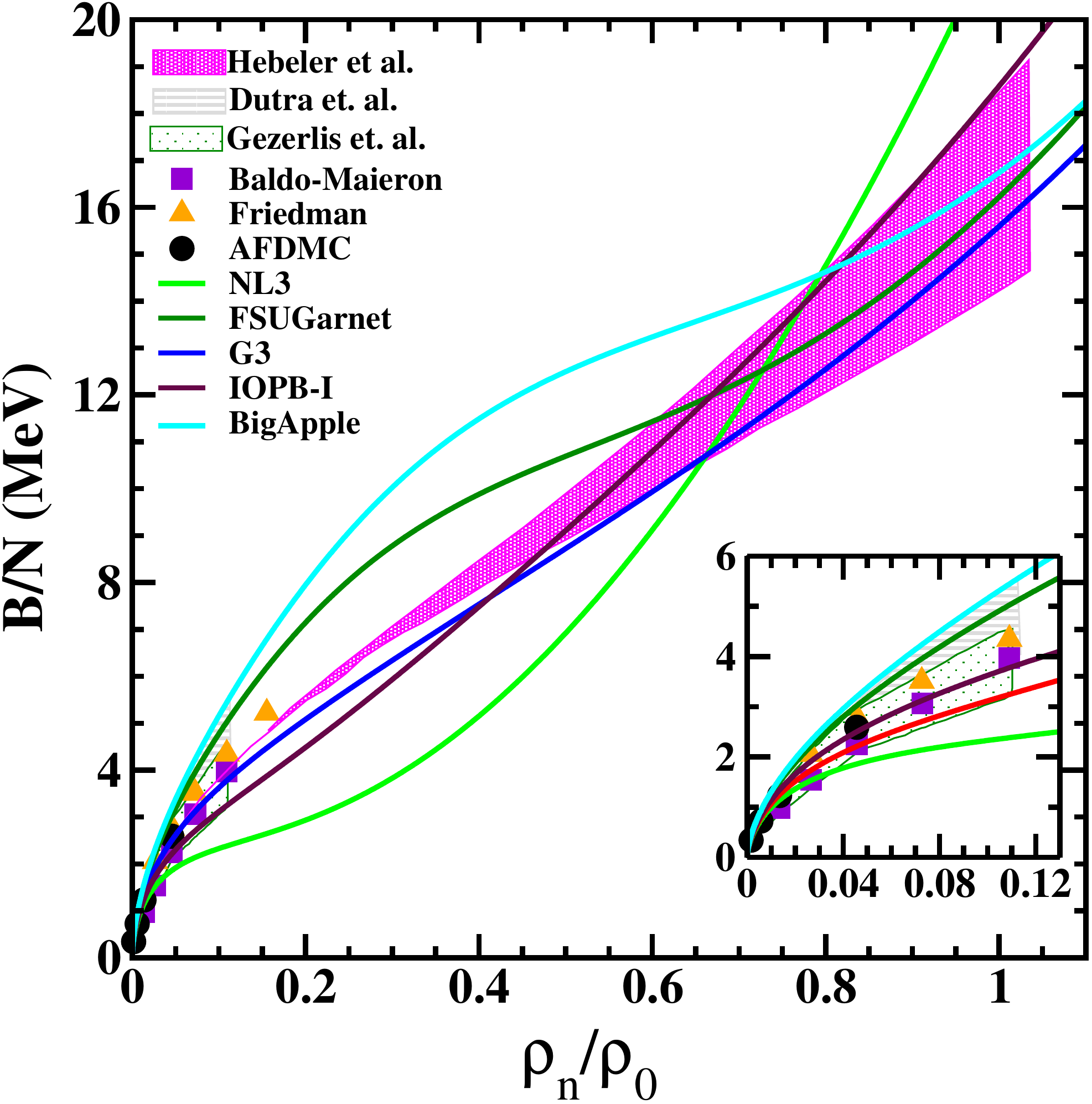}
\caption{The $B/N$ as a function of density with  NL3~\cite{Lalazissis_1997}, FSUGarnet~\cite{Chen_2015}, G3~\cite{Kumar_2017}, IOPB-I~\cite{Kumar_2018} and BigApple~\cite{Fattoyev_2020} parameter sets. The other results are from Hebeler {\it et al.}~\cite{Hebeler_2013}, Dutra {\it et al.}~\cite{Dutra_2012}, Gezerlis {\it et al.}~\cite{Gezerlis_2010}, Baldo-Maieron~\cite{Baldo_2008}, Friedman~\cite{Friedman_1981} and Auxiliary-field diffusion Monte Carlo (AFDMC)~\cite{Gandolfi_2008}.}
\label{fig:be}
\end{figure}

Next, our focus is on the $B/A$ at the saturation density. The variation of $B/A$ with baryon density ($\rho/\rho_0$) for the PNM system is shown in Fig. \ref{fig:be} for the BigApple parameter set along with NL3, FSUGarnet, G3, and IOPB-I. Some experimental data are also put for comparison. From this plot, one can see that in the low-density regions, except BigApple and NL3, other parameter sets are in harmony with the results of the microscopic calculations. The BigApple, G3, IOPB-I, and FSUGarnet parameter sets pass through the shaded regions near the saturation density. These parameter sets are qualitatively consistent with the results of Hebeler {\it et al.} data~\cite{Hebeler_2013}. The $B/A$ at the saturation density for considered parameter sets lies in the practical limit given in Table \ref{tab:nuclear_matter_properties}.
%%%%%%%%%%%%%%%
%%%%%%%%%%%%%%%
\begin{figure}
\centering
\includegraphics[width=0.5\textwidth]{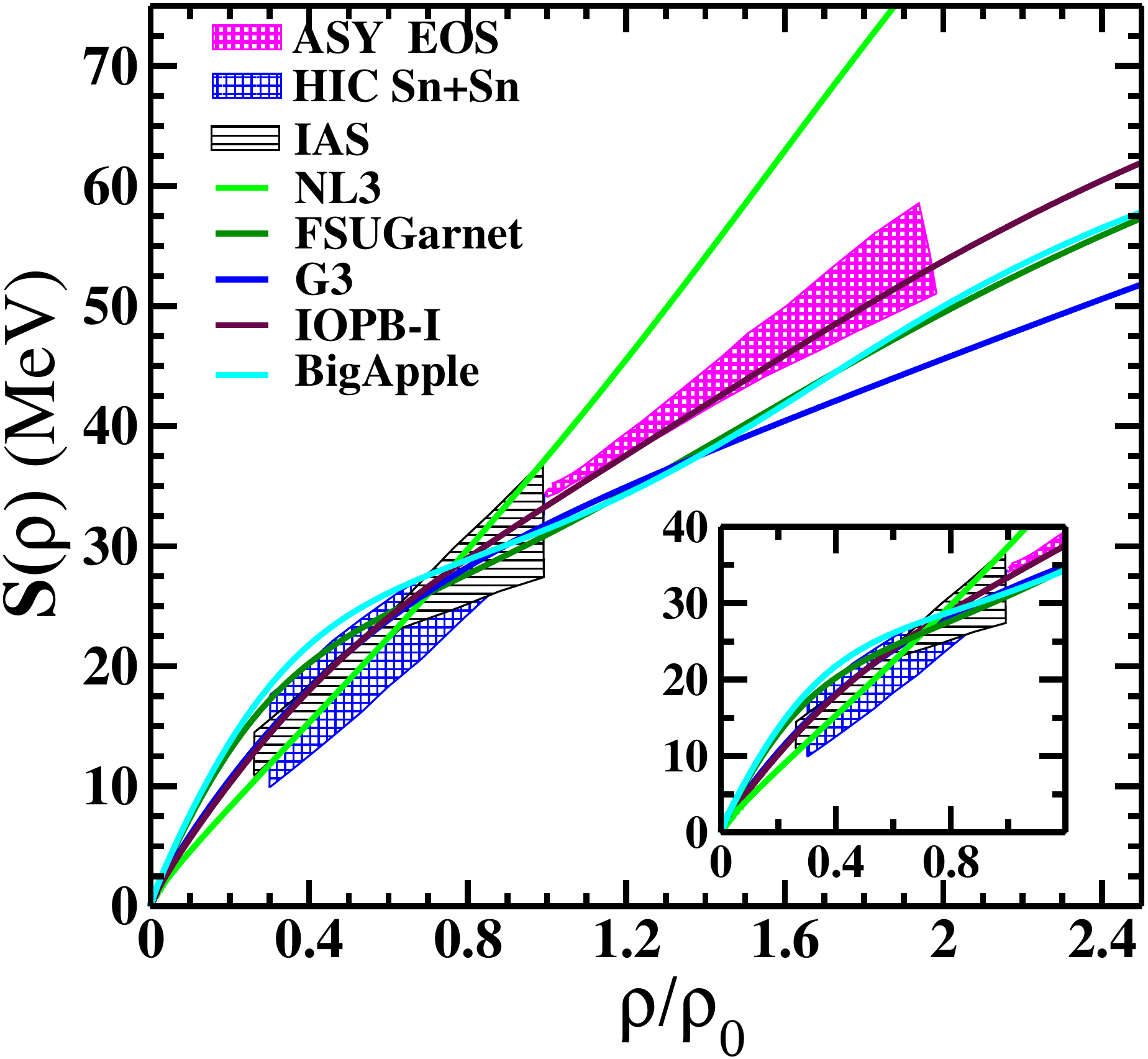}
\caption{The density-dependent symmetry energy with the baryon density for five different parameter sets. The shaded region is the symmetry energy from IAS~\cite{Danielewicz_2014}, HIC Sn+Sn~\cite{Tsang_2009, Tsang_2010} and ASY-EOS experimental data~\cite{Russotto_2016}. The zoomed pattern of the symmetry energy at low densities is shown in the inset.}
\label{fig:sym}
\end{figure}
%%%%%%%%%%%%

{\it Symmetry energy} is defined as the difference between energy per particle of PNM and SNM. The value of the symmetry energy at saturation density is known to some extent. However, its density dependence variation is still uncertain, i.e., the value of the symmetry energy at saturation density is better constrained than its density dependence. It has diverse behavior at the different densities regions~\cite{BaoLi_2019}. Also, the symmetry energy has broad relations with some properties of the NS~\cite{Dorso_2019, Zhang_2020, Lattimer_2014, Gandolfi_2016}.

In Fig. \ref{fig:sym}, we show the density-dependent symmetry energy with baryon density for five parameter sets. G3, IOPB-I, and FSUGarnet's symmetry energy predict soft behavior at the low density due to cross-coupling between $\omega$ and $\rho$ meson. Nevertheless, for the BigApple case, the values of $S(\rho)$ in low-density regions are too high. Also, at higher density, it predicts softer $S(\rho)$, which does not pass through the ASY data~\cite{Russotto_2016}. It shows a poor density dependence of the symmetry energy for the BigApple case.

In summary, we check the status of the stiff EOSs like NL3 and BigApple parameter sets to reproduce different constraints given by PNM, B/A, and $S(\rho)$,  which are given shown in Figs. \ref{fig:eosnm}--\ref{fig:sym} respectively. We find that the BigApple set does not satisfy those constraints, such as flow data and symmetry energy constraints, due to (i) its stiff behavior in low-density regions. (ii) the saturation density (0.155 fm$^{-3}$) for the BigApple parameter set is more as compared to the other two parameter sets, see Fig. \ref{fig:den}. The density distributions for NL3 and IOPB-I are almost the same, but there is a substantial shift for the BigApple case at the center. Hence in this sub-section, we examine the predictive capacity of the BigApple parameter set, i.e., we check how safe to take the BigApple parameter set for the study of NM properties. We find that it does not look too safe to take the BigApple set for the calculation.
%%%%%%%%%%%%%%%%%%%%%%%%
%%%%%%%%%%%%%%%%%%%%%%%%%
\subsection{Neutron Star}
\label{form:NS}
%%%%%%%%%%%%%%%%%%%%%%%%%
Many particles like nucleons, hyperons, kaons, quarks, and leptons are inside the NS. The neutron decays to proton, electron, and anti-neutrino inside the NS~\citep{NKGb_1997}. This process is called $\beta$--decay. The inverse $\beta$--decay process occurred to maintain the charge neutrality condition, mainly at the core part where neutron stars drifted from the nuclei. The process can be expressed as
%%%%%%%%%%%%%
\begin{align}
n \rightarrow p+e^-+\bar\nu, \nonumber\\ 
p+e^-\rightarrow n+\nu .
\label{eq:beta_equil}
\end{align}
%%%%%%%%%%%
%%%%%%%%%%%%%
\begin{align}
\mu_n &= \mu_p +\mu_e \, ,
\nonumber \\
&
\mu_e = \mu_\mu \, ,
\label{eq:chem_pot}
\end{align}
%%%%%%%%%%%
%
where $\mu_n$, $\mu_p$, $\mu_e$, and $\mu_\mu$ are the chemical potentials of neutrons, protons, electrons, and muons, respectively, and the charge neutrality conditions are
%%%%%%%%%%%%%
\begin{align}
\rho_p = \rho_e +\rho_\mu \,. 
\label{eq:total_den}
\end{align}
%%%%%%%%%%
The chemical potentials $\mu_n$, $\mu_p$, $\mu_e$, and $\mu_\mu$ are given by 
%%%%%%%%%%%%%
\begin{align}
\mu_{n,p} = W_0 \pm R_0+\sqrt{k_{n,p}^2+ M_{n,p}^{\star 2}} \, ,
\label{eq:chem_nucleon}
\end{align} 
%%%%%%%%%%%
\begin{align}
\mu_{e,\mu} = \sqrt{k_{e,\mu}^2+ m_{e,\mu}^2} \, ,
\label{eq:chem_lepton}   
\end{align}
%%%%%%%%%%%
where $m_e$ and $m_\mu$ are the masses of the leptons 0.511 and 105.66 MeV, respectively. Therefore, the total Lagrangian density for the NS is the addition of NM and lepton part, which is as follow~\cite{Dasfmode_2021}
%%%%%%%%%%%%%
\begin{align}
{\cal L}_{\rm NS} = {\cal L}_{\rm NM}
+\sum_{l}\bar\psi_{l}\Big(i\gamma^{\mu}\partial_\mu-m_l\Big)\psi_l
\label{eq:NS_lag}
\end{align}
%%%%%%%%%%%%%
%%%%%%%%%%%%%
\begin{figure}
\centering
\includegraphics[width=0.6\textwidth]{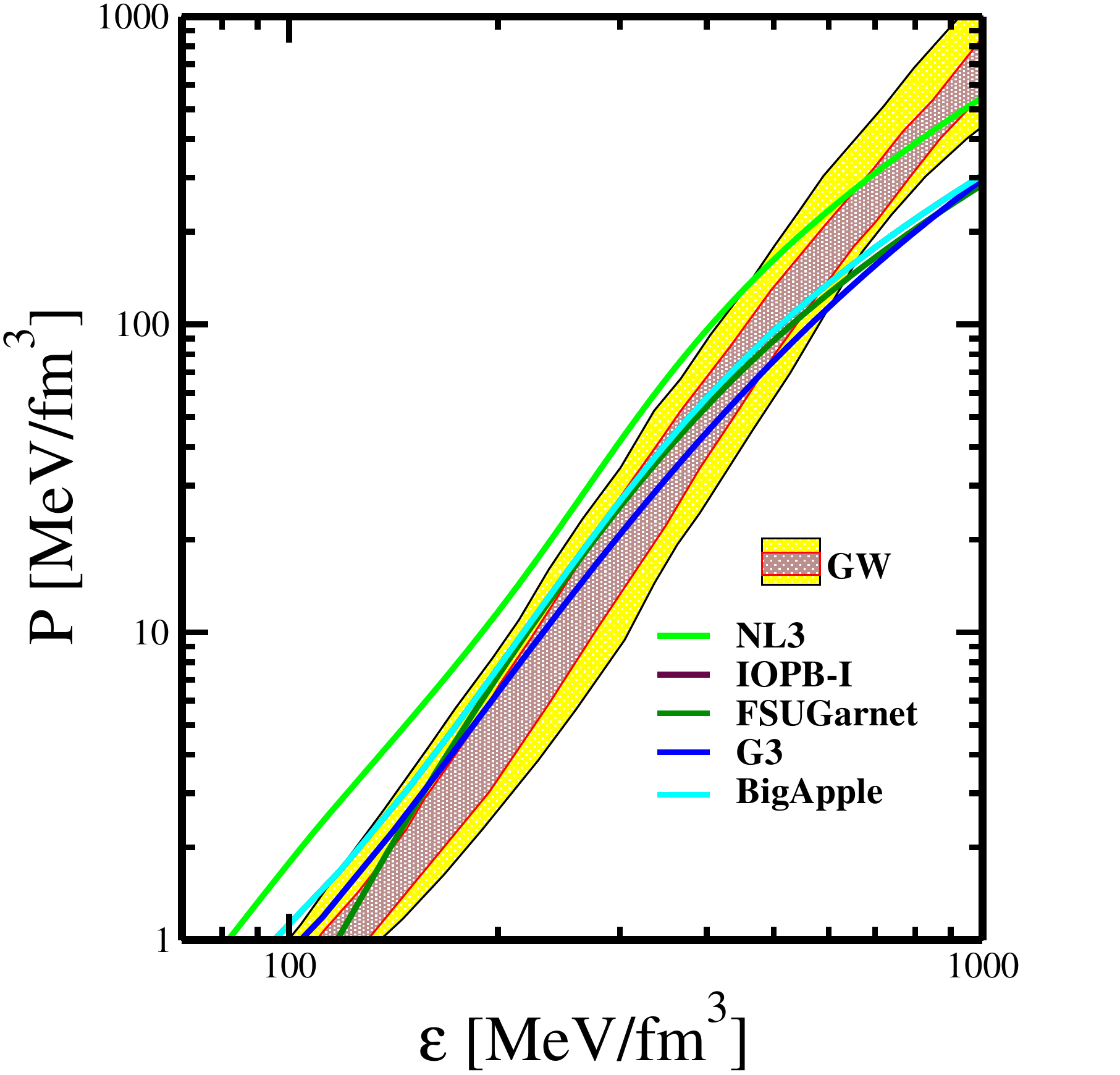}
\caption{The equations of states of $\beta$-equilibrated matter for NL3, FSUGarnet, G3, IOPB-I, and BigApple parameter sets. The shaded regions are for 50 \% (orange) and 90\% (grey) posterior credible limit given by the GW170817 data~\cite{Abbott_2018}.}
\label{EOSns}
\end{figure}
%%%%%%%%%%%%%%%
The energy density and pressure of the NS are found to be~\cite{Das_2020, Das_2021, Dasfmode_2021}
%%%%%%%%%%%%%
\begin{align}
{\cal {E}}_{{\rm NS}} = {\cal {E}_{{\rm NM}}} + {\cal {E}}_{l}, \,\,\,
P_{{\rm NS}} = P_{{\rm NM}} + P_l \, ,
\label{eq:eos_NS}
\end{align}
%%%%%%%%%%%
where, 
%%%%%%%%%%%%%
\begin{align}
{\cal {E}}_{l} = \sum_l\frac{1}{\pi^2}\int_0^{k_{F{_l}}} dk \ \sqrt{k^2+m_l^2},
\label{eq:energy_electon}
\end{align}
%%%%%%%%%%%
and
%%%%%%%%%%%%%
\begin{align}
P_{l} = \sum_l\frac{1}{3\pi^2}\int_0^{k_{F{_l}}} \frac{k^2 dk}{\sqrt{k^2+m_l^2}}\, ,
\label{eq:press_lepton}
\end{align}
%%%%%%%%%%%
where ${\cal{E}}_{l}$, $P_{l}$, and $k_l$ are the energy density, pressure, and Fermi momentum for leptons, respectively.

Nuclear EOS is the main ingredient in studying NS properties. The predicted EOS for five models alongside extracted recent GW170817 observational data are shown in Fig.  \ref{EOSns}. The shaded regions are deduced from GW170817 data with 50\% (grey) and 90\% (yellow) credible limit ~\cite{Abbott_2018}. For the crust part, we use the BCPM crust EOS~\cite{BKS_2015} and join it with the uniform liquid core to form a unified EOS. Having the EOS, we can calculate the NS properties, such as its mass, radius, tidal deformability, and moment of inertia, in the following chapters. 

\begin{figure}[H]
\centering
\includegraphics[width=0.5\textwidth]{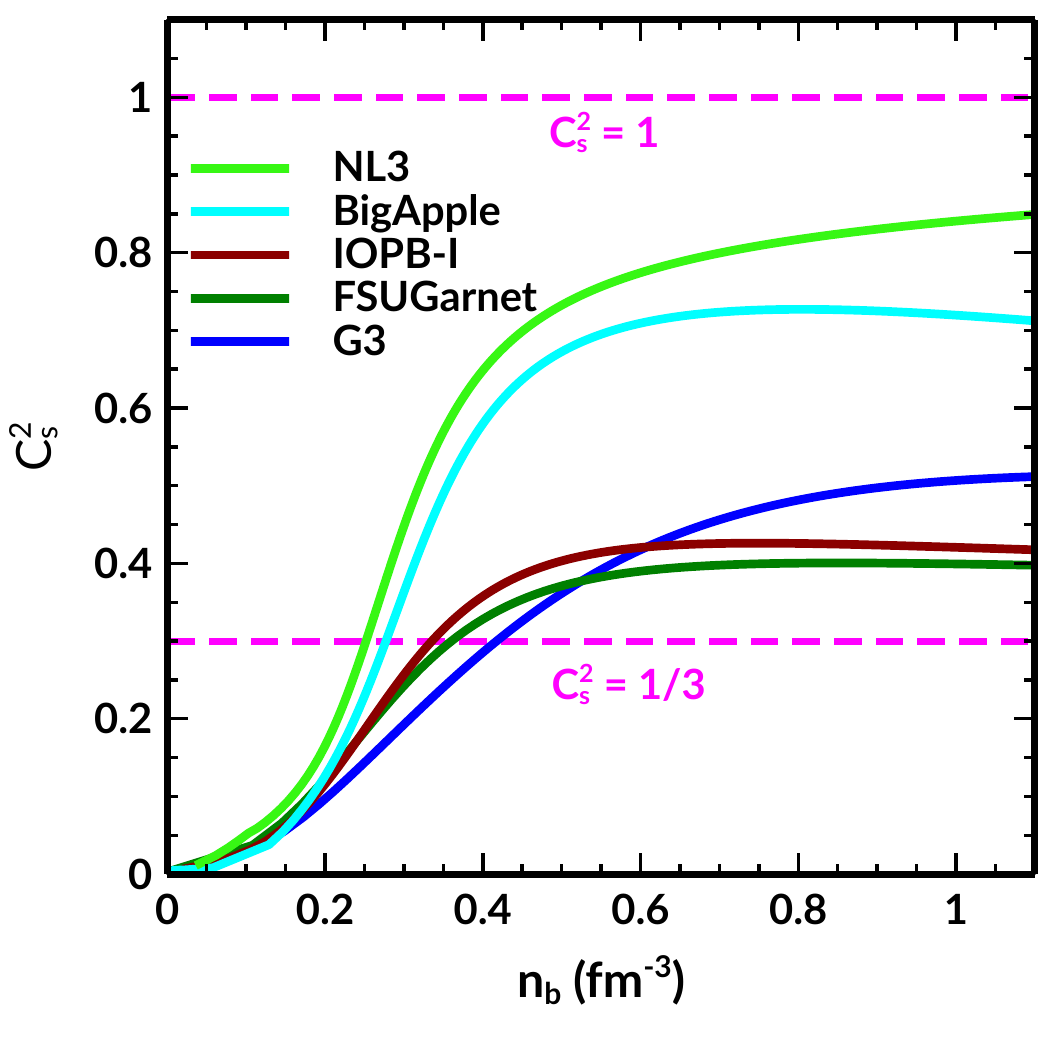}
\caption{Sound speed as a function of baryon density ($n_b$) for five considered parameter sets. The dashed magenta lines represent the causality bound ($C_s^2=1$) and QCD conformal limit ($C_s^2=1/3$).}
\label{s_speed}
\end{figure}

The speed of sound inside the NS is calculated using the equation $C_s^2=\partial P/\partial {\cal E}$. We plot the variation of $C_s^2$ as a function of baryon density in Fig. \ref{s_speed}. In Ref. \cite{Rhoades_1974}, it has been suggested that the speed of sound is always less than one ($C_s^2<1$) in the dense matter, which is defined as the ``Causality limit". Also, according to Le Chatelier's principle, the pressure is a monotonically non-decreasing function of the energy density ($dP/d{\cal E}\geq 0$), which means that the velocity of sound must be greater than zero ($C_s^2>0$). In conformal QCD symmetry, the trace of energy-momentum vanishes, which implies that the pressure is $1/3^{\rm rd}$ of the energy density. This implies the limit on the $C_s^2=1/3$ for such strongly interacting systems \cite{Bedaque_2015}. Here, all the parameter sets respect the causality in the whole density region. The value of $C_s^2$ increases up to 0.4 fm$^{-3}$, becoming constant beyond that. Hence, the causality obeys by the considered parameter sets well for the whole density regions. However, the $c/3$ limit is only valid at the higher density part where quarks play a role. In this case, we only consider the nucleons. Therefore, QCD conformal limit doesn't obey the considered parameter sets.
%%%%%%%%%%%%%
%%%%%%%%%%%%%%%
%%%%%%%%%%%%%%%%%%%%%
\section*{Conclusion}
%%%%%%%%%%%%%%%%%%%%%
In conclusion, we enumerated the details and applications of the RMF and E-RMF models by applying them to different systems. The differences between various RMF/E-RMF models have been explained in detail. The coupling constants for various forces were obtained by fitting them with different experimental/empirical data. In this study, we have taken various parameter sets, such as NL3, G3, IOPB-I, and FSUGarnet, along with BigApple, to calculate the properties of different systems. We observed that it predicts almost all properties of finite nuclei, NM, and NS well. Therefore, in the subsequent chapters, we choose those RMF and E-RMF models to explore the systems such as NM and NS.
%%%%%%%%%%%%%%%%%%%%%%%%%%%%%%%%%%%%%%%%% END %%%%%%%%%%%%%%%%%%%%%%%%%%%%%%%%%%%%%%%%%%%%
%\blankpage
%%%%%%%%%%%%%%%%%%%%%%%%%% Chapter-3 %%%%%%%%%%%%%%%%%%%%%%%%%%%%%%
%%%%%%%%%%%%%%%%%%%%%%% CHAPTER - 3 %%%%%%%%%%%%%%%%%%%%%%%%%%
\chapter{Dark matter interaction and its effects on nuclear matter properties}
\label{C3}
%%%%%%%%%%%%%%%%%%%%%%%%%%%%%%%%%%%%%%%%%%%%%%%%%%%%%%%%%%%%%%
In this chapter, we briefly give an introduction to DM. NSs are prone to the accretion of DM due to their internal compositions and high baryon density. The DM confines to the star may affect its properties. To understand the NS, one has to study nuclear matter because it is an infinite system of nucleons with $\sim 90\%$ neutrons and a few percentages of protons and leptons. Hence, it is imperative to study the NM observable, such as effective mass, binding energy, incompressibility, symmetry energy, slope, and curvature parameters, to determine the nature of the system. It is reported in Ref.  \cite{Roberts_2012, Alam_2016} that the symmetry energy and slope parameters affect the NS cooling and radii of the star. Therefore, we calculate these observables for DM admixed NM with different fractions and analyze its effects by changing the neutron and proton asymmetry.
%%%%%%%%%%%%%%%%%%%%%%%%%%%%%%%%%%%%%
\section{Dark matter in the Universe}
\label{dark_matter}
%%%%%%%%%%%%%%%%%%%%%%%%%%%%%%%%%%%%%
Around 27\% of the matter in the universe is believed to be composed of an unobservable and speculative type of matter known as DM. Additionally, some DM particles are thought to be weakly interacting; hence it is expected that this will have some influence on the NS properties \citep{Joglekar_2020}. It has also been reported that when a compact star rotates in the Galaxy through a DM halo, it captures some of the DM particles \cite{Kouvaris_2011}. The enormous gravitational force and the immense baryonic density inside the NS are responsible for the incarceration of DM particles. The efficacy of the NS observational properties depends on the amount of DM captured by it \cite{Kouvaris_2011}. 

Theoretically, several types of DM particles have been hypothesized and reported to date, like, weakly interacting massive particles (WIMP) and feebly interacting massive particles (FIMP). The WIMPs are the most abundant DM particles in the early Universe due to their freeze-out mechanism \citep{Kouvaris_2011, Ruppin_2014}. They equilibrated with the environment at freeze-out temperature and annihilated to form different standard model particles and leptons. Recently, some approaches have been dedicated to explaining the heating of NS due to the deposition of kinetic energy by the DM \cite{Baryakhtar_2017, Acevedo_2019}. The annihilation of the DM particles enhances the cooling of the NSs  \citep{Kouvaris_2008, Ding_2019, Bhat_2019}. The resurrection of the DM develops its interaction with baryon, which affects the structure of the NS \citep{De_Lavallaz_2010, Ciarcelluti_2011}. The Chandrasekhar limit constrains the accumulation of the DM inside the NS. If the accumulated quantity of the DM exceeds this limit, it can evolve into a tiny black hole and destroy the star \cite{Goldman_1989, Kouvaris_2008}. Different approaches have been used to calculate the NS properties with the inclusion of DM inside the NS \citep{Sandin_2009, De_Lavallaz_2010, Kouvaris_2010, Ciarcelluti_2011, Leung_2011, AngLi_2012, Panotopoulos_2017, Ellis_2018, Bhat_2019, Ivanytskyi_2020, Das_2019, Quddus_2020, Das_2020}. However, in the present scenario, we take the non-annihilating WIMP as a DM candidate inside the NS. We observed that the addition of DM softens the EOS, which results in the reduction of the mass-radius of the NS.
%%%%%%%%%%%%%%%%%%%%%%%%%%%%%%%%%%%
\section{Dark matter interaction}
\label{DM_lagrangian}
%%%%%%%%%%%%%%%%%%%%%%%%%%%%%%%%%%%
As it is stated earlier in this thesis, when the NS evolves in the Universe, the DM particles are accreted inside it due to its huge gravitational potential and immense baryonic density \cite{Goldman_1989, Kouvaris_2008}. In our investigations, we assume the DM particles are already inside the NS. With this assumption, we modeled the interaction Lagrangian for DM admixed NS. We consider the Neutralino as the DM particle having a mass of 200 GeV \cite{Panotopoulos_2017, Das_2019}. The DM particles are interacted with baryons by exchanging standard model (SM) Higgs. The interacting Lagrangian is given in the form \cite{Panotopoulos_2017, Das_2019, Quddus_2020, Das_2020, Das_2021, DasMNRAS_2021, DasPRD_2021}:
%%%%%%%%%%%%%
\begin{align}
	{\cal{L}}_{{\rm DM}}=\bar \chi \Big[ i \gamma^\mu \partial_\mu - M_\chi + y h \Big] \chi +  \frac{1}{2}\partial_\mu h \partial^\mu h - \frac{1}{2} M_h^2 h^2 +\sum_\alpha \frac{f M_\alpha}{v} \bar\varphi_\alpha h \varphi_\alpha , 
	\label{eq:lagrangan_DM}
\end{align}
%%%%%%%%%%%
where $\psi_{\alpha}$ and $\chi$ are the nucleons and DM wave functions, respectively. The parameters $y$ is DM-Higgs coupling, $f$ is the proton-Higgs form factors, and $v$ is the vacuum expectation value of the Higgs field. The values of $y$ and $v$ are 0.07 and 246 GeV, respectively, taken from the Refs. ~\cite{Das_2020, Das_2021}. The value of $f$ is 0.35.
%%%%%%%%%%%%%%%%%%%%%%%%%%%%%%%%%%%
\subsection{Experimental evidences}
%%%%%%%%%%%%%%%%%%%%%%%%%%%%%%%%%%%
In the present formalism, two coupling constants have major significance (i) DM-Higgs coupling ($y$) and (ii) nucleon-Higgs form factor ($f$). The detailed prescription is given as follows \\
(i) The direct detection experiment did not show any collision events till now, but they gave upper bounds on the WIMP-nucleon scattering cross-section, which is the function of the DM mass. The Higgs exchange undergoes elastic collision with the detector nucleus (at the quark level). Therefore the interaction Lagrangian, which contains both DM wave function ($\chi$) and quark wave function ($q$), can be written as \cite{Bhat_2019}
%%%%%%%%%%%%%%%%
\begin{equation}
	{\cal{L}}_{\rm int}=\alpha_q \Bar{\chi}\chi\Bar{q}q,
\end{equation}
%%%%%%%%%%%%%%%
where $\alpha_q=\frac{y f m_q}{vM_h^2}$. $q$ is the valence quark, $f$ is the nucleon-Higgs form factor, and $m_q$ is the quark's mass. In these calculations, the values of $y$ and $f$ are taken as 0.07 and 0.35. The spin-independent cross-section for the fermionic DM can be written as \cite{Bhat_2019}
%%%%%%%%%%%%%%%%
\begin{equation}
	\sigma_{\rm SI}=\frac{y^2f^2M_n^2}{4\pi}\frac{\mu_r^2}{v^2M_h^4} ,
\end{equation}
%%%%%%%%%%%%%%%
where $M_n$ (= 939 MeV) is the nucleon mass and $\mu_r$ is the reduced mass $\frac{M_nM_\chi}{M_n+M_\chi}$, $M_\chi$ is the mass of the DM particle. We calculated the $\sigma_{\rm SI}$ for three different masses of DM 50, 100, and 200 GeV, and their corresponding cross-section was found to be 9.43, 9.60, and 9.70 of the order of ($10^{-46}$ cm$^2$) respectively. That means the predicted values are consistent with the direct detection experiment like XENON-1T \cite{Xenon1T_2016}, PandaX-II \cite{PandaX_2016}, PandaX-4T \cite{Meng_2021}, and LUX \cite{LUX_2017} with 90\% confidence level. In the case of LHC, which produced various WIMP-nucleon cross-section limits in the range from $10^{-40}$ to $10^{-50}$ cm$^2$ depending on the DM production models \cite{PandaX_2016}. Thus our model also satisfies the LHC limit. Therefore in the present calculations, we constrained the value of $y$ from the direct detection experiments and the LHC results.\\
(ii) Nucleon-Higgs form factor ($f$) had been calculated in Ref. \cite{Djouadi_2012} using the implication of both lattice QCD \cite{Czarnecki_2010} and MILC results \cite{MILC_2009} whose value is $0.33_{-0.07}^{+0.30}$ \cite{Aad_2015}. The taken value of $f$ (= 0.35 ) in this calculation lies in the region.
%%%%%%%%%%%%%%%%%%%%%%%%%%%%%%%%
\subsection{Equation of motions}
%%%%%%%%%%%%%%%%%%%%%%%%%%%%%%%%
The Euler Lagrange equation of motion for DM particle ($\chi$) and Higgs boson ($h$) can be derived from Lagrangian in Eq. (\ref{eq:lagrangan_DM}) as,
%%%%%%%%%%%%%
\begin{align}
	&\bigg(i\gamma^{\mu}\partial_{\mu}-M_{\chi}+yh \bigg)\chi {} =0 \, ,
	\nonumber \\
	&\partial_{\mu}\partial^{\mu}h+M_{h}^2h=y\bar{\chi}\chi+\sum_\alpha\frac{fM_\alpha}{v}\bar\varphi_\alpha\varphi_\alpha
	\label{eq:field_EOS_DM}
\end{align}
%%%%%%%%%%%
respectively. Applying mean-field approximation, we get, 
%%%%%%%%%%%%%
\begin{align}
	&h_0=\frac{y\langle\bar{\chi}\chi\rangle+\sum_\alpha \frac{f M_\alpha}{v}\langle\bar\varphi_\alpha\varphi_\alpha\rangle}{M_h^2} \, ,
	\nonumber \\
	&\bigg(i\gamma^{\mu}\partial_{\mu}-M^{\star}_{\chi} \bigg)\chi {}=0 \, ,
	\label{eq:higgs_field}
\end{align}
%%%%%%%%%%%
where $M_\chi^{\star}$ is the DM effective mass, can be given as,
%%%%%%%%%%%%%
\begin{align}
	M_{\chi}^{\star} {}=M_{\chi}-yh_0. 
	\label{eq:efm_DM}
\end{align}
%%%%%%%%%%%
The DM scalar density ($\rho_s^{{\rm DM}}$) is
%%%%%%%%%%%%%
\begin{align}
	\rho_s^{{\rm DM}}=\langle\bar{\chi}\chi\rangle =  \frac{\gamma}{2 \pi^2}\int_0^{k_f^{{\rm DM}}} k^2 dk \  \frac{M_\chi^\star}{\sqrt{k^2+M_\chi^\star{^2}}},
	\label{eq:dm_density}
\end{align}
%%%%%%%%%%%
where $k_f^{{\rm DM}}$ is the Fermi momentum for DM, and $ \gamma$ is the spin degeneracy factor with a value of 2 for neutron and proton.

Assuming the average number density of nucleons ($n_b$) is $10^3$ times larger than the average DM density ($n_{{\rm DM}}$), which implies the ratio of the DM and the NS mass to be $\sim \frac{1}{6}$. The nuclear saturation density $n_0 \sim 0.16$ fm$^{-3}$, therefore, the DM number density becomes $n_{{\rm DM}} \sim 10^{-3}n_0\sim 0.16\times 10^{-3}$ fm$^{-3}$. Using the $n_{{\rm DM}}$, the $k_{f}^{{\rm DM}}$ is obtained from the equation $k_f^{{\rm DM}}=(3\pi^2 n_{{\rm DM}})^{1/3}$. Hence the value of $k_{f}^{{\rm DM}}$ is $\sim 0.033$ GeV. Therefore, in our case, we vary the DM momenta from 0 to 0.06 GeV.
The energy density (${\cal{E}}_{{\rm DM}}$) and pressure ($P_{{\rm DM}}$) for DM can be obtained by solving equation (\ref{eq:lagrangan_DM})
%%%%%%%%%%%%%
\begin{align}
	{\cal{E}}_{{\rm DM}}=\frac{1}{\pi^2}\int_0^{k_f^{\rm DM}} k^2 \ dk \sqrt{k^2 + M_\chi^\star{^2} } +\frac{1}{2}M_h^2 h_0^2 ,
	\label{eq:energyden_DM}
\end{align}
%%%%%%%%%%
and
%%%%%%%%%%%%%
\begin{align}
	P_{{\rm DM}}=\frac{1}{3\pi^2}\int_0^{k_f^{{\rm DM}}} \frac{ k^4 \ dk} {\sqrt{k^2 + M_\chi^\star{^2}}} - \frac{1}{2}M_h^2 h_0^2 ,
	\label{eq:presure_DM}
\end{align} 
%%%%%%%%%%%
%%%%%%%%%%%%%%
$h_0$ is the Higgs field calculated by applying the mean-field approximation. $k_f^{{\rm DM}}$ is the Fermi momentum for DM. The Higgs field's contribution to energy density and pressure is minimal. With the Higgs contribution, the system effective mass (in Eq. \ref{eq:effm_nucleons}) is redefined as 
%%%%%%%%%%%%%
\begin{align}
	M_j^\star=M- \Phi_0 \mp D_0 - \sum_j\frac{f M_j}{v}h_0 \, ,
	\label{eq:effm_NM_DM}
\end{align}
%%%%%%%%%%%
%%%%%%%%%%%%%%%%
The total EOS of the DM admixed NM system is written as \cite{Das_2020}
%%%%%%%%%%%%%
\begin{align}
	{\cal E}  = {\cal E}_{{\rm NM}} + {\cal E}_{{\rm DM}}, \ {\rm and} \  P = P_{{\rm NM}} + P_{{\rm DM}}.
	\label{eq:total_EOS_NM_DM}
\end{align}
%%%%%%%%%%%
Similarly, the total EOS of DM admixed NS is written as \cite{Das_2020, Das_2021, Dasfmode_2021, DasMNRAS_2021}
%%%%%%%%%%%%%
\begin{align}
	{\cal E}  = {\cal E}_{{\rm NS}} + {\cal E}_{{\rm DM}}, \ {\rm and} \  P = P_{{\rm NS}} + P_{{\rm DM}}.
	\label{eq:total_EOS_NS_DM}
\end{align}
%%%%%%%%%%%
%%%%%%%%%%%%%%%%%%%%%%%%%%
\section{NM Properties}
\label{NM}
%%%%%%%%%%%%%%%%%%%%%%%%%%
This section represents some of the NM properties crucial for exploring the NS. The NM properties such as binding energy per nucleon (BE/A), effective mass, incompressibility, symmetry energy, and its different co-efficient are calculated with RMF models. We also admix the DM inside the NM to see its effects on various properties.
%%%%%%%%%%%%%%%%%%%%%%%%%%%%%%%%%%%%%%%%%%%%%%%%%%%%%%%%%%%%%
\subsection{EOSs, BE/A, and effective mass for DM admixed NM}
\begin{figure}
	\centering
	\includegraphics[width=0.7\columnwidth]{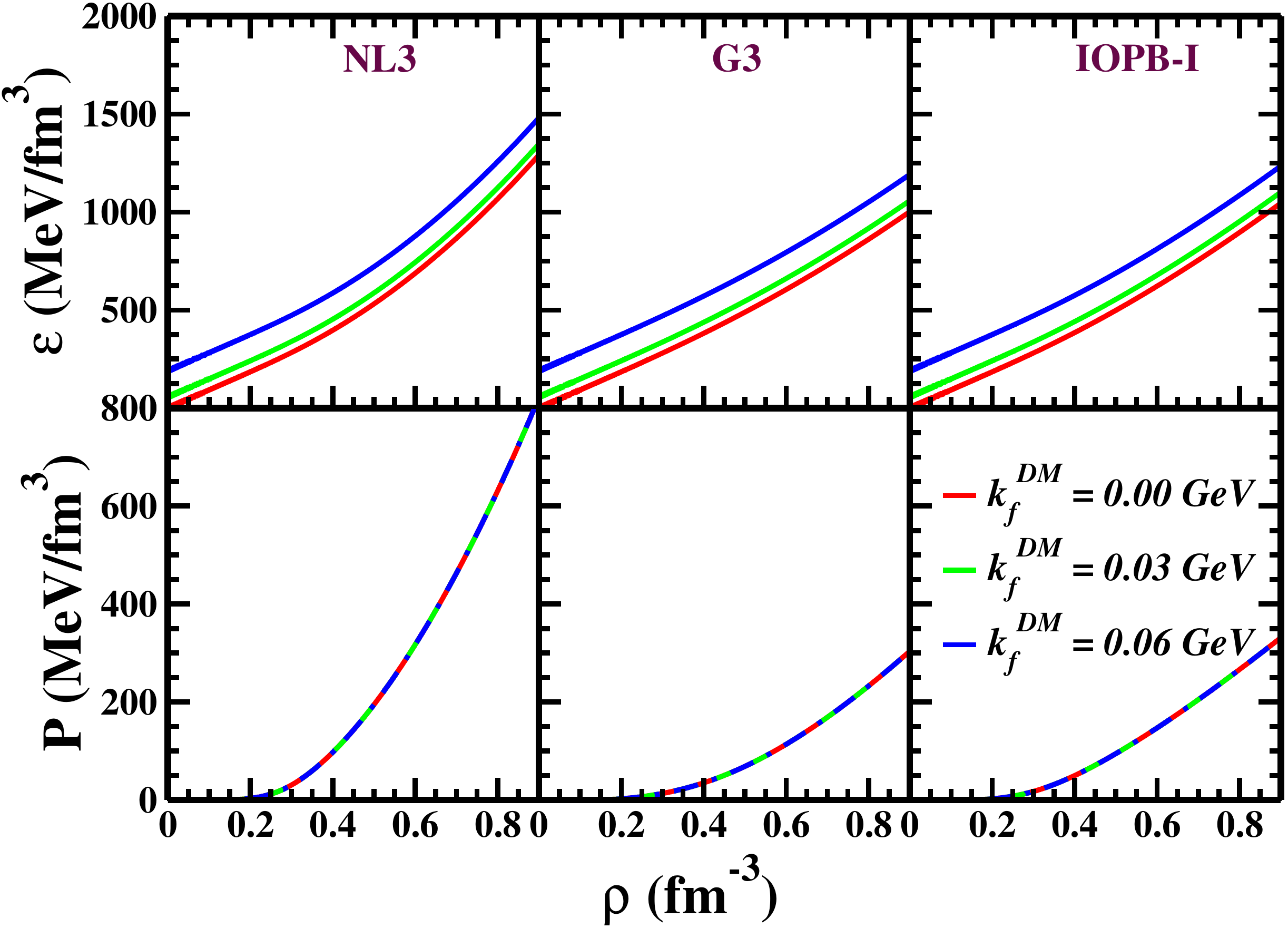}
	\caption{The energy density and pressure (Eq. \ref{eq:total_EOS_NM_DM}) for SNM with total NM density at $k_f^{\rm DM}$= 0.0, 0.03 and 0.06 GeV.}
	\label{fig:NM_DM_EOS}
\end{figure}

%%%%%%%%%%%%%%%%%%%%%%%%%%%%%%%%%%%%%%%%%%%%%%%%%%%%%%%%%%%%%
The energy density and pressure are obtained for NM admixed with DM (as given in Eq. \ref{eq:total_EOS_NM_DM}) and shown as a function of total NM density ($\rho$) in Fig. \ref{fig:NM_DM_EOS}. The exact percentage of DM is still unknown. Therefore, in this study, we vary its values from $k_f^{\rm DM}= 0.00-0.06$ GeV (as mentioned in Sub Sec. \ref{DM_lagrangian}). It is noticed that the magnitude of ${\cal{E}}$ changes significantly without affecting the $P$. This is because the DM energy density contribution is more significant than pressure. Also, it is noticed that the denominator part in Eq. (\ref{eq:presure_DM}) significantly decreases the value of $P$ in comparison with Eq. (\ref{eq:energyden_DM}). Therefore, the total contribution of the pressure to the DM admixed NM system is very less.
%%%%%%%%%%%%%%
%%%%%%%%%%%%

The parametric dependence is also shown in Fig. \ref{fig:NM_DM_EOS}. For example, NL3 gives the stiffest EOS compared to IOPB-I and G3 for the SNM case. Here also, G3 predicts the softest EOS. Thus, the qualitative nature of the EOS is similar with and without DM as far as stiffness or softness is concerned. 
%%%%%%%%%%%%%%
\begin{figure}
	\centering
	\includegraphics[width=0.7\columnwidth]{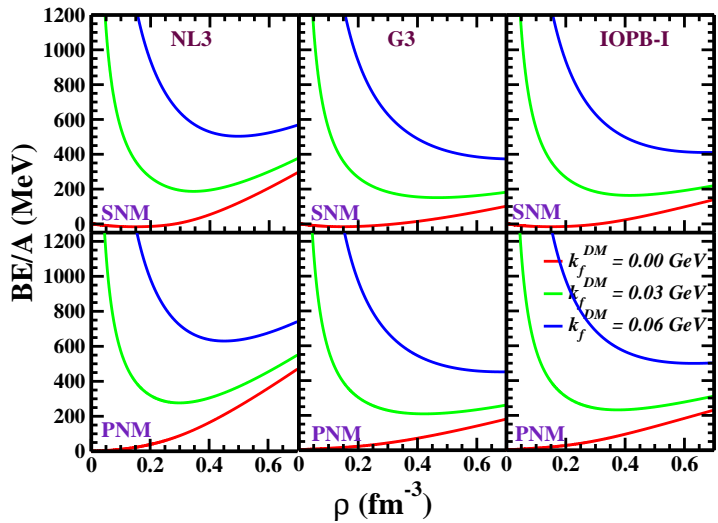}
	\caption{The BE/A of NM in the presence of DM at $k_f^{\rm DM}$= 0.0, 0.03 and 0.06 GeV.}
	\label{fig:be_NM_DM}
\end{figure}
%%%%%%%%%%%%

The BE/A is defined as ${\cal{E}}/\rho-M_n$, where the $\cal{E}$ is the total energy density in Eq. (\ref{eq:total_EOS_NM_DM}) and $\rho$ is the total NM density. The BE/A as a function of $\rho$ for different $k_f^{\rm DM}$ is shown in Fig. \ref{fig:be_NM_DM}. The effects of DM on BE/A are significant both for SNM and PNM. In the case of SNM with no DM case, the BE/A is negative, and its values are given in Table \ref{tab:NM_properties} at the saturation density. With the inclusion of DM, the BE/A goes positive. The same trend is also followed by PNM with the addition of DM.

\begin{figure}
	\centering
	\includegraphics[width=0.7\columnwidth]{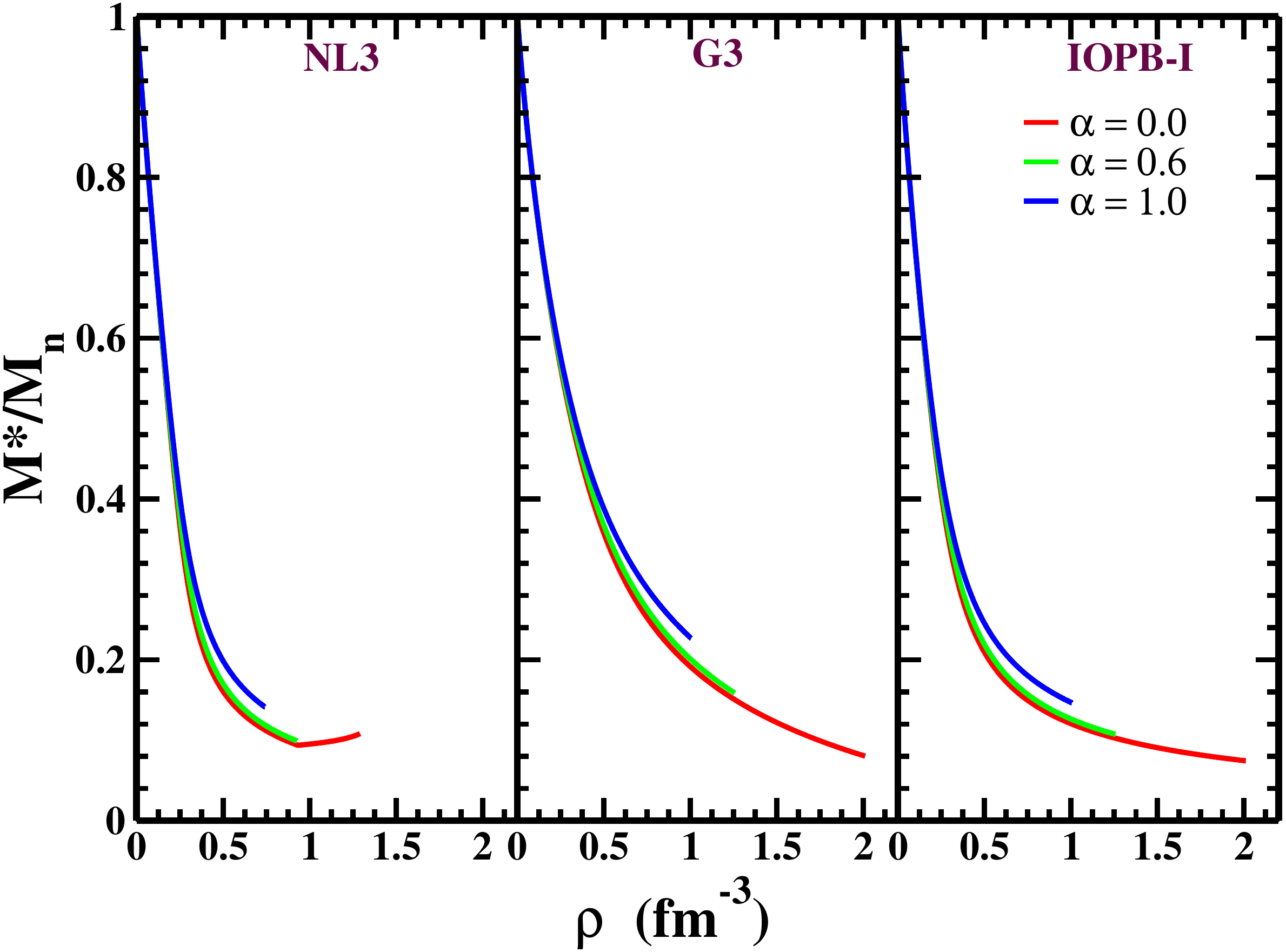}
	\caption{(colour online) The variation of effective mass in Eq. (\ref{eq:effm_NM_DM}) for different $\alpha$ with baryon density with $k_f^{\rm DM}$ = 0.06 GeV.}
	\label{fig:EFFM_NM_DM}
\end{figure}

Here, we calculate an important NM parameter, the effective mass of the NM system. The effective mass 
(see Eq. \ref{eq:effm_NM_DM}) variation as a function of total NM density is shown in Fig. \ref{fig:EFFM_NM_DM}. Here it is imperative to mention that $k_f^{\rm DM}$ = 0 GeV means $\rho_{\chi}$ is zero, but the effect on $M^\star/M_{n}$ is very less due to non-zero Higgs-nucleon Yukawa coupling as mentioned in Eq. (\ref{eq:effm_NM_DM}). The contribution of Higgs field is very small ${\cal{O}}$($10^{-6}-10^{-8}$) even for $k_f^{\rm DM}$ is 0.06 GeV. Therefore, in this case, we vary the neutron-to-proton ratio rather than the DM percentage. The $M^\star/M_{n}$ decreases with total NM density $\rho$, similar to the normal nuclear medium. As far as the neutron to proton ratio ($N/Z$) increases, the $M^\star/M_{n}$ value increases mostly in the high-density region. However, there is practically no effect of $\alpha$ in the low-density region of the NM system.
%%%%%%%%%%%%%%
%%%%%%%%%%%%
%%%%%%%%%%%%%%%%%%%%%%%%%%%%%%%%%%%%%%%%%%%%%%%%%%%%%%%%%%%
\subsection{Incompressibility, $C_{s}^{2}$ for DM admixed NM}
%%%%%%%%%%%%%%%%%%%%%%%%%%%%%%%%%%%%%%%%%%%%%%%%%%%%%%%%%%%
Another important NM parameter is incompressibility ($K$). This value tells us how much one can compress the NM system. It is a standard quantity at the saturation point, and one can put constraints on its value using isoscalar giant monopole resonance \cite{Colo_2014, Stone_2014}. The latest experimental value is $K = 240\pm20$ MeV. However, in an astronomical object like the NS, its density varies from the star's center to the crust with a variation of $\rho$ from $10-0.1\rho_0$ \cite{Lattimer_2004}. Thus, to better understand the compression mode or monopole vibration mode, we have to calculate the $K$ for all the density ranges of NM with different $\alpha$, including 0 and 1. Since we see the earlier case, DM does not affect the pressure of either SNM or PNM also in NS; hence the DM does not affect $K$.
%%%%%%%%%%%%%%
\begin{figure}
	\centering
	\includegraphics[width=0.7\columnwidth]{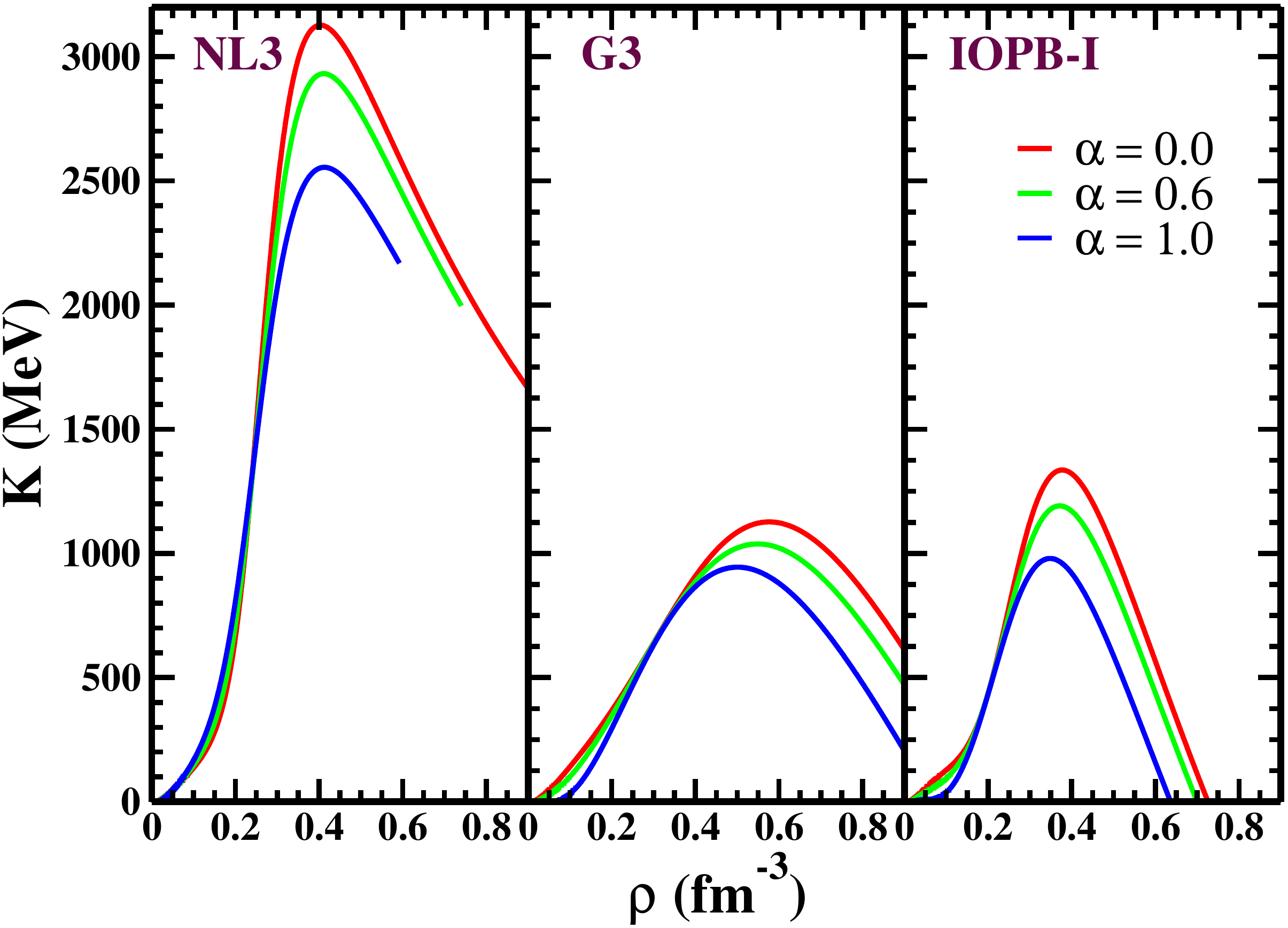}
	\caption{(colour online) The variation of incompressibility $K$ with different $\alpha$ as a function of baryon density $\rho$ with $k_f^{\rm DM}$ = 0.06 GeV.}
	\label{fig:ICOMP_NM_DM} 
\end{figure}
%%%%%%%%%%%%

The variation of $K$ with baryon density for different $\alpha$ is displayed in Fig. \ref{fig:ICOMP_NM_DM}. One can find the value of $K$ at the saturation density in Table \ref{tab:NM_properties} for the SNM system are 271.38, 243.96, and 222.65 MeV for NL3, G3, and IOPB-I, respectively. It is worth mentioning that DM does not affect incompressibility. That means the $K$ values remain unaffected with the variation of $k_f^{\rm DM}$. On the other hand, substantial variation is seen with the different $\alpha$. We found that the value of $K$ increases up to a certain maximum and then gradually decreases, as shown in Fig. \ref{fig:ICOMP_NM_DM}. The calculation shows that the incompressibility decreases with increasing $\alpha$ irrespective of the parameter sets. The values of $K$ for G3 and IOPB-I parameter sets lie in the region (except for NL3) given by the experimental value in Table \ref{tab:NM_properties}. Since the NL3 gives very stiff EOS, all NM parameters like $K$, $J$, and $L$ provide large values and do not lie in the range.

The recent gravitational wave observation from the merger of binary NSs event GW170817 \cite{Abbott_2017, Abbott_2018}, constraints the upper limit on the tidal deformability $\Lambda$ and predicts a small radius. Also, the recent discovery of the three highly massive stars $\sim$ 2 $M_\odot$ \cite{Antoniadis_2013, Fonseca_2016, Cromartie_2020} predicts that the pressure in the inner core of the star is significant, where the typical baryon number density relatively high $\rho$ $>$ 3$\rho_0$ in this region. The pressure in the outer core of the massive NS is considered to be small in the density range 1 to 3 $\rho_0$ \cite{McLerran_2019}. Combining these observations of large masses and the smaller radii of the massive NSs, one can infer that the causality \cite{Rhoades_1974, Bedaque_2015, Kojo_2015, Moustakidis_2017, McLerran_2019} of the NM inside the inner core of the NS can violate  \cite{McLerran_2019}. 
%%%%%%%%%%%%%%
\begin{figure}
	\centering
	\includegraphics[width=0.7\columnwidth]{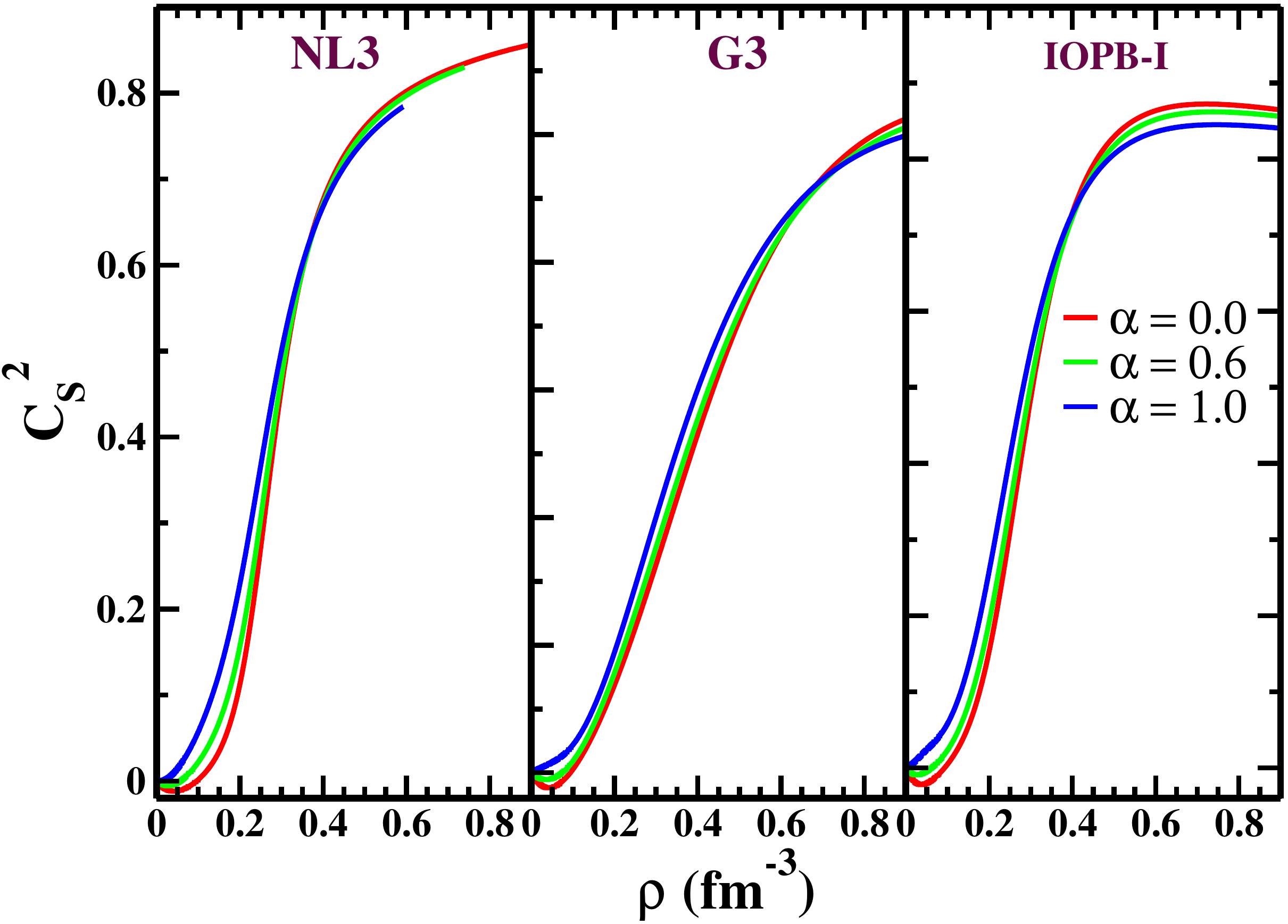}
	\caption{The variation of the speed of the sound with baryon density for NL3, G3, and IOPB-I parametrizations at different $\alpha$ with $k_f^{\rm DM}$ = 0.06 GeV.}
	\label{fig:VEL_NM_DM}
\end{figure}
%%%%%%%%%%%%

It is conjectured that the speed of the sound $C_s\leq c/\sqrt{3}$, where $C_s^2=\frac{\partial P}{\partial {\cal{E}}}$ with $C_s$ and $c$ are the speed of the sound and light, respectively. To see the causality condition for the NM case with an admixture of DM, we plot $C_s^2$ as a function of total NM density $\rho$ in Fig. \ref{fig:VEL_NM_DM} at different $\alpha$ for NL3, G3, and IOPB-I parameter sets. We find that the $C_s^2$ increases approximately up to 0.8 fm$^{-3}$, which is constant for high-density regions. It is clear from the values of $C_s^2$ that the causality remains intact for a wide range of density for all three parameter sets, as shown in Fig. \ref{fig:VEL_NM_DM}. However, the density at the core of the NS is so high that the nuclei break into quarks. In that medium, the sound speed is basically slower in comparison to the nucleonic medium. This is the main tension that the sound speed
violates the QCD conformal limit for only nucleonic cases as discussed in Refs. \cite{Bedaque_2015, McLerran_2019}. In this study, we include only nucleons. Therefore, it violates that limit in the whole region of the stars.
%%%%%%%%%%%%%%%%%%%%%%%%%%%%%%%%%%%%%%%%%%%%%%%%%%%%%%%%%%%%%%%%%%%%%%%%%%%%%%%%%%%%%%%%%%%%%%%
\subsection{Symmetry energy and its higher order coefficients for DM admixed NM}
%%%%%%%%%%%%%%%%%%%%%%%%%%%%%%%%%%%%%%%%%%%%%%%%%%%%%%%%%%%%%%%%%%%%%%%%%%%%%%%%%%%%%%%%%%%%%%%
The symmetry energy $S$ and its coefficients $L$, $K_{\rm sym}$, and $Q_{\rm sym}$ are defined in Eqs. (\ref{eq:L_sym} -- \ref{eq:Q_sym}), play a crucial role in the EOS for symmetric and asymmetric NM.
%%%%%%%%%%%%%%
\begin{figure}
	\centering
	\includegraphics[width=0.6\columnwidth]{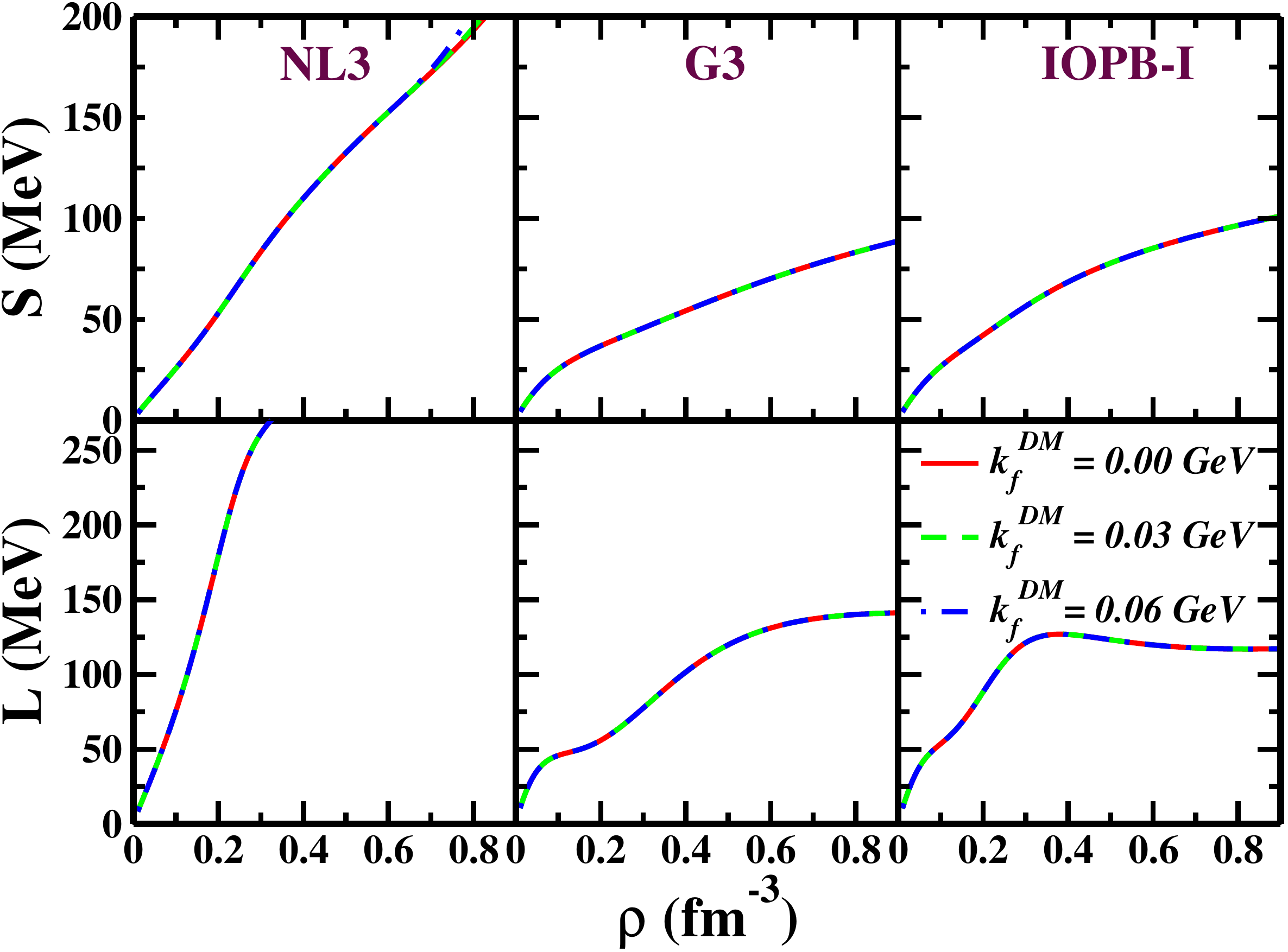}
	\caption{The symmetry energy $S$ and its slope parameter $L$ are plotted with the varying $k_f^{\rm DM}$ as a function of $\rho$. We found almost the same results (Table \ref{tab:NM_properties}) with and without DM for all the three-parameter sets.}
	\label{fig:SYMM_NM_DM}
\end{figure}
%%%%%%%%%%%%
As we have mentioned earlier, these parameters are important quantities to determine the nature of the EOS. Just after the supernova explosion, its remnants, which lead towards the formation of a NS, are in a high temperature ($\sim 200$ MeV) state \cite{Gendin_2001,Page_2004,Yakovlev_2005,Yakovlev_2010}. Soon after the formation of the NS, it starts cooling via the direct URCA processes to attain stable charge neutrality and $\beta$-equilibrium condition. The dynamic process of NS cooling is affected heavily by these NM parameters. Thus, it is very much intuitive to study these parameters more rigorously.
%%%%%%%%%%%%%%
\begin{table}[t]
	\centering
	%\captionsetup{width=1.0\textwidth}
	\caption{The NM properties such as BE/A, symmetry energy, and its derivatives, etc., are given at saturation density for three-parameters sets with $k_f^{\rm DM}$ = 0 GeV (without DM), 0.03 GeV and 0.06 GeV respectively for SNM. Similarly the effective mass ($M^\star/M_N$), incompressibility ($K$) varying with $\alpha$ = 0, 0.6, 1.0 at $\rho_0$ (not in the variations of $k_f^{\rm DM}$) in last three-rows. The empirical/experimental value for $\rho_0$, BE/A, $J$, $L$, $K_{\rm sym}$ and $K$ are also given at saturation density.}
	\renewcommand{\arraystretch}{1.3}
	\scalebox{0.7}{
		\begin{tabular}{|l|lll|lll|lll|l|}
			\hline
			\multirow{2}{*}{\begin{tabular}[c]{@{}l@{}}NM\\ Parameters\end{tabular}} & \multicolumn{3}{l|}{\hspace{1.2cm} NL3} & \multicolumn{3}{l|}{ \hspace{1.4cm}G3} & \multicolumn{3}{l|}{\hspace{1.2cm}IOPB-I}&
			\multirow{2}{*}{\begin{tabular}[c]{@{}l@{}}Empirical/\\ Experimental \end{tabular}} \\ \cline{2-10}  
			&  0.0   &  0.03     &  0.06     &   0.0    &  0.03     & 0.06      &   0.0    &  0.03  & 0.06  &          \\ \hline
			$\rho_0$ (fm$^{-3}$)        &  0.148  &  0.148    &  0.148    &   0.148   &  0.148    & 0.148     &   0.149   &  0.149    &  0.149 &  0.148 -- 0.185 \cite{Bethe_1971}   \\ 
			BE/A (MeV)                    & -16.35  &  143.95   &  1266.11  &   -16.02  &  143.28   & 1266.44   &   -16.10  &  143.09   & 1257.51  &  -15 -- -17 \cite{Bethe_1971} \\ 
			$J$ (MeV)                     &  37.43  &  38.36    &  38.36    &   31.84   &  31.62    & 31.62     &   33.30   &  34.45    &  34.45   &  30.20 -- 33.70  \cite{Danielewicz_2014} \\ 
			$L$ (MeV)                     &  118.65 & 121.44    &  121.45   &   49.31   &  49.64    & 49.64     &   63.58   &  67.16    &  67.67   &  35.00 -- 70.00 \cite{Danielewicz_2014}   \\ 
			$K_{\rm sym}$ (MeV)             &  101.34 & 101.05    &  100.32   &   -106.07 &  -110.38  & -111.10 & -37.09  &  -45.94   & -46.67   &  -174 -- -31  \cite{Zimmerman_2020}  \\ 
			$Q_{\rm sym}$ (MeV)             &  177.90 & 115.56    &  531.30   &   915.47  &  929.67   & 1345.40   &   868.45  &  927.84   &  1343.58  & ---------------  \\ \hline
			$\alpha$ =                  &  \hspace{0.2cm}0       & \hspace{0.1cm}0.6      & \hspace{0.1cm} 1.0      &   \hspace{0.2cm} 0       & \hspace{0.1cm} 0.6      &  \hspace{0.1cm}1.0      &  \hspace{0.2cm} 0       &  \hspace{0.1cm}0.6      &  \hspace{0.1cm}1.0   &   \\ \hline 
			$M^\star/M_n$                        &  0.595   & 0.596    &  0.606    &  0.699    &  0.700    &  0.704    &   0.593   &  0.599    & 0.604  & ---------------  \\ 
			$K$ (MeV)                     &  271.38  & 312.45   &  372.13   &  243.96   &  206.88   & 133.04    &   222.65  &  204.00   & 176.28  & 220 -- 260 \cite{Stone_2014}    \\ \hline
	\end{tabular}}
	\label{tab:NM_properties}
\end{table}
%%%%%%%%%%%%

The symmetry energy ($S$) and its coefficient ($L$) for the whole density range for three parameter sets NL3, G3, and IOPB-I with different $k_f^{\rm DM}$ are displayed in Fig. \ref{fig:SYMM_NM_DM}. The effect of DM on $S$ and $L$ is minimal, and it is not easy to notice in the figure. We enumerated their values at the saturation point in Table \ref{tab:NM_properties}. For example, with NL3 set, $S(\rho_0$) = $J$ = 37.43 MeV for PNM, and it increases to a value $J$ = 38.36 MeV with DM. Similarly, $L$ = 118.65 MeV without DM, $L$ = 121.44 and 121.45 MeV in the presence of DM with different momenta. For other sets, one can see their values in Table \ref{tab:NM_properties}. This is because the effect of DM does not change NM asymmetry to a significant extent. A careful inspection of Fig. \ref{fig:SYMM_NM_DM} and Table \ref{tab:NM_properties} cleared that $S$ and $L$ are force-dependent. It is maximum for NL3 and minimum for G3 sets. The experimental value of $J$ and $L$ at $\rho_0$ is between 30.2 -- 33.7 MeV and 35.0 -- 70.0 MeV, respectively. With the addition of DM, the values of $J$ and $L$ lie (except for NL3) in the experimental limit.
%%%%%%%%%%%%%%
\begin{figure}[H]
	\centering
	\includegraphics[width=0.6\columnwidth]{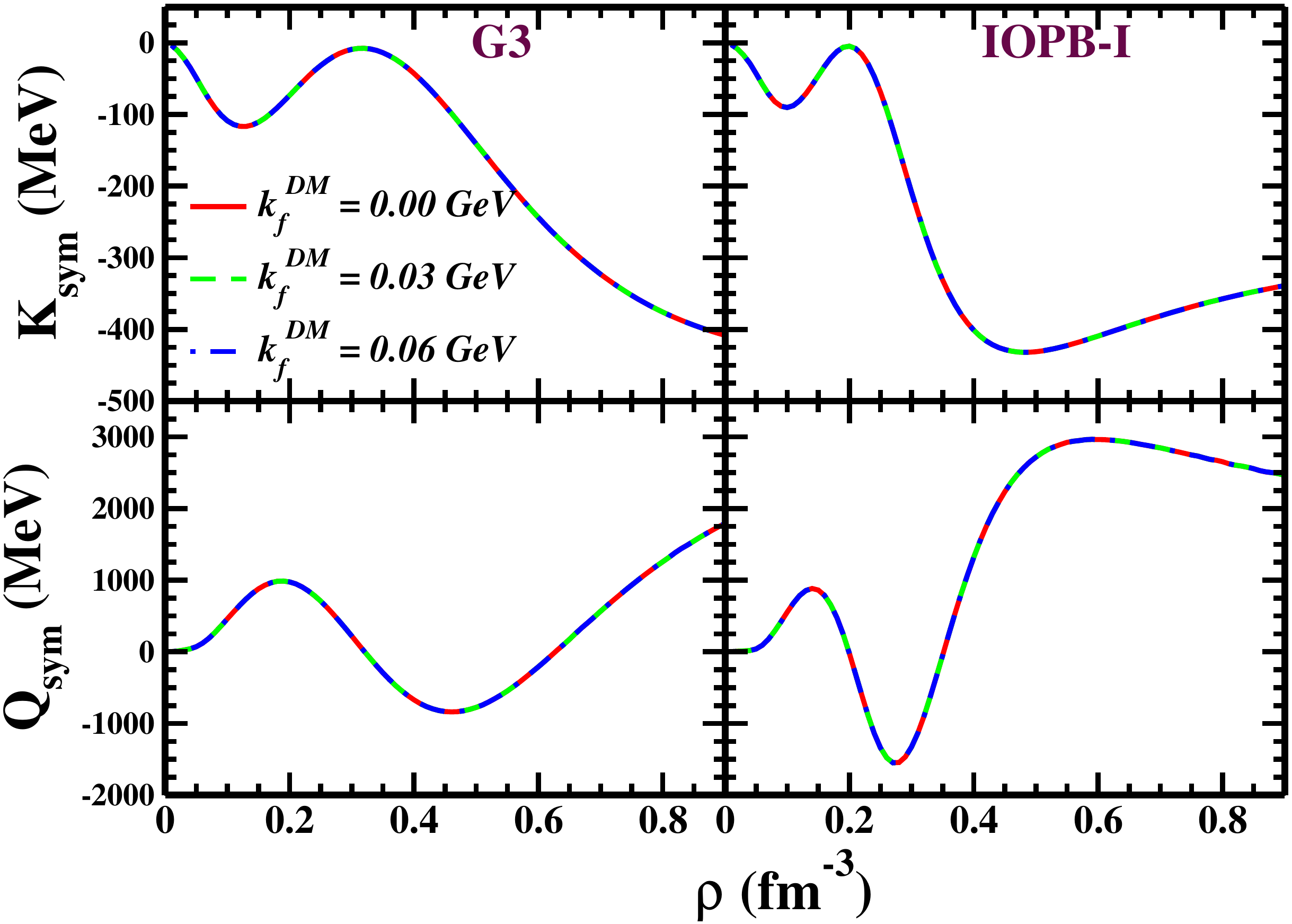}
	\caption{The variation of $K_{\rm sym}$ and $Q_{\rm sym}$ with baryon density is plotted for G3 and IOPB-I at different  $k_f^{\rm DM}$. The $K_{\rm sym}$ and $Q_{\rm sym}$ are opposite to each other both for G3 and IOPB-I.}
	\label{fig:KQSYM_NM_DM}
\end{figure} 
%%%%%%%%%%%%

The other higher-order derivatives of symmetry energy like $K_{\rm sym}$ and $Q_{\rm sym}$ are also calculated in this section. The results are displayed in Fig. \ref{fig:KQSYM_NM_DM}, and their numerical values are tabulated in Table \ref{tab:NM_properties}. The $K_{\rm sym}$ is a parameter that tells a lot not only about the surface properties of the astrophysical object (such as NS and white dwarf) but also the surface properties of finite nuclei. The whole density range of $K_{\rm sym}$ and $Q_{\rm sym}$ for G3 and IOPB-I sets are shown. The $K_{\rm sym}$ affected marginally, but the parameter $Q_{\rm sym}$ was influenced by DM (see Table \ref{tab:NM_properties}) for these values at saturation. At low density, $K_{\rm sym}$ initially decreases slightly, then it increases up to $\rho$ ($\sim$ 0.2 fm$^{-3}$) and after that decreases the value almost like an exponential function. Recently the value of $K_{\rm sym}$ is calculated by Zimmerman et al. \cite{Zimmerman_2020} with the help of NICER \cite{Miller_2019, Riley_2019, Bogdanov_2019, Bilous_2019, Raaijmakers_2019, Guillot_2019} and GW170817 data and its values lie in the range -174 to -31 MeV as given in Table \ref{tab:NM_properties}. Our $K_{\rm sym}$ values lie in this region at the saturation density.
%%%%%%%%%%%%%%%%%%%%%
\section*{Conclusion}
%%%%%%%%%%%%%%%%%%%%%
In this chapter, we calculated the DM admixed NM properties for different RMF and E-RMF sets. Since the NS is a highly asymmetric NM system, it has direct relations with the properties of nuclear matter. Therefore, in this study, we explored the effects of DM on the nuclear matter. The DM model has been constructed by assuming that they interact with nucleons by exchanging Higgs. Therefore, the system energy density is the addition of both nucleons and DM. It was observed that the EOS ($P\sim {\cal{E}}$) becomes softer with the increasing DM momentum. The energy density increases with $k_f^{\rm DM}$ without adding much to the pressure. We also found that the DM has marginal effects on the NM properties, except the EOSs, the binding energy per particle, and $Q_{\rm sym}$. This is because it has significantly less effect on the pressure of the system due to the contribution of the Higgs field in the order of $10^{-6}-10^{-8}$ MeV. One can choose other DM models by varying their mass, and their fraction to explore the DM admixed NM properties.
%%%%%%%%%%%%%%%%%%%%%%%%%%%%%%%%%% END %%%%%%%%%%%%%%%%%%%%%%%%%%%%%%%%%%%%%%%%%%%%%%%%	
%\blankpage 
%%%%%%%%%%%%%%%%%%%%%%%%%% Chapter-4 %%%%%%%%%%%%%%%%%%%%%%%%%%%%%%
%%%%%%%%%%%%%%%%%%%%%%% CHAPTER - 4 %%%%%%%%%%%%%%%%%%%%%%%%%%
\chapter{Impacts of dark matter on isolated and static/rotating neutron star properties}
\label{C4} 
%%%%%%%%%%%%%%%%%%%%%%%%%%%%%%%%%%%%%%%%%%%%%%%%%%%%%%%%%%%%%%%
This chapter deals with the static and rotating isolated NS for the DM admixed NS cases. First, we calculate the EOS for the DM admixed NS by changing the percentage. In addition to this, the speed of sound is also studied. The Tolman-Oppenheimer-Volkoff equations for the static and spherically symmetric stars are solved with RMF and E-RMF equation of states that provides the mass, radius, and gravitational profile of the star. The moment of inertia for slowly rotating NS is obtained with the Hartle-Throne approximations. For rotating NS (RNS), we use Cook, Shapiro, and Teukolsky method, which has been implemented in the RNS code developed by Stergioulas \cite{NikolaosStergioulas_1999}. We compare the magnitude of observables for different fractions of DM and constrain its percentage with various pulsars data. The Keplerian frequency is also compared with different observational data. The curvature formed by the static compact objects is also estimated with different DM content. Next, we extend our study to explore the secondary components of the GW190814 event \cite{RAbbott_2020}. There are many debates regarding the secondary component, whether it is the heaviest NS or the lightest black hole? In this study, we provide a possibility that the secondary component might be a DM admixed NS. 
%%%%%%%%%%%%%%%%%%%%%%%%%%%%%%%%%%%%%%%%%%%%%%%%%%%%%%%%%%%
\section{Mass, Radius, and the Moment of Inertia of the NS}
\label{DM_NS_mri}
%%%%%%%%%%%%%%%%%%%%%%%%%%%%%%%%%%%%%%%%%%%%%%%%%%%%%%%%%%%
The mass and radius are the important macroscopic properties of the NS. Other properties such as the moment of inertia, tidal deformability, curvatures, and oscillations directly rely on them. This section provides a formalism for static, slowly rotating, and rapidly rotating cases by solving the following Einstein equation in different conditions.
%%%%%%%%%%%%%%%%%%%%%%%%%%%%%%
\subsection{Einstein equation}
%%%%%%%%%%%%%%%%%%%%%%%%%%%%%%
The Einstein equation simply describes that matter tells space how to curve, and space tells matter how to move, and the equation is given as \cite{Einstein_1916}
%%%%%%%%%%%%%
\begin{align}
G^{\mu\nu}=R^{\mu\nu}-\frac{1}{2}g^{\mu\nu}R=8\pi T^{\mu\nu},
\label{eq:Einstein_eq}
\end{align}
%%%%%%%%%%%
where $G^{\mu\nu}, R^{\mu\nu}, R$, and $T^{\mu\nu}$ are the Einstein tensor, Riemann tensor, Ricci scalar, and stress-energy tensor, respectively. The stress-energy tensor is \cite{NKGb_1997}
%%%%%%%%%%%%%
\begin{align}
T^{\mu\nu}=({\cal E}+P)u^{\mu}u^{\nu}+Pg^{\mu\nu},
\label{eq:stress_tensor}
\end{align}
%%%%%%%%%%%%%%%%%%%%%%%%%%%
where ${\cal E}$, $P$, and $u^\mu$ are the energy density, pressure, and 4-velocity (which satisfy the condition $u^\mu u_\mu=-1$), respectively. The stress-energy tensor is directly connected to the EOS of the star. 
%%%%%%%%%%%%%%%%%%%%%%%%%%%%%%%%%%%%%%%%%%%%%
\subsection{TOV equation for the static star}
%%%%%%%%%%%%%%%%%%%%%%%%%%%%%%%%%%%%%%%%%%%%%
In the case of static, spherically symmetric stars, the metric is in the form of 
%%%%%%%%%%%%%
\begin{align}
ds^2= -e^{2\nu(r)}dt^2+e^{2\lambda(r)}dr^2+r^2d\theta^2+r^2sin^2\theta d\phi^2,
\label{eq:spherical_metric}
\end{align}
%%%%%%%%%%%%%
where $r, \theta, {\rm and}\, \phi$ are the spherical co-ordinates. The terms $\nu(r)$ and $\lambda(r)$ are the metric potentials given as \cite{NKGb_1997}
%%%%%%%%%%%%%
\begin{align}
e^{2\lambda(r)} = [1-\gamma(r)]^{-1},
\label{eq:pot_lambda}
\end{align}
%%%%%%%%%%%%%
\begin{align}
e^{2\nu(r)}=e^{-2\lambda(r)} = [1-\gamma(r)], \qquad r>R_{star}
\label{eq:pot_nu}
\end{align}
%%%%%%%%%%%
with
%%%%%%%%%%%%%
\begin{align}
\gamma(r)=\left\{
\begin{array}{l l}
\frac{2m(r)}{r}, & \quad \mbox{if $r<R_{star}$}\\\\
\frac{2M}{r}, & \quad \mbox{if $r>R_{star}$}
\end{array}
\right.
\label{eq:fun_gamma}
\end{align}
%%%%%%%%%%%
By solving the Einstein equation using the metric Eq. (\ref{eq:spherical_metric}), one can reproduce the Tolman-Oppenheimer-Volkoff (TOV) equations for the static, spherically symmetric star given by \cite{TOV1, TOV2}
%%%%%%%%%%%%%
\begin{align}
\frac{d\nu(r)}{dr}&=\frac{m(r)+4\pi r^3P(r)}{r[r-2m(r)]},
\\
\frac{dP(r)}{dr}&=-\Big[P(r)+{\cal E}(r)\Big] \frac{d\nu(r)}{dr},
\\
\frac{dm(r)}{dr}&=4\pi r^2 {\cal E}(r).
\label{eq:TOV}
\end{align}
%%%%%%%%%%%
Here, ${\cal E}(r)$ and $P(r)$ are the energy density and pressure, respectively. The enclosing mass $m(r)$ at a distance $r$ from the center of the star is obtained by solving the TOV equations with boundary conditions $r=0$ and $P=P_c$ at a fixed central density. The maximum mass ($M$) of the NS and the corresponding radius ($R$) are obtained from the coupled differential equations assuming the pressure vanishes at the surface of the star, i.e., $P(r)=0$ at $r=R$.

The solution of $\nu(r)$ must match with the exterior solution as given in Eq. (\ref{eq:pot_nu}). If $\nu(r)$ is the solution, then any constant added to it is also a solution. Hence, we have to correct the solution as follows \cite{NKGb_1997}
%%%%%%%%%%%%%
\begin{align}
    \nu(r) \rightarrow \nu(r) - \nu(R) + \frac{1}{2} \ln\left(1 - \frac{2M}{R}\right)
\end{align}
%%%%%%%%%%%
%%%%%%%%%%%%%%%%%%%%%%%%%%%%%%%%%%%%%%%%%
\subsection{Slowly rotating neutron star}
%%%%%%%%%%%%%%%%%%%%%%%%%%%%%%%%%%%%%%%%%
The moment of inertia (MI) of the slow rotation is calculated using the Hartle-Throne approximation \cite{Hartle_1967, Hartle_1968}. In this case, we assume that the NS is rotating uniformly with a frequency $\Omega$, which is significantly smaller than the Keplerian frequency $\Omega_k ( = \sqrt{M/R^3})$. The complete discussion on the MI can be found in Ref. \cite{Hartle_1973}. Here, we briefly describe some of the equations as given in the following \cite{Fattoyev_2010}. The metric for slowly, uniformly rotating NS is given as \cite{Hartle_1967}
%%%%%%%%%%%%%
\begin{align}
ds^2= -e^{2\nu}dt^2+e^{2\psi}(d\phi - \omega dt^2)+e^{2\alpha}(r^2d\theta^2+ d\phi^2),
\label{eq:metric_rotation}
\end{align}
%%%%%%%%%%%
where the $\psi$ and $\alpha$ are the metric functions due to slow rotation as the functions on $r$ and $\theta$. The 4-velocity changes due to the rotation, and it becomes 
%%%%%%%%%%%%%
\begin{align}
u^{\mu}=\frac{e^{\mu}}{\sqrt{1-v^2}}(t^{\mu}+\Omega\phi^{\mu}),
\label{eq:four_velocity}
\end{align}
%%%%%%%%%%%
where $t^\mu$ and $\phi^\mu$ are the killing vectors. $v$ is the spatial velocity represented as $(\Omega-\omega)e^{\psi-\nu}$. 

In the slow rotation approximation, the MI of the uniformly rotating, axially symmetric NS is given as \cite{Lattimer_2000, Krastev_2008, Fattoyev_2010} 
%%%%%%%%%%%%%
\begin{align}
I \approx \frac{8\pi}{3}\int_{0}^{R}({\cal E}+P)\ e^{-\nu(r)}\Big[1-\frac{2m(r)}{r}\Big]^{-1}\frac{\Bar{\omega}}{\Omega}r^4 dr,
\label{eq:slow_rot_mom}
\end{align}
%%%%%%%%%%%
where the $\Bar{\omega}$ is the dragging angular velocity for a uniformly rotating star. The $\Bar{\omega}$ satisfying the boundary conditions are 
%%%%%%%%%%%%%
\begin{align}
\Bar{\omega}(r=R)=1-\frac{2I}{R^3},\qquad \frac{d\Bar{\omega}}{dr}\bigg|_{r=0}=0 .
\label{eq:slow_mom_bc}
\end{align}
%%%%%%%%%%%
 The Keplerian frequency is defined as  \cite{Komatsu_I_1989,Komatsu_II_1989,Stergioulas_2003,Dhiman_2007,Jha_2008,Krastev_2008,Worley_2008,Haensel_2009,Koliogiannis_2020}
%%%%%%%%%%%%%
\begin{align}
\nu_k(M) \approx \chi \Big(\frac{M}{M_{\odot}}\Big)^{1/2}\Big(\frac{R}{10\ \ {\rm km}}\Big)^{-3/2},
\label{eq:kep_freq}
\end{align}
%%%%%%%%%%%
The value of $\chi$ can be obtained either by fitting or empirically, as given in Refs. \cite{Lattimer_2004, Haensel_2009}. In Ref. \cite{Haensel_2009}, it is found that the value of $\chi=1.08$ kHz is empirically for static NS. For rotating NS, we obtained the value of $\nu_k$ directly from RNS code \citep{NikolaosStergioulas_1999} and show in Fig. \ref{fig:FREQ_NS_DM}.
%%%%%%%%%%%%%%%%%%%%%%%%%%%%%%%%%%%%%%%%%%
\subsection{Rapidly rotating neutron star}
%%%%%%%%%%%%%%%%%%%%%%%%%%%%%%%%%%%%%%%%%%
Many formalisms have already been established to calculate the rapidly rotating NS properties. For details, we refer the reader to see the Ref. \cite{Stergioulas_2003} and references therein. Mainly people used the Komatsu, Eriguchi, and Hachis (KEH) \cite{Komatsu_I_1989, Komatsu_II_1989} and Cook, Shapiro, and Teukolsky (CST) improved version of the KEH scheme. These formalisms are well implemented in RNS code written by Stergioulas \cite{NikolaosStergioulas_1999}.
%%%%%%%%%%%%%%%%%%%%%%%%%%%%%%%%%%%%
\subsection{Results and Discussions} 
%%%%%%%%%%%%%%%%%%%%%%%%%%%%%%%%%%%%
%%%%%%%%%%%%%%
\begin{figure}
\centering
\includegraphics[width=0.6\columnwidth]{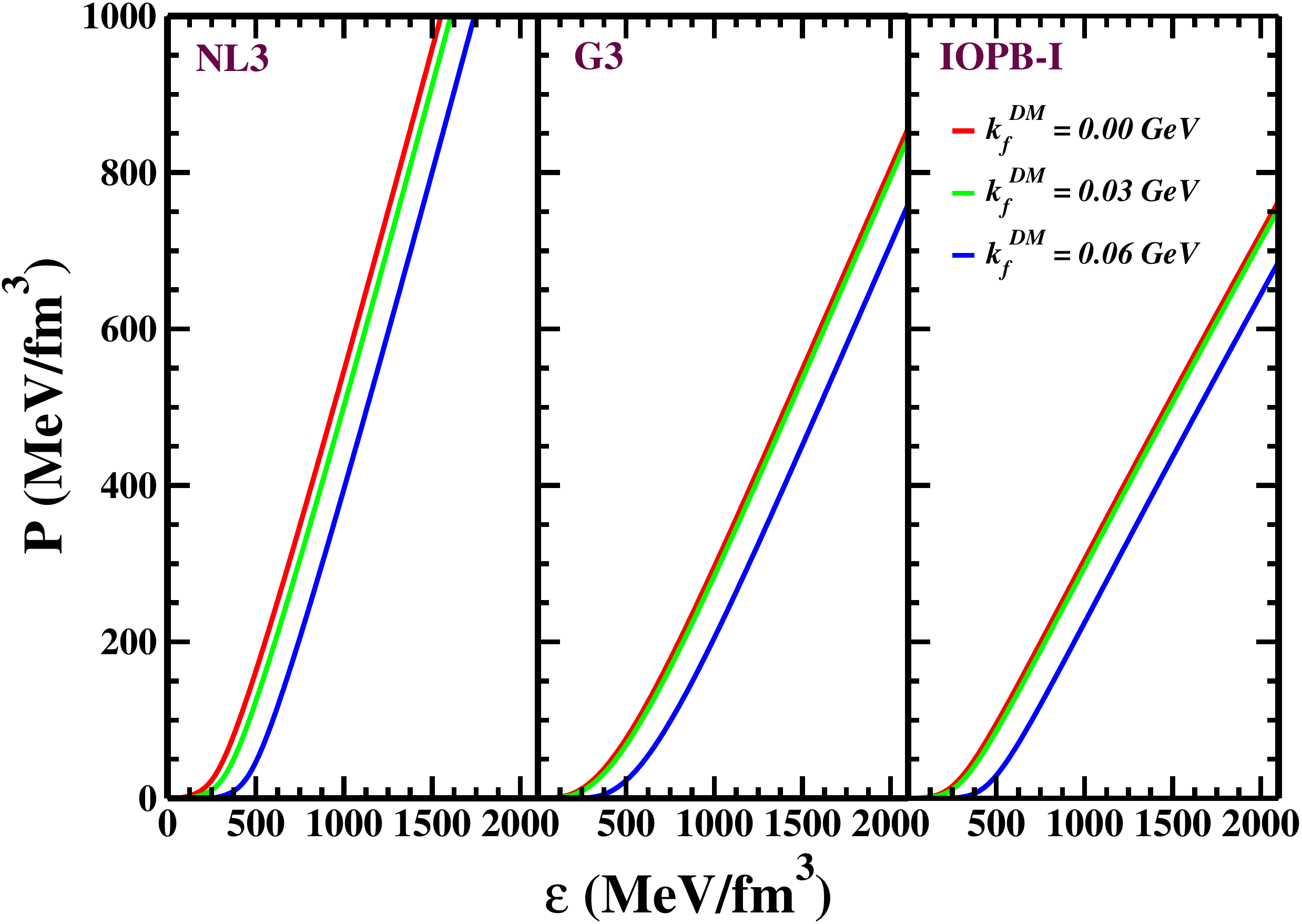}
\caption{The pressure as the function of energy density (given in Eq. \ref{eq:total_EOS_NS_DM}) with different $k_f^{\rm DM}$. NL3, G3, and IOPB are the stiffest, softest, and moderate stiff EOSs.}
\label{fig:EOS_NS_DM}
\end{figure}
%%%%%%%%%%%%%%
In this study, we take the mixed EOS, i.e., hadron matter with DM. This EOS is calculated with the RMF model by maintaining both $\beta-$equilibrium and charge neutrality conditions within the NS. Also, we vary the DM percentage inside the NS to see its effects on the NS properties such as $M$, $R$, and $I$. The EOS for NL3, G3 and IOPB-I with $k_f^{\rm DM}$= 0.0, 0.03 and 0.06 GeV are shown in Fig. \ref{fig:EOS_NS_DM}. In the previous chapter, we mentioned that the EOS is very sensitive to $k_f^{\rm DM}$ both for PNM and SNM. The same behavior is also seen for NS with DM. We find that the EOS becomes softer with $k_f^{\rm DM}$ \cite{AngLi_2012, Panotopoulos_2017, Bhat_2019, Das_2019, Quddus_2020}. 

To calculate the macroscopic/gross properties of the NS, one should use unified EOS, i.e., the EOS for both core and crust. For the core part, we use the RMF model EOSs like NL3, G3, and IOPB. For the crust part, we use Baym, Pethick, Sutherland (BPS) \cite{BPS_1971} EOS. Construction of the BPS EOS is based on the minimization of the Gibbs function and effects of the Coulomb lattice, which gives a suitable combination of the A and Z. Once we know the entire EOS, we can calculate the properties of both static and rotating NS. 

\begin{figure}
\centering
\includegraphics[width=0.6\columnwidth]{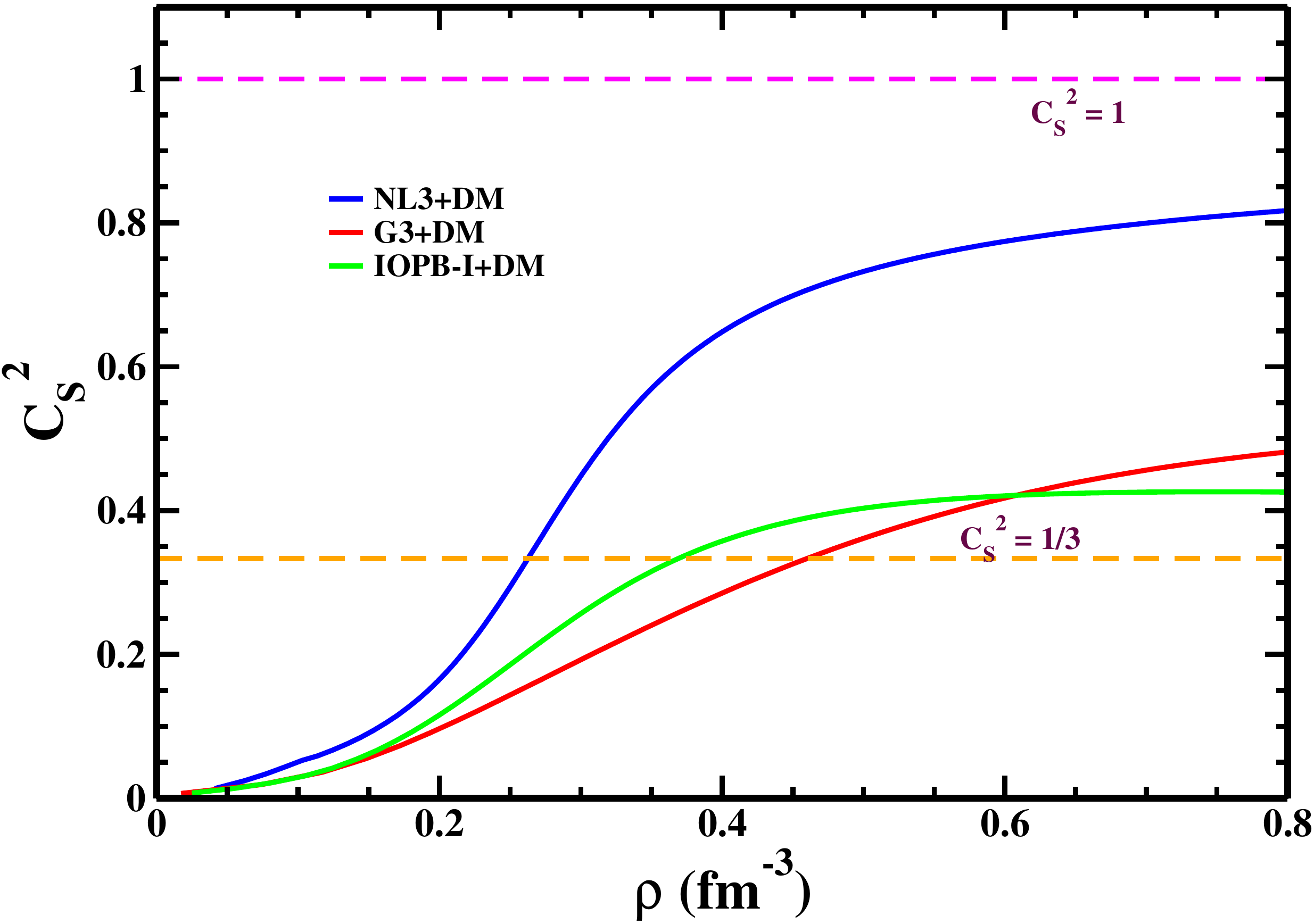}
\caption{The speed of the sound as a function of total baryon density with $k_f^{\rm DM}$ = 0.06 GeV.}
\label{fig:VEL_NS_DM}
\end{figure}

Here, we have also tested the causality condition similar to the case of nuclear matter. We have seen that both in NM and NS, the causality is not violated throughout the region, as shown in Fig. \ref{fig:VEL_NM_DM} and \ref{fig:VEL_NS_DM} respectively. The dashed horizontal line is the conjectured $C_s^2=1/3$ value given in Fig. \ref{fig:VEL_NS_DM}. The NL3 set predicts a stiff rise in $C_s^2$ compared to G3 and IOPB-I. But in all the cases $C_s^2$ is less than 1/3 for very low density region ($\rho <0.4$ fm$^{-3}$). Compared to the NM, the NS contains nucleons, electrons, and muons, which is a completely different stable system survived by the balancing force due to the attractive gravitation and the repulsive degenerate neutrons with short-range repulsive nuclear force. The causality is also well tested inside the DM admixed NS.
%%%%%%%%%%%%%%
%%%%%%%%%%%%%%
\begin{figure}
\centering
\includegraphics[width=0.6\columnwidth]{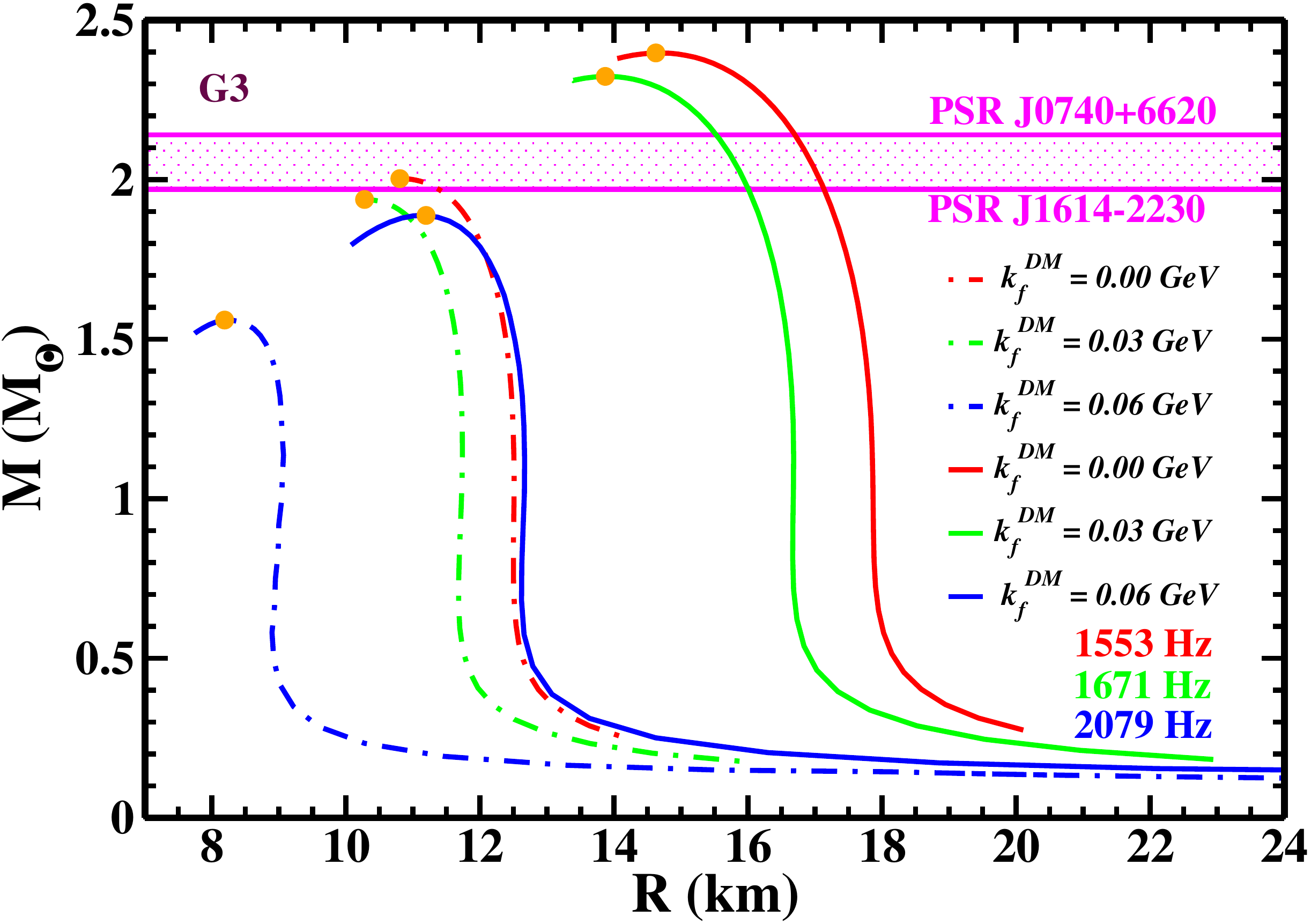}
\caption{The variation of mass as a function of the equatorial radius with different $k_f^{\rm DM}$. The orange dot shows the maximum mass of the corresponding $k_f^{\rm DM}$ for the G3 parameter set. The dotted-dashed lines represent the SNS, and the solid lines represent the RNS. The maximum rotational frequencies of NS are also shown. The recent observational constraints on NS masses \cite{Demorest_2010, Cromartie_2020} are also shown.}
\label{fig:MR_NS_DM}
\end{figure}
%%%%%%%%%%%%%

The mass-radius ($M-R$) relation is calculated by solving the TOV Eqs. \ref{eq:TOV} for G3 EOS as input with different DM percentages, and the relation for both static NS (SNS) and RNS is shown in Fig. \ref{fig:MR_NS_DM}. The precisely measured NSs masses, such as PSR J1614-2230 \cite{Demorest_2010} and PSR J0740+6620  \cite{Cromartie_2020}, are shown in the horizontal lines with pink colors. These observations suggest that the maximum mass predicted by any theoretical model should reach the limit $\sim$ 2.0 $M_{\odot}$, and this condition is satisfied in all of the EOSs, which are taken into consideration. We noticed that the increase in $k_f^{\rm DM}$ lowers the maximum mass and the equatorial radius of the SNS and RNS. For the RNS case, the maximum mass is increased by $\sim$ 20\%, and the radius increased by $\sim$ 26\% for the given EOS, which is approximately equal to the increase in mass due to the rapid rotation of the NS \cite{Stergioulas_2003, Worley_2008}.
%%%%%%%%%%%%%%
\begin{figure}
\centering
\includegraphics[width=0.6\columnwidth]{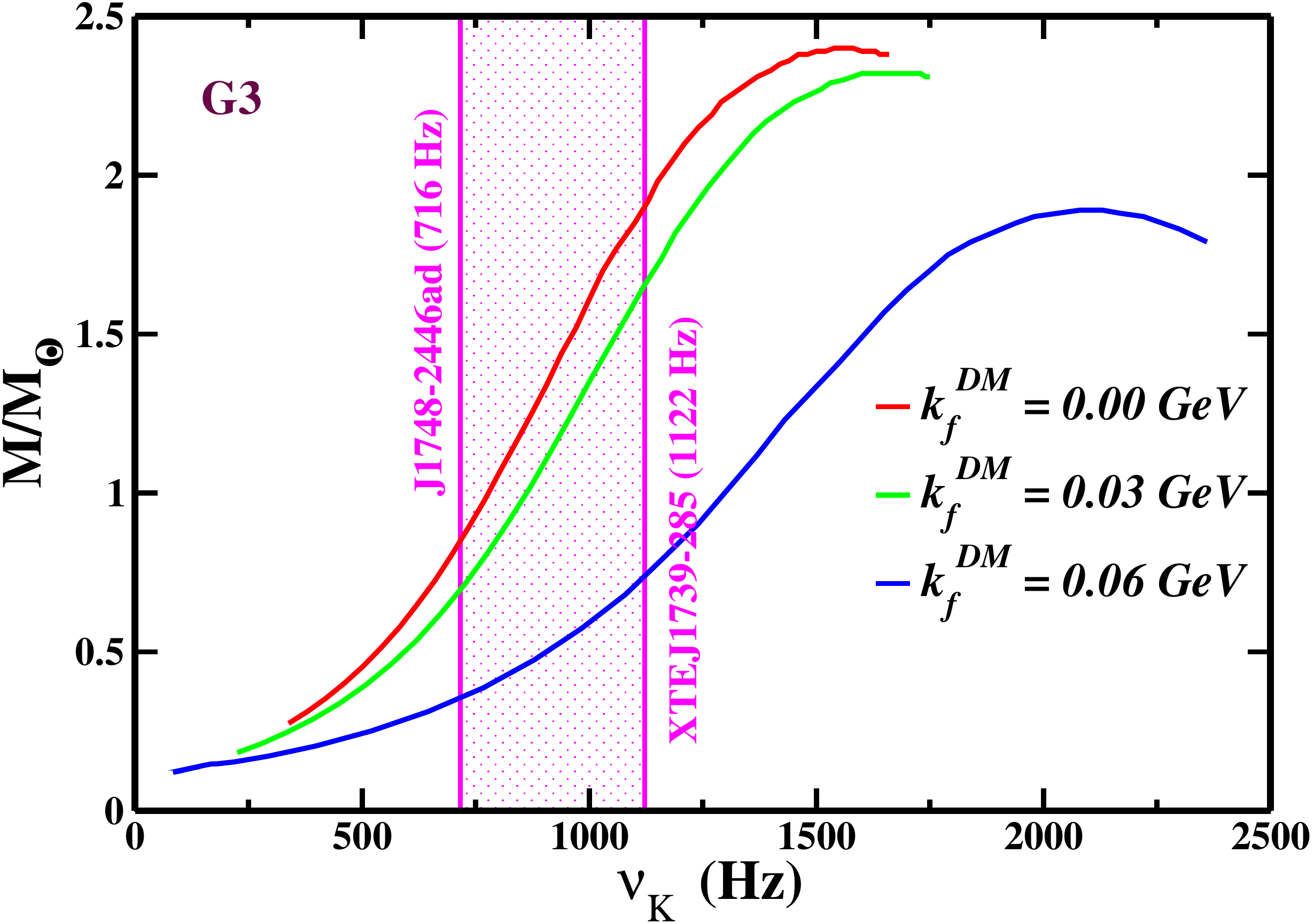}
\caption{The variation NS mass with Keplerian frequency $\nu_K$ is shown for the G3 parameter set. The two vertical magenta line represents the frequencies of the fastest NSs J1748-2446ad \cite{Hessels_2006} and XTE J1739-285 \cite{Kaaret_2007}. }
\label{fig:FREQ_NS_DM}
\end{figure}
%%%%%%%%%%%%%%
%%%%%%%%%%%%%%
\begin{figure}
\centering
\includegraphics[width=0.6\columnwidth]{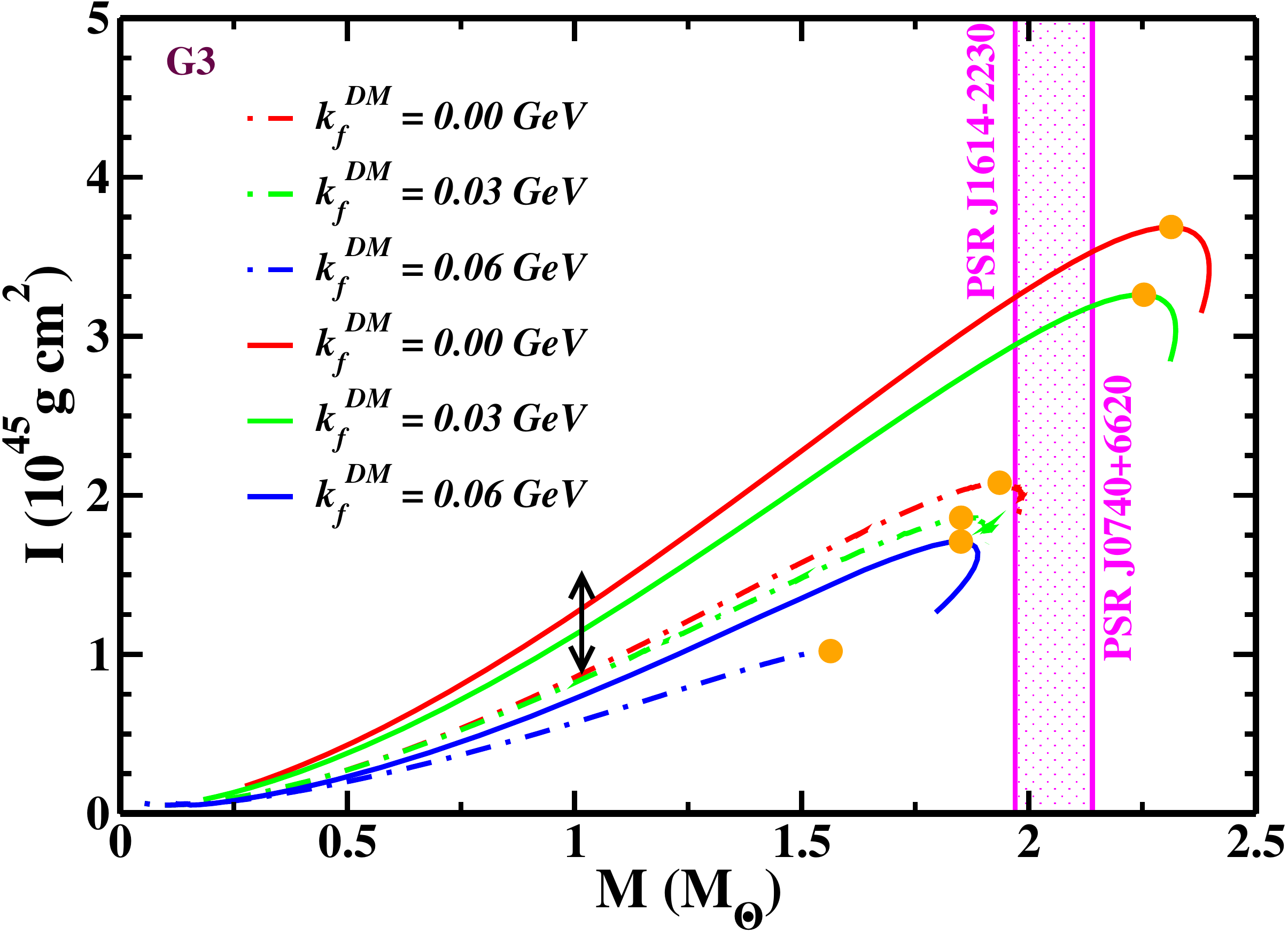}
\caption{The variation of $I$ with the mass of NS for different $k_f^{\rm DM}$ using the G3 parameter set. The orange dot shows the maximum mass of the corresponding $k_f^{\rm DM}$. The dotted-dashed line represents the SNS, and the bold line represents the RNS. The overlaid arrows represent the constraints on $I$, of PSR J0737-3039A set \cite{Landry_2018,Kumar_2019} from the analysis of GW170817 data \cite{Abbott_2017,Abbott_2018}}
\label{fig:MOM_NS_DM}
\end{figure}
%%%%%%%%%%%%%%    

Here, we examine the effect of Kepler frequency ($\nu_K$) on the mass of the NS. In Fig. \ref{fig:MR_NS_DM}, the mass of the RNS is increased due to the rapid rotation of the NS due to its high $\nu_K$, i.e., the mass of NS directly depends on the $\nu_K$. Theoretical calculations allow the value of $\nu_K$ to be more than 2000 Hz \cite{Koliogiannis_2020}, but till now, the two fastest pulsars have been detected having frequencies 716 Hz \cite{Hessels_2006} and 1122 Hz \cite{Kaaret_2007}. From Fig. \ref{fig:FREQ_NS_DM}, one can conclude that the NS mass, approximately more than 1.7 $M_{\odot}$, is rotated more than the fastest pulsar as of today. In our case, we find the $\nu_K$ are 1553, 1671, and 2079 Hz for the DM momentum 0, 0.03, and 0.06 GeV, respectively, at the maximum mass. The spherical NS leads to a deformed shape with increasing mass (or frequency) 
    
A measurement of the MI of PSR J0737-3039A is expected via optical observation of the orbital precision in the double pulsar system soon \cite{Burgay_2003}. As the MI depends on the internal structure of the NS, its measurement will constrain the unknown EOS of supra-nuclear matter, which is believed to be universal for all NS \cite{Landry_2018, Kumar_2019}. Here we show the variation of $I$  with $M (M_\odot$) in Fig. \ref{fig:MOM_NS_DM}. The change of $I$ with $M$ is almost linear for different values of $k_f^{\rm DM}$ up to the maximum mass of the star. Then there is a drop of $I$, as shown in Fig. \ref{fig:MOM_NS_DM}. Since the increase of DM momentum leads to softer EOS, hence the decrease of $I$. This is because $I$ ($\sim$ MR$^2$) is directly proportional to the mass and square of the equatorial radius of the rotating object. The moment of inertia is larger for the stiffer EOS as it predicts a larger radius and vice-versa.
%%%%%%%%%%%%%%%%%%%%%%%%%%%%%%%%%%%%%%%%
\section{Curvatures of the neutron star}
\label{form:curvature}
%%%%%%%%%%%%%%%%%%%%%%%%%%%%%%%%%%%%%%%%
To study the curvature of space-time, we have Ricci scalar and tensor, Riemann tensor, and Weyl tensor are already calculated \cite{NKGb_1997, Carroll_2019}
We describe the different curvature quantities briefly in the concept of the general theory of relativity (GR) from Ref. \cite{Carroll_2019}. The Riemann tensor is a necessary quantity to measure curvature, and it has twenty independent components in four dimensions. The Kretschmann scalar is defined as the square root of the full contraction of the Riemann tensor and has the same property as the Riemann tensor. The Ricci tensor is the contraction of the Riemann tensor, and the trace of the Ricci tensor is known as the Ricci scalar or curvature scalar. The Ricci scalar and the Ricci tensor contain all the information about the Riemann tensor leaving only the trace-less part. Weyl tensor can be formed by removing all the contraction terms of the Riemann tensor. Physically, the Ricci scalar and the Ricci tensor measure the volumetric change of the body in the presence of the tidal effect, and the Weyl tensor gives information about the shape distortion of the body. However, the Riemann tensor measures both the distortion of shape and the volumetric change of the body in the presence of the tidal force.

The robust field regime of the NS can be probed more profoundly with the help of modern observational instruments. The core is around 15 times denser than its surface, which means most of the matter is concentrated in the core, which makes the accurate measurement of the $M-R$ profile of the NS more complicated. However, the compactness and surface curvature within the star is increasing radially towards the surface \cite{Kazim_2014}. This is probably why the measurement of the maximum mass-radius is more prominent for the EOS rather than gravity. We explain the results of some recent approaches to studying the curvature of the NS. In  Ref. \cite{Kazim_2014}, it is quantified the NS's unconstrained gravity in the GR framework. They have calculated the curvature of the NS and noticed that GR is not well tested in the whole range of the star than EOS. Also, the Weyl tensor radial variation follows the power law for the large part of the star. 
Further, Xiao \textit{et al.} \cite{Xiao_2015} have taken both RMF and SHF EOSs and concluded that the symmetry energy affects the curvature of lighter NS significantly and has minimal effects on the massive NS. Moreover, to quantify the deviations from GR in the strong-field regime, a detailed understanding is required to study the properties of DM in the Universe \cite{Psaltis_2008}. Therefore, in the present work, we quantitatively investigate the NS's curvature in the presence of DM by using RMF EOSs.
%%%%%%%%%%%%%%%%%%%%%%%%%%%%%%%%%%%%%%%%%%%%%%%%%%
\subsection{Mathematical formulation for different curvatures}
%%%%%%%%%%%%%%%%%%%%%%%%%%%%%%%%%%%%%%%%%%%%%%%%%%
We adopt the mathematical form of different curvature quantities from Ref. \cite{Kazim_2014}, which measures the curvatures for both inside and outside the star. The curvatures are Ricci scalar, Ricci tensor, Riemann tensor, and Weyl tensor, which are formulated as 
\\
The Ricci scalar
%%%%%%%%%%%%%
\begin{align}
{\cal R}(r)=8\pi\left[{\cal{E}}(r) -3 P(r)\right],
\label{eq:RS}
\end{align}
%%%%%%%%%%%
the full contraction of the Ricci tensor
%%%%%%%%%%%%%
\begin{align}
{\cal J}(r) \equiv \sqrt{{\cal R}_{\mu \nu} {\cal R}^{\mu \nu}} = 8\pi \left[ {\cal{E}}^2(r) + 3P^2(r)\right]^{1/2},
\label{eq:RT}
\end{align}
%%%%%%%%%%%
the Kretschmann scalar (full contraction of the Riemann tensor)
%%%%%%%%%%%%%%
\begin{align}
{\cal{K}}(r)&\equiv\sqrt{{\cal{R}}^{\mu\nu\rho\sigma}{\cal{R}}_{\mu\nu\rho\sigma}}
\nonumber\\
&
=8\pi \left[\left(3{\cal{E}}^2(r)+3P^2(r)
+2P(r){\cal{E}}(r)\right)-\frac{128{\cal{E}}(r)m(r)}{r^3}
% \nonumber\\
% &
+\frac{48m^2(r)}{r^6}\right]^{1/2},
\label{eq:KS}
\end{align}
%%%%%%%%%%%
and the full contraction of the Weyl tensor
%%%%%%%%%%%%%
\begin{align}
{\cal W}(r) \equiv \sqrt{{\cal C}^{\mu \nu \rho \sigma }{\cal C}_{\mu \nu \rho \sigma}} = \bigg[\frac43 \left( \frac{6m(r)}{r^3} - 8\pi {\cal{E}}(r) \right)^2\bigg]^{1/2}.
\label{eq:WT}
\end{align}
%%%%%%%%%%%
Where ${\cal{E}}(r)$, $P(r)$, $m(r)$, and $r$ are the energy density, pressure, mass, and radius of the NS, respectively. At the surface $m\rightarrow M$ and $r$ becomes $R$. The Ricci tensor and Ricci scalar vanish outside the star because they depend on the ${\cal{E}}(r)$, $P(r)$, which are zero outside the star. However, there is a non-vanishing component of the Riemann tensor that does not vanish; $\tensor{{\cal R}}{^1_{010}}=-\frac{2M}{R^3}=- \xi$, even in the outside of the star \citep{Kazim_2014, Xiao_2015}. Therefore, the Riemann tensor is a more relevant quantity to measure the curvature of the stars than others. Kretschmann scalar is the square root of the full contraction of the Riemann tensor. The vacuum value for both $\cal{K}$ and $\cal{W}$ is $\frac{4\sqrt{3}M}{R^3}$ as easily can see Eqs. (\ref{eq:KS}) and (\ref{eq:WT}). Therefore, one can take $\cal K$ and $\cal W$ as two reasonable measures of the curvature within the star.
%%%%%%%%%%%%%%
\begin{figure}
\centering
\includegraphics[width=0.7\textwidth]{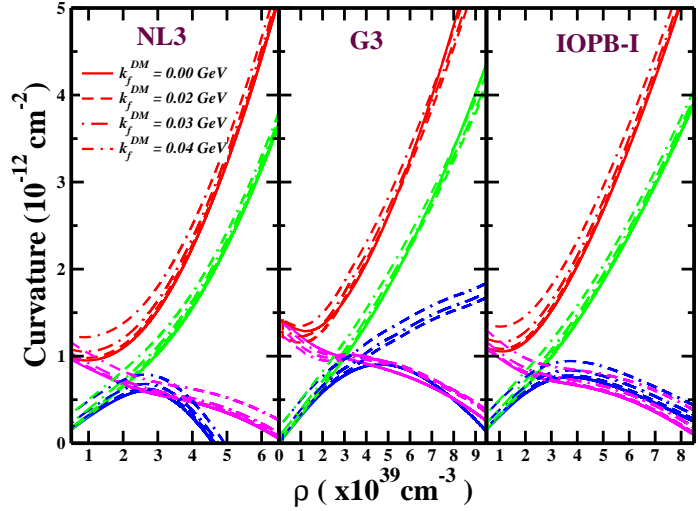}
\caption{The variation of different curvatures  $\cal{K}$ (red), $\cal{J}$ (green), $\cal{R}$ (blue) and $\cal{W}$ (magenta) with baryon density for NL3 (left), G3 (middle) and IOPB-I (right) in the presence of DM for corresponding maximum mass.}
\label{fig:density_curv}
\end{figure}
%%%%%%%%%%%%%
%%%%%%%%%%%%%%
\begin{figure}[t]
\centering
\includegraphics[width=0.7\textwidth]{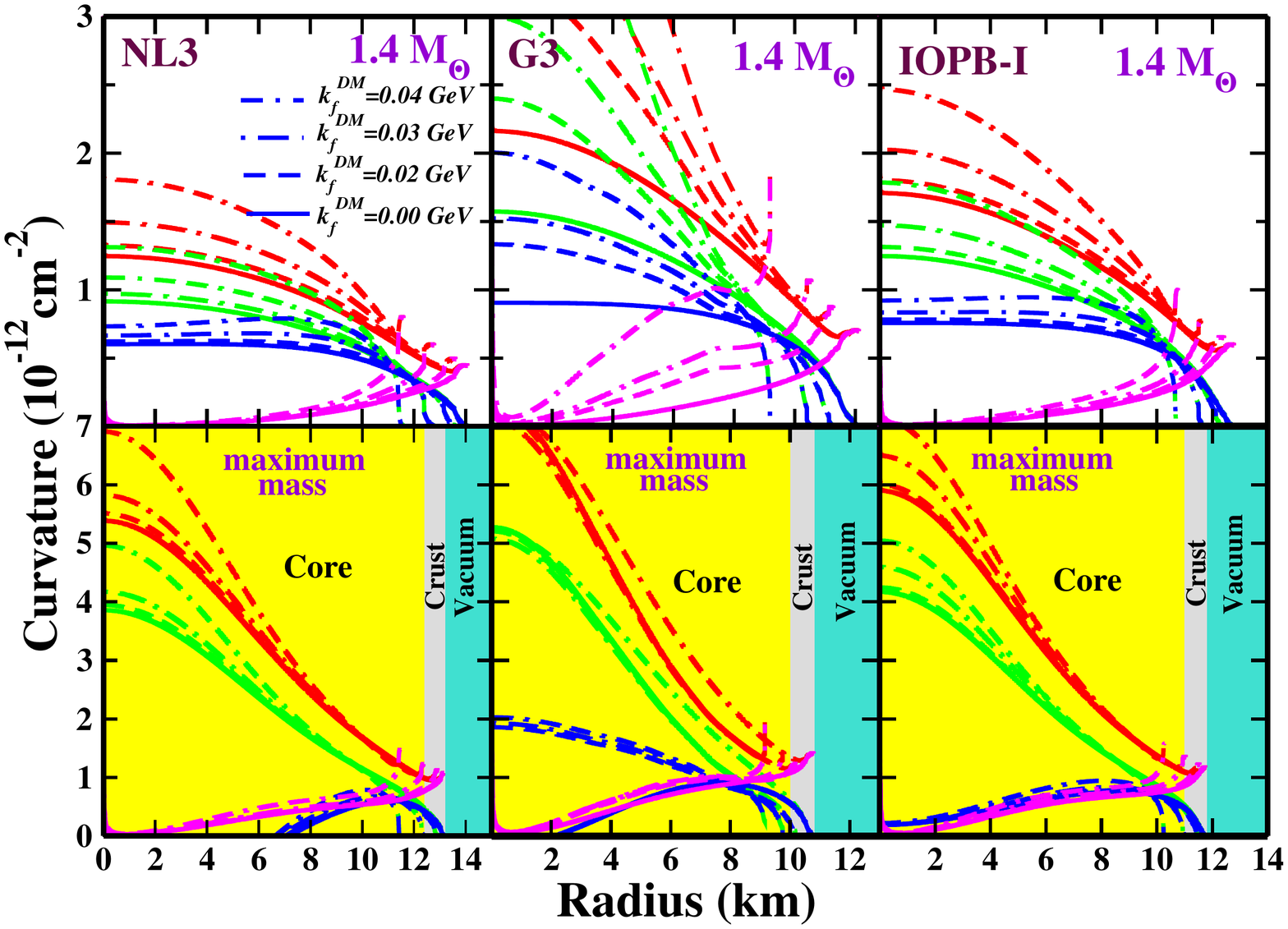}
\caption{The radial variation of all the curvatures  $\cal{K}$ (red), $\cal{J}$ (green), $\cal{R}$ (blue) and $\cal{W}$ (magenta) for NL3 (left), G3 (middle) and IOPB-I (right) in the presence of DM. The yellow, grey, and cyan color regions represent the core, the crust, and the vacuum of the NS, respectively. The curvatures $\cal R$ become negative for a maximum mass star without DM. This negative value is about 6.5 km, 2 km, and 1.2 km for NL3, G3, and IOPB-I, respectively. This is due to the limitations of EOSs obtained from the assumed parameter sets, which are not fully compatible with special relativity. The EOS violates ${\cal{E}}\geq3P$ in higher densities limit \cite{Kazim_2014}.}
\label{fig:curv_kr_all}
\end{figure}
%%%%%%%%%%%%
%%%%%%%%%%%%%%%%%%%%%%%%%%%%%%%%%%%%
\subsection{Results and Discussions}
%%%%%%%%%%%%%%%%%%%%%%%%%%%%%%%%%%%%
We calculate various curvatures like $\cal{K}, \cal{J}, \cal{R}$ and $\cal{W}$ of the NS in the presence of DM. The curvature of the NS as the function of baryon density for different DM momenta is shown in Fig. \ref{fig:density_curv}. At the low-density region (near the surface), the curvatures $\cal{J}$ and $\cal{R}$ almost vanish due to their zero vacuum value (see Eqs. \ref{eq:RS} and \ref{eq:RT}) but the curvatures $\cal{K}$ and $\cal{W}$ approach each other at the local maximum $\frac{4\sqrt{3} M}{R^3}$. In the high dense portion, the $\cal K$ and $\cal J$ procure larger curvature than others. All the curvatures increase with the $k_f^{\rm DM}$ as shown in Fig. \ref{fig:density_curv}. This is because the EOS becomes softer with $k_f^{\rm DM}$, which gives more curvature compared to a stiffer one. G3 gives large curvature compared to IOPB-I and NL3 due to its soft nature. 
    
The radial variation of the curvatures with the addition of DM is shown in Fig \ref{fig:curv_kr_all}. All the curvatures are maximum at the star's center except the full contraction Weyl tensor. However, the Ricci scalar is negative within the star (for maximum mass), as shown in Fig. \ref{fig:curv_kr_all}. At the surface of the star, ${\cal{E}}=0$ and $P=0$, so $\cal{K}$ and $\cal{W}$ are equal (see Eqs. (\ref{eq:KS}) and (\ref{eq:WT})). Near the surface of the NS, the $\cal{J}$ and $\cal{R}$ approaches zero. If we assume that the NS has uniform density, i.e., $m=\frac{4}{3}\pi r^3\rho$, then the Eq. (\ref{eq:WT}) is equal to zero. Therefore, the $\cal{W}$ tends to zero at the core. As we go from the outer crust to the surface, the density in this region is like a diffuse state so that $\cal{W}$ is maximum at the surface. Thus, it can be concluded that $\cal K$ and $\cal W$ attain different values within the star and approach each other at the crust and give identical values in a vacuum, as shown in Fig. \ref{fig:curv_kr_all}. The radial variation of curvatures follows the same trends in the presence of the DM, but the magnitude of the curvature is more. 
    
\begin{figure}
\centering
\includegraphics[width=0.7\columnwidth]{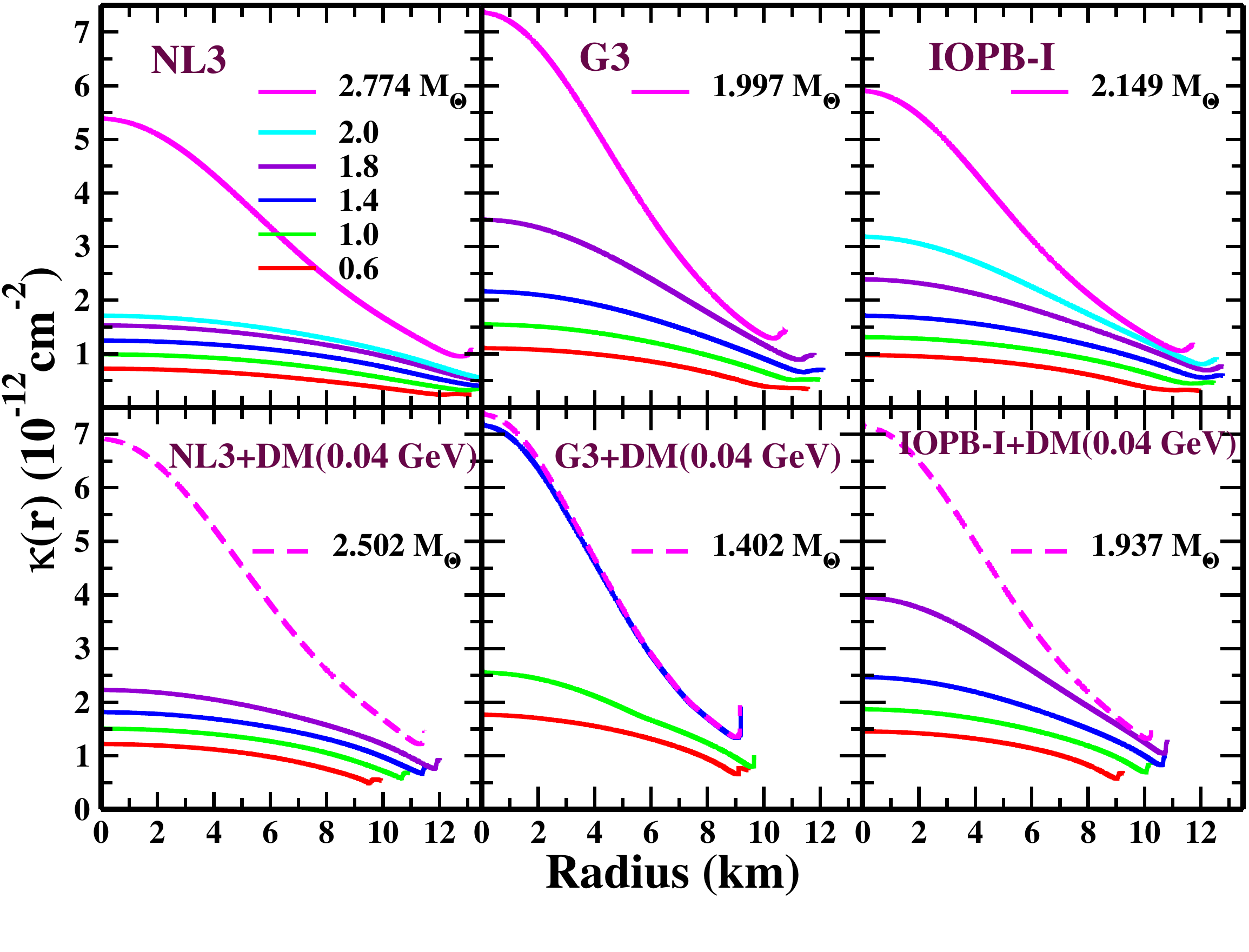}
\caption{The radial variation of ${\cal{K}}(r)$ without and with DM having momentum 0.04 GeV. The corresponding maximum mass is shown in bold line (without DM) and dashed line (with DM).}
\label{fig:curv_kr}
\end{figure}

Here, we also calculate the ${\cal{K}}(r)$ within the NS. To see the parametric dependence of the curvature with radius, we fix the DM momentum at 0.04 GeV as shown in Fig. \ref{fig:curv_kr} for different masses of the star. If one sees carefully, the change of ${\cal{K}}(r)$ is not significant up to the canonical mass as compared to the maximum mass of the star. We calculate the change of ${\cal{K}}(r)$ with and without DM is $\approx$ 33 \%, and the change increases for the maximum mass star. Hence, we inferred that the DM affects all the curvature parameters $\cal{K}$, $\cal{J}$, $\cal{R}$, and $\cal{W}$ of the NS. 
%%%%%%%%
%%%%%%%%%%%%%
\begin{table}
\centering
\caption{The central density ${\cal{E}}_c$, mass $M$, radius $R$, surface curvature ${\cal{K}}(R)$, binding energy $B/M$ of the NS are tabulated with the variation of $k_f^{\rm DM}$ both for canonical (1.4 $M_{\odot}$) and maximum mass star for NL3, G3 and IOPB-I parameter sets.}
\label{tab:curvature_table}
\renewcommand{\arraystretch}{1.6}
\scalebox{0.6}{
\begin{tabular}{|l|l|l|l|l|l|l|l|l|l|l|l|l|l|l|l|l|}
\hline
\multirow{2}{*}{\begin{tabular}[c]{@{}l@{}}$k_f^{\rm DM}$\\(GeV)\end{tabular}} &
  \multirow{2}{*}{\begin{tabular}[c]{@{}l@{}}Star\\ type\end{tabular}} &
  \multicolumn{3}{l|}{\begin{tabular}[c]{@{}l@{}} \hspace{1.5cm} ${\cal{E}}_c$\\ \hspace{0.8cm} (MeV fm$^{-3}$)\end{tabular}} &
  \multicolumn{3}{l|}{\begin{tabular}[c]{@{}l@{}}\hspace{1.6cm}$M$\\ \hspace{1.5cm}($M_\odot$)\end{tabular}} &
  \multicolumn{3}{l|}{\begin{tabular}[c]{@{}l@{}}\hspace{1.6cm}R\\ \hspace{1.5cm}(km)\end{tabular}} &
  \multicolumn{3}{l|}{\begin{tabular}[c]{@{}l@{}}\hspace{1.5cm}${\cal{K}}(R)$\\ \hspace{1.5cm}($10^{14}\cal{K}_{\odot}$)\end{tabular}} &
  \multicolumn{3}{l|}{\hspace{1.5cm}$B/M$} \\ \cline{3-17} 
                      &      & NL3 & G3 & IOPB-I & NL3 & G3 & IOPB-I & NL3 & G3 & IOPB-I & NL3 & G3 & IOPB-I & NL3 & G3 & IOPB-I \\ \hline
\multirow{2}{*}{0.00} & Cano &270     & 460   & 366       & 1.400    &1.400    &1.400        & 14.08    &12.11    &12.78        &1.477     &2.320    &1.977        &-0.084     &-0.098    &-0.092        \\ \cline{2-17} 
                      & Max  &870     &1340    &1100        &2.774     &1.997    &2.149        &13.16     &10.78       &11.76     &3.584   &4.695        &3.894     &-0.207    &-0.162 & -0.165      \\ \hline
\multirow{2}{*}{0.02} & Cano &286     &482    &385        & 1.400    &1.400    &1.400        & 13.63    & 11.77   &12.42        & 1.626    &2.534    &2.153        &-0.038     &-0.066    &-0.057        \\ \cline{2-17} 
                      & Max  &890     &1400    &1120        &2.734     &1.974    &2.118        &12.91     &10.55    & 11.54       &3.741     &4.957    &4.061        &-0.178     &-0.141    &-0.139        \\ \hline
\multirow{2}{*}{0.03} & Cano & 320    &530    &430        &1.400     &1.400    &1.400        &12.78     &11.09    &11.75        &1.976     &3.034    &2.546        &0.045     &-0.006    &0.016        \\ \cline{2-17} 
                      & Max  &940     &1480    &1190        &2.646     &1.923    &2.050        &12.39     &10.13    & 11.06       &4.097     &5.468    &4.492        &-0.116     &-0.096    &-0.009        \\ \hline
\multirow{2}{*}{0.04} & Cano &383     &640    &518        &1.400     &1.400    &1.400        &11.60     &10.27    &10.76        &2.638     &3.839    &3.317        &0.016     &0.076    & 0.105       \\ \cline{2-17} 
                      & Max  &1100     &1600    &1390        &2.502     &1.837    &1.937        &11.46     &9.53    &10.23        &4.093     &6.274    &5.341        &-0.023     &-0.026    & -0.002       \\ \hline
\end{tabular}
}
\end{table}
%%%%%%%%%%%%%%
\begin{figure}
\centering
\includegraphics[width=0.7\columnwidth]{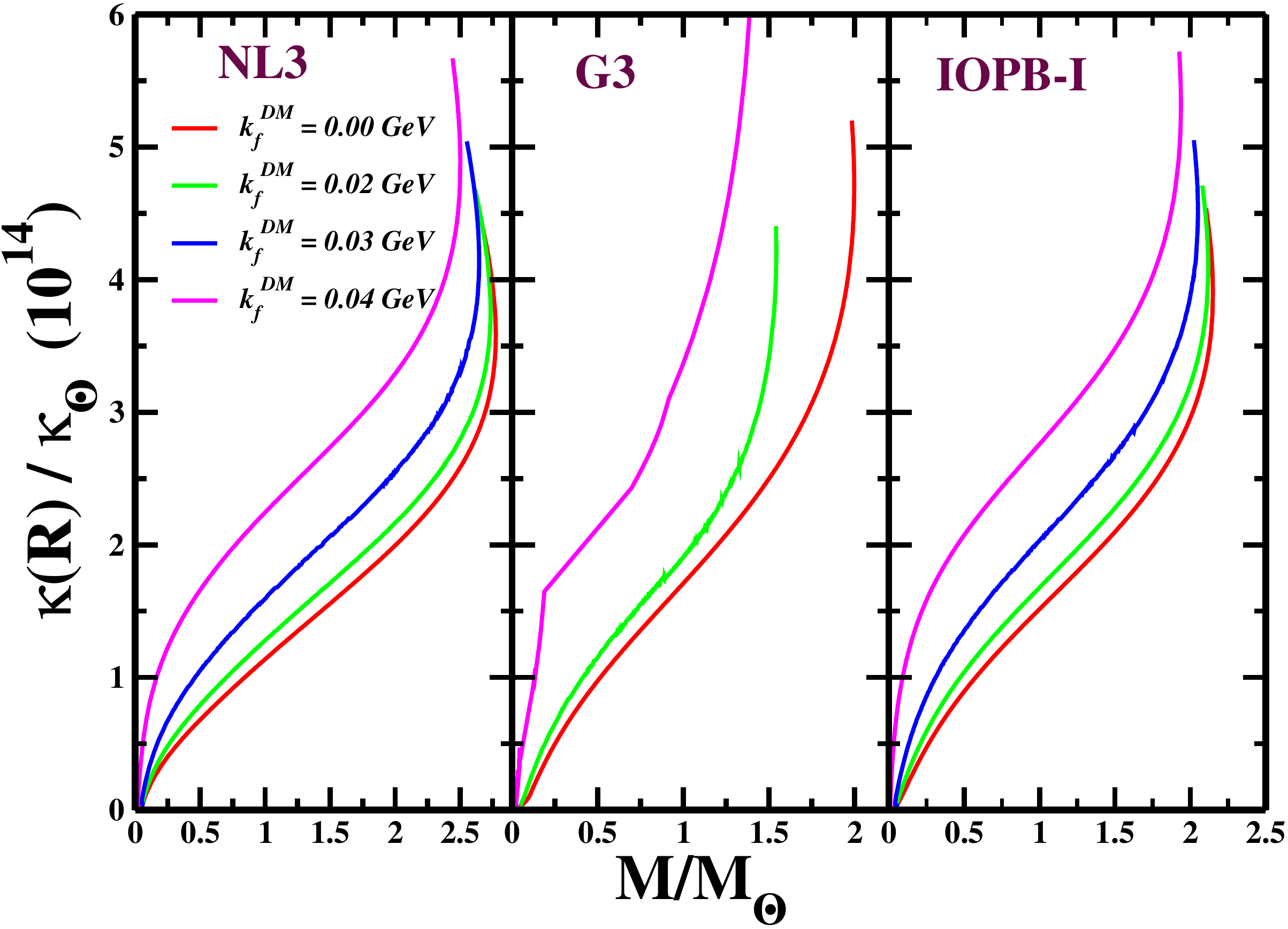}
\caption{The ratio of the surface curvature of NS and the Sun with the variation of NS mass with DM for NL3, G3, and IOPB-I.}
\label{fig:surf_curv}
\end{figure}
%%%%%%%%%
   
The curvature at the surface of the NS is more prominent to quantify the space-time wrap in the Universe. The variation of ${\cal{K}}(R)$/$\cal{K}_{\odot}$ with the mass of the NS is shown in Fig. \ref{fig:surf_curv}. In our calculations, we find that the surface curvature of NS with respect to the Sun for NL3, G3, and IOPB-I parameter sets are 3.584, 4.695, and 3.894, respectively, without including the DM, and the values are 4.093, 6.478 and 5.341 with DM momentum (0.04 GeV). The surface curvature of the Sun  ($\cal{K}_{\odot}$) is $3.06\times 10^{-27}$~cm$^{-2}$  \cite{Kazim_2014}. The comparison of our results with the Sun gives the  ${\cal{K}}(R)$/$\cal{K}_{\odot}$ $\approx$ 10$^{14}$. The ratio ${\cal{K}}(R)$/$\cal{K}_{\odot}$ increases with the inclusion of DM. The G3 parameter set provides a softer EOS than IOPB-I, which indicates that the softer EOS facilitates us with larger surface curvature. The numerical values for both the canonical and maximum mass star are given in Table \ref{tab:curvature_table}. If the DM density is very high inside the NS, then the EOS becomes softer, which affects the curvatures significantly at the surface. More curvature means the space-time curve is more, depending on the amount of DM percentage inside the NS.
    
From the above discussions, we summarized that the prediction of the strength of gravity within the NS is $\sim 10^{15}$ times more than the Sun. It increases a few times with the addition of DM. In our calculations, the curvature $\cal K$ within the star decreases a few times towards the crust of the star, and that order is $\sim 15$ times than ${\cal{K}}_\odot$ without DM (shown in Fig. \ref{fig:curv_kr_all}). The $\cal K$ increases with the increase of DM density. The value of $\cal W$ is zero at the core and almost radially increases towards the crust. On the other hand, $\cal K$ is maximum at the core, which comes mainly from the unconstrained parts of the NS. However, the $\cal W$ has maximum value almost at the crust where GR plays a significant role. That means GR breaks down in the strong-field gravity while it retains its nobleness at the surface. From this analysis, one can say that GR is not well tested on the whole part of the NS as the EOS.
    
\begin{figure}
\centering
\includegraphics[width=0.7\textwidth]{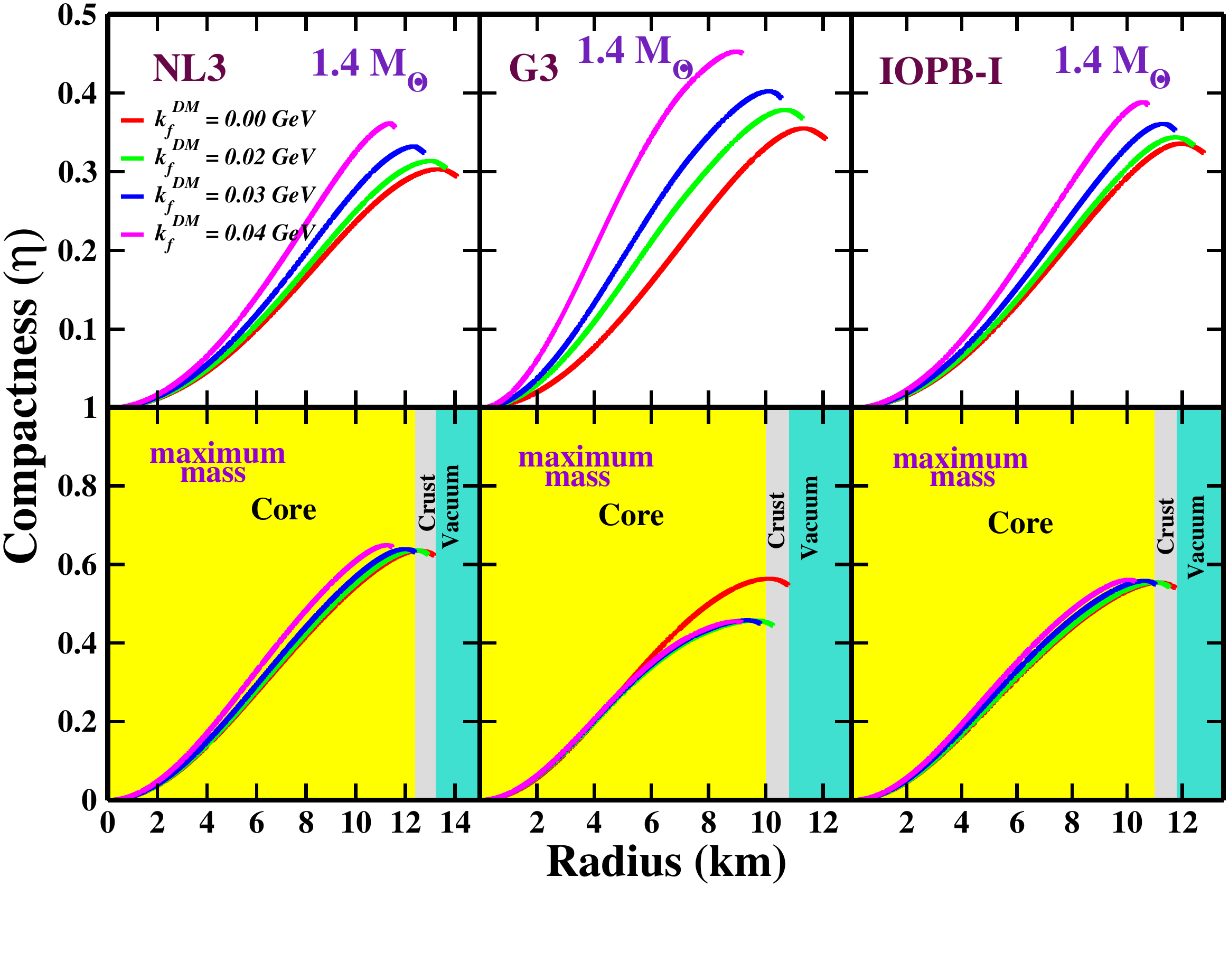}
\caption{The radial variation of compactness ($\eta$) for NL3 (left), G3 (middle), and IOPB-I (right) in the presence of DM.}
\label{fig:eta_curv}
\end{figure}

The compactness ($\eta=\frac{m}{r}$) measures the degree of the denseness of a star. The NS has a larger mass and smaller radius than the Sun, so its compactness is $10^5$ times larger than our Sun. Therefore to study the compactness of the NS, we plot the radial variation of the compactness of the NS in the presence of the DM, which is depicted in Fig. \ref{fig:eta_curv}. The star's compactness increases with the DM momentum for different parameter sets, shown in Fig. \ref{fig:eta_curv}. It has a large magnitude for the maximum mass NS than the canonical star. With the increase of the DM percentage, the EOS becomes softer and has less compactness than the stiff EOS. The compactness is maximum at the surface of the star.
%%%%%%%
%%%%%%%%%

The gravitational binding energy ($B$) is defined as the difference between the gravitational mass ($M$) and  baryonic mass ($M_B$) of the NS;  $B=M-M_B$, where $M$ is calculated as \citep{NKGb_1997, Xiao_2015}
%%%%%%%%%%%%%%%%
\begin{equation}
M=\int_{0}^{R} dr \  4\pi r^2 {\cal E}(r),
\label{eq:GM_mass}
\end{equation}
%%%%%%%%%%%%%%
and $M_B=Nm_b$, where $m_b$ is the mass of baryons (931.5 MeV) and $N$ is the number of baryons calculated by integrating over the whole volume in the Schwarzchild limit as 
%%%%%%%%%%%%%%%%
\begin{eqnarray}
N=\int_{0}^{R} dr \ 4\pi r^2 \Big[1-\frac{2m(r)}{r}\Big]^{-1/2}.
\label{eq:number_nucleons}
\end{eqnarray}
%%%%%%%%%%%%%%%
The $N$ is found to be $\approx10^{57}$ same as given in the Ref. \cite{NKGb_1997}. In our work, binding energy $B$ originally corresponds to $B/M$, which is more convenient for comparison purposes. The binding energy per particle of the symmetric NM is $\approx$ -16 MeV, i.e., it needs 16 MeV to unbound the system. For the pure neutron matter (PNM) system, it is positive \cite{Serot_1986}. That means the PNM system is already unstable. It is well acknowledged that the nuclear force is state-dependent, and the nucleon-nucleon interaction is divided into three categories- singlet-singlet, triplet-triplet, and singlet-triplet. The singlet-singlet and the triplet-triplet interaction are repulsive in nature, while the singlet-triplet interaction is attractive \cite{Patra_1992,Satpathy_2004,Kaur_2020}. Due to the excess number of neutrons, the repulsive part adds instability to NS. However, its enormous gravitational force balances the repulsive nuclear force. Thus for the whole NS, then $B$ is negative.
    
\begin{figure}
\centering
\includegraphics[width=0.7\columnwidth]{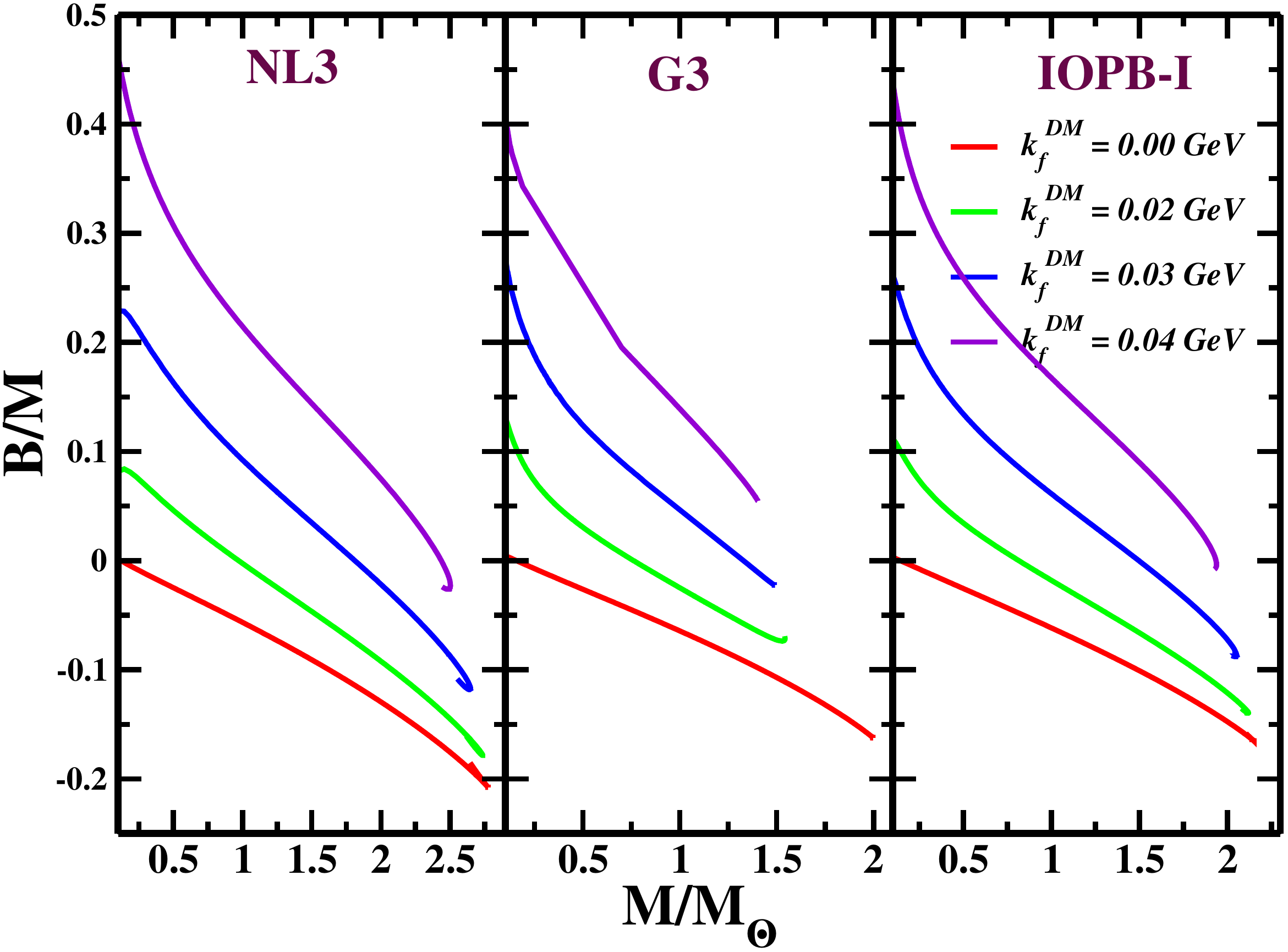}
\caption{The variation of $B/M$ with the M/$M_{\odot}$ of the NS with and without DM.}
\label{fig:BE_curv}
\end{figure}

With the addition of DM inside the NS, the $B$ steps up toward positive, which means it will be unstable. However, the instability of NS depends on the DM percentage. The variation of $B/M$ with $k_f^{\rm DM}$ is depicted in Fig. \ref{fig:BE_curv} for assumed parameter sets. The numerical values are given in Table \ref{tab:curvature_table}. A careful inspection of Table \ref{tab:curvature_table} shows that up to 0.02 GeV, the $B/M$ of the canonical and maximum mass NS are negative. It indicates that both the canonical and maximum mass NS are bound systems with this amount of DM. However, if we increase the DM momentum, the canonical NS system becomes unbound with positive $B$. For example, with a DM momentum of 0.04 GeV, the binding energy for the canonical star becomes positive for different parameter sets. However, the maximum mass NS still shows a bound system with negative binding energy. From this, we conclude that one can constrain the DM percentage inside the NS. If the DM contained is more than the canonical star, it forms a mini black hole at the core and destroys the NS \cite{Goldman_1989, De_Lavallaz_2010, Kouvaris_2011, Kouvaris_2012}. The cooling of NS is also faster with the increasing DM mass \cite{Ding_2019, Bhat_2019}. That means the positive $B$ may have a relation with the cooling properties of the NS; in other words, it may accelerate the Urca process. 

%%%%%%%%%%%%
%%%%%%%%%%%%
%%%%%%%%%%%%%%%%%%%%%%%%%%%%%%%%%%%%%%%%%%%%%%%%%%%
\section{The secondary component in GW190814 event}
%%%%%%%%%%%%%%%%%%%%%%%%%%%%%%%%%%%%%%%%%%%%%%%%%%%
The LIGO and Virgo collaboration detected the most enigmatic compact binary coalescence event (GW190814) involving a  black hole and a compact object with a mass range of $22.2-24.3$ and $2.50-2.67 \ M_\odot$ respectively \cite{RAbbott_2020}. Since the electromagnetic counterpart has not been found and has no measurable signature of tidal deformations from the GW, the secondary component might be the lightest black hole or the heaviest NS. Therefore, it is a big challenge to explain the nature of such a compact object. There is a lot of debate about understanding the mystery of the secondary component of the GW190814 event \cite{Most_2020, Vattis_2020, Tews_2021, ZhangAAS_2020, Lim_2021, Godzieba_2021, Huang_2020, Tan_2020, Fattoyev_2020, Roupas_2021, Biswas_2021}. The analysis of GW190814 data implies the possible nature of such a compact object as a NS only when (i) the EOS is very stiff \cite{Huang_2020} or (ii) it is a rapidly rotating compact object below the mass shedding frequency \cite{ZhangAAS_2020, Biswas_2021}. Thus, one can consider either of the two facts, which can produce the mass of the secondary component of GW190814 around 2.50 $M_\odot$.
	
The binary NS merger event GW170817 \cite{Abbott_2017} has evoked to put an upper bound on the maximum mass of nonrotating NS. By combining the total binary mass of GW170817 inferred from GW signal with electromagnetic observations, an upper limit of $M_{{\rm max}}\leq2.17 \ M_\odot$ is predicted by Margalit {\it et al.} \cite{Margalit_2017}. Rezzolla {\it et al.} have put the upper bound by combining the GW observations and quasi-universal relations as $M_{{\rm max}}\leq 2.16_{-0.15}^{+0.17}\ M_\odot$ \cite{Rezzolla_2018}. Further analysis by employing energy and momentum conservation laws and the numerical-relativity simulations show that the maximum mass of a cold NS is bound to be less than $2.3\ M_\odot$ \cite{Shibata_2019}. Also, various massive pulsar discoveries constrained the EOSs of the supra-nuclear matter inside the core of the NS \cite{Demorest_2010, Antoniadis_2013, Cromartie_2020}. These observational data also suggest strong constraints on the maximum mass of the slowly rotating NS with a lower bound of $\sim 2\ M_\odot$, which discarded many EOSs. Recently, the NICER data also imposed stringent constraints on the mass and radius of a canonical star from the analysis of PSR J0030+0451 data \cite{Miller_2019, Riley_2019, Raaijmakers_2019}. However, one cannot exclude the existence of a super-massive NS as the secondary component of GW190814. 
	
Many energy density functionals have been formulated to study the finite nuclei, the nuclear matter, and the NS properties, out of which only a few EOSs can reproduce the properties consistently with empirical/experimental data \cite{Dutra_2012, Dutra_2014}. To achieve the mass of the secondary component of the GW190814 event, one has to include stiff EOS, which may or may not satisfy different constraints such as flow \cite{Danielewicz_2002}, GW170817 \cite{Abbott_2017, Abbott_2018}, x-ray \cite{Antoniadis_2013, Cromartie_2020}, and NICER data \cite{Miller_2019, Riley_2019}. For this, we search for stiff RMF EOSs and find that the linear Walecka models (L-RMF) \cite{Serot_1986}, the standard nonlinear $\sigma$-self-coupling  interactions (NL-RMF) \cite{Boguta_1977} and the density-dependent RMF (DD-RMF) \cite{Huang_2020, Rather_2021} sets generally reproduce the maximum  mass of the NS more than 2.50 $M_\odot$. The BigApple force is designed to reproduce the maximum mass of the NS more than 2.50 $M_\odot$ \cite{Fattoyev_2020, DasBig_2021}. Although the L-RMF model predicts larger mass, it fails to reproduce the finite nuclei and nuclear matter properties. The recent work of Huang {\it et al.} indicates the possibility of the secondary object in the GW190814 event as pure hadronic NS with DD-RMF parameter sets  \cite{Huang_2020}. In the DD-RMF family, the DD-LZ1 and DD-MEX forces reproduce the NS masses as 2.554 and 2.556 $M_\odot$, respectively. Thus, the DD-RMF sets could only predict the secondary component's lower limit mass in the GW190814 event. Hence, it ruled out the possibility of the hybrid star or any admixture of exotic components, such as hyperons and DM inside the NS. Therefore, the L-RMF and DD-RMF forces are not very useful for the investigation.
%%%%
\begin{figure}
\centering
\includegraphics[width=0.8\textwidth]{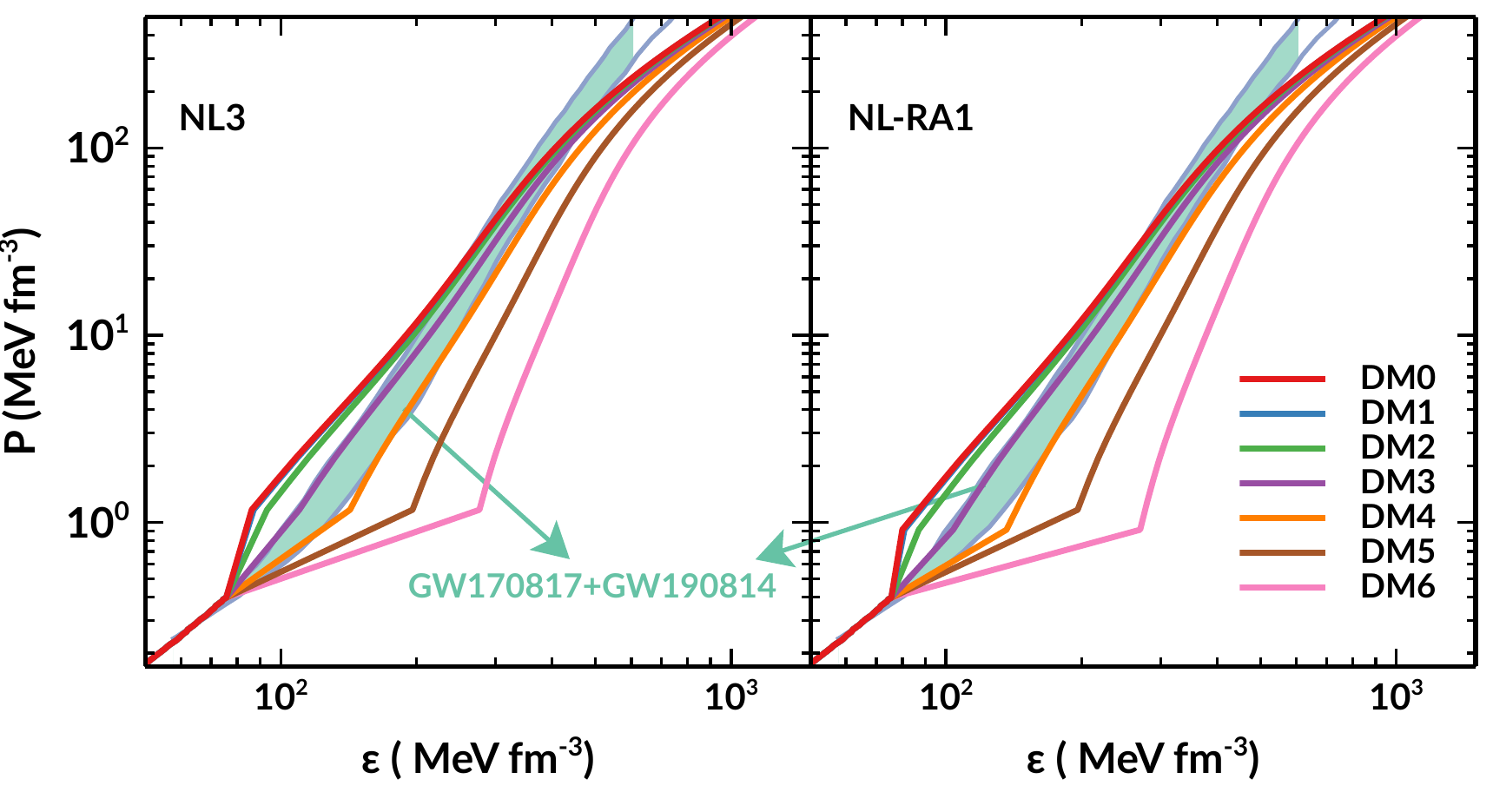}
\caption{The EOS are shown for NL3 and NL-RA1 with DM Fermi momenta 0--0.06 GeV. The joint constraints from the gravitational wave data GW170817 and GW190814 in the shaded region are adopted from Ref. \cite{RAbbott_2020}.}
\label{fig:EOS_dm_GW190814}
\end{figure}
%%%%
	
In the present study, we consider the NL-RMF parameter sets as our primary input to achieve the mass of the secondary component. Some of these forces are  NL3 \cite{Lalazissis_1997}, NL3* \cite{Lalazissis_2009}, NL1 \cite{Reinhard_1986}, NL-SH \cite{Sharma_1993}, NL3-II \cite{Lalazissis_1997} and NL-RA1 \cite{Rashdan_2001}, which are generally considered to be stiff EOSs. The maximum masses of all these parametrizations fall in the $2.7-2.8$ $M_\odot$, which are slightly more massive than the secondary component of the GW190814 event. Therefore, we have two possible scenarios to reduce the maximum mass and radius of the NS (i) with the addition of hyperons or (ii) with the addition of DM. With the addition of exotic particles, the EOS becomes softer, which reduces the mass, radius, and tidal deformability of the NS \cite{Weissenborn_2012, Biswal_2019, Biswalaip_2019, Das_2019, Quddus_2020, Das_2020, DasBig_2021}. However, in the case of hyperons inside the NS, the predicted canonical radius for NL-RMF type sets was found to be $\sim15$ km, which does not lie in the range given by the NICER ($11.96-14.26$ km) \cite{Miller_2019} and GW190814 ($12.2-13.7$ km) \cite{RAbbott_2020} predictions. Therefore, we exclude the hyperons in this case and involve the DM as an extra candidate inside the NS.
%%%%%%%%%%%%%%%%%
\begin{figure}[ht]
\centering
\includegraphics[width=0.5\textwidth]{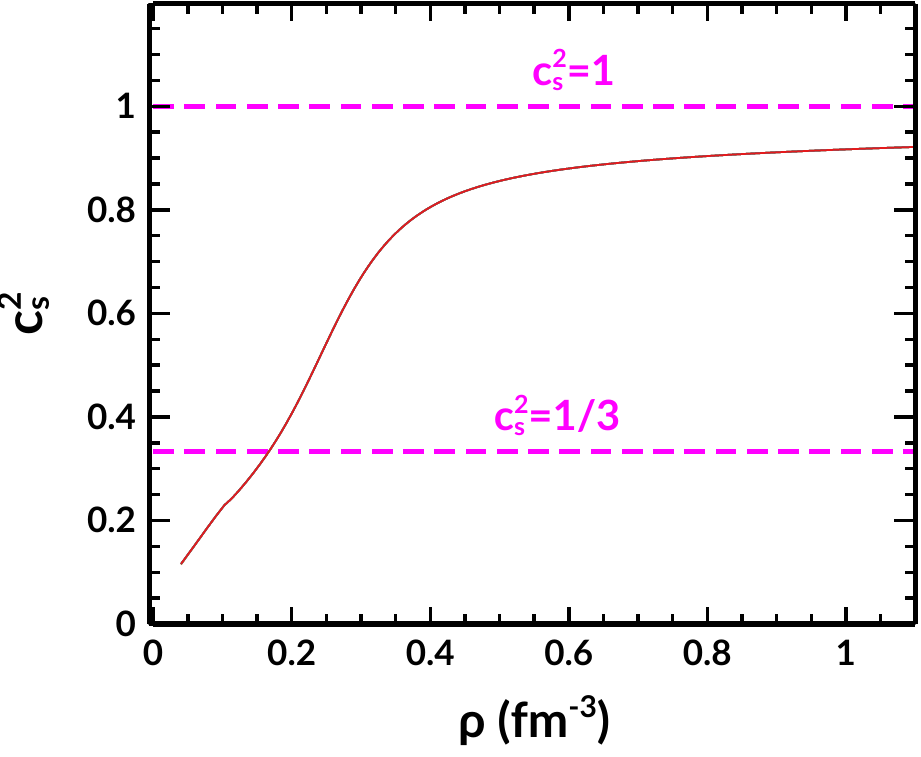}
\caption{The speed of sound for NL3 parameter set with DM Fermi momenta $0-0.06$ GeV. The QCD conformal limit ($c_s^2=1/3$) and maximum speed limit ($c_s^2=1$) are shown with a dashed magenta line.}
\label{fig:causa_dm_GW190814}
\end{figure}
%%%%%%%%%%%%%%
\begin{figure}
\centering
\includegraphics[width=0.8\textwidth]{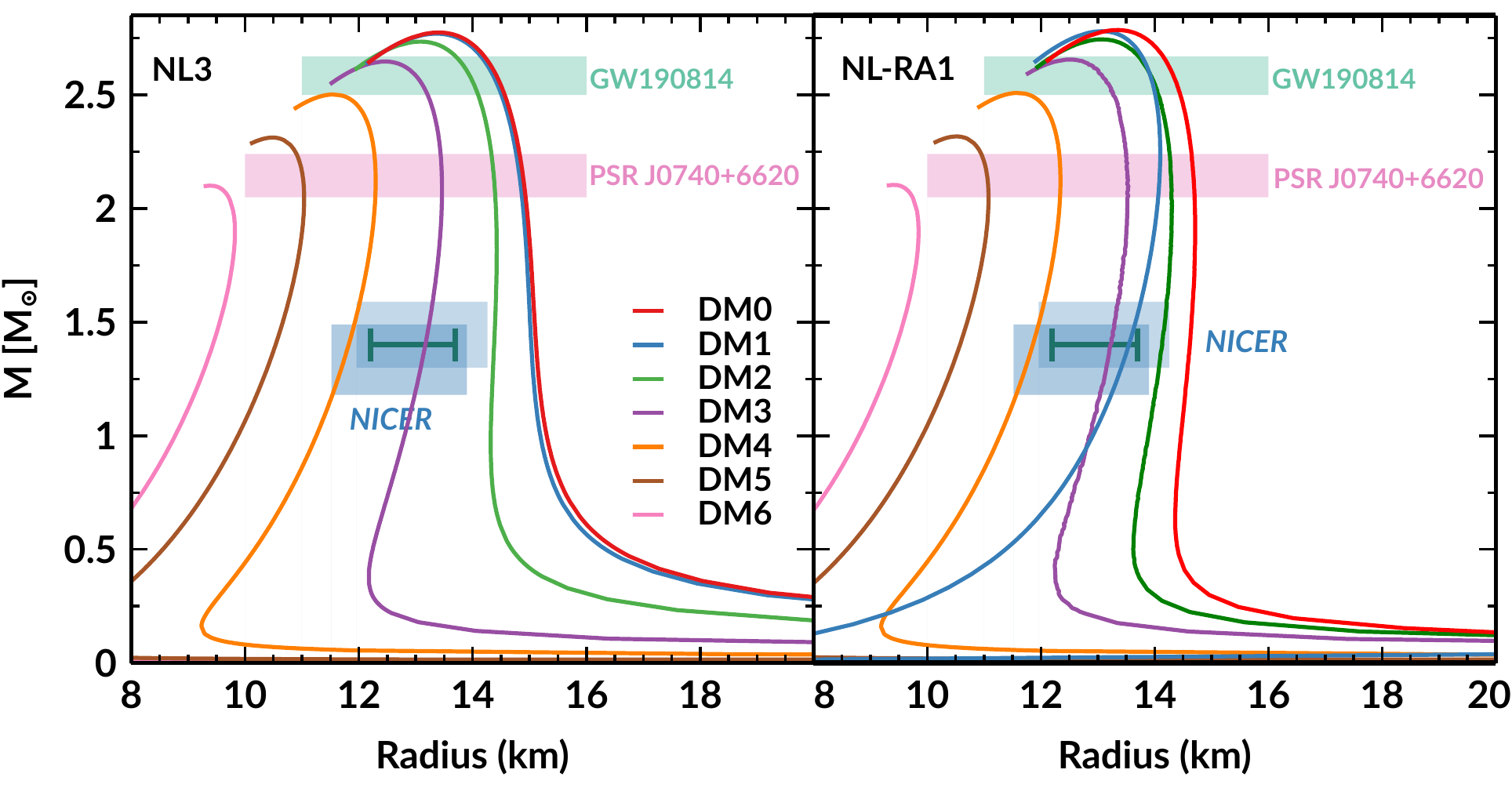}
\caption{The $M-R$ relations for NL3 and NL-RA1 with different DM Fermi momenta. The horizontal bars represent the maximum mass constraints from the PSR J0740+6620 (light pink) and the GW190814 event (dark cyan). The NICER data are also shown with two boxes from two different analyses \cite{Miller_2019, Riley_2019}. The double-headed green line represents the radius constraints by the GW190814 \cite{RAbbott_2020} for 1.4 $M_\odot$ NS.}
\label{fig:mr_dm_GW190814}
\end{figure}
%%%%%%%%%%%%%
%%%%%%%%%%%%%%%%%%%%%%%%%%%%%%%%%%%%%%%%%%%%%%%%%%
\subsection{Results and Discussions}
%%%%%%%%%%%%%%%%%%%%%%%%%%%%%%%%%%%%%%%%%%%%%%%%%%
In this section, we calculate the NS properties by using NL-RMF-type EOSs. All these sets predict the maximum NS mass of more than $2.67 \ M_\odot$. In Fig. \ref{fig:EOS_dm_GW190814}, the EOSs (such as NL3 and NL-RA1) are depicted with $k_f^{\rm DM}$ in the range of $0-0.06$ GeV, which are shortly represented as DM0-DM6. The EOSs become softer with the increase of $k_f^{\rm DM}$. The GW190814 discovery provides a joint constraint for NS-black hole (NSBH) merger  \cite{RAbbott_2020}. It is assumed to be a NS only when the maximum mass is not less than the secondary component of GW190814. The EOSs that correspond to DM3 and DM4 almost pass through the joint constraints imposed by GW170817 and GW190814. This amount of DM ($0.03-0.04$ GeV) may be available inside the NS. To test the validity of the EOS, we calculate the speed of sound ($c_s^2=\partial P/\partial \mathcal{E}$) for DM admixed NS. We find that our model respects the causality, which is shown in Fig. \ref{fig:causa_dm_GW190814}. The value of $c_s^2$ increases up to 0.4 fm$^{-3}$, becoming constant beyond that.
%%%%%%%%%%%%%%%
\begin{figure}
\centering
\includegraphics[width=0.8\textwidth]{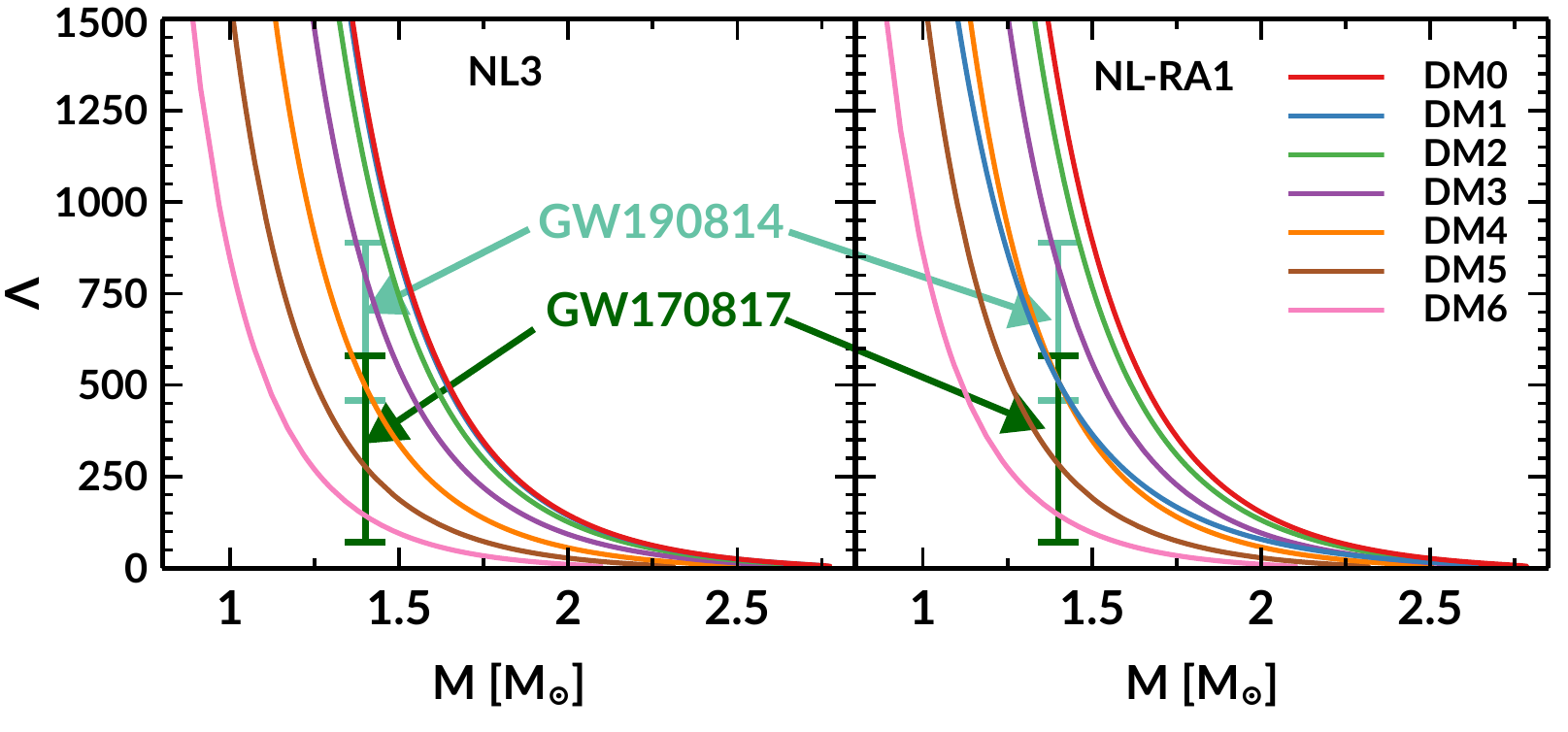}
\caption{The tidal deformability of the NS with the mass for NL3 and NL-RA1 parameter set for different DM Fermi momenta. The dark cyan (green) line implies the constraint given by GW190814 (GW170817) data for 1.4 $M_\odot$ NS.}
\label{fig:tidal_dm_GW190814}
\end{figure}
%%%%%%%%%%%%%%
	
The macroscopic properties such as mass ($M$), radius ($R$), and dimensionless tidal deformability ($\Lambda$) are also calculated with different DM fractions. By solving TOV equations \cite{TOV1, TOV2}; one can find the $M$-$R$ relation for each EOSs by assuming the $P(0)=P_C$ and $P(R)=0$ as the boundary conditions. We find that the NS's mass and radius reduce with DM's addition, as shown in Fig. \ref{fig:mr_dm_GW190814}. The maximum mass of the secondary object and the canonical radius constraints are $2.50-2.67$ $M_\odot$ and $12.20-13.70$ km as given in GW190814 \cite{RAbbott_2020}. The $M_{{\rm max}}$ and $R_{1.4}$ range correspond to DM3 are found to be $2.65-2.67$ $M_\odot$ and $12.38-12.75$ km respectively for the considered parameter sets (see Table \ref{tab:table_dm_GW190814}), which is consistent with GW190814 data. Under the assumption of a stiff nuclear EOS, the value of maximum mass with the assumed parameter set hints toward the possibility of the secondary component of the GW190814 event being a DM admixed NS. The NICER \cite{Miller_2019, Riley_2019} constraints are also respected by DM3 and DM4. Thus we conclude that given a specific model of the nuclear EOS, one can fix the fraction of DM inside the NS from the observational data.
	
When the NS presents in the tidal field created by its companion star, it is deformed. The relation between quadruple deformation $Q_{ij}$ and tidal field $\epsilon_{ij}$ is written as \cite{Hinderer_2008, Kumartidal_2017}
%%%%%%%%%%%%%%%%
\begin{equation}
Q_{ij}=-\lambda \epsilon_{ij},
\end{equation}
%%%%%%%%%%%%%%%
where the $\lambda$ is defined as tidal deformability. The dimensionless tidal deformability can be defined as $\Lambda=\lambda/M^5$. We calculate the $\Lambda$ for different DM Fermi momenta as shown in Fig. \ref{fig:tidal_dm_GW190814}. The values of $\Lambda_{1.4}$ for DM3 are 842.12, 784.59, 932.99, 804.36, and 818.48 for NL3, NL3*, NL-SH, NL3-II, and NL-RA1 respectively. The predictions of $\Lambda_{1.4}$ from GW170817 is $190_{-120}^{+390}$ at 90\% confidence level \cite{Abbott_2018}. The GW190814 also put a constraint on $\Lambda_{1.4}$ under NSBH scenario, $\Lambda_{1.4}=616_{-158}^{+273}$ \cite{RAbbott_2020}. The calculated values of $\Lambda_{1.4}$ for DM3 are well matched with GW190814 data, and, for the DM4 case, it is well consistent with GW170817 data. This also gives sufficient hints to constrain the amount of DM inside the NS.  

In summary, the mass gap of $2.5-5.0 \ M_\odot$ in the secondary component of GW190814 motivates us to explore its true nature, and we find the possibility that the compact object could be a DM admixed NS. To know the mystery of DM inside the NS, we modeled a Lagrangian density by assuming that the DM particles interact with nucleons via SM Higgs. We adopt the RMF model with NL-RMF  parameter sets to calculate NS EOS. The DM admixed NS properties are computed with different DM fractions inside it. 
    
To achieve the mass of the secondary object, one should take stiff EOSs. In RMF parametrizations, only NL-RMF and DD-RMF sets are stiff, which predicts the maximum masses of more than 2.50 $M_\odot$. Huang {\it et al.} \cite{Huang_2020} also suggested the possibility of the secondary object as a NS (with DD-RMF sets without any admixture of DM or strange particles). However, the NL-RMF forces with NS EOSs predict masses more than 2.67 $M_\odot$ are candidates for DM admixed NS. This scenario is only seen for NL-RMF type sets because no DD-RMF forces predict the mass of more than 2.67 $M_\odot$. Hence, we assume that those stars corresponding to DD-RMF sets are only NS without DM inside.
%%%%%
\begin{table*}
\centering
\caption{The quantities such as $M_{{\rm max}}$, $R_{{\rm max}}$, $R_{1.4}$ and $\Lambda_{1.4}$ are given with $k_f^{\rm DM}=$ 0, 0.03 and 0.04 GeV for NL3, NL3*, NL1, NL-SH, NL3-II and NL-RA1 respectively.}
\label{tab:table_dm_GW190814}
\renewcommand{\tabcolsep}{0.02cm}
\renewcommand{\arraystretch}{1.2}
\scalebox{0.63}{
\begin{tabular}{|l|l|l|l|l|l|l|l|l|l|l|l|l|l|l|l|l|l|l|}
\hline
\multirow{2}{*}{\begin{tabular}[c]{@{}l@{}}$k_f^{\rm DM}$\\   (GeV)\end{tabular}} & \multicolumn{3}{l}{\hspace{1.1cm} NL3} & \multicolumn{3}{|l}{\hspace{1.1cm} NL3*} & \multicolumn{3}{|l}{\hspace{1.1cm} NL1} & \multicolumn{3}{|l}{\hspace{1.1cm} NL-SH} & \multicolumn{3}{|l}{\hspace{1.1cm} NL3-II} & \multicolumn{3}{|l|}{\hspace{1.1cm} NL-RA1} \\ \cline{2-19}
&0.00 &0.03 &0.04 &0.00 &0.03 &0.04 &0.00 &0.03 &0.04 &0.00 &0.03 &0.04 &0.00 &0.03 &0.04 &0.00 &0.03 &0.04 \\ \hline
\begin{tabular}[c]{@{}l@{}}$M_{{\rm max}}$\\ $(M_\odot)$\end{tabular} &
2.78 &2.65 &2.50 &2.76 &2.64 &2.49 &2.84 &2.71 &2.55 &2.79 &2.66 &2.51 &2.77 &2.65 &2.50 &2.78 &2.66 &2.51 \\ \hline
\begin{tabular}[c]{@{}l@{}}$R_{{\rm max}}$\\   (km)\end{tabular} &
13.40 &12.46 &11.53 &13.10 &12.38 &11.60 &13.63 &12.75 &11.67 &13.53 &12.63 &11.60 &13.14 &12.42 &11.52 &13.19 &12.47 &11.57 \\ \hline
\begin{tabular}[c]{@{}l@{}}$R_{1.4}$\\   (km)\end{tabular} &
14.08 &13.16 &11.82 &14.03 &13.13 &12.13 &14.73 &13.46 &11.90 &14.35 &13.41 &11.90 &14.05 &13.16 &
11.81 &14.11 &13.22 &11.85 \\ \hline
$\Lambda_{1.4}$ &
1311.16 &842.12 &499.04 &1250.45 &784.59 &492.40 &1503.77 &912.65 &548.83 &1528.42 &932.99 &563.90 &1282.90 &804.36 &503.89 &1338.98 &818.48 & 516.96 \\ \hline
\end{tabular}}
\end{table*}
%%%%%%

With the addition of DM, the EOS becomes softer, $M_{{\rm max}}$, $R_{{\rm max}}$, $R_{1.4}$ and $\Lambda_{1.4}$ reduce with increase of DM percentage as given in Table \ref{tab:table_dm_GW190814}. The EOS corresponds to DM3, and DM4 well passes through the joint constraint inferred from GW170817+GW190814 by assuming that the secondary component is a NS only when its maximum mass is not less than the 2.50 $M_\odot$. This model also respects the causality for all values of DM Fermi momenta. Other properties such as $M_{{\rm max}}$ and $R_{1.4}$ are well consistent with the GW190814 data for the DM3 case. The values of $R_{1.4}$ for DM3 and DM4 are well constrained with the NICER results in almost all assumed sets. The calculated $\Lambda_{1.4}$ for DM3 and DM4 lie in the range of GW190814 and GW170817 data. Thus, one can constrain the DM percentage inside the NS from these observational data.
    
Therefore, the possibility of the secondary object in GW190814 as a DM admixed NS only when the maximum mass should be more than 2.67 $M_\odot$. We hope future GW and x-ray observations may answer the presence of DM inside the NS and constrain its properties. Thus, we suggest that the LIGO/Virgo include DM inside the compact objects when they infer some properties, such as mass, radius, and tidal deformability of the binary NS.
%%%%%%%%%%%%%%%%%%%%%
\section*{Conclusion}
%%%%%%%%%%%%%%%%%%%%%
In conclusion, we have calculated the properties of isolated, static, and rotating DM admixed NS. The simple DM model is taken as discussed in Chapter-\ref{C3}. It has been noticed that DM has significant effects on both static and rotating NS. For example, the mass and radius decrease by increasing the DM percentage. When we compared the static and rotating cases, the change was found to be $\sim 20 \% $ for different macroscopic properties. The change of curvature with and without DM is obtained as $\sim33\%$ for the canonical star, and the value increases with mass. The magnitude of surface curvature and compactness are increased. The binding energy for the DM admixed NS increased towards a positive value with the increase of DM momentum, which makes the NS unstable. From this study, we concluded that a tiny amount of DM can accumulate inside the NS. If a star has more DM contains, it heats the NS, and results in the acceleration of the Urca process. Due to this phenomenon, the cooling of NS enhances and makes the NS unstable. With the DM admixed EOSs, we have also suggested that the secondary component might be a NS with DM content if the underlying nuclear EOS is sufficiently stiff.
%%%%%%%%%%%%%%%%%%%%%%%%%%%%%%%% END %%%%%%%%%%%%%%%%%%%%%%%%%%%%%%%%%%%%%%%%%%%%%%%%%%%%%%
%\blankpage 
%%%%%%%%%%%%%%%%%%%%%%%%%% Chapter-5 %%%%%%%%%%%%%%%%%%%%%%%%%%%%%%
%%%%%%%%%%%%%%%%%%%%%%%%%%%%%%% CHAPTER - 5 %%%%%%%%%%%%%%%%%%%%%%%%%%%%%%%%%%%%
\chapter{Dark matter effects on properties of the oscillating neutron/hyperon stars}
\label{C5} 
%%%%%%%%%%%%%%%%%%%%%%%%%%%%%%%%%%%%%%%%%%%%%%%%%%%%%%%%%%%%%%%%%%%%%%%%%%%%%%%%
The NS oscillates due to vibrant explosion in the type-II supernovae. Due to its complex internal compositions, it oscillates with different ranges of frequencies and emits gravitational waves. In this chapter, we explore the fundamental mode frequency of the NS since it emits a large amount of energy through this mode. We use the relativistic mean-field EOS for the DM admixed hyperons star within the relativistic Cowling approximations to get the eigen frequencies of the star. Various oscillating properties of hyperon stars are calculated with different fractions of DM for well-known EOSs. We enumerate some empirical relations, which can be used to constrain the internal properties of the NS if we detect the emitted GWs in the future. 
%%%%%%%%%%%%%%%%%%%%%%%%%%%%%%%%%%%%%%%%%%
\section{Hyperons inside the neutron star}
%%%%%%%%%%%%%%%%%%%%%%%%%%%%%%%%%%%%%%%%%%
It is well known that hyperons appear in the NS interior at densities around $2-3\rho_0$. The presence of the hyperons makes the EOS softer; consequently, the magnitude of the maximum mass substantially reduces, which is incompatible with the observation known as the ``hyperon puzzle". In addition to DM, we also include hyperons production inside the NS. Hyperons appear at the higher-density region or core of the NS. The idea behind the appearance of hyperons inside the NS is not brand-new. In 1959, Cameron {\it et al.} first put forward the concept of hyperon puzzle ~\cite{Cameron_1959}. According to Ambartsumyan \& Saakyan, the core of a massive NS consists of an inner hyperon core and an outer nucleon shell ~\cite{Ambartsumyan_1960}. The detailed discussion on hyperons production and interaction with mesons and their coupling parameters have been studied in Refs. ~\cite{Pais_1966, Millener_1988, NKGfp_1992, NKGPRL_1991, Schaffner_1994, Schaffner_1996, Batty_1997, Schaffner_2000, Harada_2005, Harada_2006, Kohno_2006, Friedman_2007, Weissenborn_2012, WeissenbornE_2014, Lopes_2014, Bhuyan_2017, Biswal_2019, Biswalaip_2019}. Although different approaches have been tried to solve the issues, such as (i) potential depths for each hyperon, (ii) coupling between hyperons-mesons, and (iii) hyperon puzzles, however, these are still open problems. This is because only some hypernuclei have been discovered in the terrestrial laboratory ~\cite{HAYANO_1989, Nagae_1998, Saha_2004}. More hypernuclei will help us to constrain the coupling parameters by fitting the potential depth. In this study, the hyperons-scalar mesons coupling constants are calculated by fitting with hyperon potential depth, while for hyperons-vector mesons, SU(6) method is used, as mentioned in Refs. ~\cite{Lopes_2018, Weissenborn_2012, WeissenbornNPA_2012}. Another efficient approach to fix the hyperons-mesons coupling parameters is SU(3) group theory method, as given in Refs. ~\cite{Weissenborn_2012, Tsuyoshi_2013, Lopes_2014, Lopes_2021}. The coupling constants for scalar and vector mesons are constrained using the SU(3) method. One can calculate NM and NS properties by using both SU(3) and SU(6) methods and compare their results as done in Refs. ~\cite{Weissenborn_2012, Lopes_2014}. In this chapter, our aim is to calculate the $f$-mode frequency of the DM admixed hyperon star as the function of different astrophysical quantities.
%%%%%%%%%%%%%%%%%%%%%%%%%%%%%%%%%%%%%%%%%%%%%%%%%%%%%%%%%%%%%%%%%%%%%%%%%%%
\subsection{Interaction between baryons-mesons and mesons-mesons}
\label{form:hyp}
%%%%%%%%%%%%%%%%%%%%%%%%%%%%%%%%%%%%%%%%%%%%%%%%%%%%%%%%%%%%%%%%%%%%%%%%%%%
In this chapter, we have taken baryon octets ($p, n, \Lambda, \Sigma^{+,-,0}, \, {\rm and} \, \Xi^{+,-}$) along with other two strange mesons such as $\sigma^*$, and $\phi$ ~\cite{Schaffner_1996,Biswal_2019,Pradhan_2021}. Hence, the Lagrangian density in Eq. (\ref{eq:NS_lag}) is almost the same, but an extra part is added for the strange meson. Hence the Lagrangian density is modified as follows \cite{Dasfmode_2021}
%%%%%%%%%%%%%%%%
\begin{eqnarray}
{\cal L}_{\rm NS} = {\cal L}_{\rm NM} + {\cal L}_{\rm YY} + \sum_{l}\bar\psi_{l}\Big(i\gamma^{\mu}\partial_\mu-m_l\Big)\psi_l, 
\end{eqnarray}
where
%%%%%%%%%%%%%%%%
\begin{eqnarray}
{\cal L}_{\rm YY}&=&\sum_Y  \bar{\psi}_{_Y}  (g_{\sigma^*_Y} \sigma^*-g_{\phi_ Y}\gamma_{\mu}\phi^{\mu})\psi_{_Y}+\frac{1}{2}m_{\phi}^2 \phi_{\mu}\phi^{\mu}-\frac{1}{4} \phi_{\mu \nu}\phi^{\mu \nu}  
\nonumber\\
&+&\frac{1}{2}  (\partial_{\mu} \sigma^* \partial^{\mu}\sigma^* - m_{\sigma^*}^2 {\sigma^*}^2).
\end{eqnarray}
%%%%%%%%%%%%%%
Also, it is imperative to mention that the summation over $\Sigma_{p, n}$  in Eq. (\ref{rmflag}) is now modified as $\Sigma_{B}$, where $B$ represents the baryon octets. The energy density and pressure are also modified with the addition of other baryons and strange mesons and given as ~\cite{Kumar_2018, Biswal_2019, Kumar_2020}:
%%%%%%%%%%%%%
\begin{align}
{\cal E}_{\rm NS} = & \sum_B \frac{\gamma_{_B}}{2\pi^2}\int_{0}^{k_{F_B}} k^2\ dk \sqrt{k^2+m^{*2}_{_B}}+n_{_B} g_{\omega_B}\omega_0\frac{n_{3B}}{2}g_{\rho{_B}}\rho_0
\nonumber\\
-&\frac{1}{3!}\zeta_{0}{g_{\omega_B}^2}\omega_0^4-\Lambda_{\omega}g_{\rho_B}^2g_{\omega_B}^2\rho_{03}^2\omega_0^2+m_{\sigma}^2{\sigma_0}^2\Bigg(\frac{1}{2}+\frac{\kappa_{3}}{3!}\frac{g_{\sigma_B}\sigma_0}{m_{_B}}+\frac{\kappa_4}{4!}\frac{g_{\sigma_B}^2\sigma_0^2}{m_{_B}^2}\Bigg)
\nonumber\\
-&\frac{1}{2}m_{\omega}^2\omega_0^2\Bigg(1+\eta_{1}\frac{g_{\sigma_B}\sigma_0}{m_{_B}}+\frac{\eta_{2}}{2}\frac{g_{\sigma_B}^2\sigma_0^2}{m_{_B}^2}\Bigg)-\frac{1}{2}\Bigg(1+\frac{\eta_{\rho}g_{\sigma_B}\sigma_0}{m_{_B}}\Bigg)m_{\rho}^2\rho_{03}^{2}
\nonumber\\
+&\frac{1}{2}m_{\delta}^2 \delta_0^{2}+\frac{1}{2}m_\phi^2\phi_0^2
%\nonumber\\
+\frac{1}{2}m_{\sigma^*}^2\sigma_0^{*2}+\sum_l\frac{\gamma_{_l}}{2\pi^2}\int_0^{k_{F{_l}}} dk \ \sqrt{k^2+m_l^2},
\label{eq:ENS}
\end{align}
%%%%%%%%%%%
and
%%%%%%%%%%%%%
\begin{align}
P_{\rm NS} = & \sum_B \frac{\gamma_{_B}}{6\pi^2}\int_{0}^{k_{F_B}} \frac{k^4\ dk}{\sqrt{k^2+m^{*2}_{_B}}} +\frac{1}{3!}\zeta_{0}{g_{\omega_B}^2}\omega_0^4+\Lambda_{\omega}g_{\rho_B}^2g_{\omega_B}^2\rho_{03}^2\omega_0^2
\nonumber\\
-&m_{\sigma}^2{\sigma_0}^2\Bigg(\frac{1}{2}+\frac{\kappa_{3}}{3!}\frac{g_{\sigma_B}\sigma_0}{m_{_B}}+\frac{\kappa_4}{4!}\frac{g_{\sigma_B}^2\sigma_0^2}{m_{_B}^2}\Bigg)
+\frac{1}{2}m_{\omega}^2\omega_0^2\Bigg(1+\eta_{1}\frac{g_{\sigma_B}\sigma_0}{m_{_B}}+\frac{\eta_{2}}{2}\frac{g_{\sigma_B}^2\sigma_0^2}{m_{_B}^2}\Bigg)
\nonumber\\
+&\frac{1}{2}\Bigg(1+\frac{\eta_{\rho}g_{\sigma_B}\sigma_0}{m_{_B}}\Bigg)m_{\rho}^2\rho_{03}^{2}-\frac{1}{2}m_{\delta}^2 \delta_0^{2}+\frac{1}{2}m_{\sigma^*}^2\sigma_0^{*2}+\frac{1}{2}m_\phi^2\phi_0^2
\nonumber\\
+&\sum_l\frac{\gamma_{_l}}{6\pi^2}\int_0^{k_{F{_l}}} \frac{k^2 dk}{\sqrt{k^2+m_l^2}},
\label{eq:PNS}
\end{align}
%%%%%%%%%%%
where $\gamma_{_B}$ and $\gamma_{l}$ are the spin degeneracy factor for baryons and leptons, respectively. $k_{F_B}$ and $k_{F_l}$ are the baryons and leptons Fermi momentum, respectively. $m^*_{_B}$ is the effective masses of the baryons, which is written by 
%%%%%%%%%%%%%%%%
\begin{eqnarray}
m^*_B &=& m_B-g_{\sigma_B}\sigma_0-g_{\delta_B}\tau_B\delta_0-g_{\sigma^*_B}\sigma^*.
\label{eq:effm}
\end{eqnarray}
%%%%%%%%%%%%%%
The last term in Eq. (\ref{eq:effm}) doesn't contribute to nucleons. This is because the value of $g_{\sigma^*_B}$ and $g_{\phi_B}$is zero for the nucleons. The mass, charge, isospin, and strangeness of each baryon are taken in these calculations are given in Table \ref{tab:qty_baryons}
%%%%%%%%%%%%%
\begin{table}
	\centering
	\caption{Mass, charge, isospin, and strangeness of all baryon octet.}
	\label{tab:qty_baryons}
	\scalebox{1.0}{
		\begin{tabular}{|l|l|l|l|l|l|l|l|l|}
			\hline
			\begin{tabular}[c]{@{}l@{}}Baryons \\ Name\end{tabular} & $p$ & $n$ & $\Lambda$ & $\Sigma^+$ & $\Sigma^0$ & $\Sigma^-$ & $\Xi^0$ & $\Xi^-$ \\ \hline
			\begin{tabular}[c]{@{}l@{}}Mass\\ (MeV)\end{tabular}    & 939 & 939 & 1116      & 1193       & 1193       & 1193       & 1318    & 1318    \\ \hline
			Charge ($q$)      & +1  & 0    & 0  & +1 & 0  & -1 & 0   & -1   \\ \hline
			Isospin ($I_3$)   & 1/2 & -1/2 & 0  & +1 & 0  & -1 & 1/2 & -1/2 \\ \hline
			Strangeness ($s$) & 0   & 0    & -1 & -1 & -1 & -1 & -2  & -2  \\ \hline
		\end{tabular}%
	}
\end{table}
%%%%%%%%%%%
%%%%%%%%%%%%%%%%%%%%%%%%%%%%%%%%%%%%%%%%%%%%%%%%%%%%%%%%%%%%%%%%%%%%%%%%%%%%%
\subsubsection{1. Coupling constants for Nucleons-Mesons and Mesons-Mesons:-}
%%%%%%%%%%%%%%%%%%%%%%%%%%%%%%%%%%%%%%%%%%%%%%%%%%%%%%%%%%%%%%%%%%%%%%%%%%%%%
The coupling between nucleons-mesons and meson-mesons differs and depends on the parameter sets. For example, the G3 parameter set has different nucleons-mesons and mesons-mesons coupling compared to IOPB-I forces (see Table \ref{tab:rmf_mass_coupling}). Many RMF parameter sets have been developed, including different types of interaction between nucleons-mesons and mesons-mesons (both self and cross). Therefore, to study how the coupling parameters affect both NM and NS properties, we take four different types of parameter sets such as NL3 ~\cite{Lalazissis_1997}, IOPB-I ~\cite{Kumar_2018}, FSUGarnet ~\cite{Chen_2014}, and G3 ~\cite{Kumar_2017}. We tabulate the masses of different mesons and coupling constants in Table \ref{tab:rmf_mass_coupling} for these parameter sets. The NM properties corresponding to four parameter sets are also reported in Table \ref{tab:NM_properties} ~\cite{Kumar_2018, DasBig_2021}. Table \ref{tab:NM_properties} shows that all three parameter sets reproduce the NM properties well except NL3. For example, the value of incompressibility, $K=$ 271.38 MeV, is higher than other sets. This is because NL3 is a stiff EOS compared to G3, which is a softer one.
%%%%%%%%%%%%%%%%%%%%%%%%%%%%%%%%%%%%%%%%%%%%%%%%%%%%%%%%%%%%%
\subsubsection{2. Coupling constants for Hyperons-Mesons:-}
%%%%%%%%%%%%%%%%%%%%%%%%%%%%%%%%%%%%%%%%%%%%%%%%%%%%%%%%%%%%%
Until now, we do not know exactly how the hyperons interact with each other compared to the nucleons-nucleons interaction because of the limitation of the hyper-nuclei data ~\cite{NKGPRC_2001}. Therefore, fixing the couplings between hyperons-mesons is quite difficult. Generally, the SU(6) symmetry group is used to fix the couplings between hyperons and vector mesons, while the couplings with scalar mesons are fixed through the hyperon potential depth ~\cite{Pais_1966, Lopes_2014, Biswal_2019, Biswalaip_2019}. The potential depth of the $\Lambda$-hyperon ($U_\Lambda$) is known to be -28 MeV ~\cite{Millener_1988,NKGPRL_1991,Schaffner_1994,Batty_1997, Torres_2017, Lopes_2018} but the $\Sigma$ and $\Xi$ potentials ($U_\Sigma$, $U_\Xi$) provide a large uncertainty and not even the signs of the potentials are well defined ~\cite{Dover_1984, Schaffner_1994}. But some hyper-nuclear experiments show that the $U_\Sigma$ and $U_\Xi$ potential depth close to +30 MeV and -18 MeV respectively ~\cite{Schaffner_2000,Harada_2005, Harada_2006, Kohno_2006, Friedman_2007}.

The hyperon potential depth is defined as ~\cite{NKGb_1997}
%%%%%%%%%%%%%%%%
\begin{equation}
U_Y = -g_{\sigma{_N}}x_{\sigma{_Y}}\sigma_0+g_{\omega {_N}}x_{\omega{_Y}}\omega_0,
\label{eq:uYN}
\end{equation}
%%%%%%%%%%%%%%
where $x_{\sigma{_Y}}$ and $x_{\omega{_Y}}$ is defined as $g_{\sigma{_Y}}/g_{\sigma{_N}}$ and $g_{\omega{_Y}}/g_{\omega{_N}}$ respectively. For symmetric NM the quantities $g_{\sigma{_Y}}/g_{\sigma{_N}}$ and $g_{\omega{_Y}}/g_{\omega{_N}}$ are given in Refs. ~\cite{NKGb_1997, Biswal_2019}
%%%%%%%%%%%%%
\begin{align}
&&g_{\sigma{_N}}\sigma_0=M-M^*=M\Big(1-\frac{M^*}{M}\Big),  \, 
g_{\omega{_N}}\omega_0=\Big(\frac{g_{\omega{_N}}}{m_\omega}\Big)^2\rho_0,
\label{eq:NKGY}
\end{align}
%%%%%%%%%%%
Rewriting Eq. (\ref{eq:uYN}) by inserting Eq. (\ref{eq:NKGY}), we get
%%%%%%%%%%%%%%%%
\begin{eqnarray}
U_Y = M\Big(\frac{M^*}{M}-1\Big)x_{\sigma{_Y}}+\Big(\frac{g_{\omega{_N}}}{m_\omega}\Big)^2\rho_0x_{\omega{_Y}}.
\end{eqnarray}
%%%%%%%%%%%%%%%
In this case, we choose the values of $U_\Lambda =-28$ MeV, $U_\Sigma=+30$ MeV, and $U_\Xi = -18$ MeV. To get the desired potential, one has to choose the values of $x_{\sigma{_Y}}$ and $x_{\omega{_Y}}$ for a fixed parameter set. For example NL3 case, we fix,  $x_{\sigma{_\Lambda}}=0.8$, and the value of $x_{\omega{_\Lambda}}$ is found to be 0.8982. Similarly for IOPB-I set, we fix, $x_{\sigma{_\Lambda}}$ = 0.8 which gives $x_{\omega{_\Lambda}}=0.8338$. For $\Sigma$ and $\Xi-$hyperons the values are  $x_{\sigma_\Sigma}=0.7$, $x_{\omega_\Sigma}=0.8932$, $x_{\sigma_\Xi}=0.8$, and $x_{\omega_\Xi}=0.8639$ for IOPB-I set. Hence, one can get different combinations to fit the potential depth. The hyperon interactions with $\rho-$meson are fitted according to SU(6) symmetry method ~\cite{Dover_1984, Schaffner_1994} and their couplings with hyperons are given as: $x_{\rho{_\Lambda}}=0.0$, $x_{\rho{_\Sigma}}=2.0$ and $x_{\rho{_\Xi}}=1.0$. 

The couplings constants for hyperons and $\sigma^*-$meson are $x_{\sigma^*\Lambda}=0.69$, $x_{\sigma^*\Sigma}=0.69$, and $x_{\sigma^*\Xi}=1.25$. Similarly, for hyperons and $\phi-$mesons coupling constants are $x_{\phi\Lambda}=-\frac{\sqrt{2}}{3}g_{\omega_N}$, $x_{\phi\Sigma}=-\frac{\sqrt{2}}{3}g_{\omega_N}$, and $x_{\phi\Xi}=-\frac{2\sqrt{2}}{3}g_{\omega_N}$ respectively for $\Lambda$, $\Sigma$ and $\Xi-$hyperons ~\cite{Biswal_2019}. In this calculation, we neglected the $\sigma^*$ and $\phi$ mesons-nucleons interaction for numerical simplicity.
%%%%%%%%%%%%%%%%%%%%%%%%%%%%%%%%%%%%%%%%%%%%%%%%%%%%%%
\section{Dark matter admixed hyperon star}
\label{form:EOS}
%%%%%%%%%%%%%%%%%%%%%%%%%%%%%%%%%%%%%%%%%%%%%%%%%%%%%%
%%%%%%%%%%%%%%%%%%%%%%%%%%%%%%%%%%%%%%%%%%%%%%%%%%%%%%%%%%%
\subsection{Interaction between Baryons-DM}
\label{form:DM}
%%%%%%%%%%%%%%%%%%%%%%%%%%%%%%%%%%%%%%%%%%%%%%%%%%%%%%%%%%%
In this sub-sec, our aim is to calculate the DM interactions with all baryons. We have already formulated the DM-nucleons interaction in chapter-\ref{C2} (see sub-section \ref{DM_lagrangian}). With the addition of a baryon octet, the interaction Lagrangian can be modified as 
%%%%%%%%%%%%%%%%%
\begin{align}
{\cal{L}}_{\rm DM}=  \bar \chi \Big[ i \gamma^\mu \partial_\mu - M_\chi + y h \Big] \chi +  \frac{1}{2}\partial_\mu h \partial^\mu h - \frac{1}{2} M_h^2 h^2 +\sum_B f_{_B} \frac{m_{_B}}{v} \bar{\psi}_{_B} h \psi_{_B}, 
\label{eq:ldm}
\end{align}
%%%%%%%%%%%%%%
where $\psi_{_B}$ and $\chi$ are the baryons and DM wave functions, respectively. The parameters $y$ is DM-Higgs coupling, $f_{_B}$ is the baryons-Higgs form factors, and $v$ is the vacuum expectation value of the Higgs field. The values of $y$ and $v$ are 0.07 and 246 GeV, respectively, taken from the Refs. ~\cite{Das_2020, Das_2021}. The value of $f_{_B} (=$ 0.35) for all baryons is taken to be the same because we do not know the form factor for hyperons except nucleons with Higgs. 
%%%%%%%%%%%%%%
\begin{figure}
	\centering
	\includegraphics[width=0.7\textwidth]{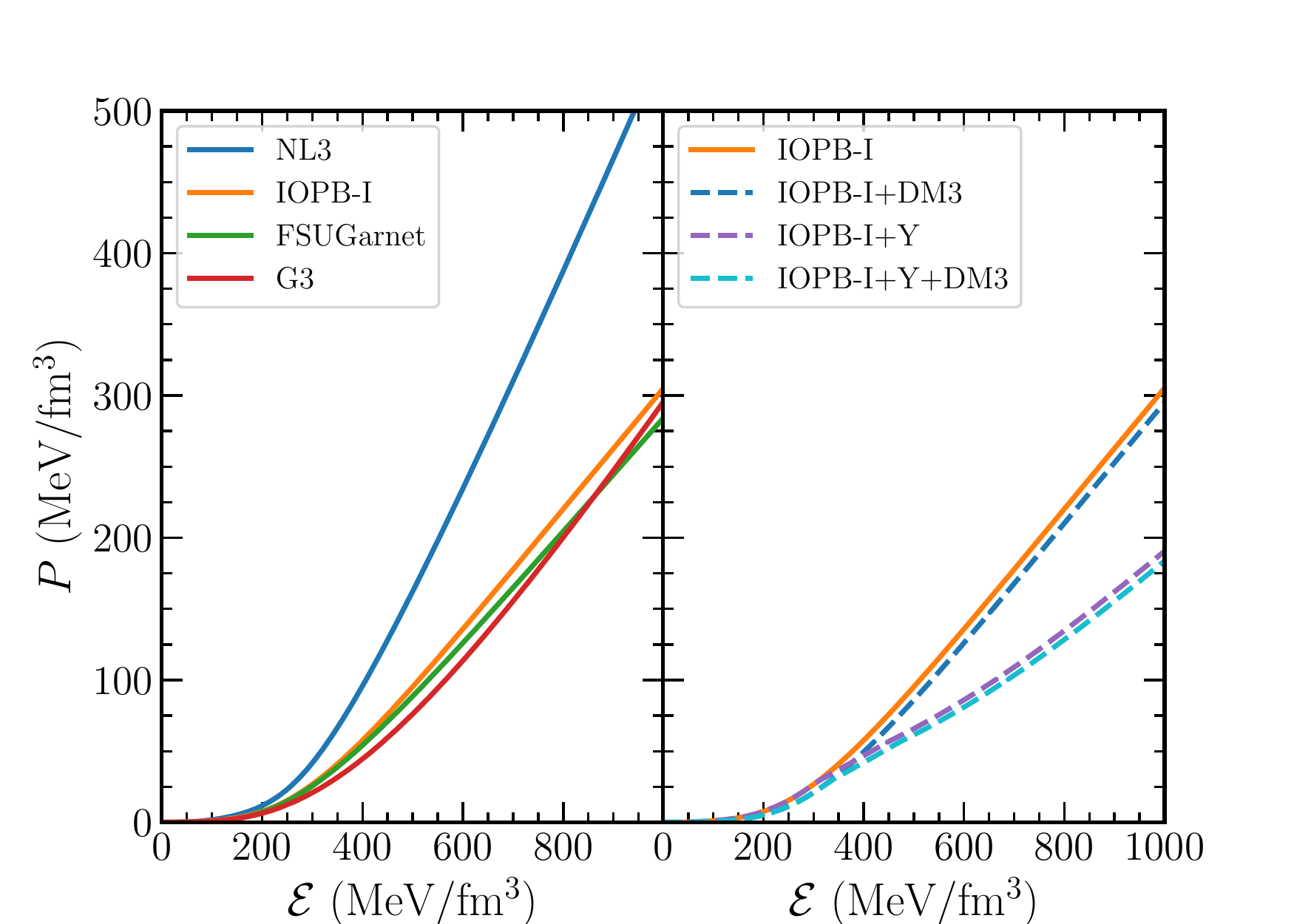}
	\caption{{\it Left:} EOSs for different RMF parameter sets. {\it Right}: EOSs are for IOPB-I with a yellow line (only nucleons), IOPB-I+DM3 with a blue dashed line (nucleons+DM3), IOPB-I+Y with a purple dashed line (nucleons+hyperons), IOPB-I+Y+DM with dashed cyan line (nucleons+hyperons+DM3). }
	\label{fig:EOS}
\end{figure}	
%%%%%%%%%%%%
%%%%%%%%%%%%%%%%%%%%%%%%%%%%%%%%%
\subsection{EOSs, $M-R$ profiles}
%%%%%%%%%%%%%%%%%%%%%%%%%%%%%%%%%
The total energy density (${\cal E}$) and pressure ($P$) for the DM admixed hyperonic NS are as follows:
%%%
\begin{eqnarray}
{\cal{E}} =   {\cal{E}}_{\rm NS}+ {\cal{E}}_{\rm DM}, \ {\rm and }\ P =   P_{\rm NS}+ P_{\rm DM}.
\label{eq:EOS}
\end{eqnarray}
%%%%

In Fig. \ref{fig:EOS}, we plot $P$ with ${\cal E}$ for four considered parameter sets. It is clear from the figure that the NL3 is the stiffest EOS as compared to others, while G3 is the softer EOS. On the right side of the plot, we show the EOS for the IOPB-I parameter set for nucleons (IOPB-I), nucleons with DM (IOPB-I+DM3), nucleons with hyperons (IOPB-I+Y), and nucleons with both hyperons and DM (IOPB-I+Y+DM3). The EOS becomes softer with the addition of DM/hyperons or DM+hyperons, but the softness is more for DM+hyperons (dashed cyan line) than others, which is visible in the figure. 

In the presence of hyperons/DM particles, the EOS of the NS becomes softer. This is because every system would like to minimize its energy. As the density increases, the Fermi momenta or the Fermi energy increases. Because the nucleons are fermions and need to be placed in a higher orbit with increasing density, it is well known that the density is proportional to the cube of the Fermi momenta. At sufficiently high density, the energy of the nucleon increases, which is the rest of mass energy plus the kinetic energy or $E=\sqrt{k_f^2+M^2}$. When this energy exceeds the mass of the hyperons/kaons/DM, the decay of the nucleon to these particles. In other words, it is economical energy-wise for the system to have the hyperons/kaons/DM in the lower energy states rather than nucleons at higher Fermi energy at higher density, though the hyperons are heavier than the nucleons. In other words, the nucleons are replaced by hyperons/kaons/DM depending on the system's density. As a result, a fraction of the gravitational mass is converted to kinetic energy and decreases the gravitational mass. With these EOSs, we calculate the $M-R$ profiles and the $f$-mode frequencies of the NS.
%%%%%%%%%%%%%%%%%%%%%%%%%%%%%%%%
\subsection{Mass-Radius profile}
%%%%%%%%%%%%%%%%%%%%%%%%%%%%%%%%
%%%%%%%%%%%%%%
\begin{figure}
	\centering
	\includegraphics[width=0.6\textwidth]{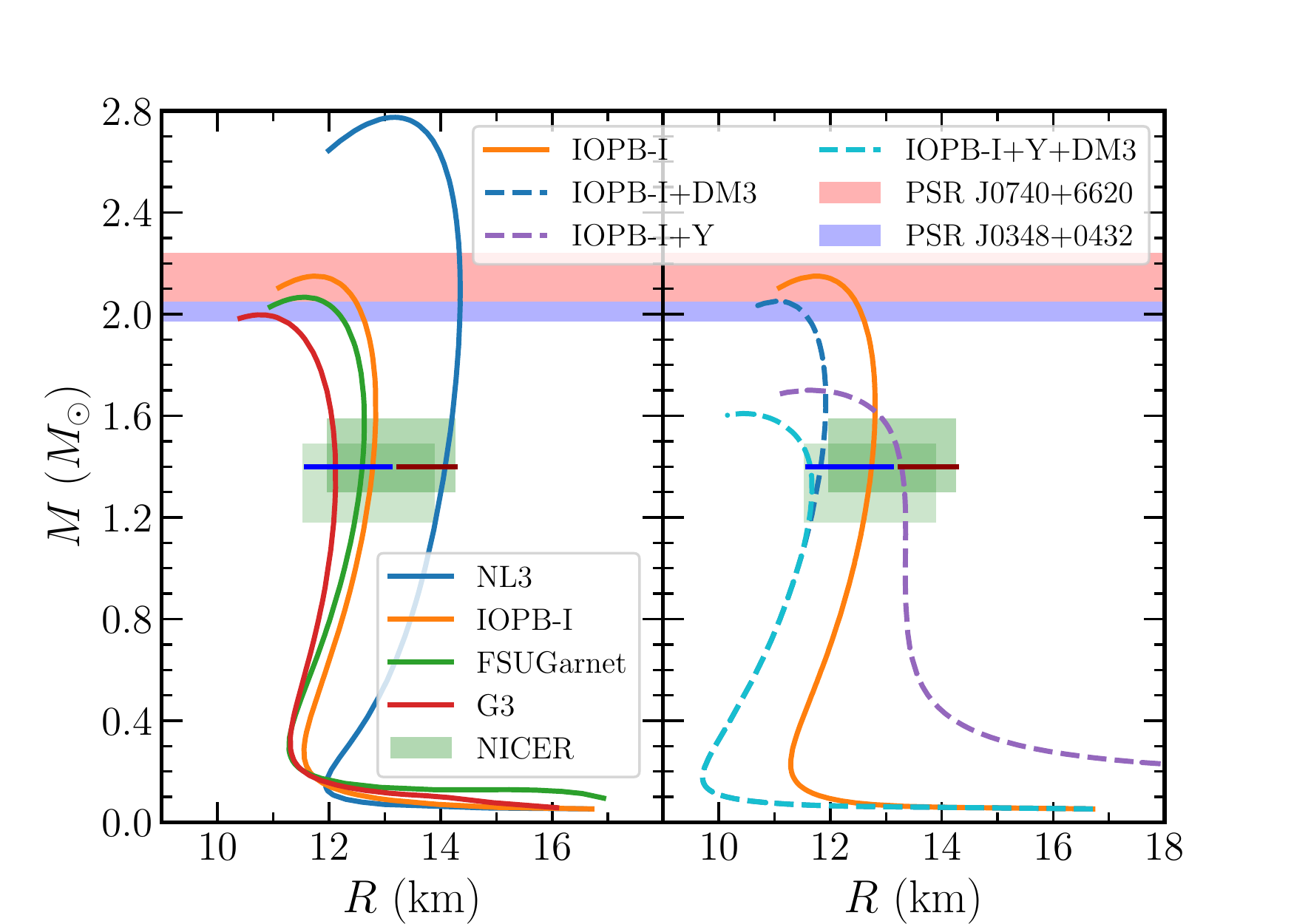}
	\caption{{\it Left:} $M-R$ relations for four different RMF parameter sets. {\it Right}: $M$--$R$ relations for IOPB-I, IOPB-I+DM3, IOPB-I+Y, and IOPB-I+Y+DM3. The red and blue band represents the masses of the massive pulsars observed by Cromartie {\it et al.} ~\cite{Cromartie_2020} and Antoniadis {\it et al.} ~\cite{Antoniadis_2013}. The old NICER results are shown with two green boxes from two different analyses ~\cite{Miller_2019, Riley_2019}. The new NICER result is shown for canonical radius (blue horizontal line) given by  Miller {\it et al.} ~\cite{Miller_2021}. Reed {\it et al.} radius range is also shown with a dark red line for canonical star \cite{Reed_2021}.}
	\label{fig:mr}
\end{figure}
%%%%%%%%%%%%

We calculate the mass-radius profile of the DM admixed hyperon star by solving the TOV equations (see Eq. \ref{eq:TOV}) and shown in Fig. \ref{fig:mr}. We observe that the NL3 EOS predicts the maximum mass around 2.77 $M_\odot$ and radius $\sim$14 km. Since it is a stiffer EOS than others, its maximum mass is higher than the other three sets. We put maximum mass constraints measured from the different pulsars such as PSR \ J0740+6620 ~\cite{Cromartie_2020}, and PSR J0348+0432 ~\cite{Antoniadis_2013}. Except for NL3, the other three EOSs can support maximum NS mass 2.0 $M_{\odot}$, which are compatible with the precise measured NS masses by Cromartie {\it et al.} ($2.14_{-0.09}^{+0.10}\ M_\odot$ with 68.3\% credibility interval) ~\cite{Cromartie_2020} and Antoniadis {\it et al.} ($2.01\pm0.04 \ M_\odot$) ~\cite{Antoniadis_2013}. The simultaneous measurement of mass and radius by the NICER give the $M = 1.44_{-0.14}^{+0.15}\ M_\odot$ ($M = 1.34_{-0.16}^{+0.15}\ M_\odot$), and $R = 13.02_{-1.06}^{+1.24}$ \ km ($R = 12.71_{-1.19}^{+1.14}$ \ km) respectively from the analysis of PSR J0030+0451 by Miller {\it et al.} (Riley {\it et al.}) ~\cite{Miller_2019, Riley_2019}. We called it old NICER data. We put the old NICER data with two green boxes from two different analyses in the same figure ~\cite{Miller_2019, Riley_2019}. The predicted radius by all the EOSs is compatible with old NICER data, as shown in Fig. \ref{fig:mr}. On the left side of Fig. \ref{fig:mr}, we noticed that IOPB-I and IOPB-I+DM3 could able to reproduce the maximum mass constraints. The maximum mass and radius corresponding to IOPB-I EOS are 2.15 $M_\odot$ and 11.75 km, respectively. The maximum masses and radii are (2.05 $M_\odot$, 11.02 \ km), (1.70 $M_\odot$, 11.39\ km), and (1.61 $M_\odot$, 10.40\ km) for IOPB-I+DM3, IOPB-I+Y, and IOPB-I+Y+DM3 respectively. Hence, these models satisfy the old NICER data. Recently, Miller {\it et al.} put another radius constraint on both for the canonical ($R_{1.4} = 12.45 \pm 0.65$ \ km) and maximum NS ($R_{2.08} = 12.35 \pm 0.75$ \ km) from the NICER and X-ray Multi-Mirror (XMM) Newton data are termed as new NICER data ~\cite{Miller_2021}. We depicted the new NICER data as a horizontal blue line in Fig. \ref{fig:mr}. All considered EOSs satisfy the new NICER data except NL3 and IOPB-I+Y sets.

Recently, the PREX-2 experiment has given the updated neutron skin thickness of $^{208}$Pb as $R_{\rm skin}=0.284\pm0.071$ fm, within an $1\%$ error \cite{Adhikari_2021}. Based on this data, Reed {\it et al.} inferred the symmetry energy and slope parameter as $J=38.1\pm4.7$ MeV and $L=106\pm37$ MeV, respectively, with the help of limited relativistic mean-field forces (using PREX-2 data) \cite{Reed_2021}. These $J$ and $L$ values are larger in magnitude than the old PREX data (PREX-I). The latest $J$ and $L$ can be reproduced mostly from the stiff EOS. That means the PREX-2 result allows a stiff EOS. Reed {\it et al.} also predicted the radius for canonical NS is $13.25<R_{1.4}<14.26$ km. In this case, only NL3 satisfies the Reed {\it et al.} data as shown in Fig. \ref{fig:mr}. Recently, Miller {\it et al.} also gave new NICER constraints both for canonical and maximum mass NS from the X-ray study of PSR J0030+0451 ~\cite{Miller_2021}. The improved radius estimate for a canonical star is $11.8<R_{1.4}<13.1$ km. This new NICER constraint allows a narrow radius range as compared to old NICER data ($ 11.52<R_{1.4}<14.26$ km). From the radii constraint, we find that the new NICER data allows a narrow radius range contrary to a large range of PREX-2 and the old NICER data, leaving us an inconclusive determination of the NS radius.
%%%%%%%%%%%%%%%%%%%%%%%%%%%%%%%%%%%%%%%%
\section{$f$-mode oscillation of the NS}
\label{form:fmode}
%%%%%%%%%%%%%%%%%%%%%%%%%%%%%%%%%%%%%%%%
NS oscillates when it is disturbed by an external/internal event. Hence, it emits gravitational waves with different modes of frequencies. The most important modes are fundamental mode ($f$), first and second pressure modes ($p$), and the first gravitational mode ($g$) ~\cite{Anderson_1996, Sandoval_2018, Pradhan_2021}. Almost all the energy of the NS is emitted as GWs radiations with these modes. To study different modes of oscillations, one has to solve perturbed fluid equations in the vicinity of GR. Throne and Campollataro ~\cite{Throne_1967} first calculated the non-radial oscillation of NS in the framework of general relativity. Lindblom and Detweiler gave the first integrated numerical solution of the NS ~\cite{Lindblom_1983}. To solve the non-radial oscillations, one can also use the Cowling approximation, which is simpler than the Lindblom and Detweiler methods. In Cowling approximations, the space perturbations are neglected ~\cite{Cowling_1941}. The obtained frequencies within Cowling approximations differ by 10-30\% as compared to full linearized equations of GR. 

In this work, we want to calculate the non-radial oscillations of the NS using the Cowling approximations for a DM admixed hyperon star. To find different oscillations mode  frequencies, one has to solve the following coupled differential equations ~\cite{Sotani_2011, Flores_2014, Sandoval_2018, Pradhan_2021} 
%%%%%%%%%%%%%%%%%
\begin{eqnarray}
\frac{d W(r)}{dr}&=&\frac{d {\cal E}}{dP}\left[\omega^2r^2e^{\Lambda (r)-2\Phi (r)}V(r)+\frac{d \Phi(r)}{dr} W (r)\right] -l(l+1)e^{\Lambda (r)}V (r)
\nonumber \\
\frac{d V(r)}{dr} &=& 2\frac{d\Phi (r)}{dr} V (r)-\frac{1}{r^2}e^{\Lambda (r)}W (r).
\label{eq:wv}
\end{eqnarray}
%%%%%%%%%%%%%%
The functions $V (r)$ and $W (r)$ along with frequency $\omega$, characterize the Lagrange displacement vector  ($\eta$) associate to perturbed fluid,
%%%%%%%%%%%%%%%
\begin{eqnarray}
\eta=\frac{1}{r^2}\Big(e^{-\Lambda (r)}W (r),-V (r)\partial_{\theta},
-\frac{V(r)}{ \sin^{2}{\theta}}\  \partial _{\phi}\Big)Y_{lm},
\end{eqnarray}
%%%%%%%%%%%%%%%
where $Y_{lm}$ is the spherical harmonic, which is the function of $\theta$ and $\phi$. Solution of Eq. (\ref{eq:wv}) with the fixed background metric in Eq. (\ref{eq:spherical_metric}) near origin will behave as:
%%%%%%%%%%%%%%%%
\begin{equation}
W (r)=Br^{l+1}, \ V (r)=-\frac{B}{l} r^l,
\label{eq:bc1}
\end{equation}
%%%%%%%%%%%%%%
where $B$ is an arbitrary constant. At the surface of the star, the perturbation pressure must vanish, which provides another boundary condition as follows
%%%%%%%%%%%%%%%
\begin{equation} 
\omega^2 e^{\Lambda (R)-2\Phi (R)}V (R)+\frac{1}{R^2}\frac{d\Phi (r)}{dr}\Big|_{r=R}W (R)=0.
\label{eq:bc2}
\end{equation}
%%%%%%%%%%%%%
After solving Eq. (\ref{eq:wv}) with two boundary conditions (Eqs. (\ref{eq:bc1}) \& (\ref{eq:bc2})), one can get eigenfrequencies of the NS. In this work, we solve the oscillations equations Eq. (\ref{eq:wv}) by using the shooting method with some initial guess for $\omega^2$. The equations are integrated from the center to the surface and try to match the surface boundary conditions. After each integration, the initial guess of $\omega^2$ is corrected through the Secant method to get the desired precision, improving the initial guess.
%%%%%%%%%%%%%%%%%%%%%%%%%%%%%%%%%%%%%%%%%%%%%%%%%%%%%%%%%%%%%%%%%%%%%%%%%%%%%%%%%%%%
\subsection{Calculation of $f$-mode frequency as functions of different observables}
\begin{figure}
	\centering
	\includegraphics[width=0.7\textwidth]{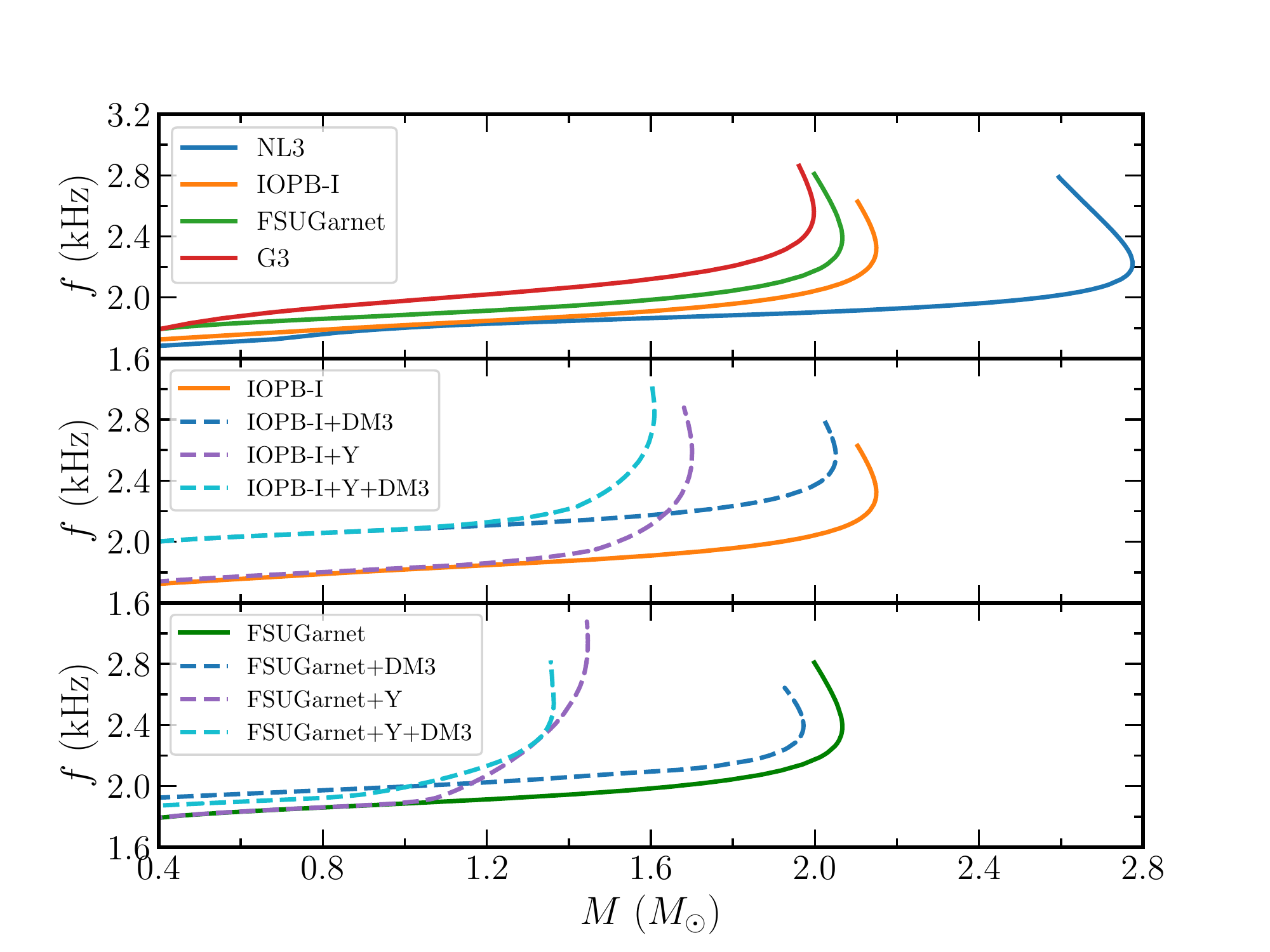}
	\caption{{\it Upper:} $f$-mode frequencies as a function of mass for four different RMF parameter sets. {\it Middle}: $f$-mode frequencies for IOPB-I set with DM3/hyperons and hyperons+DM3. {\it Lower}: Same as the middle one but for FSUGarnet.}
	\label{fig:fm2}
\end{figure}

%%%%%%%%%%%%%%%%%%%%%%%%%%%%%%%%%%%%%%%%%%%%%%%%%%%%%%%%%%%%%%%%%%%%%%%%%%%%%%%%%%%%
The $f$-mode frequencies (only for $l=2$)for different RMF EOSs are shown in Fig. \ref{fig:fm2}. In addition to this, the $f$-mode frequencies for IOPB+DM3, IOPB+Y, and IOPB-I+Y+DM are also shown in the middle panel of the figure. To see the parametric dependence of either DM/Y or DM+Y, we repeat the calculations with FSUGarnet as shown in the lower panel of Fig. \ref{fig:fm2}. 
%%%%%%%%%%%%%%
%%%%%%%%%%%%%
%%%%%%%%%%%%%%
\begin{figure}
	\centering
	\includegraphics[width=0.7\textwidth]{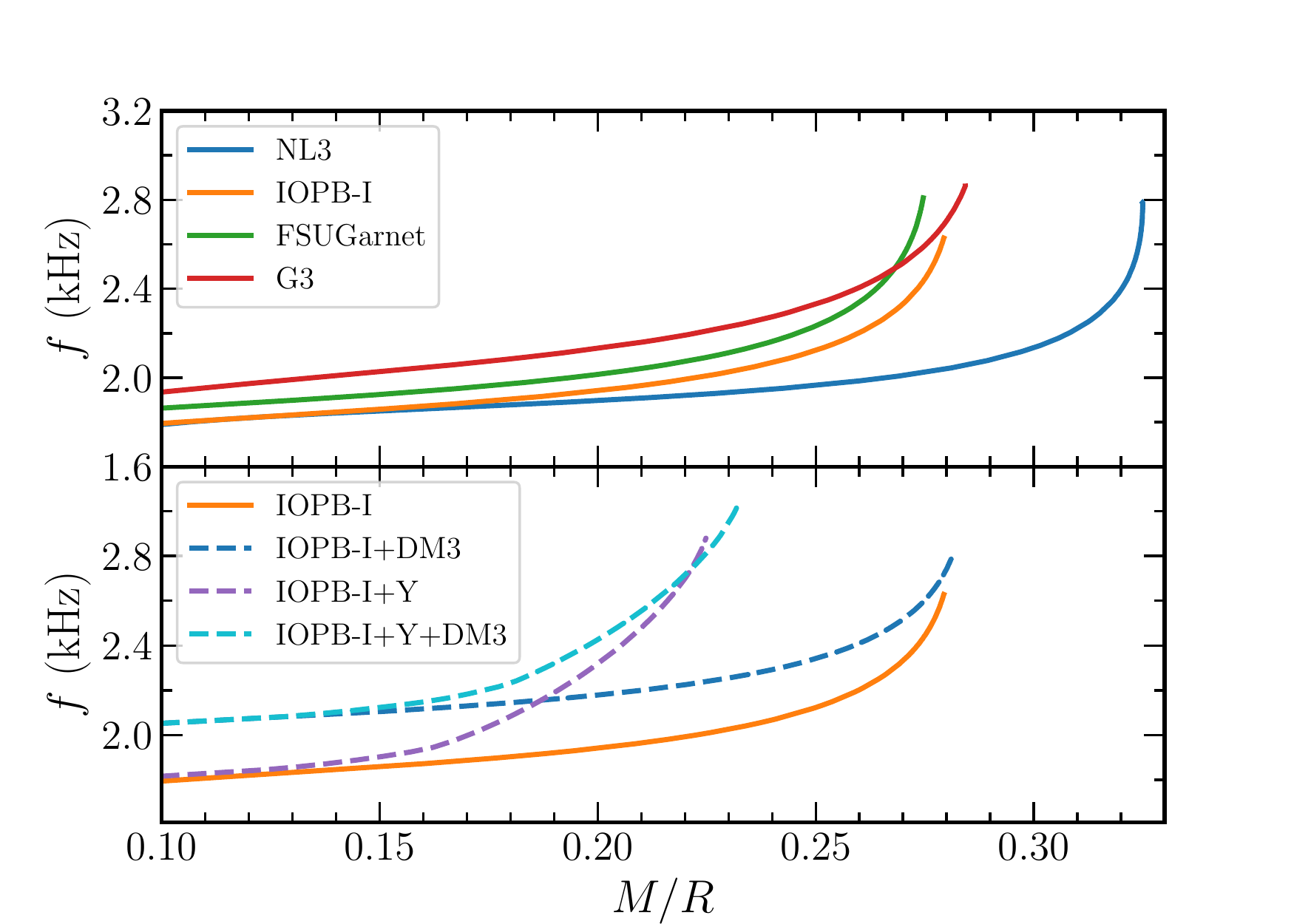}
	\caption{{\it Upper:} $f$-mode frequencies are shown for four different RMF parameter sets as a function of compactness. {\it Lower}: $f$-mode frequencies for IOPB-I set with DM3/hyperons and hyperons+DM3.}
	\label{fig:fc2}
\end{figure}
%%%%%%%%%%%%

The $f$-mode frequency corresponds to maximum mass ($f_{\rm max}$) for the four EOSs are shown in the upper panel of Fig. \ref{fig:fm2}. The $f_{\rm max}$ are 2.16, 2.32, 2.38, and 2.55 kHz for NL3, IOPB-I, FSUGarnet, and G3, respectively. The magnitude of $f_{\rm max}$ is maximum for G3 and minimum for NL3, which are in-between for IOPB-I and FSUGarnet sets. This is because G3 is a soft EOS, and NL3 is the stiffest EOS. In the middle panel of Fig. \ref{fig:fm2}, we observe a minor change of $f$-mode frequencies up to 1.3 $M_\odot$ for the IOPB-I case. There is no change in the $f$-mode frequencies up to $1 \ M_\odot$ for FSUGarnet as shown in the lower panel of Fig. \ref{fig:fm2}. The change in the $f$-mode frequencies is seen mainly at the core part. This is due to the presence of hyperons and DM particles in the dense region of the NS, which generally occurs in the central region. It is well known that the EOS is model-dependent, mainly in the core region. Also, the appearance of hyperons/DM is not possible in the lower-density region. Therefore, the change in $f-$mode frequency due to hyperon/DM particle is seen mainly at the core part rather than the crust.

The $f$-mode frequencies differ considerably with force parametrizations. The values of $f_{\rm max}$ are 2.32, 2.57, 2.58, and 2.85 kHz for IOPB-I, IOPB-I+DM3, IOPB-I+Y, and IOPB-I+Y+DM3, respectively. The EOSs are softer with the addition of either DM3/hyperons or DM3+hyperons than the original IOPB-I EOS. Therefore, the $f_{\rm max}$ value is more for IOPB-I+Y+DM3 in comparison to others. We tabulate the value of $f_{\rm max}$ and $f_{1.4}$ for four different RMF forces along with different NS observables in Table \ref{tab:table4}. The mass variation of frequency changes within interval $1.75-2.55$ kHz for $l=2$, which is almost consistent with Pradhan {\it et al.} ~\cite{Pradhan_2021} for the considered parameter sets. For hyperons/DM3 and hyperons+DM3 cases, the frequency range is within the interval of $1.8-2.85$ kHz. This range is higher in magnitude as compared without hyperons/DM cases. 

The compactness of a star is defined as ($C=M/R$), where $M$ and $R$ are the mass and radius of the star ~\cite{NKGb_1997, schaffner-bielich_2020}. Like $C$, another significant quantity is surface red-shift $Z_s$ is defined as ~\cite{NKGb_1997}
%%%%%%%%%%%%%%%%
\begin{equation}
Z_s =\frac{1}{\sqrt{1-\frac{2M}{R}}} -1 = \frac{1}{\sqrt{1-2C}}-1.
\end{equation}
%%%%%%%%%%%%%%
If we find the value of $Z_s$, one can constrain the mass and radius of the star. Till now, only one value of $Z_s=0.35$ is reported in the Ref. ~\cite{Cottam_2002} from the analysis of stacked bursts in the low-mass x-ray binary EXO 0748-676, which is also discarded by the subsequent observation ~\cite{Cottam_2008}.

\begin{figure}
	\centering
	\includegraphics[width=0.7\textwidth]{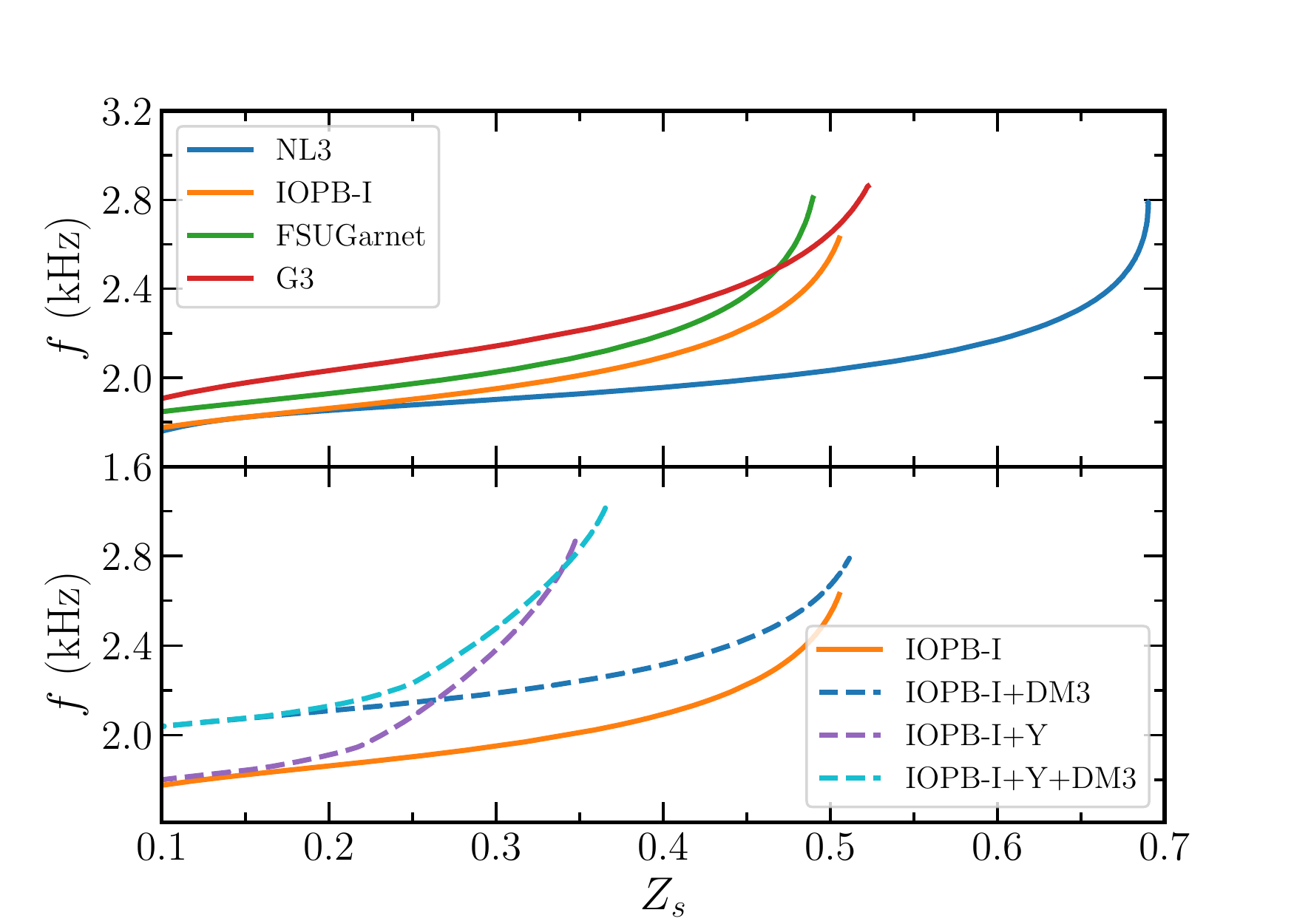}
	\caption{Same as Fig. \ref{fig:fc2}, but for $f$ as a function of $Z_s$.}
	\label{fig:fr2}
\end{figure}

We calculate $C$ and $Z_s$, which are shown in Figs. \ref{fig:fc2} and \ref{fig:fr2} with $f$-mode frequencies. We observed that the variation of $f$-mode frequencies for $C$ and $Z_s$ looks almost identical. This is because the $Z_s$ is the function of stellar compactness. The numerical values of $C$ and $Z_s$ correspond to both canonical and maximum mass NS in Table \ref{tab:table4}. In our case, we find the range of $C_{\rm max}$ and $Z_{s_{\rm max}}$ are $0.22-0.31$ and $0.33-0.62$, respectively, for the considered EOSs. The values of $C_{\rm max}$ and $Z_{s_{\rm max}}$ decrease as compared to only nucleonic EOSs.

The average density of a star is defined as $\bar{\rho}= \sqrt{\bar{M}/\bar{R^3}}$, where $\bar{M} = \frac{M}{1.4\ M_\odot} $ and $\bar{R}=\frac{R}{10 \ \mathrm{km}}$. We plot the $f$-mode frequencies with the variation of $\bar{\rho}$. On top of that, we plot some empirical fit relations (see Table \ref{tab:table3}) from previous calculation ~\cite{Anderson_1998, Benhar_2004, Pradhan_2021} including our fitting results.  
%%%%%%%%%%%%%%
%%%%%%%%%%%%%
%%%%%%%%%%%%%%%%%%%%%%%%%%%%%%%
\subsubsection{Fitting formula:-}
%%%%%%%%%%%%%%%%%%%%%%%%%%%%%%%
Andersson and Kokkotas (AK) first calculated the relation between $f$-mode frequency as a function of average density ~\cite{Anderson_1996}. They got an empirical relation by taking the polytropic EOSs as follows:
%%%%%%%%%%%%%%%%%
\begin{equation}
f(\mathrm{kHz})\approx 0.17+2.30\sqrt{\left(\frac{M}{1.4 \ M_\odot}\right)\left(\frac{10 \ \mathrm{km }}{R}\right)^3}.
\label{eq:AK1}
\end{equation}
%%%%%%%%%%%%%%%%
This empirical relation is again modified by their subsequent paper for the realistic EOSs ~\cite{Anderson_1998}, which is given as 
\begin{equation}
f(\mathrm{kHz})\approx 0.78+1.635\sqrt{\frac{\bar{M}}{\bar{R^3}}}.
\label{eq:AK2}
\end{equation}
%%%%%%%%%%%%%%
%%%%%%%%%%%%%%
\begin{figure}
	\centering
	\includegraphics[width=0.7\textwidth]{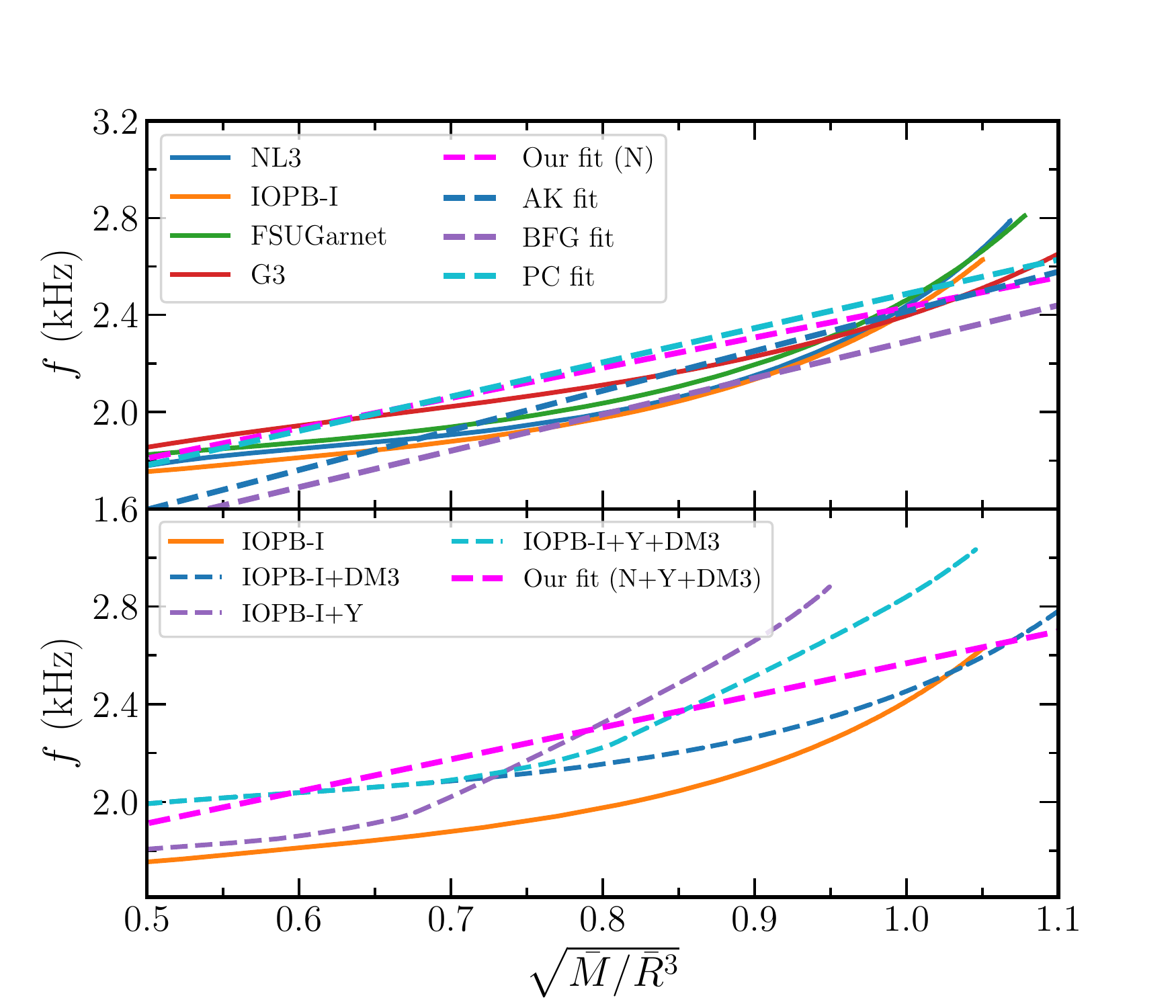}
	\caption{{\it Upper:} $f$-mode frequency as a function of average density for four parameter sets. The dashed lines with different colors are the fitted relations taken from different analyses, including ours. {\it Lower:} The same relations but for nucleonic EOSs along with DM3/Y and DM3+Y.}
	\label{fig:fd2}
\end{figure}
%%%%%%%%%%%%%
%%%%%%%%%%%%%%%%%
\begin{figure}
	\centering
	\includegraphics[width=0.7\textwidth]{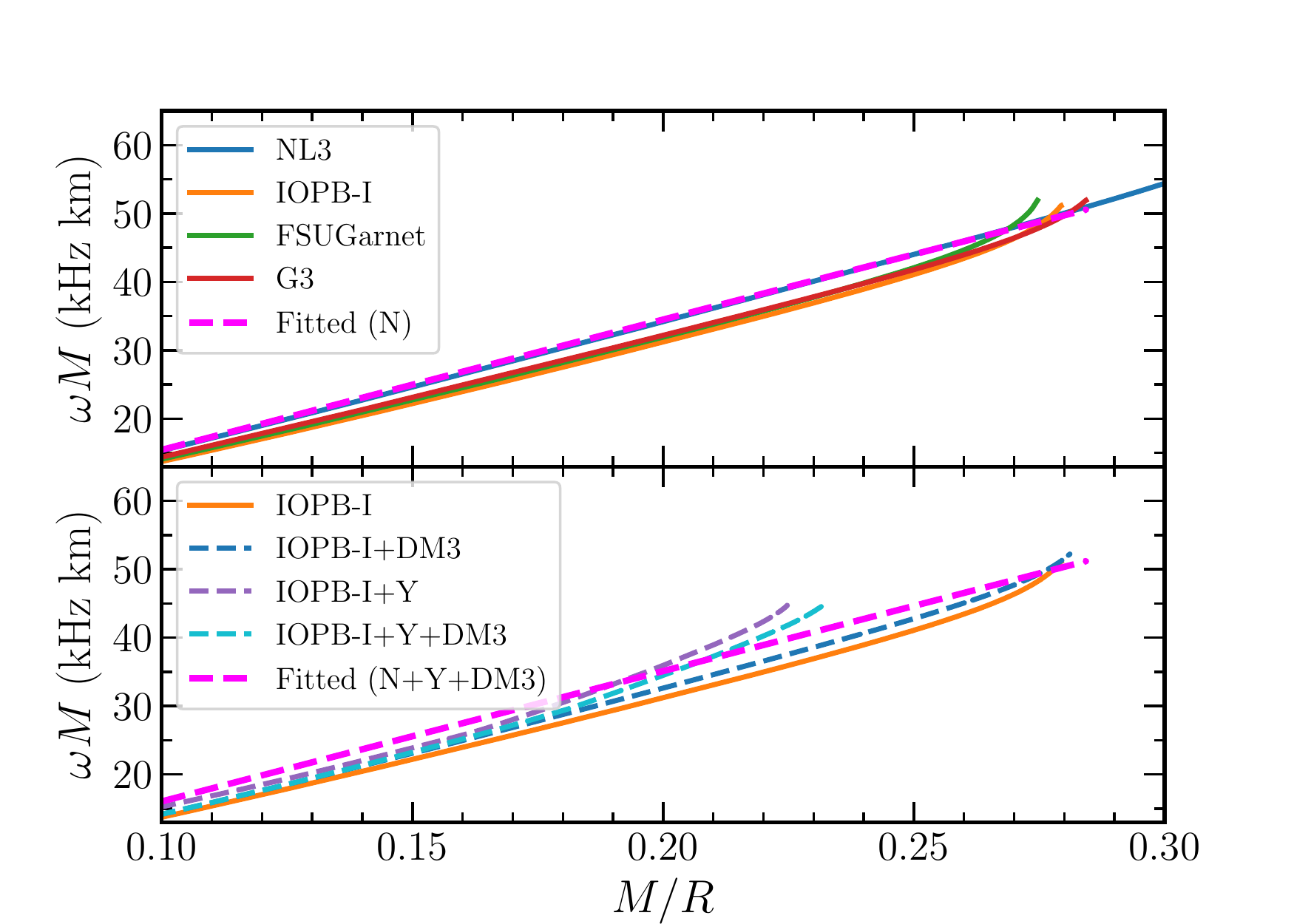}
	\caption{{\it Upper:} Angular frequencies ($\omega=2\pi f$) scaled by mass ($\omega M$) are shown for four different RMF parameter sets. {\it Lower}: $\omega M$ for only IOPB-I with a yellow line, IOPB-I+DM3 with a blue dashed line, IOPB-I+Y with a purple dashed line, IOPB-I+Y+DM with a cyan dashed line.}
	\label{fig:fom2}
\end{figure}	
%%%%%%%%%%%%%
Later, these empirical relations were modified by Benhar, Ferrari, and Gualtieri (BFG) ~\cite{Benhar_2004} by using hybrid EOSs. Pradhan and Chatterjee (PC) ~\cite{Pradhan_2021} have put an empirical relation using hyperonic EOSs. We tabulate the coefficients of the empirical relations from different analyses in Table \ref{tab:table3}. We also find the fitting relations (i) for only nucleon EOSs (N) and (ii) for nucleon EOSs with DM/Y and Y+DM (N+Y+DM). The fitting coefficients $a$ and $b$ for different analyses are given in Table \ref{tab:table3} for comparison.

If we measure the $f$-mode frequency of a star, then one can infer the mass and radius of the NS using the fit relations, which can be used to constrain the NS EOS ~\cite{Anderson_1996}. Several approaches have already tried to find the universal relations between $f$-mode frequency and compactness/average density ~\cite{Anderson_1996, Anderson_1998, Benhar_2004, Tsui_2005, Lau_2010, Wen_2019}. However, in this study, our aim is to explore those relations for DM admixed hyperon stars. 
%%%%%%%%%%%%%
\begin{table}
	\centering
	\caption{Empirical relations, $f(\mathrm{kHz})=a+b\sqrt{\frac{\bar{M}}{\bar{R^3}}}$ between $f$-mode frequency and average density from different works for $l=2$ mode, where $a$ and $b$ are fitting coefficients.}
	\label{tab:table3}
	\scalebox{1.0}{
		\begin{tabular}{|l|l|l|}
			\hline
			Different works & a(kHz) & b(kHz) \\ \hline
			Our fit (N)      & 1.185       & 1.246      \\ \hline
			Our fit (N+Y+DM) & 1.256       & 1.311  \\ \hline
			AK  fit      & 0.78        & 1.635      \\ \hline
			BFG fit      & 0.79        & 1.500      \\ \hline
			PC fit       & 1.075       & 1.412      \\ \hline 
	\end{tabular}}
\end{table}
%%%%%%%%%%%%%
%%%%%%%%%%%%%%%%%%%%%%%%%%%%%%%%%%%%%%
\subsubsection{Universal relations:-}
\begin{figure}
	\centering
	\includegraphics[width=0.7\textwidth]{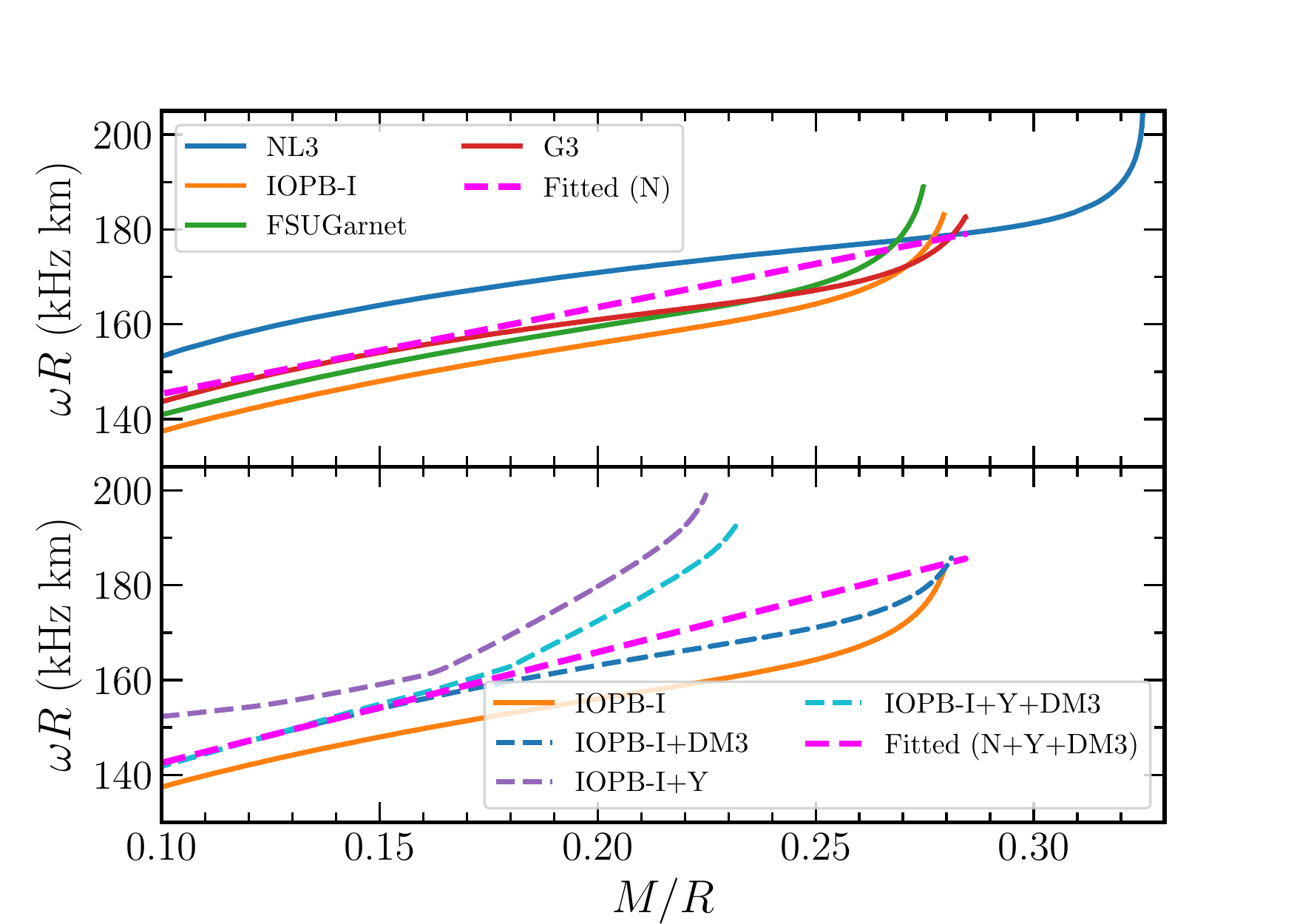}
	\caption{Same as Fig. \ref{fig:fom2}, but for $\omega R$ with $M/R$.}
	\label{fig:for2}
\end{figure}

%%%%%%%%%%%%%%%%%%%%%%%%%%%%%%%%%%%%%%
Different correlations have been seen between different modes, such as $f$, $p$, and $g$, with compactness/average density. All these relations are relatively independent of the EOSs. In Ref. ~\cite{Sotani_2011}, it is observed that the mass-scaled angular frequency ($\omega M$) as a function of compactness was found to be universal for $g$-mode frequency. The same type of correlation between $\omega M$ as a function of compactness for $p$ and $w$ modes have been proposed in Ref. ~\cite{Salcedo_2014} and for $f$-modes see Ref. ~\cite{Wen_2019}. Thus, we have checked these universal relations as a function of compactness, as shown in Fig. \ref{fig:fom2} and Fig. \ref{fig:for2}. We find a universal relation between $\omega M$ and $M/R$. However, the correlation is weaker in the case of $\omega R$ with $M/R$. The Universal relations are given as follows:
%%%%%%%%%%%%%%%%
\begin{eqnarray}
\omega M (\mathrm{kHz \ km}) &=& a_i\left(\frac{M}{R}\right)-b_i,
\label{eq:uni1}
\\
\omega R (\mathrm{kHz \ km}) &=& c_j\left(\frac{M}{R}\right)+d_j,
\label{eq:uni2}
\end{eqnarray}
%%%%%%%%%%%%%%
where $a_i$, $b_i$, $c_j$ and $d_j$ are the fitting co-efficients in kHz km. The value of $i=1$ for nucleon (N) and $i=2$ for N+Y+DM3. The coefficients are $a_1=190.447$, $b_1=4.538$ and $a_2=190.475$, $b_2=2.984$ for $\omega M\sim M/R$. The values of $c_1=182.585$, $d_1=127.1$ and $c_2=233.596$, $d_2=119.18$ for $\omega R\sim M/R$.
%%%%%%%%%%%%%%
%%%%%%%%%%%%%%
%%%%%%%%%%%%%%
\begin{figure}
	\centering
	\includegraphics[width=0.7\textwidth]{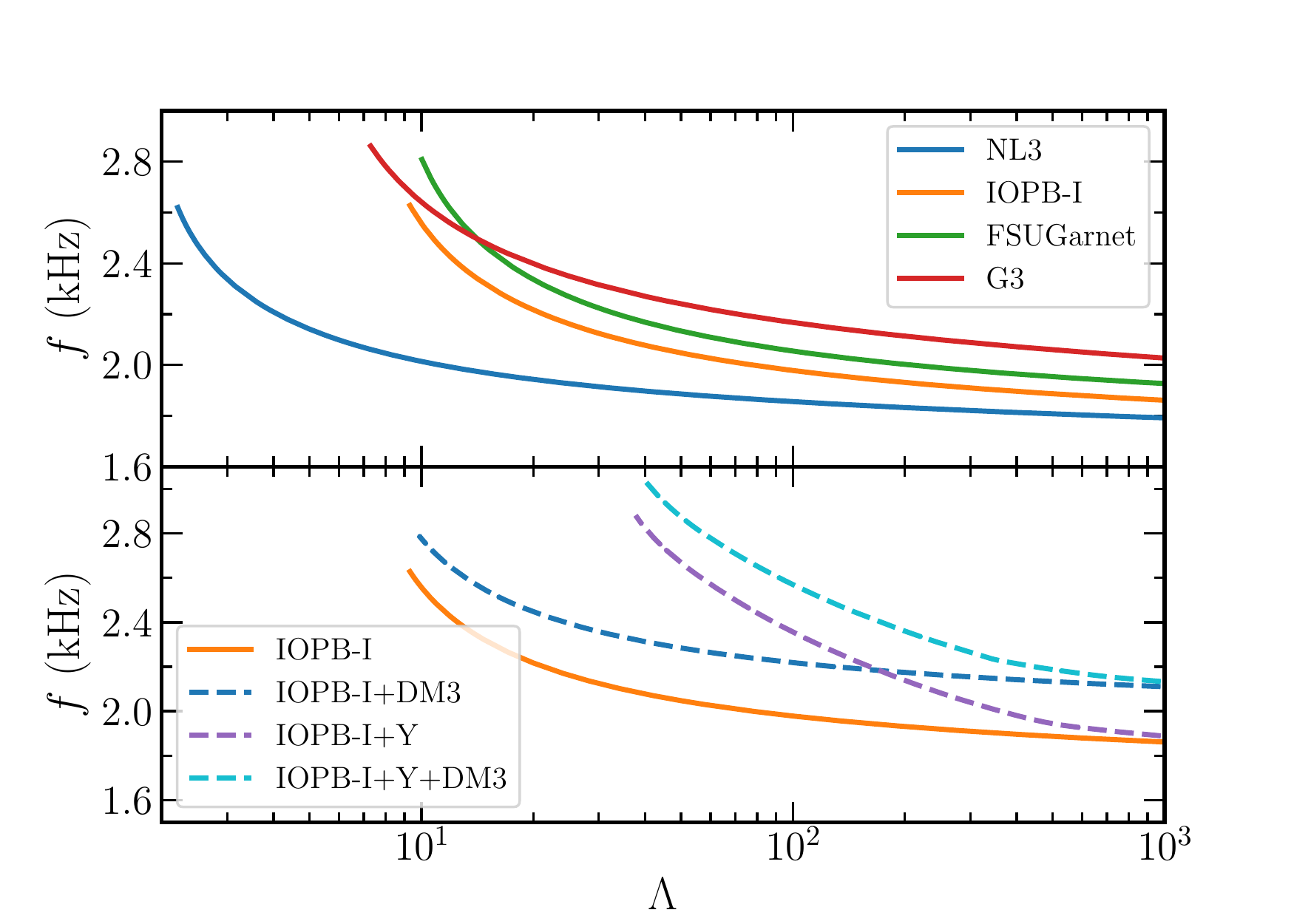}
	\caption{{\it Upper:} $f$-mode frequencies are shown with the variation of $\Lambda$ for four different RMF parameter sets. {\it Lower}: $f$-mode frequencies with $\Lambda$ for IOPB-I, IOPB-I+DM3, IOPB-I+Y, and IOPB-I+Y+DM.}
	\label{fig:tidal}
\end{figure}
%%%%%%%%%%%

The tidal deformability is the most important quantity, which gives important information about the NS EOSs. The GW170817 event has limited the tidal deformability of the canonical NS, which is used to constrain the EOS of neutron-rich matter at 2--3 times the nuclear saturation densities ~\cite{Abbott_2017, Abbott_2018}. Due to a strong dependence on the tidal deformability with radius ($\Lambda \sim R^5$), it can put stringent constraints on the EOS. Several approaches ~\cite{Bauswein_2017,Annala_2018,Fattoyev_2018,Radice_2018,Mallik_2018,Most_2018,Tews_2018,Nandi_2019,Capano_2020} have been tried to constraint the EOS on the tidal deformability bound given by the GW170817. The dimensionless tidal deformability $\Lambda$ is defined as~\cite{Hinderer_2008,Hinderer_2009, Hinderer_2010}
%%%%%%%%%%%%%%%%
\begin{eqnarray}
\Lambda = \frac{2}{3}k_2 \ C^{-5},
\label{eq:tidal}
\end{eqnarray}
%%%%%%%%%%%%%%
where $k_2$ is the second Love number, which depends on the internal structure and mass and radius of a star ~\cite{Hinderer_2008, Hinderer_2009, Kumartidal_2017, DasBig_2021}. 

We calculate the $f$-mode frequencies as a function of $\Lambda$ for different RMF EOSs, including hyperons and DM. The value of $\Lambda$ decreases when one goes from stiff to soft EOSs, and the corresponding $f$-mode frequencies increase contrary to $\Lambda$. The numerical values of $\Lambda_{1.4}$ are given in Table \ref{tab:table4} for different EOSs. The GW170817 event put constraint on $\Lambda_{1.4}=190_{-120}^{+390}$ ~\cite{Abbott_2018}. In Ref. ~\cite{Wen_2019}, they have put a limit on the value of $f$-mode frequencies for $1.4 M_\odot$ are 1.67--2.18 kHz by combining two constraints: the EOS parameter space allowed by terrestrial nuclear experiments and the tidal deformability data from GW170817. Our predicted results of $\Lambda$ are consistent with Wen {\it et al.} \cite{Wen_2019}, except for the case of IOPB-I+Y+DM3. More NS mergers are expected to be measured in the future, which may constrain the $f$-mode frequency more tightly.
%%%%%%%%%%%%%%
\begin{table}
	\centering
	\caption{The observables such as $M_{\rm max}$ ($M_\odot$), $R_{\rm max}$ (km), $R_{1.4}$ (km), $f_{\rm max}$ (kHz), $f_{1.4}$  (kHz), $C_{\rm max}$, $C_{1.4}$, $Z_{s_{\rm max}}$, $Z_{s_{1.4}}$, $\Lambda_{\rm max}$, $\Lambda_{1.4}$ are given for NL3, G3, FSUGarnet, IOPB-I, IOPB-I+DM3, IOPB-I+Y, IOPB-I+Y+DM3 both for maximum mass and canonical NS.}
	\label{tab:table4}
	\renewcommand{\tabcolsep}{0.2cm}
	\renewcommand{\arraystretch}{1.2}
	\scalebox{0.8}{
		\begin{tabular}{|l|l|l|l|l|l|l|l|}
			\hline 
			Observable      & NL3   & G3    & FSUGarnet & IOPB-I & IOPB-I+DM3 & IOPB-I+Y & IOPB-I+Y+DM3 \\ \hline
			$M_{\rm max}$       & 2.77  & 1.99   & 2.07  & 2.15  & 2.05  & 1.70  & 1.61  \\ \hline
			$R_{\rm max}$       & 13.17 & 10.79  & 11.57 & 11.76 & 11.04 & 11.41 & 10.38 \\ \hline
			$R_{1.4}$       & 14.08 & 12.11  & 12.59 & 12.78 & 11.76 & 12.81 & 11.57 \\ \hline
			$f_{\rm max}$       & 2.16  & 2.55   & 2.38  & 2.32  & 2.57  & 2.58  & 2.85  \\ \hline
			$f_{1.4}$       & 1.78  & 2.06   & 1.94  & 1.87  & 2.14  & 1.92  & 2.22  \\ \hline
			$C_{\rm max}$       & 0.31  & 0.27   & 0.26  & 0.27  & 0.27  & 0.22  & 0.23  \\ \hline
			$C_{1.4}$       & 0.15  & 0.17   & 0.16  & 0.16  & 0.18  & 0.16  & 0.18  \\ \hline
			$Z_{s_{\rm max}}$   & 0.62  & 0.48   & 0.45  & 0.47  & 0.49  & 0.33  & 0.36  \\ \hline
			$Z_{s_{1.4}}$   & 0.18  & 0.23   & 0.22  & 0.21  & 0.24  & 0.21  & 0.25  \\ \hline
			$\Lambda_{\rm max}$ & 4.48  & 12.12  & 17.67 & 14.78 & 14.39 & 59.58 & 50.58 \\ \hline
			$\Lambda_{1.4}$ & 1267.79 & 461.28& 624.81 & 681.27 & 471.06 & 650.55 & 391.68 \\ \hline
	\end{tabular}}
\end{table}
%%%%%%%%%%%%%%
%%%%%%%%%%%%%%
\begin{figure}
	\centering
	\includegraphics[width=0.7\textwidth]{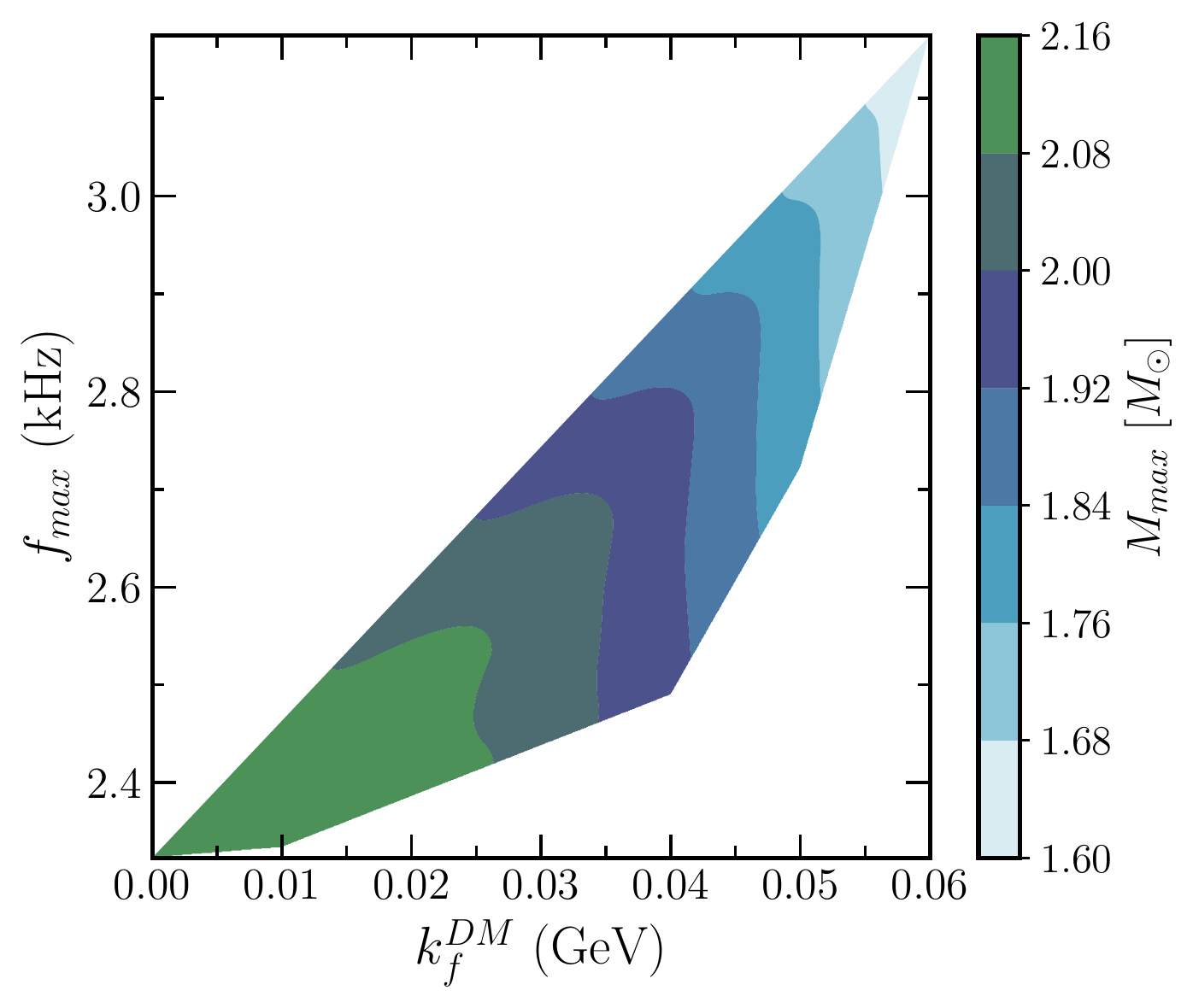}
	\caption{$f_{\rm max}$ are shown with different DM momenta for IOPB-I parameter sets. The color scheme represents the maximum masses corresponding to different $k_f^{\rm DM}$. }
	\label{fig:fm_dm2}
\end{figure}
%%%%%%%%%%%%%%
\begin{figure}
	\centering
	\includegraphics[width=0.7\textwidth]{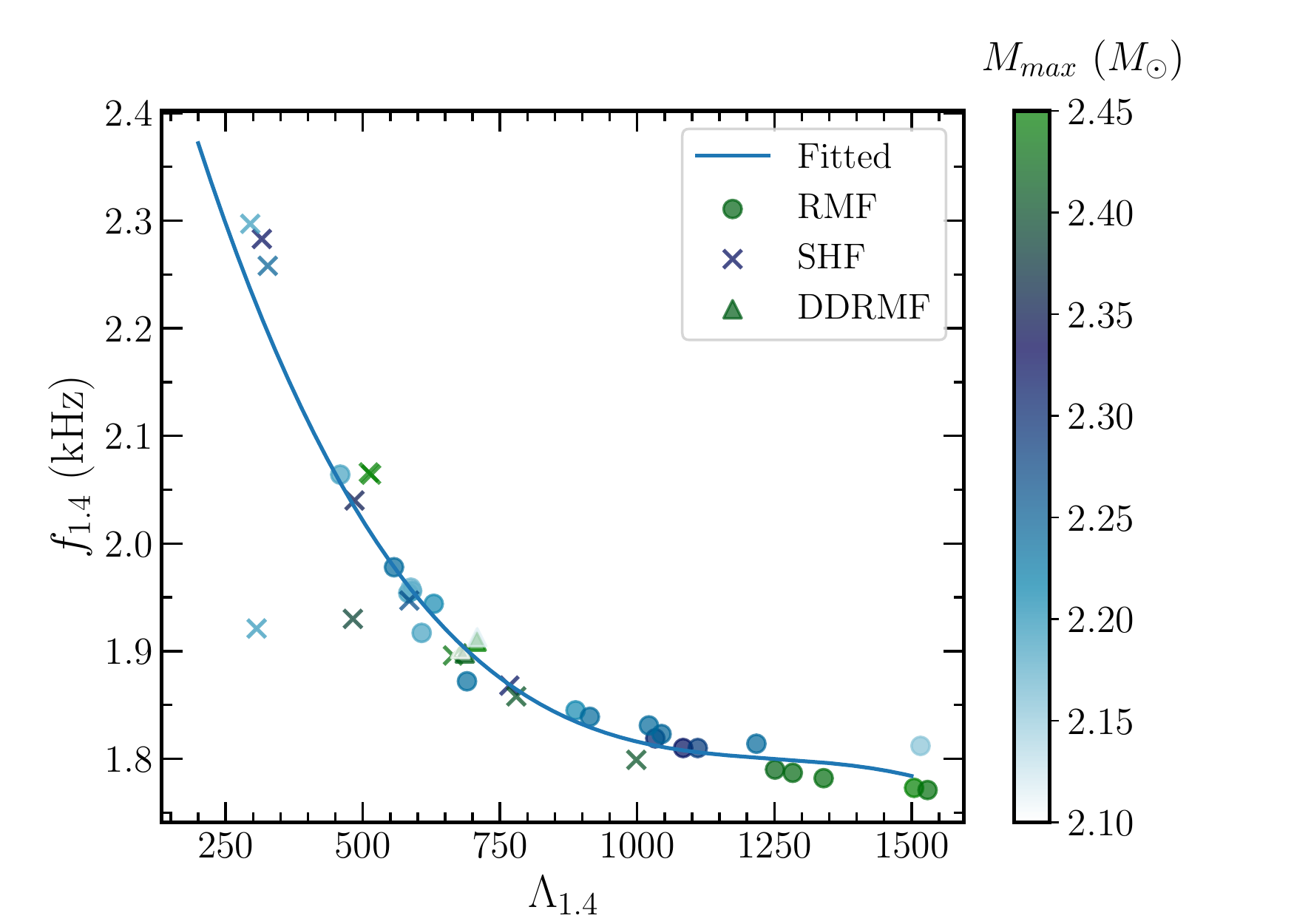}%
	\caption{Correlation between $f_{1.4}$ and $\Lambda_{1.4}$ are shown for 41 parameter sets, including (RMF, SHF, and DDRMF). The color bar represents the maximum masses of the corresponding parameter sets.}
	\label{fig:fCLC}
\end{figure}
%%%%%%%%%%%%
%%%%%%%%%%%%%%
\begin{figure}
	\centering
	\includegraphics[width=0.7\textwidth]{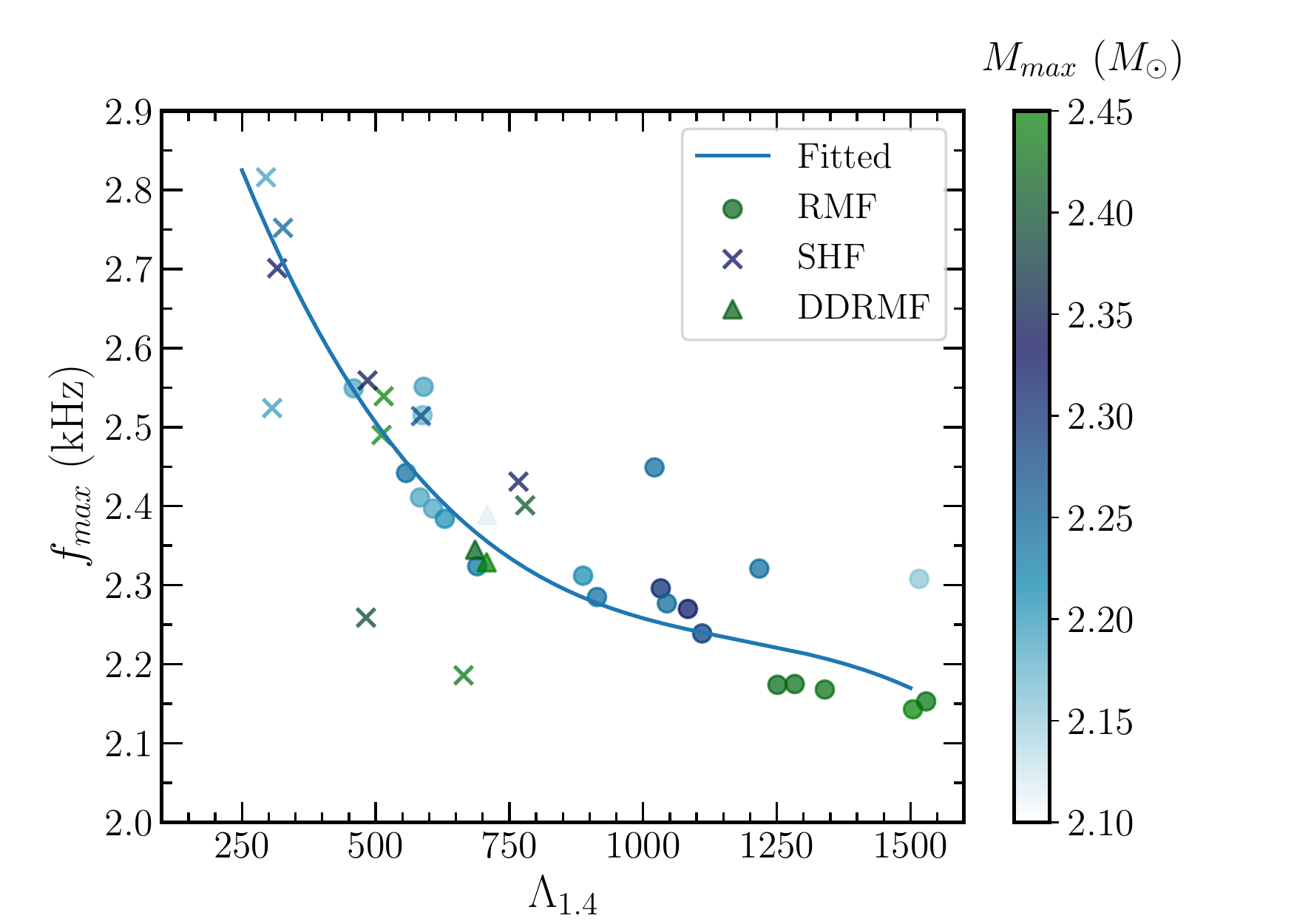}
	\caption{Same as Fig. \ref{fig:fCLC}, but for $f_{\rm max}$ and $\Lambda_{1.4}$. }
	\label{fig:fmLC}
\end{figure}
%%%%%%%%%%%%%%

In Fig. \ref{fig:fm_dm2}, we plot the values of $f_{\rm max}$ with different DM Fermi momenta. The $f$-mode frequencies increase with increasing $k_f^{\rm DM}$. The range of $f_{\rm max}$ is 2.3--2.4 kHz corresponds to $k_f^{\rm DM}=0.00-0.025$ GeV; after that, it increases with more proportion beyond $k_f^{\rm DM}=0.03$ GeV. The color bar represents the maximum masses ($M_{\rm max}$) of the NS for different $k_f^{\rm DM}$. The area of $M_{\rm max}$ at lower $k_f^{\rm DM}$ (green color) is found to be more as compared to the higher $k_f^{\rm DM}$, and the area slightly reduces from lower $k_f^{\rm DM}$ to a higher one. This is because the slight increase in DM percentage reduces the value $M_{\rm max}$ a little bit. When we increase the DM percentage by more than 0.03 GeV, the maximum mass reduces significantly. For example, for $k_f^{\rm DM}=$ 0.00, 0.01, and 0.02 GeV, the values of $M_{\rm max}$ are 2.149, 2.146, and 2.119 $M_\odot$ respectively. If we increase the DM Fermi momenta $k_f^{\rm DM}=$ 0.03, 0.04, 0.05, and 0.06 GeV, the values of $M_{\rm max}$ are 2.051, 1.938, 1.788, and 1.614 $M_\odot$ respectively.
%%%%%%%%%%%%%%%%%%%%%%%%%%%%%%%%%%%%%%%%%%%%%%%%%%%%%%%%%%%%%%%%%%
\subsection{Correlations between $f$-mode and tidal deformability}
%%%%%%%%%%%%%%%%%%%%%%%%%%%%%%%%%%%%%%%%%%%%%%%%%%%%%%%%%%%%%%%%%%
We also find some correlations between $f_{1.4}$--$\Lambda_{1.4}$, and  $f_{\rm max}$--$\Lambda_{1.4}$ which are shown in Figs. \ref{fig:fCLC} and \ref{fig:fmLC}. This correlation is first calculated by Wen {\it et al.} ~\cite{Wen_2019} by taking 23000 phenomenological EOSs, 11 microscopic EOSs, and two quark EOSs. Here, in our calculation, we use 23 RMF EOSs, 14 Skyrme-Hartree-Fock (SHF) forces, and four density-dependent (DDRMF) sets are given in Ref. ~\cite{Biswal_2020} having mass more than $\sim 2 \ M_\odot$. We get a correlation between $f_{1.4}$--$\Lambda_{1.4}$ and  $f_{\rm max}$--$\Lambda_{1.4}$ as shown in figure. The accurately measured tidal deformability of the star from the GW analysis can be used as a constraint on the $f$-mode frequency of a star. We expect that future GW observation may answer this type of correlation.
%%%%%%%%%%%%%%%%%%%%%
\section*{Conclusion}
%%%%%%%%%%%%%%%%%%%%%
In this chapter, we have calculated the $f$-mode oscillations of the DM admixed hyperon stars within the relativistic Cowling approximations. Hyperons are appeared inside the NS, mainly at the core part. To compute the hyperon interactions mediated by the mesons, we used SU(6) model to obtain the coupling between hyperons-vector mesons. However, the coupling between hyperons-scalar mesons has been obtained by fitting with the depth of the hyperon potential. For DM admixed hyperons stars, the DM particles interact with nucleons as well as hyperons by exchanging the Higgs. The hyperons-Higgs form factor is assumed to be the same as nucleons, and the available theoretical data constrain its value.

The $f$-mode frequencies of the DM admixed hyperon star for various EOSs are computed. The magnitude of $f$-mode frequencies decreases for softer EOSs because it predicts smaller mass as compared to stiffer ones. The angular frequencies scaled by mass and radius are also calculated as the function of compactness. It has been found that both quantities follow a linear relationship. The variation of $f$-mode frequencies as a function of tidal deformability is calculated for a large variety of EOSs. We obtained the correlations between $f_{1.4} – \Lambda_{1.4}$ and $f_{\rm max} – \Lambda_{\rm max}$ which are almost consistent with the earlier prediction \cite{Wen_2019}. The predicted range of $f_{1.4}$ is $1.78-2.22$ kHz for the considered EOSs. We hope the discovery of more BNS merger events will open up tight constraints on the $\Lambda$ in the future, which will put more stringent constraints on $f$-mode frequency.
%%%%%%%%%%%%%%%%%%%%%%%%%%%%% END %%%%%%%%%%%%%%%%%%%%%%%%%%%%%%%
%\blankpage 
%%%%%%%%%%%%%%%%%%%%%%%%% Chapter-6 %%%%%%%%%%%%%%%%%%%%%%%%%%%%%%%
%%%%%%%%%%%%%%%%%%%%%%% CHAPTER - 6 %%%%%%%%%%%%%%%%%%%%
\chapter{Effects of dark matter on binary neutron star properties}
\label{C6} 
%%%%%%%%%%%%%%%%%%%%%%%%%%%%%%%%%%%%%%%%%%%%%%%%%%%%%%%%%%
Neutron stars (NSs) in binary systems can interact through their gravitational fields, resulting in tidal deformability, a measure of the degree to which the tidal forces distort the shape of the NS. In this study, we calculate the tidal Love numbers and corresponding deformabilities for NSs containing dark matter (DM) for different levels of deformation. During the late inspiral stage of the binary, the NSs emit large amounts of gravitational waves as the separation distance decreases. Using the post-Newtonian formalism, we compute various inspiral properties such as gravitational wave frequency, inspiral phase, last time in the inspiral stage, strain amplitudes, and post-Newtonian parameters. With the detection of gravitational waves from the merger of binary NS in 2017 (GW170817 event), this study offers theoretical insights into exploring the properties of binary NS, especially if they contain some amount of DM.
%%%%%%%%%%%%%%%%%%%%%%%%%%%%%%%%%%%%%%%%
\section{Binary Neutron Star Properties}
%%%%%%%%%%%%%%%%%%%%%%%%%%%%%%%%%%%%%%%%
%%%%%%%%%%%%%%%%%%%%%%%%%%%%%%%%%%%%%%%%%%%%%%%%
\subsection{Tidal Love number and deformability}
\label{form:tidal}
%%%%%%%%%%%%%%%%%%%%%%%%%%%%%%%%%%%%%%%%%%%%%%%%
In a BNS system, tidal interactions come into the picture when the two stars are in close orbit with each other. The mutual gravitational interaction increases with time because the inspiralling stars emit gravitational radiation due to their acceleration in the gravitational field. The shape of the star becomes increasingly tidally deformed. Analogous to electromagnetic radiation, here also, the gravitational field is generated from the two types of tidal fields, i.e., (i) electric (even parity or polar) ${\cal E}_L$ and (ii) magnetic (odd parity or axial) ${\cal M}_L$ fields,  where $L$ represents the space indices. These two fields induce a mass multipole moment ($Q_{L}$) and a current multipole moment ($S_{L}$), which are defined \mbox{as \cite{Damour_2009, Chamel_2021}}:
%%%%%%%%%%%%%
\begin{align}
Q_L=\lambda_l {\cal E}_L,
\label{eq:Q_L}
\end{align}
%%%%%%%%%%%
%%%%%%%%%%%%%
\begin{align}
S_L=\sigma_l {\cal M}_L,
\label{eq:S_L}
\end{align}
%%%%%%%%%%%
where $\lambda_l$ and $\sigma_l$ are the gravitoelectric and gravitomagnetic tidal deformability of order $l$. The mass and current multipole moments come into the picture in general relativity compared to Newtonian dynamics, where only the mass multipole moment plays a role. The gravitoelectric ($k_l$) and gravitomagnetic Love numbers ($j_l$) are related with the deformability parameters $\lambda_l$ and $\sigma_l$, respectively, as \cite{Chamel_2021}:
%%%%%%%%%%%%%
\begin{align}
k_l = \frac{(2l-1)!!}{2}\frac{\lambda_l}{R^{2l+1}},
\label{eq:k_l}
\end{align}
%%%%%%%%%%%
%%%%%%%%%%%%%
\begin{align}
j_l = 4(2l-1)!!\frac{\sigma_l}{R^{2l+1}}.
\label{eq:j_l}
\end{align}
%%%%%%%%%%%
The dimensionless tidal deformability parameters correspond to electric $\Lambda_l$ and magnetic $\Sigma_l$ multipole moments, which help express the tidal deformability of the BNS. These quantities are often used conveniently in various measurements and related to the Love numbers and compactness of the star $C$ with the relation \cite{Chamel_2021}:
%%%%%%%%%%%%% 
\begin{align}
\Lambda_l=\frac{\lambda_l}{M^{2l+1}}=\frac{2}{(2l-1)!!}\frac{k_l}{C^{2l+1}}
\label{eq:Lambda_l}
\end{align}
%%%%%%%%%%%    
%%%%%%%%%%%%%    
\begin{align}
\Sigma_l= \frac{\sigma_l}{M^{2l+1}}=\frac{1}{4(2l-1)!!}\frac{j_l}{C^{2l+1}}.
\label{eq:Sigma_l}
\end{align}
%%%%%%%%%%%
As discussed earlier in this paper, the leading order perturbation of the tidal deformations is the gravitoelectric and gravitomagnetic quadrupole deformations $\lambda_2$ and $\sigma_2$. However, the higher-order deformations are also quite substantial due to the close orbital separation of the binary system and at the merging condition. All these tidal deformations can be expressed with the various Love numbers $k_l$ and $j_l$ as defined in Eqs. (\ref{eq:k_l}) and (\ref{eq:j_l}).  To calculate these tidal Love numbers along with the dimensionless tidal deformability $\Lambda_l$ and $\Sigma_l$, we adopt the formalism developed in \mbox{Ref. \cite{Chamel_2021}}. The gravitoelectric perturbations are evaluated from the differential equation in terms of metric functions ($H$), which are written as:
%%%%%%%%%%%%%
\begin{align}
H_\ell''(r)&+ H_\ell'(r) \biggl[1-\frac{2m(r)}{r}\biggr]^{-1} \Bigg\{\frac{2}{r} - \frac{2 m(r)}{r^2} - 4\pi r \left[{\cal E}(r) -  P(r)\right] \Bigg\}
\nonumber  \\
&
+H_\ell(r) \biggl[1-\frac{2m(r)}{r}\biggr]^{-1} \times \Bigg\{4\pi \bigg[5{\cal E}(r)+ 9P(r) +\dfrac{d{\cal E}}{dP} \left[{\cal E}(r)  + P(r)\right] \bigg]
\nonumber  \\
&
-\frac{\ell(\ell+1)}{r^2} - 4\biggl[1-\frac{2m(r)}{r}\biggr]^{-1} \biggl[\frac{m(r)}{r^2} + 4\pi\, r\,  P(r)\biggr]^2 \Bigg\} = 0, \;
\label{eq:H_eq}
\end{align}
%%%%%%%%%%%
and the gravitomagnetic perturbations are obtained by solving the differential alignment, which is given as:
%%%%%%%%%%%%%
\begin{align}
\widetilde{H}_{\ell}''(r)&-\widetilde{H}_{\ell}'(r) \biggl[1-\frac{2 m(r)}{r}\biggr]^{-1} \times 4\pi r  \left[ P(r) + {\cal E} (r) \right]
\nonumber \\
&
-\widetilde{H}_{\ell}(r) \biggl[1-\frac{2 m(r)}{r}\biggr]^{-1} \times \left\{ \frac{\ell(\ell+1)}{r^2} - \frac{4m(r)}{r^3} +8 \pi \theta  \left[ P(r) + {\cal E} (r) \right] \right\} = 0,
\label{eq:Ht_eq}
\end{align}
%%%%%%%%%%%
%%%%%%%%%%%%%
\begin{align}
y(R)=\frac{R\beta(R)}{H_{\ell}(R)}, \ {\rm and} \ \widetilde{y}(R)=\frac{R\widetilde{\beta}(R)}{\widetilde{H}_{\ell}(R)}.
\end{align}
%%%%%%%%%%%
The gravitoelectric Love numbers $k_2$, $k_3$ and $k_4$ are given as \cite{Damour_2009, Hinderer_2008, Hinderer_2010, Kumar_2017, Chamel_2021}:
%%%%%%%%%%%%%
\begin{align}
k_2 &=  \frac{8}{5} C^5 (1-2C)^2 \big[ 2(y_2-1)C - y_2 + 2 \big]
\nonumber \\
&
\times \Big\{ 2C \big[ 4(y_2+1)C^4 + 2(3y_2-2)C^3 - 2(11y_2-13)C^2 + 3(5y_2-8)C 
\nonumber \\
&
- 3(y_2-2) \big] + 3(1-2C)^2 \big[ 2(y_2-1)C-y_2+2 \big] \log(1-2C) \Big\}^{-1} \, ,
\label{eq:k2}
\end{align}
%%%%%%%%%%%
%%%%%%%%%%%%%
\begin{align}
k_3 &=  \frac{8}{7} C^7 (1-2C)^2 \big[ 2(y_3-1)C^2 - 3(y_3-2)C + y_3 - 3 \big]
\nonumber \\
&
\times \Big\{ 2C \big[ 4(y_3+1)C^5 + 2(9y_3-2)C^4 - 20(7y_3-9)C^3 + 5(37y_3-72)C^2 
\nonumber \\ 
&
- 45(2y_3-5)C + 15(y_3-3) \big] + 15(1-2C)^2 \big[ 2(y_3-1)C^2 - 3(y_3-2)C 
\nonumber \\ 
&
+ y_3 - 3 \big] \log(1-2C) \Big\}^{-1}  \, ,
\label{eq:k3}
\end{align}
%%%%%%%%%%
and
%%%%%%%%%%%%%
\begin{align}
k_4 &= \frac{32}{147} C^9 (1-2C)^2 \big[ 12(y_4-1)C^3 - 34(y_4-2)C^2 + 28(y_4-3)C - 7(y_4-4) \big]
\nonumber \\
&
\times \Big\{ 2C \big[ 8(y_4+1)C^6 + 4(17y_4-2)C^5 - 12(83y_4-107)C^4 + 40(55y_4-116)C^3 \nonumber \\ &- 10(191y_4-536)C^2 + 105(7y_4-24)C - 105(y_4-4) \big]
\nonumber \\
&
+ 15(1-2C)^2\times \big[ 12(y_4-1)C^3 - 34(y_4-2)C^2 + 28(y_4-3)C 
\nonumber \\
&
- 7(y_4-4) \big] \log(1-2C) \Big\}^{-1}. \,
\label{eq:k4}
\end{align}
%%%%%%%%%%
Similarly, the gravitomagnetic Love numbers $j_2$, $j_3$ and $j_4$ are given as \cite{Chamel_2021}:
%%%%%%%%%%%%%
\begin{align}
j_2 &= \frac{24}{5} C^5 \big[ 2(\ym_2-2)C - \ym_2 + 3 \big] \times \Big\{ 2C \big[ 2(\ym_2+1)C^3 + 2\ym_2 C^2 + 3(\ym_2-1)C 
\nonumber \\
&
- 3(\ym_2-3) \big] + 3 \big[ 2(\ym_2-2)C - \ym_2 + 3 \big] \log(1-2C) \Big\}^{-1} \, ,
\label{eq:j2}
\end{align}
%%%%%%%%%%%
%%%%%%%%%%%%%
\begin{align}
j_3 &= \, \frac{64}{21} C^7 \big[ 8(\ym_3-2)C^2 - 10(\ym_3-3)C + 3(\ym_3-4) \big]
\nonumber \\
&
\times \Big\{ 2C \big[ 4(\ym_3+1)C^4 + 10\ym_3C^3 + 30(\ym_3-1)C^2 - 15(7\ym_3-18)C + 45(\ym_3-4) \big]
\nonumber \\
&
+ 15 \big[ 8(\ym_3-2)C^2 - 10(\ym_3-3)C + 3(\ym_3-4) \big] \log(1-2C) \Big\}^{-1} \, ,
\label{eq:j3}
\end{align}
%%%%%%%%%%%
%%%%%%%%%%%%%
\begin{align}
j_4 &= \frac{80}{147} C^9 \big[ 40(\ym_4-2)C^3 - 90(\ym_4-3)C^2 + 63(\ym_4-4)C - 14(\ym_4-5) \big]
\nonumber \\ &\times \Big\{ 2C \big[ 4(\ym_4+1)C^5 + 18\ym_4C^4 + 90(\ym_4-1)C^3 - 5(137\ym_4-334)C^2
\nonumber \\
&
+ 105(7\ym_4-26)C - 210(\ym_4-5) \big] + 15 \big[ 40(\ym_4-2)C^3 - 90(\ym_4-3)C^2
\nonumber \\ &+ 63(\ym_4-4)C - 14(\ym_4-5) \big] \log(1-2C) \Big\}^{-1}\, .
\label{eq:j4}
\end{align}
%%%%%%%%%%%
The dimensionless tidal deformability $\Lambda_l$ and $\Sigma_l$ are calculated using Eq. (\ref{eq:Lambda_l}) and (\ref{eq:Sigma_l}) for different values of $l$. The Love numbers play a major role in the inspiral phases of the BNS merger. The dominant contribution comes mainly from the $l=2$ mode compared to higher order polarities, which will be discussed in sub-section \ref{rd:el}.
%%%%%%%%%%%%%%%%%%%%%%%%%%%%%%%%%%%%%%%%%
\subsection{Shape/Surficial Love Numbers}
%%%%%%%%%%%%%%%%%%%%%%%%%%%%%%%%%%%%%%%%%
The surficial Love numbers $h_l$ are associated with the surface deformation of a body. Damour and Nagar gave a relativistic theory of $h_l$ of a body in the influence of tidal forces~\cite{Damour_2009}. It is further extended in Ref. \cite{Landry_2014} to understand it in terms of a deformed curvature of the body's surface. For a perfect fluid, the surficial Love numbers are related with $k_l$ as $h_l=1+2k_l$ in Newtonian case \cite{Damour_2009, Landry_2014}. We calculate the $h_l$ in the limit of a perfect fluid using the relation given in \cite{Damour_2009, Landry_2014, Kumartidal_2017}:
%%%%%%%%%%%%%
\begin{align}
h_\ell = \Gamma_1 + 2 \Gamma_2\, k_{\ell},
\label{h_vs_k1}
\end{align}
%%%%%%%%%%%
where $\Gamma_1$ and $\Gamma_2$ are defined as
%%%%%%%%%%%%%
\begin{align}
\Gamma_1 &= \frac{\ell+1}{\ell-1} (1-M/R) F(-\ell,-\ell;-2\ell;2M/R)
- \frac{2}{\ell-1} F(-\ell,-\ell-1;-2\ell;2M/R), \\
\Gamma_2 &= \frac{\ell}{\ell+2} (1-M/R) F(\ell+1,\ell+1;2\ell+2;2M/R)
+ \frac{2}{\ell+2} F(\ell+1,\ell;2\ell+2;2M/R),
\end{align}
%%%%%%%%%%%
The $F(a,b;c;z)$ is the hypergeometric function and
$\Gamma_1$ and $\Gamma_2$ can be approximated as follow for $C<<1$ \cite{Landry_2014}:
%%%%%%%%%%%%%
\begin{align}
\Gamma_1 &= 1 - (\ell+1) (M/R)
+ \frac{ \ell(\ell+1)(\ell^2-2\ell+2) }{ (\ell-1)(2\ell-1) } (M/R)^2
+ \cdots, \\
\Gamma_2 &= 1 + \ell (M/R)
+ \frac{ \ell(\ell+1)(\ell^2+4\ell+5) }{ (\ell+2)(2\ell+3) } (M/R)^2
+ \cdots.
\end{align} 
%%%%%%%%%%
When compactness goes to zero, the surficial Love number attains the Newtonian limit $h_l=1+2k_l$. In the case of a non-rotating black hole, the surficial Love numbers also play a crucial role, as discussed in Refs. \cite{DamourBH_2009, Damour_2009, Landry_2014}. The surficial Love number is independent of the star's internal structure, which is a function of the mass and radius, i.e., the compactness of the body. 
%%%%%%%%%%%%%%%%%%%%%%%%%%%%%%%%%%%%%%%%%%%%%%%%%%%%%%%
\subsection{Tidal effects on the inspiral phase of BNS}
\label{form:tidal_BNS}
%%%%%%%%%%%%%%%%%%%%%%%%%%%%%%%%%%%%%%%%%%%%%%%%%%%%%%%
During different phases of BNS systems, including the inspiral, merger, and post-merger phases, GWs are emitted. The tidal interaction between the two NSs during the inspiral phase significantly affects the waveforms of the GWs. This study explores the effects of DM on the inspiral properties of BNS and calculates the tidal corrections to GWs waveforms for DM-admixed BNS, including the higher-order tidal properties. The effects of DM on the inspiral properties mainly depend on (i) the type of DM, (ii) the percentage of DM present in the star, and (iii) the lifetime of the BNS evolution. We also calculate the effects of tidal corrections to GWs waveforms for DM admixed BNS in the last part of this chapter.

To calculate the effects of DM on the inspiral/emitted properties of GWs, we use the PN formalism \cite{Blanchet_2006, Boyle_2007, Blanchet_2008, Baiooti_2011, Hotokezaka_2013, Hotokezaka_2016}. The detailed methods can be found in our previous study \cite{DasMNRAS_2021}. We solve the energy balance equation with energy and luminosity up to 3  and 3.5 PN order, respectively. From the energy balance equation, the PN parameter ($x$) is evaluated, which is again used for amplitude, and phase calculations as in Ref. \cite{DasMNRAS_2021}. The gravitoelectric Love numbers $k_l$ lead to a correction of order ($2l+1$)PN to the phase of the gravitational-wave signal \cite{Yagi_2014, Chamel_2021}:
%%%%%%%%%%%%%
\begin{align}
\Psi_{\ell} &= - \sum_{i=1}^{2} \Biggl[\frac{5}{16}\frac{(2\ell-1)!!(4\ell+3)(\ell+1)}{(4\ell-3)(2\ell-3)} \Lambda_{\ell,i} X_i^{2\ell-1} x^{2\ell-3/2}+ \frac{9}{16}\delta_{\ell 2} \Lambda_{2,i} \frac{X_i^4}{\eta} x^{5/2} \Biggr]
\nonumber \\
&
+\mathcal{O}(x^{2\ell-1/2})\, ,
\label{eq:Psil}
\end{align}
%%%%%%%%%%%
with $X_i=M_i/M$ and $\eta=M_1M_2/M^2$. The $\delta_{\ell\ell'}$ is the Kronecker delta, and $i=1,2$ is used to distinguish the two stars of the binary system.
%%%%%%%%%%%%%%%%%%%%%%%%%%%%%%%%%
\section{Results and Discussions}
\label{R&D}
%%%%%%%%%%%%%%%%%%%%%%%%%%%%%%%%%
This section presents our calculated results for the NS's EOS, mass, radius, various Love numbers, and deformabilities in Figs. \ref{fig:EOS_mr}--\ref{fig:Sigma}. The tidal deformability is expressed in the gravitoelectric and gravitomagnetic Love numbers, and the surficial Love numbers describe the surface deformation of the star. The detailed discussions are outlined in the subsequent subsections.
%%%%%%%%%%
\subsection{Equation of State and Mass-Radius Relations}
\begin{figure}
	\centering
	\includegraphics[width=0.9\textwidth]{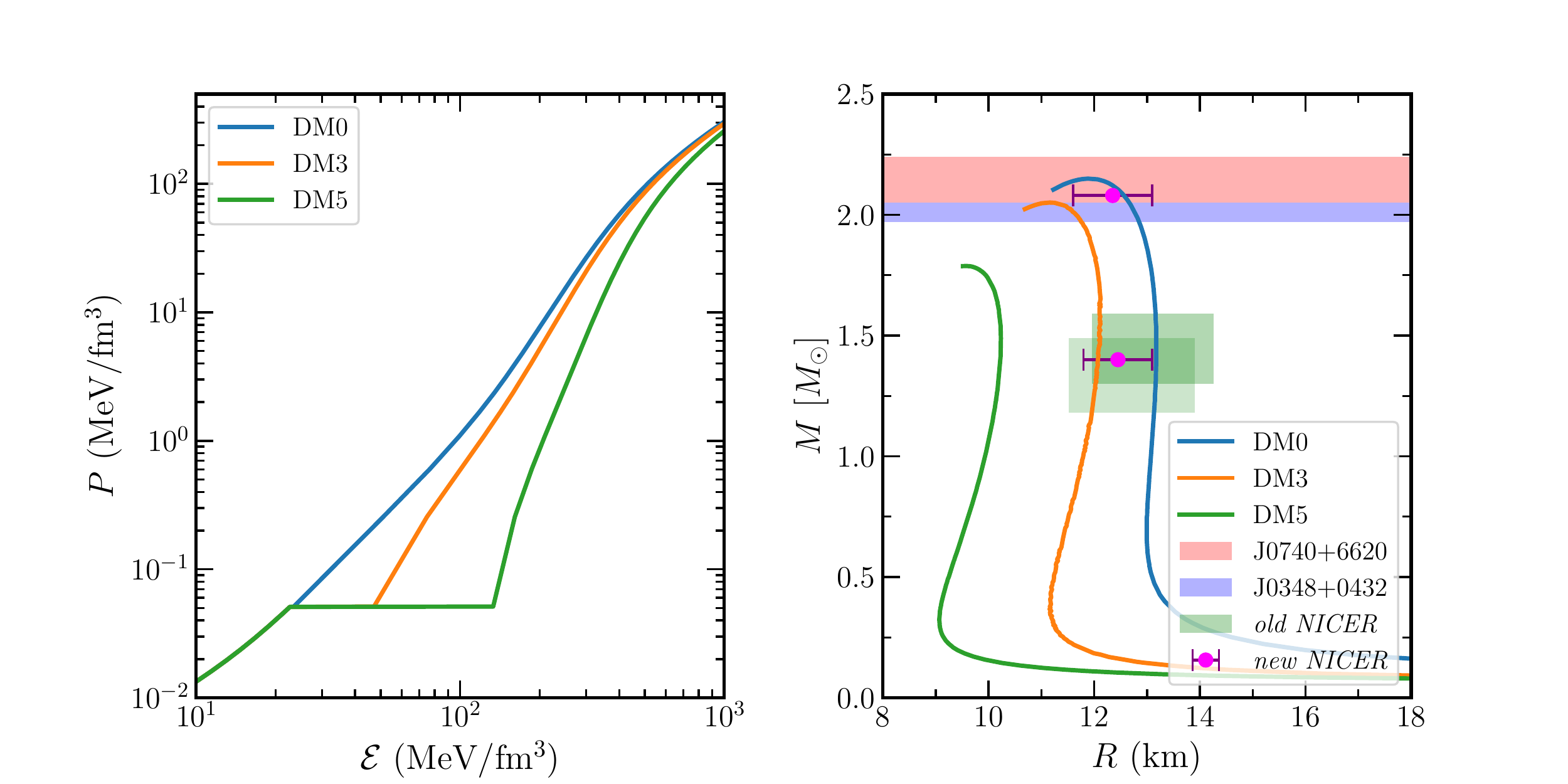}
	\caption{{\it Left}: EOSs for unified IOPB-I (IOPB-I-U) with different DM percentages. The red line represents the IOPB-I crust calculated in Ref. \cite{Parmar_2021}. {\it Right}: $M-R$ profiles for DM admixed NS. Different massive pulsars constraints, such as MSP J0740+6620 by Cromartie et al. \cite{Cromartie_2020} and PSR J0348+0432 by Antoniadis et al. \cite{Antoniadis_2013} are overlaid with different color bars. Both old NICER \cite{Miller_2019, Riley_2019} constraints for canonical stars and new NICER limits for both 1.40 and 2.08 $M_\odot$ are also depicted \cite{Miller_2021}.}
	\label{fig:EOS_mr}
\end{figure}

%%%%%%%%%%
The IOPB-I unified EOS (IOPB-I-U), represented as DM0, and two DM admixed EOSs, such as DM3 and DM5 (DM0, DM3, and DM5 represent the DM Fermi momenta 0.00, 0.03 and 0.05 GeV, respectively.), are shown on the left side of Fig. \ref{fig:EOS_mr}. At the lower density region, all three EOSs coincide into a single line representing the crust part, as shown in the figure. This means the crust EOS is the same for all three cases. With the addition of DM, the EOSs become softer, which is clearly seen in the figure. The softening of the EOS depends on the amount of DM contained inside the NS. Therefore, EOS with 0.05 GeV DM momentum is the softest compared to the other two cases, DM3 and DM0.
%%%%%%%%%%%%%%	
%%%%%%%%%%%%
On the right side of Fig. \ref{fig:EOS_mr}, we depict the mass $M$ and radius $R$ of the DM admixed NS with the IOPB-I-U equation of state. The IOPB-I-U predicts a mass of $M=2.149 \ M_\odot$ and radius $R=11.748$ km. Both the mass and radius are well within the known observational constraints, as shown in the figure. With the addition of DM, the $M$ and $R$ of the NS decreases as the presence of DM softens the EOSs. We noticed a certain range for DM momentum that produces mass--radius values compatible with the latest massive pulsar and NICER limit. Hence, from the observational data, such as advanced LIGO and NICER, one can fix the percentage of DM inside the NS. Theoretically, by generating many such EOSs and using Bayesian analysis, one could also fix the DM percentage inside the NS.
%%%%%%%%%%%%%%%%%%%%%%%%%%%%%%%%%%%%%%%%%%%%%%%%%%%%%%%%%
\subsection{Electric Love Number and Tidal Deformability}
\label{rd:el}
%%%%%%%%%%%%%%%%%%%%%%%%%%%%%%%%%%%%%%%%%%%%%%%%%%%%%%%%%
%%%%%%%%%%%%%%
\begin{figure}
	\centering
	\includegraphics[width=1.0\textwidth]{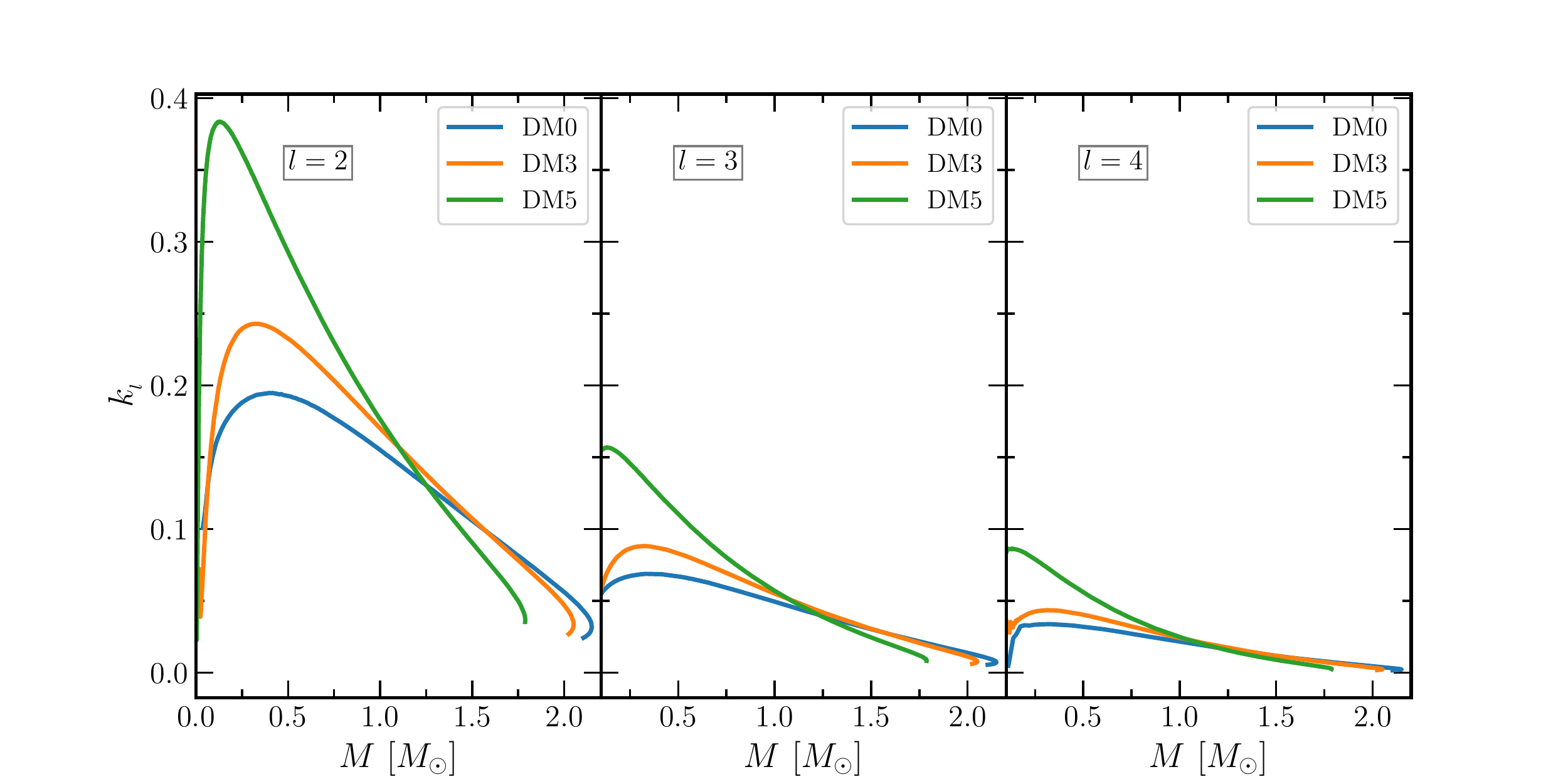}
	\caption{Gravitoelectric tidal Love numbers for IOPB-I-U EOS with DM Fermi momenta 0.00, 0.03 and 0.05 GeV with $l=2, 3, {\rm and } \  4 $.}
	\label{fig:kl}
\end{figure}
%%%%%%%%%%%%
The gravitoelectric Love numbers with different values of $l$ are calculated using Eqs. (\ref{eq:k2})--(\ref{eq:k4}), and the results are shown in Fig. \ref{fig:kl}. As expected, the value of $k_2$ is more pronounced than the $k_3$ and $k_4$. The effects of crust on the $k_l$ are more significant than the EOS without considering the crust, as observed in Ref. \cite{Chamel_2020}. Hence, we take the unified EOSs to calculate the $k_l$. The effects of DM on the gravitoelectric Love numbers are also shown in the figure. The maximum value of $k_l$ increases with DM percentage and the differences are seen mainly in the lower mass regions, where the crust plays an important role. Hence, choosing suitable crust EOSs for Love numbers and tidal deformability calculations is crucial.

With the increase of $l$, the magnitude of $k_l$ decreases more profoundly. However, the impacts of DM on $k_3$ and $k_4$ are non-negligible. It is discussed in Ref. \cite{Chamel_2021} that the symmetry energy is less important on the crust than the stiffness of the EOS. We also found similar results for the DM admixed NS, as the EOSs softened with adding DM.

\begin{figure}
	\centering
	\includegraphics[width=1.0\textwidth]{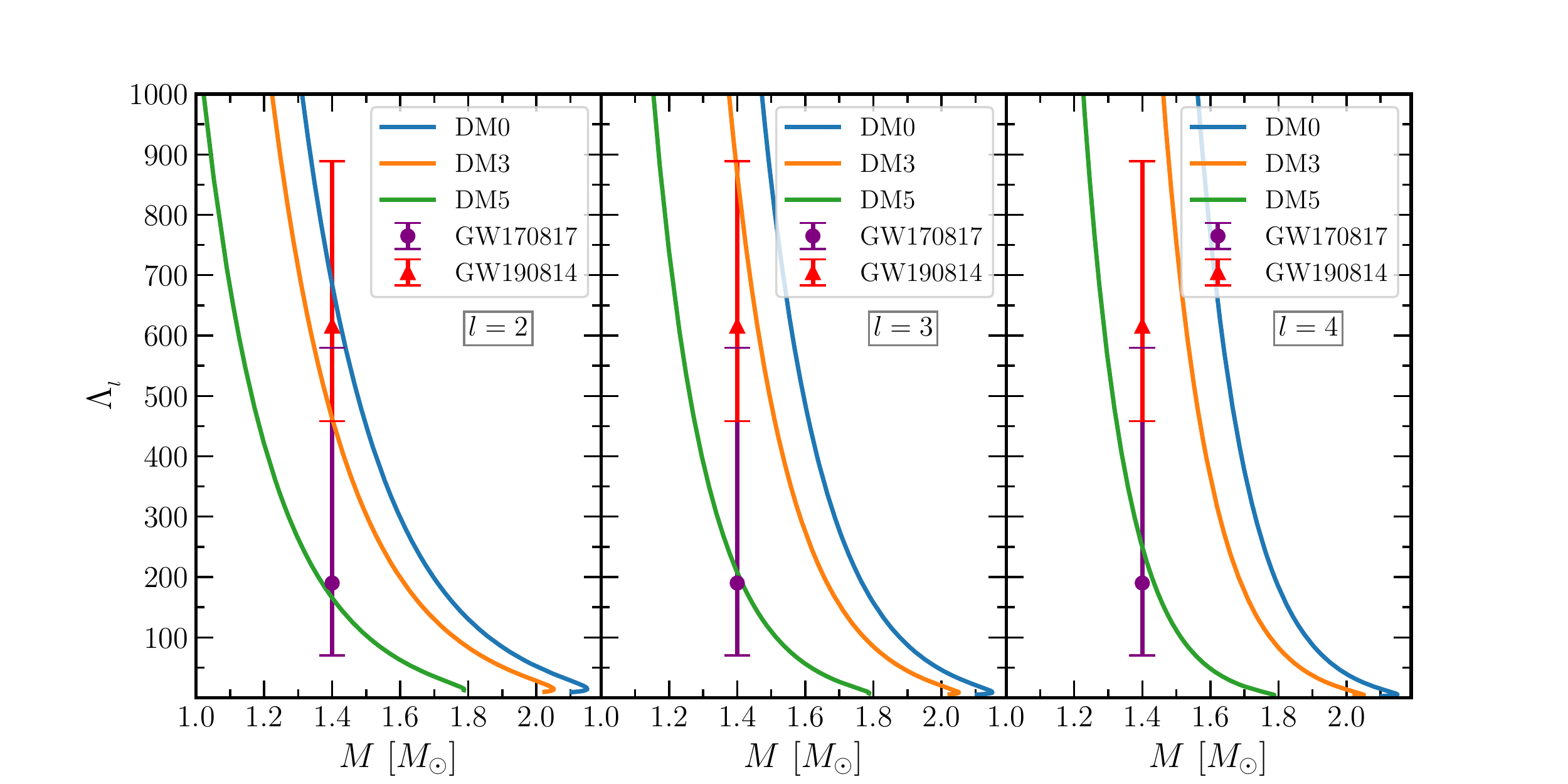}
	\caption{Same as Figure \ref{fig:kl} but for dimensionless electric tidal deformability. The purple colour error bar represents the constraint on $\Lambda_{1.4}$ given by LIGO/Virgo \cite{Abbott_2017, Abbott_2018} from the BNS merger event GW170817 with, $\Lambda_{1.4}=190_{-70}^{+390}$. The red colour error bar represents the $\Lambda_{1.4}$ constraints from the GW190814 event under the assumption of NSBH scenario with, $\Lambda_{1.4}= 616_{-158}^{+273}$ \cite{RAbbott_2020}.}
	\label{fig:Lambda}
\end{figure}

The dimensionless tidal deformability $\Lambda_l$ was calculated for DM admixed NS within the IOPB-I-U equation of state, and the results are shown in Fig. \ref{fig:Lambda}. All the three EOSs, such as DM0, DM3, and DM5, well satisfy the $\Lambda_{1.4}$ constraints for $l=2$. On the other hand, we found a shifting on the $\Lambda_l$ with the increased $l$ values, which do not satisfy the GW170817 limit except for higher DM momenta.
%%%%%%%%%%%%%%	
%%%%%%%%%%%%%	
The secondary component of the GW190814 event is very much in debate, ``{\it whether it is a supermassive NS or the lightest black hole?}". Several attempts have been made to explore this object \cite{Huang_2020, DasPRD_2021}. Some of the suggestions are as follows: (i) It is a heavy NS with deconﬁned QCD core \cite{Tan_2020}, (ii) a super-fast pulsar \cite{Zhang_2020}, (iii) a binary black-hole merger \cite{Fattoyev_2020, DasBig_2021} and (iv) the DM admixed NS with EOS is sufficiently stiff \cite{DasPRD_2021} etc. If this event is the merger of the NS and black hole (NSBH), then the tidal deformability of the NS for the canonical star is $\Lambda_{1.4}= 616_{-158}^{+273}$. 

Hence, one can impose a constraint on the NS mass, radius, and tidal deformability from both GW170817 and GW190814 data. We noticed that using IOPB-I EOS, DM0, and DM3 satisfies the values of $\Lambda_l$ given by GW190814 for the lowest multipole moment. For higher values of $l$, the values of $\Lambda_l$ are shifted and do not satisfy the GW190814 limit.
%%%%%%%%%%%%%%%%%%%%%%%%%%%%%%%%%%%%%%%%%%%%%%%%%%%%%%%%%
\subsection{Magnetic Love Number and Tidal Deformability}
\begin{figure}
	\centering
	\includegraphics[width=1.0\textwidth]{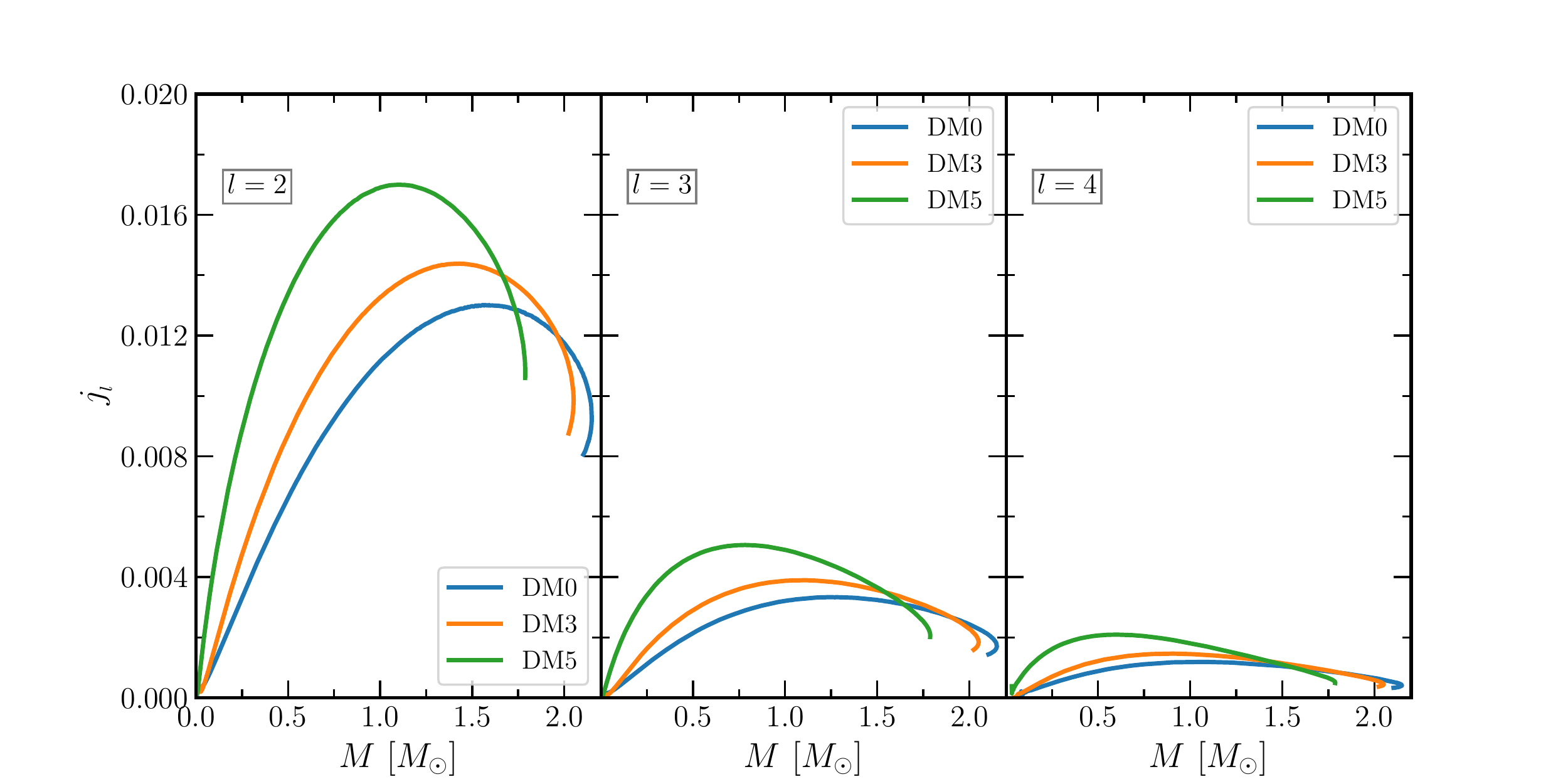}
	\caption{Magnetic tidal Love numbers for IOPB-I EOS with DM Fermi momenta 0.00, 0.03 and 0.05 GeV with three different $l$ values for static fluid.}
	\label{fig:jl_static}
\end{figure}

\begin{figure}
	\centering
	\includegraphics[width=1.0\textwidth]{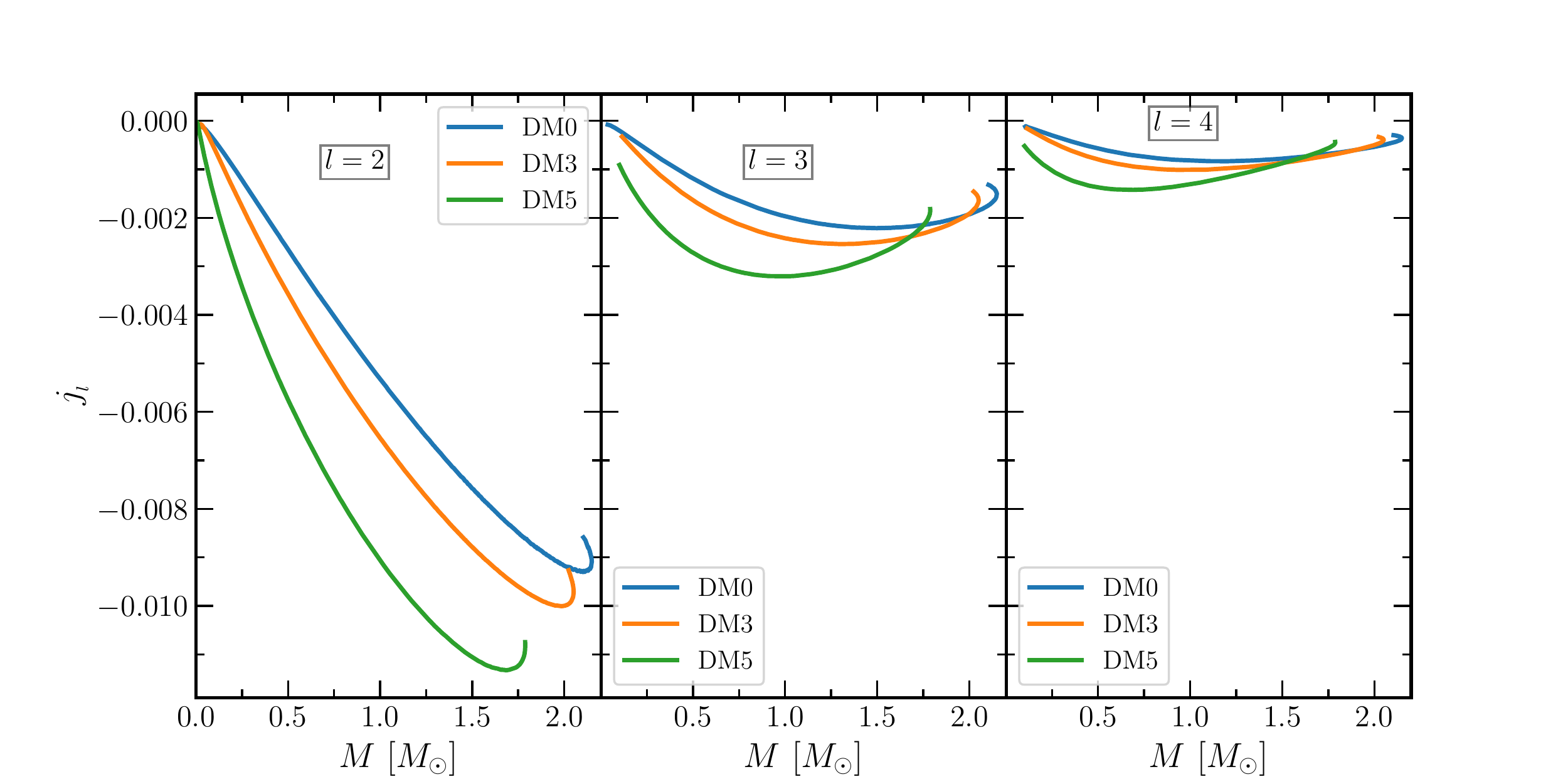}
	\caption{Same as Fig. \ref{fig:jl_static} but for irrotational fluid.}
	\label{fig:jl_irrot}
\end{figure}

\begin{figure}
	\centering
	\includegraphics[width=1.0\textwidth]{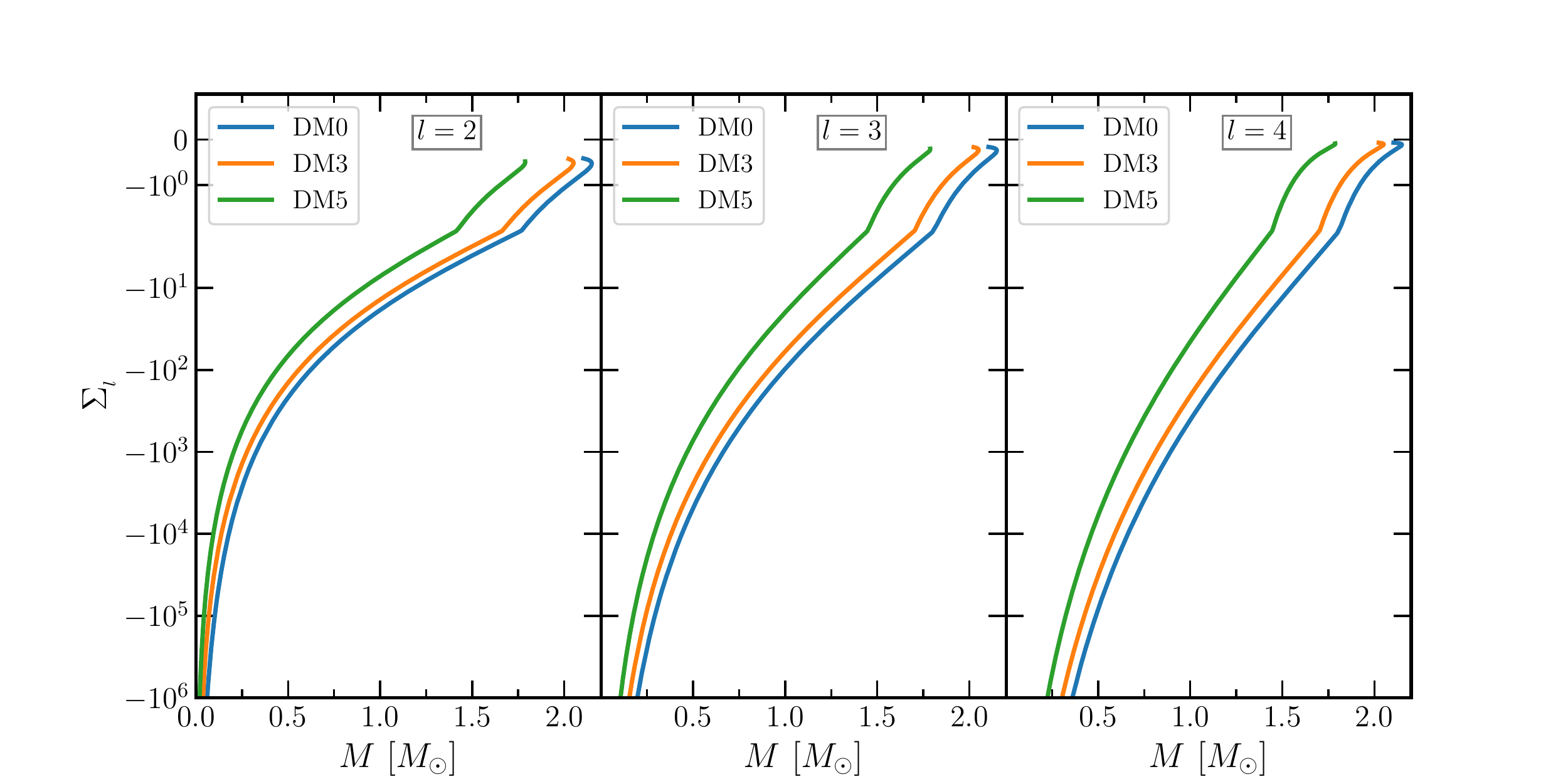}
	\caption{Same as Fig. \ref{fig:jl_irrot} but for dimensionless magnetic tidal deformability.}
	\label{fig:Sigma}
\end{figure}

%%%%%%%%%%%%%%%%%%%%%%%%%%%%%%%%%%%%%%%%%%%%%%%%%%%%%%%%%
We calculate the gravitomagnetic Love numbers and tidal deformabilities for \mbox{$l=2, 3, 4$} using Eqs. (\ref{eq:Sigma_l}) and (\ref{eq:j2})--(\ref{eq:j4}), and the results are depicted in Figs. \ref{fig:jl_static} and  \ref{fig:jl_irrot}. The magnitude of $j_2$ is more negative than both $j_3$ and $j_4$, i.e.,  with the addition of DM in the EOSs, the $j_l$ values correspond to DM3 and DM5 decrease as compared to the DM0 case. Hence, the effects of DM on the $j_l$ are significant. This also depends on the DM contained inside the NS. The magnitudes of the gravitomagnetic tidal Love numbers for both static and irrotational fluids are almost comparable; however,  their values are in opposite signs (as seen from the figures). These tidal Love numbers also impinge less effect on the waveform of GWs, which might be detected in the upcoming modern detectors. 
%%%%%%%%%%%%%%
%%%%%%%%%%%%
%%%%%%%%%%%%%%	
%%%%%%%%%%%%
The gravitomagnetic tidal deformability is shown in Fig. \ref{fig:Sigma} for DM admixed NS. The values of $\Sigma_l$ increase with DM momenta, contrary to the $\Lambda_l$. However, this effect is less significant on the waveforms of the emitted GWs since it depends on the value of $j_l$ compared to $\Lambda_l$. There are also no significant changes with different values of $l$ with the DM percentage. We also observed that the changes are seen mainly in the core part of the NS, as the crust has less contribution to $j_l$ and $\Sigma_l$. Hence, the crust part does not affect the gravitomagnetic Love numbers and tidal deformabilities.
%%%%%%%%%%%%%%%%%%%%%%%%%%%%%%%%%%%
\subsection{Surficial Love Numbers}
\begin{figure}
	\centering
	\includegraphics[width=1.0\textwidth]{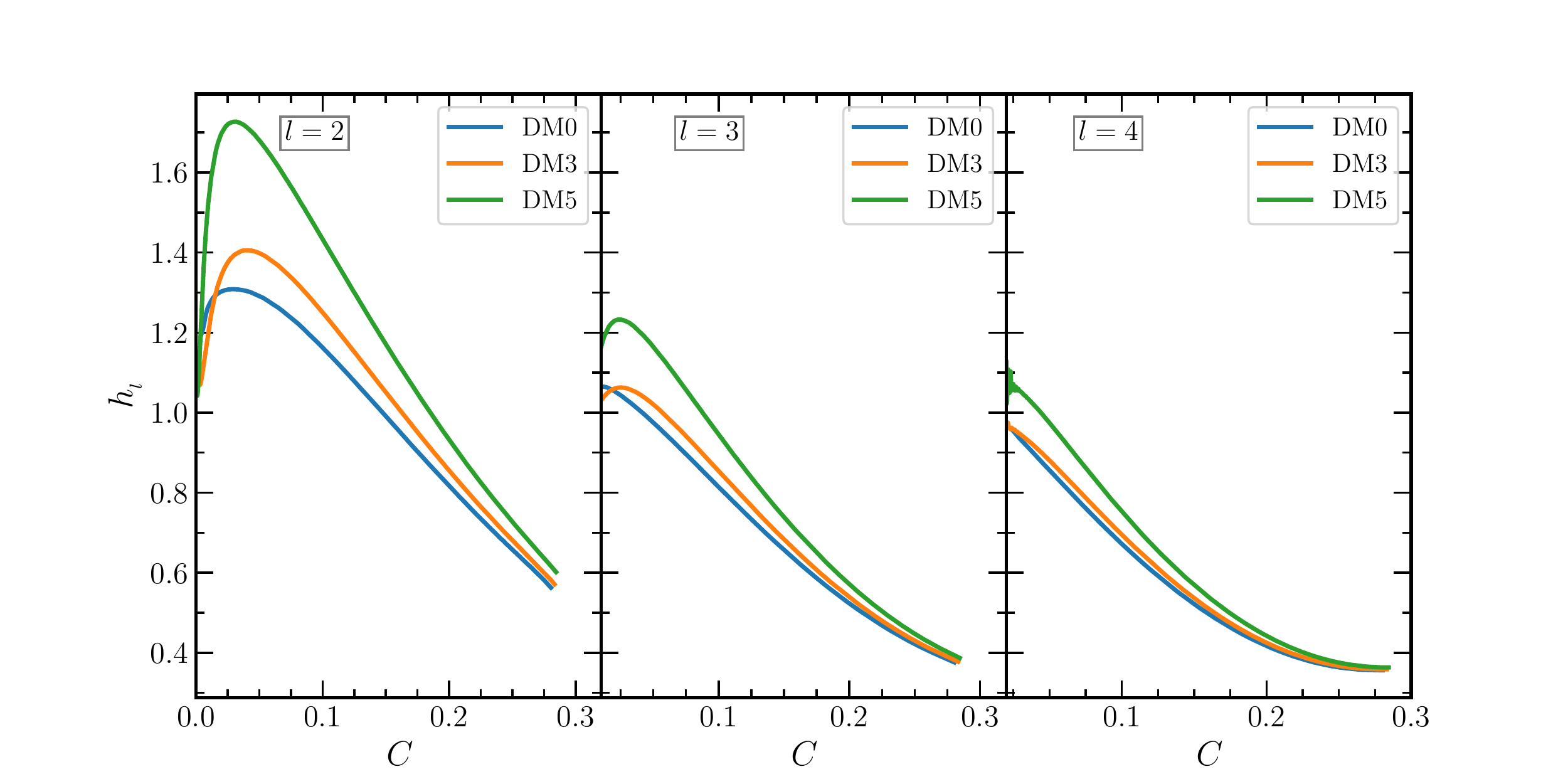}
	\caption{Shape/surficial Love numbers for IOPB-I EOS with DM Fermi momenta 0.00, 0.03, and 0.05 GeV with three different $l$.}
	\label{fig:hl}
\end{figure}

%%%%%%%%%%%%%%%%%%%%%%%%%%%%%%%%%%%
The surficial Love numbers describe the surface deformation of a body in the presence of a tidal field. These are expressed in terms of the compactness of the NS in these calculations. The calculated surficial Love numbers $h_l$ with DM admixed NS are shown in Fig. \ref{fig:hl} for different multipole moments $l$. The pattern of the variation of $h_l$ is almost similar to the $k_l$, but the magnitudes $h_l$ are higher compared to $k_l$. When the compactness $C$$\rightarrow$$0$, all the values of $h_l$, i.e., $h_2$, $h_3$ and $h_4$  approach  $\sim{1}$, irrespective of the moments (see Fig. \ref{fig:hl}). 
%%%%%%%%%%%%%%
%%%%%%%%%%%%%

The value of $h_l$ increases with the percentage of DM contained in the NS. Thus, the surface of the NS is more distorted for the DM5 case than in the other two DM3 and DM0 conditions. A comparative analysis among the multipole moments tells that the surface deformation is mainly due to the quadrupole deformation of the star. However, the contributions of $h_l$ for $l=3$ and 4 are non-negligible. In all the cases of $l$, the maximum distortion of the star takes place at around $C\sim{0.04}$. 
%%%%%%%%%%%%%%
%%%%%%%%%%%%
%%%%%%%%%%%%%%
\begin{figure}
	\centering
	\includegraphics[width=1.0\textwidth]{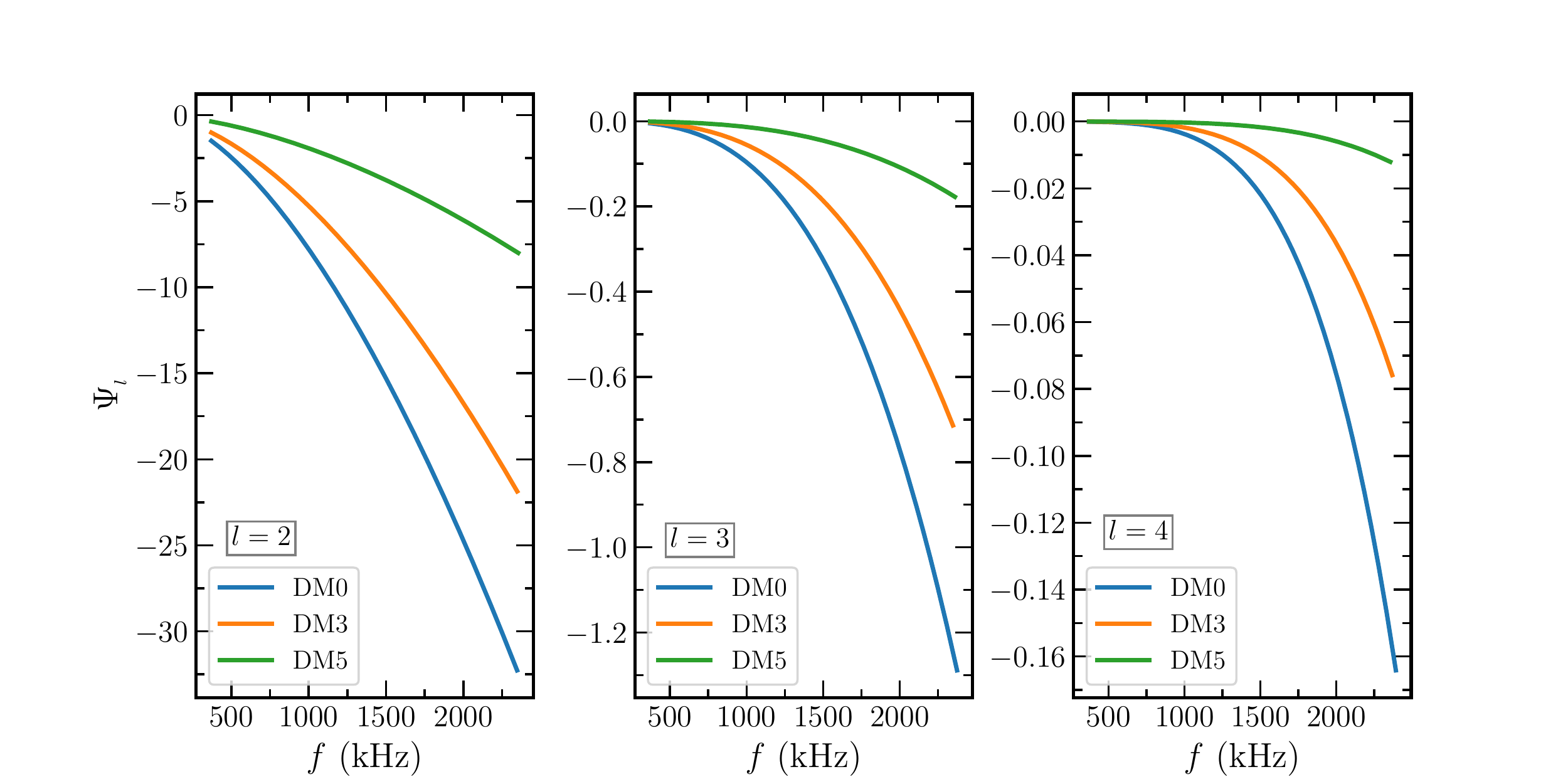}
	\caption{Tidal corrections as a function of gravitational frequency for IOPB-I EOS with DM Fermi momenta 0.00, 0.03, and 0.05 GeV with three different $l$.}
	\label{fig:psil}
\end{figure}
%%%%%%%%%%%%%
%%%%%%%%%%%%%%%%%%%%%%%%%%%%%%%%%%%%%%%%%%%
\subsection{Tidal Effects on the GW Signal}
%%%%%%%%%%%%%%%%%%%%%%%%%%%%%%%%%%%%%%%%%%%
The tidal deformations have significant effects on the gravitational wave in the later stage of the inspiral phase. In this calculation, we study the effects of DM on tidal corrections on the gravitational wave. The tidal corrections are calculated using a PN approximation. The correction is only related to NS observable through the tidal deformability parameter, which is model-dependent. In our previous study \cite{DasMNRAS_2021}, we calculated the inspiral properties with a DM equation of state and observed that the tidal deformability of the star decreased with increasing DM percentage, which slowed down the merger dynamics. The effects of DM on the GW amplitude (such as $h_+$, $h_\times$ and $h_{22}$), inspiral phase ($\Phi$), etc., are calculated. Here, we calculate the tidal corrections ($\Psi_l$) and depict these in Fig.~\ref{fig:psil} for different values of $l$ as a function of frequency ($f$). Similar to all previous cases, we observe that the leading order contribution comes from the $l=2$ case. The magnitude of the tidal contribution was found to be greater in the higher frequency region. Hence, the corrections also accelerate the merger of the BNS for a higher massive star than the less massive one.

The mass of the NS decreases with the addition of DM to the BNS. From our calculated results (see Fig. \ref{fig:psil}), it is clear that the magnitude of $\Psi_l$ decreases with the increasing percentage of DM. We can connect this behavior with the decreasing nature of $\Lambda_l$, as $\Psi_l$ is related to $\Lambda_l$ (see Eq. (\ref{eq:Psil})). Thus, the merging time of the binary system increases with the addition of DM in the NSs. Similar to other quantities, the contribution to waveform $\Psi_l$ has less significance than the higher-order deformation of the binary system. Thus, most of the GWs signal comes from the quadrupole deformations.
%%%%%%%%%%%%%%%%%%%%%%%%%%%%%%
\section{Inspiral properties}
\label{BNS:insp}
%%%%%%%%%%%%%%%%%%%%%%%%%%%%%%
%%%%%%%%%%%%%%%%%%%%%%%%%%%%%%%%%%%%%%%%%%%%%%
\subsection{Post-Newtonian expanded Taylor-T4}
\label{PN}
%%%%%%%%%%%%%%%%%%%%%%%%%%%%%%%%%%%%%%%%%%%%%%
The PN formalism is a slow-motion, weakly stressed, and weak field approximation to the general theory of relativity and is valid for any mass ratio \cite{Blanchet_2006, Buonanno_2015}. In the PN method, the equations of motion are obtained by systematically expanding the metric. Einstein's field equations are expressed in powers of the dimensionless parameter $\epsilon=\sqrt{GM_t/c^2d}\sim\frac{v}{c}$, where $M_t$ is the total mass of the system, $d$ is the distance between two NS, and $v$ is the characteristic velocity of particles. The metric is then solved iteratively in powers of $\epsilon$, and the equations of motion are evaluated from the metric using the geodesic equation. An expansion containing terms up to $\epsilon^n$ or equivalently $(1/c)^n$ is denoted an $\frac{n}{2}$PN expansion \cite{Blanchet_2006, Boyle_2007}. This calculation takes the energy and luminosity up to 3PN and 3.5PN order, respectively, for the in-spiraling binary in quasi-circular orbits. 

The BNS emit GWs in their inspiral phases. Because of the emission of the GWs for a classical system, the BNS has to rotate in quasi-circular orbits. The flux (luminosity) emitted by the system balances the rate of change of energy with time in that orbit, and the energy balance equation is given as \cite{Boyle_2007, Maggiore_2008, Lackey_2012}:
%%%%%%%%%%%%%%%%
\begin{equation}
{\cal{L}}=-\frac{dE}{dt}=-\frac{dE/dx}{dt/dx},
\end{equation}
%%%%%%%%%%%%%%
where $E$ is the energy of the system, $t$ is the time taken and $x$ is the PN parameter which is defined as \cite{Boyle_2007,Maggiore_2008,Lackey_2012}
%%%%%%%%%%%%%%%%
\begin{equation}
x=\Big(M_t\frac{d\Phi}{dt}\Big)^{2/3}=\Big(M_t\Omega\Big)^{2/3},
\end{equation}
%%%%%%%%%%%%%%
where $M_t$ is the total mass of BNS systems, $\Phi$ is the orbital phase, $\Omega=2\pi\omega$ is the orbital angular velocity ($\omega=f/2$  is the orbital angular frequency), and $f$ is the frequency of the emitted GW. The energy of the system calculated up to 3PN order in terms of $x$ is given as \cite{Blanchet_2006, Maggiore_2008, Lackey_2012}:
%%%%%%%%%%%%%%%%
\begin{eqnarray}
E&=&-\frac{1}{2}M_t\nu x\Bigg\{1+\Bigg(-\frac{3}{4}-\frac{\nu}{12}\Bigg)x+\Bigg(-\frac{27}{8}+\frac{19\nu}{8}-\frac{\nu^2}{24}\Bigg)x^2
\nonumber \\
&&
+\Bigg[-\frac{675}{64}+\Bigg(\frac{34445}{576}-\frac{205\pi^2}{96}\Bigg)\nu
-\frac{155\nu^2}{96}-\frac{35\nu^3}{5184}\Bigg]x^3\Bigg\},
\label{enerPN}
\end{eqnarray}
%%%%%%%%%%%%%%
where $\nu=m_1m_2/M_t^2$ is the symmetric mass ratio, $m_1$ and $m_2$ are the individual masses of the binary. The luminosity ${\cal{L}}$ is calculated by the time derivative of Eq. (\ref{enerPN}) using PN expansions up to 3.5 order is given as \cite{Blanchet_2006, Maggiore_2008, Lackey_2012}
%%%%%%%%%%%%%
\begin{align}
{\cal{L}}=&\frac{32}{5}\nu^2x^5\Bigg\{1+\Bigg(-\frac{1247}{336}-\frac{35\nu}{12}\Bigg)x+4\pi x^{3/2}+\Bigg(-\frac{44711}{9072}+\frac{9271\nu}{504}+\frac{65\nu^2}{18}\Bigg)x^2
\nonumber\\
+&\Bigg(-\frac{8191}{672}-\frac{583\nu}{24}\Bigg)\pi x^{5/2}
+\Bigg[\frac{6643739519}{69854400}+\frac{16\pi^2}{3}-\frac{1712\gamma_E}{015}-\frac{856}{105}\ln(16x)
\nonumber\\
+&\Bigg(-\frac{134543}{7776}+\frac{41\pi^2}{48}\Bigg)\nu-\frac{94403\nu^2}{3024}-\frac{775\nu^3}{324}\Bigg]x^3
+\Bigg(-\frac{16285}{504}+\frac{214745\nu}{1728}
\nonumber\\
+&\frac{193385\nu^2}{3024}\Bigg)\pi x^{7/2}\Bigg\},
\label{lumPN}
\end{align}
%%%%%%%%%%%
where $\gamma_E\approx 0.5772$ is the Euler's constant. The phase evolution of the binary system can be calculated by solving the following equations:
%%%%%%%%%%%%%%%%
\begin{eqnarray}
\frac{dx}{dt}=\frac{dE/dt}{dE/dx}=-\frac{{\cal{L}}}{dE/dx},
\label{dxdt}
\end{eqnarray}
%%%%%%%%%%%%%
\begin{eqnarray}
\frac{d\Phi}{dt}=\frac{x^{3/2}}{M_t}.
\label{dpdt}
\end{eqnarray}
%%%%%%%%%%%%%%
There are various methods to integrate this system of equations labeled as TaylorT1-TaylorT4 \cite{Boyle_2007, Creighton_2011, Lackey_2012}. In the TaylorT1 method, the Eqs. (\ref{enerPN}) and (\ref{lumPN}) are inserted in Eq. (\ref{dxdt}), and then the integration is performed by using the initial conditions $x_0=(M_t\Omega_0)^{2/3}$ and $\Phi_0$. In the TaylorT2 method, the equations are written by starting with a parametric solution of energy balance equations. Then, each expression integrand is re-expanded as a single PN parameter x and truncated at the appropriate order \cite{Boyle_2007, Lackey_2012}
%%%%%%%%%%%%%%%%
\begin{eqnarray}
t(x)=t_0+\int_{x}^{x_0}dx\frac{(dE/dx)}{{\cal{L}}},
\label{tpn}
\end{eqnarray}
%%%%%%%%%%%%%%
%%%%%%%%%%%%%%%%
\begin{eqnarray}
\Phi(x)=\Phi_0+\int_{x}^{x_0}dx\frac{x^{3/2}}{M_t}\frac{(dE/dx)}{{\cal{L}}}.
\end{eqnarray}
%%%%%%%%%%%%%%
In the present calculations, we take the TaylorT4 approximation. In this method, the right-hand side of the Eq. (\ref{tpn}) is re-expanded as a single series and truncated at 3.5PN order before doing the integration. After calculating $x$, $t$, and $\Phi$, one can know the properties related to the in-spiraling binaries for the point-particle system. 

In the case of BNS, the tidal interaction comes into the picture, which has a significant role in the inspiral properties. Therefore, we have to include the extra part of tidal interactions with this point-particle approximation. The system's motion is tidally interacting, which influences its internal structures. Also, the tidal interactions affect the evolution of the GW phase by a parameter $\lambda$. Hence, the BNS system's tidal contributions must be added on the right-hand side of Eq. (\ref{dxdt}), which is modified as \cite{Baiooti_2011, Hotokezaka_2013, Hotokezaka_2016}:
%%%%%%%%%%%%%%%%
\begin{equation}
\frac{dx}{dt}=\frac{64\nu}{5M_t} x^5\{F_{3.5}^{\mathrm{Taylor}}(x)+F^{\mathrm{Tidal}}(x)\},
\end{equation}
%%%%%%%%%%%%%%
where $F_{3.5}^{\mathrm{Taylor}}$is PN-expanded expression for point-mass contribution using Taylor-T4 approximation, given by \cite{Blanchet_2008,Boyle_2007,Baiooti_2011,Hotokezaka_2013,Hotokezaka_2016}
%%%%%%%%%%%%%
\begin{align}
F_{3.5}^{\mathrm{Taylor}}(x)=&1-\Bigg(\frac{743}{336}+\frac{11}{4}\nu\Bigg)x+4\pi x^{3/2}+\Bigg(\frac{34103}{18144}+\frac{13661}{2016}\nu+\frac{59}{18}\nu^2\Bigg)x^2
\nonumber\\ 
-&\Bigg(\frac{4159}{672}+\frac{189}{8}\nu\Bigg)\pi x^{5/2}
+\Bigg[\frac{16447322263}{139708800}-\frac{1712}{105}\gamma_E-\frac{56198689}{217728}\nu
\nonumber\\ 
+&\frac{541}{896}\nu^2-\frac{5605}{2592}\nu^3+\frac{\pi^2}{48}(256+451\nu)-\frac{856}{105}\ln(16x)\Bigg]x^3
\nonumber\\ 
+&\Bigg(-\frac{4415}{4032}+\frac{358675}{6048}\nu+\frac{91495}{1512}\nu^2\Bigg)\pi x^{7/2},
\end{align}
%%%%%%%%%%%
With the addition of the tidal part to the PN formalism, we adopt the method developed by Vines {\it et al.} \cite{Vines_2011} up to 1PN accuracy, which is expressed as 
%%%%%%%%%%%%%
\begin{align}
F^{\mathrm{Tidal}}(x)=&\frac{32\chi_1\lambda_2}{5M_t^6}\Big[12(1+11\chi_1)x^{10}+\Big(\frac{4421}{28}-\frac{12263}{28}\chi_2+\frac{1893}{2}\chi_2^2-661\chi_2^3\Big)x^{11}
\nonumber\\ 
+&(1\leftrightarrow2)\Big],
\label{tidalEq}
\end{align}
%%%%%%%%%%%
where $\lambda_2$ is the tidal deformability of the second star and $\chi_1$ is the mass fraction of the first star ($\chi_1=M_1/M_t$). The numerical values of $k_2$ and $R$ are given in Table \ref{table1} for different parameter sets with three DM fractions. The numerical values are different from others because they are model dependent. Baiotti {\it et al.} \cite{Baiooti_2011} re-expressed the Eq. (\ref{tidalEq}) in terms of tidal Love number $k_2$ and compactness parameter $C$ as
%%%%%%%%%%%%%%%%
\begin{equation}
F^{\mathrm{Tidal}}(x)=\sum_{I=1,2}F_{LO}(\chi_I)x^5\big(1+F_1(\chi_I)x\big),
\end{equation}
%%%%%%%%%%%%%%%
with 
%%%%%%%%%%%%%%%%
\begin{equation}
F_{LO}(\chi_I)=4\hat{k_2^I}\Bigg(\frac{12-11\chi_I}{\chi_I}\Bigg),
\end{equation}
%%%%%%%%%%%%%%%
and 
%%%%%%%%%%%%%%%%
\begin{equation}
F_1(\chi_I)=\frac{4421-12263\chi_I+26502\chi_I^2-18508\chi_I^3}{336(12-11\chi_I)}.
\end{equation}
%%%%%%%%%%%%%%%
The expression for $\hat{k_2^I}$ is defined as:
%%%%%%%%%%%%%%%%
\begin{equation}
\hat{k_2^I}=k_2^I\Bigg(\frac{\chi_I}{C_I}\Bigg)^5  \ \ I=1,2.
\end{equation}
%%%%%%%%%%%%%%%
For equal-mass binary, $\chi_1 = \chi_2=\chi=1/2$ and $C=C_1=C_2$, the Eq. (\ref{tidalEq}) can be written as \cite{Hotokezaka_2013} 
%%%%%%%%%%%%%%%%
\begin{equation}
F^{\mathrm{Tidal}}(x)= \frac{52}{5M_t}\frac{k_2}{C^5}x^{10}\Big(1+\frac{5203}{4368}x\Big).
\end{equation}
%%%%%%%%%%%%%%
Although the tidal interactions affect the NS-NS inspiral only at 5PN order, its coefficient is 10$^4$ order magnitude for a NS with a radius of 10–15 km, $k_2 \sim 0.1$, and $C \sim 0.14–0.20$. Hence, it plays a vital role in the late inspiral stage.
%%%%%%%%%%%%%%%%%%%%%%%%%%%%%%%%%%%
\subsection{Polarization waveforms}
\label{pola}
%%%%%%%%%%%%%%%%%%%%%%%%%%%%%%%%%%%
From the observational point of view, the GW detector detects the radiation in the direction of the source. Therefore, the polarization waveforms are measured in the line-of-sight of the source, which can be expressed in spherical coordinates ($R,\hat{\theta},\hat{\phi}$) as \cite{Maggiore_2008, Isoyama_2020}
%%%%%%%%%%%%%%%%
\begin{eqnarray}
&&h_+(\mathrm{t})=\frac{4}{r}{\cal{M}}^{5/3}\omega^{2/3}\Bigg(\frac{1+\cos^2\theta}{2}\Bigg)\cos(2\omega t_{\rm ret}+2\phi),\nonumber\\ \mathrm{and} \\ \nonumber
&&
h_\times(\mathrm{t})=\frac{4}{r}{\cal{M}}^{5/3}\omega^{2/3}\cos\theta\sin(2\omega t_{\rm ret}+2\phi),
\label{h+hx}
\end{eqnarray}
%%%%%%%%%%%%%%
where the chirp mass ${\cal{M}}$ is defined as ${\cal{M}}=\nu^{3/5}M$. The amplitudes of the GW polarization with a fixed $\omega$ depends on the binary masses through ${\cal{M}}$, which can also be derived as ${\cal{M}}=\frac{(m_1m_2)^{3/5}}{(m_1+m_2)^{1/5}}$. It is measured by LIGO with a good precision ${\cal{M}}=1.188_{-0.002}^{+0.004}M_\odot$ \cite{Abbott_2017}. If we see the orbit edge-on, $\theta=\pi/2$, then $h_\times$ vanishes, and the GW is linearly polarized. If we put $\theta=0$, then $h_+$ and $h_\times$ have the same amplitude, but the only difference is the phase, which is circularly polarized.

The polarization waveforms for an equal mass binary NS (EMBNS) systems along the z-axis (optimally oriented observer) calculated up to 2.5PN are given by  \cite{Arun_2004, Arun_2005, Kidder_2007, Boyle_2007}
%%%%%%%%%%%%%
\begin{align}
h_+^{(z)}=&\frac{M_t}{2D}x\Bigg(\cos2\Phi\bigg\{-2+\frac{17}{4}x-4\pi x^{3/2}+\frac{15917}{2880}x^2+9\pi x^{5/2}\bigg\}
\nonumber \\
+&\sin2\Phi\Big\{\frac{59}{5}x^{5/2}\Big\}\Bigg),
\end{align}
%%%%%%%%%%%
%%%%%%%%%%%%%
\begin{align}
h_\times^{(z)}=&\frac{M_t}{2D}x\Bigg(\sin2\Phi\bigg\{-2+\frac{17}{4}x-4\pi x^{3/2}+\frac{15917}{2880}x^2+9\pi x^{5/2}\bigg\}
\nonumber \\
+&\cos2\Phi\Big\{-\frac{59}{5}x^{5/2}\Big\}\Bigg).
\end{align}
%%%%%%%%%%%
The polarization waveforms are shown in Fig. \ref{hphc} for IOPB-I parameter sets. 
The dominant (2,2) mode up to 3PN order is written as follow \cite{Blanchet_2008,Kidder_2008, Lackey_2012}
%%%%%%%%%%%%%
\begin{align}
h_{22}=&-8\nu\sqrt{\frac{\pi}{5}}\frac{M_t}{D}e^{-2i\Phi}x  \Bigg \{1+\Bigg(-\frac{107}{42}+\frac{55\nu}{42}\Bigg)x+2\pi x^{3/2}
\nonumber\\
+&\Bigg(-\frac{2173}{1512}-\frac{1069\nu}{216}+\frac{2047\nu^2}{1512}\Bigg)x^2 
+\Bigg[\frac{-107\pi}{21}+\Bigg(\frac{34\pi}{21}-24i\Bigg)\nu\Bigg]x^{5/2}
\nonumber\\
+&\Bigg[\frac{27027409}{646800}-\frac{856\gamma_E}{105}+\frac{2\pi^2}{3}+\frac{428i\pi}{105}-\frac{428}{105}\ln(16x)
\nonumber\\
+&\Bigg(\frac{41\pi^2}{96}-\frac{278185}{33264}\nu\Bigg)-\frac{20261\nu^2}{2772}+\frac{11463\nu^3}{99792}\Bigg]x^3 \Bigg \},
\end{align}
%%%%%%%%%%%
where $D$ is the distance between source and observer, taken as 100 Mpc in our calculations. 
%%%%%%%%%%%%%%%%%%%%%%%%%%%%%%%%
\subsection{Tidal deformability}
\label{NSprop}
\begin{figure}
	\centering
	\includegraphics[width=0.7\textwidth]{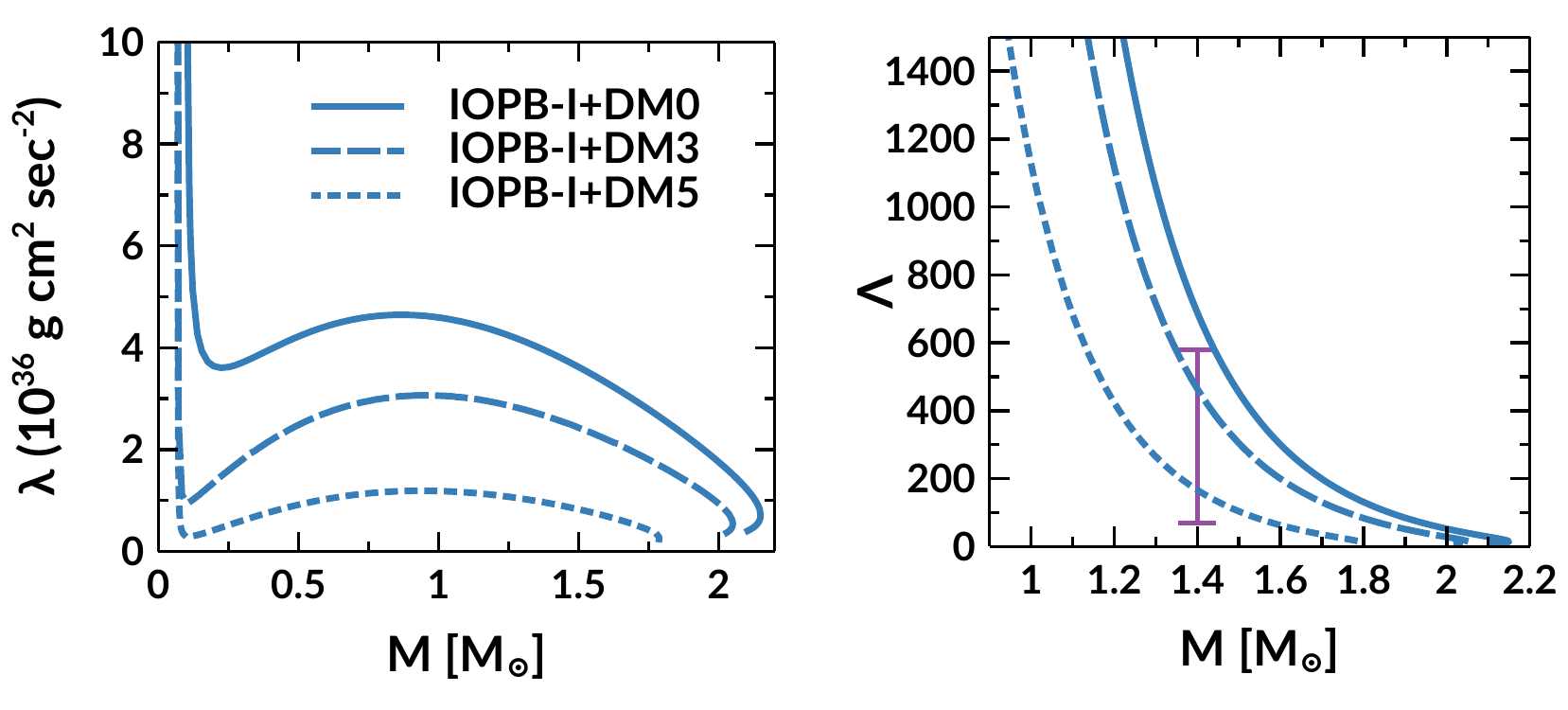}
	\caption{{\it Left}: The mass variation of tidal deformability of the NS is shown for IOPB-I set with two different DM momentum. {\it Right}: The dimensionless tidal deformability vs. masses of the NS with the addition of DM are shown. The vertical line with bar represents the GW170817 constraints on $\Lambda_{1.4}$ \cite{Abbott_2017,Abbott_2018}.}
	\label{lambda}
\end{figure}

%%%%%%%%%%%%%%%%%%%%%%%%%%%%%%%%
The tidal deformability of the NS decreases with DM's addition, as shown in Fig. \ref{lambda}. The same trend was also obtained for $\Lambda$. This is because, with the addition of DM, the EOS becomes softer, and the mass and radius change accordingly. Therefore, the softer EOS gives a lower $\lambda$ than the stiffer one. The vertical double-headed line represents the constraints from GW170817 with $\Lambda_{1.4}=190_{-120}^{+390}$ \cite{Abbott_2018}. The values of $\Lambda_{1.4}$ are 473.22, and 169.644 for DM momentum 0.03 and 0.05 GeV, respectively. This implies that the value of $\Lambda_{1.4}$ with these DM momenta satisfies the constraints given by GW170817.
%%%%%%%%%%%%%%
%%%%%%%%%%%%%%
%%%%%%%%%%%%%%
\begin{table*}
	\centering
	\caption{The maximum mass of the spherical NS $M_{\rm max}$ obtained with different EOS, the radius $R_{1.35}$, the tidal deformability $\lambda_{1.35}$, the dimensionless tidal deformability $\Lambda_{1.35}$, the tidal love number $k_{2,1.35}$, the compactness $C_{1.35}$ for NS mass 1.35 $M_\odot$. The total mass of the BNS system is $M_t=2.7 M_{\odot}$, and $M_t\Omega_0=M_t\pi f_0=0.0155$ are taken in this calculation with the minimum gravitational wave frequency $f_0$  $\approx$ 371 Hz. $\lambda_{1.35}^*$ is in $10^{36}$ g cm$^2$ sec$^{-2}$ unit.}
	\label{table1}
	\begin{tabular}{|l|l|l|l|l|l|l|l|}
		\hline %\hline
		EOSs & $M_{\rm max}$ ($M_{\odot})$ & $R_{1.35}$ (km) & $\lambda_{1.35}^*$& $\Lambda_{1.35}$ & $k_{2,1.35}$ & $C_{1.35}$ & $M_t\Omega_0$ \\ \hline
		NL3        & 2.775 & 14.568 & 7.415  & 1575.239 & 0.1130  & 0.137 & 0.0155 \\ \hline
		G3         & 1.979 & 12.476 & 2.754  & 585.182  & 0.0911  & 0.159 & 0.0155 \\ \hline
		IOPB-I     & 2.149 & 13.168 & 4.026  & 855.304  & 0.1017  & 0.151 & 0.0155 \\ \hline
		IOPB-I+DM3 & 2.051 & 12.047 & 2.717  & 577.237  & 0.1071  & 0.165 & 0.0155 \\ \hline
		IOPB-I+DM5 & 1.788 & 10.204 & 0.990  & 210.387  & 0.0896  & 0.195 & 0.0155 \\ \hline %\hline
	\end{tabular}
\end{table*}
%%%%%%%%%%%%
\subsection{Gravitational waves properties}
\label{GW}
In this sub-sec., we describe the inspiral properties of the  BNS using the PN formalism applied to a quasi-circular orbit. For simplicity, we restrict our calculations only to EMBNS. Again, we take IOPB-I as a representative EOS for the DM admixed NS since it predicts the mass of the NS as 2.149 $M_{\odot}$ which is consistent with PSR J0740+6620 \cite{Cromartie_2020}. Here, we calculate the polarization waveforms and the GW properties, such as amplitude, frequency, and orbital phase, for the EMBNS. The observables are calculated at retarded time defined by \cite{Hotokezaka_2013}: 
%%%%%%%%%%%%%%%%
\begin{eqnarray}
t_{\rm ret}=t-r_*,
\label{ret1}
\end{eqnarray}
%%%%%%%%%%%%%%%
where $r_*$ is the tortoise coordinate defined as 
%%%%%%%%%%%%%%%%
\begin{eqnarray}
r_*=r_A+2M_t\ln\Big(\frac{r_A}{2M_t}-1\Big),
\end{eqnarray}
%%%%%%%%%%%%%%
where $r_=\sqrt{A/4\pi}$ and A is the proper surface area of the sphere. In a simplified manner, $r_A$ is written as  \cite{Hotokezaka_2013}
%%%%%%%%%%%%%%%%
\begin{eqnarray}
r_A=r\Bigg(1+\frac{M_t}{2r}\Bigg)^2,
\end{eqnarray}
%%%%%%%%%%%%%%%
where $r\approx200M_t$ is the finite spherical-coordinate radius.
%%%%%%%%%%%%%%
\begin{figure}
	\centering
	\includegraphics[width=0.7\textwidth]{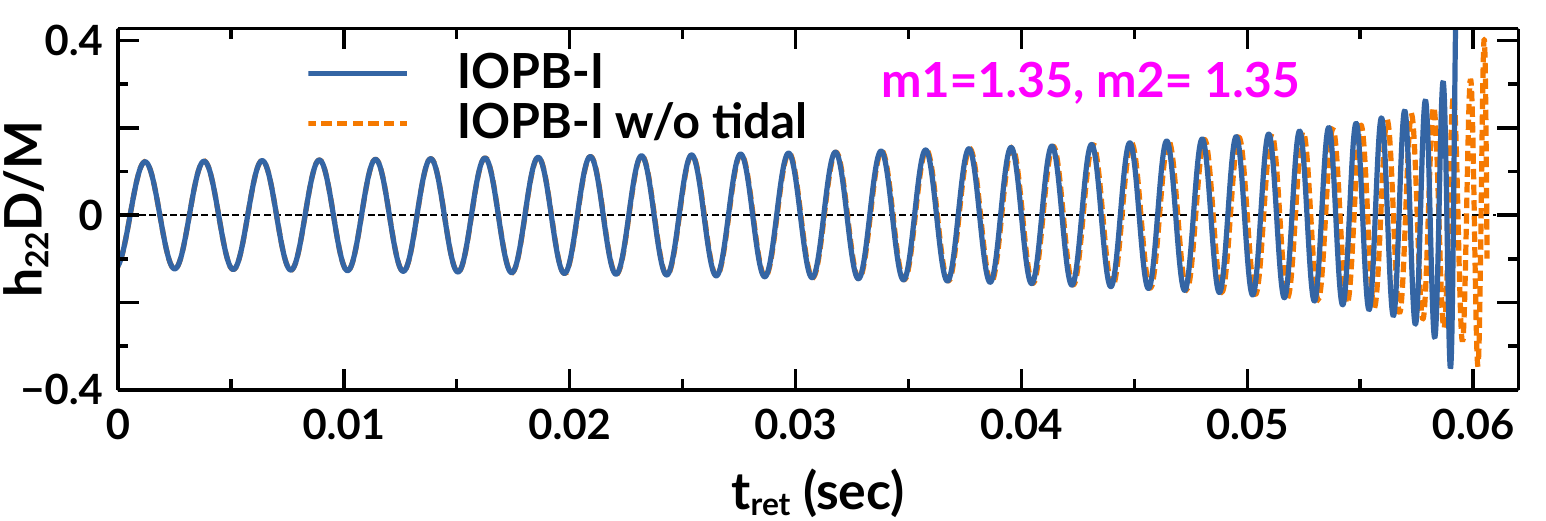}
	\caption{(2,2) mode waveforms are shown with and without tidal contributions for the IOPB-I parameter sets as a representative case.}
	\label{iopb_wo_tidal}
\end{figure}
%%%%%%%%%%%%%
%%%%%%%%%%%%%
\begin{figure}
	\centering
	\includegraphics[width=0.7\textwidth]{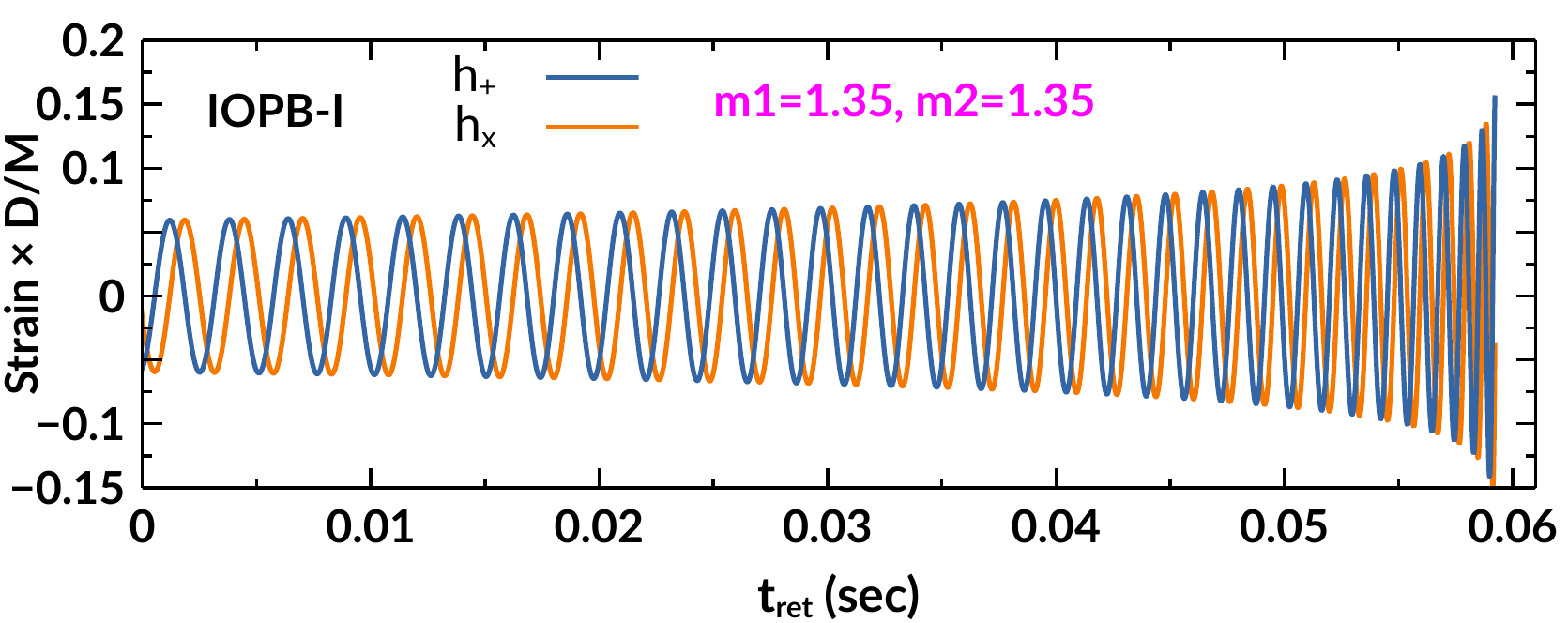}
	\caption{Plus and cross-polarization waveforms are shown for the IOPB-I parameter sets.}
	\label{hphc}
\end{figure}
%%%%%%%%%%%%%%
%%%%%%%%%%%%%%
\begin{figure}
	\centering
	\includegraphics[width=0.7\textwidth]{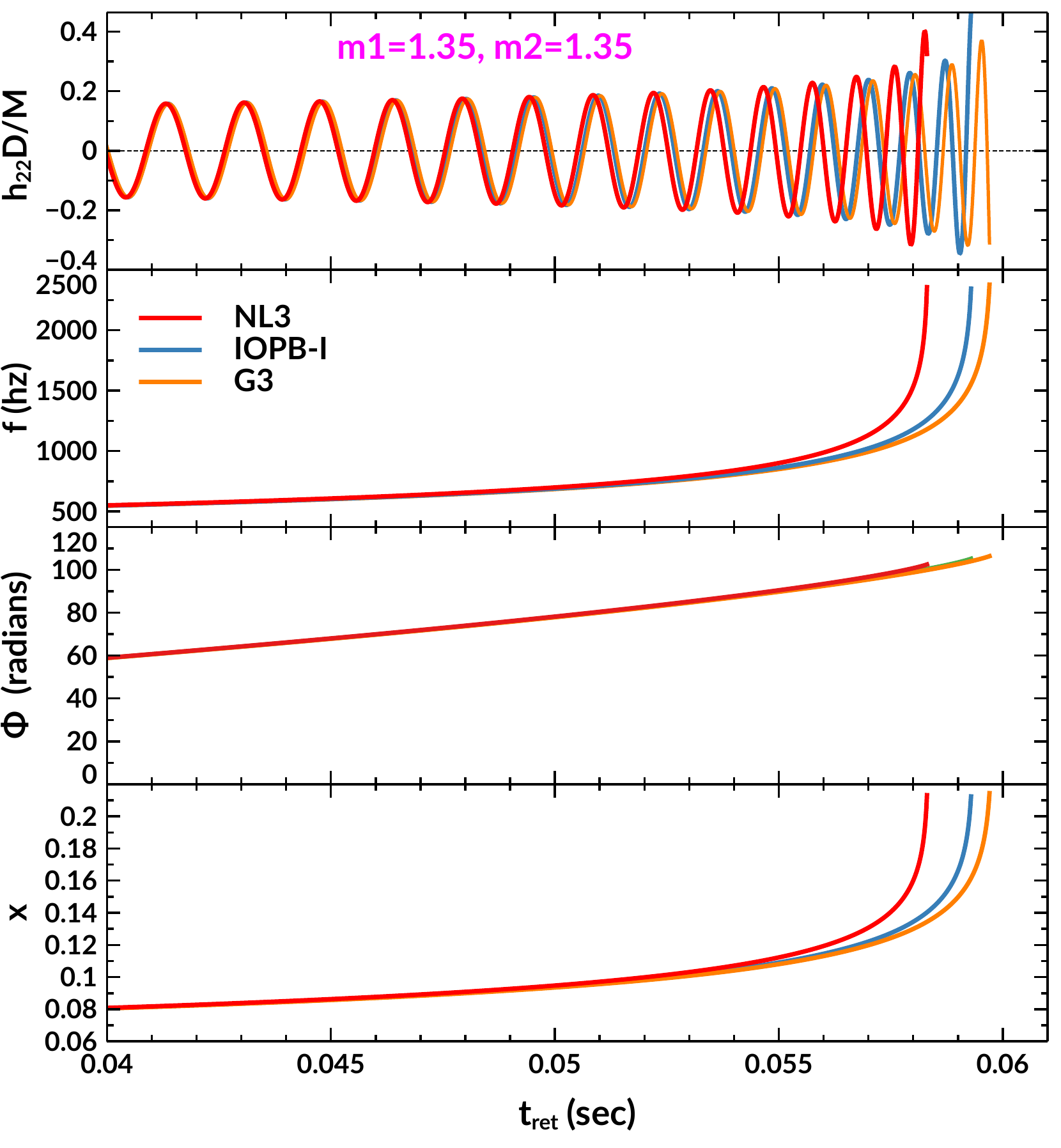}
	\caption{(2,2) mode waveform, frequency, phase, and PN-parameters are shown for three parameter sets of the EMBNS in the retarded time interval calculated for $D=100$ Mpc.}
	\label{all1}
\end{figure}
%%%%%%%%%%%%

Fig. \ref{all1} shows the waveforms, frequencies, phases, and $x$ for NL3, G3, and IOPB-I sets. The stiff EOS, like NL3, predicts higher mass and tidal deformability than the softest EOS, like G3. Therefore, the massive NS is easily deformed by the presence of tidal fields created by its companion star. Hence, in the inspiral phase, a less massive BNS system sustains a longer time than a more massive BNS system. This is because the tidal interactions accelerate the orbital evolution in the late inspiral phases due to increased interactions between two NSs. Some significant effects have been seen in Fig. \ref{iopb_wo_tidal} with and without tidal interactions.

To have a quantitative and comparative estimation for the inspiral properties of the BNS merger, we enumerate the numerical value of $t_{\rm ret}$, $\Phi$, $f$ and $x$ in Table \ref{table2} using NL3, IOPB-I, and G3. The last orbital time is 0.0583 sec. for NL3, which is less than from other two sets and confirms the model-dependent prediction. There is significant influence seen for $t_{\rm ret}$, $\Phi$ and $f$, except the value of $x$. The frequency and phase are also increased for the BNS of soft EOS. The pictorial representations of $f$, $\Phi$, and $x$ are compared in Fig. \ref{all1} with and without DM contains.
%%%%%%%%%%%%%
\begin{table}
	\centering
	\caption{The orbital phase ($\Phi$), gravitational wave frequency ($f$), and post-Newtonian parameter ($x$) are given in the last inspiral orbits of the BNS system.}
	\label{table2}
	\begin{tabular}{|l|l|l|l|l|}
		\hline %\hline
		EOSs & $t_{\rm ret}$ (sec.) & $\Phi$ (radians)& $f$ (Hz) & \quad $x$ \\ \hline
		NL3        & 0.0583 & 102.505 & 2356.47  & 0.213  \\ \hline
		G3         & 0.0597 & 106.383 & 2377.13  & 0.214  \\ \hline
		IOPB-I     & 0.0593 & 105.178 & 2343.66  & 0.212  \\ \hline
		IOPB-I+DM3 & 0.0597 & 106.429 & 2387.09  & 0.215   \\ \hline
		IOPB-I+DM5 & 0.0603 & 108.224 & 2365.90  & 0.214  \\ \hline  %\hline
	\end{tabular}
\end{table}
%%%%%%%%%%%%%%
%%%%%%%%%%%%%%
\begin{figure}
	\centering
	\includegraphics[width=0.7\textwidth]{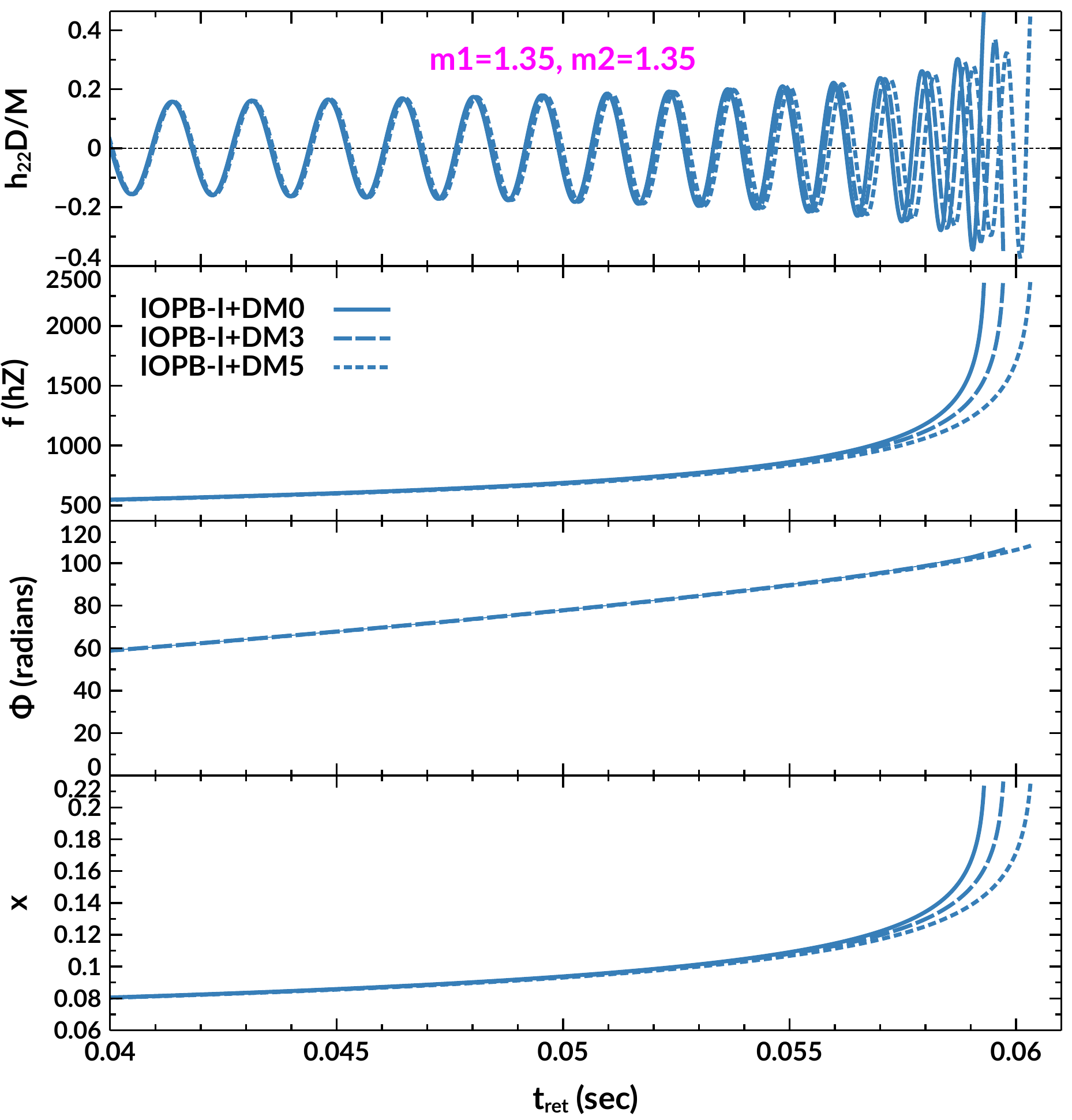}
	\caption{Same as Fig. \ref{all1}, but for DM admixed NS with IOPB-I parameter set as a representative case.}
	\label{all2}
\end{figure}
%%%%%%%%%%%%

With the addition of DM to the NS EOS, the calculated macroscopic properties ($M$, $R$, and $\lambda$) decreases. As a result, the tidal interaction between the two NSs becomes weaker compared to the normal BNS combination. For example, $M=2.149 M_\odot$ without DM and it is 2.051 $M_{\odot}$ with $k_f^{\rm DM}=0.03$ GeV with IOPB-I set. Similarly, the radius of the star is $R_{1.35}=13.168$ km without DM, and it is 12.047 km when DM is included (see Table \ref{table1}). The calculated values for $f$, $\Phi$, and $x$ are tabulated in Table \ref{table2} for IOPB-I parameter set as a representative case with DM momenta $k_f^{\rm DM}=0.03$ and $0.05$ GeV. The time period $t_{\rm ret}$ is found to be 0.0603 sec. with IOPB-I+DM5, which is larger than the other two values obtained with IOPB-I and IOPB-I+DM3. These results infer that the BNS system with a higher percentage of DM sustains more time in the inspiral phase (see Fig. \ref{all2}). The other quantities, such as $f$ and $\Phi$ also affected by DM, as shown in Fig. \ref{all2}. 

In general, with the addition of DM, the EOS becomes softer, which affects the macroscopic properties of the BNS, i.e., the mass of the NS decreases with DM. However, much ambiguity prevails in the post-merger remnants of BNS. This has been settled partially after the analysis of GW170817 and GW190425 events. Some possible pathways can be found in Ref. \cite{Sarin_2021}. The likely theoretical probabilities for a post-merger NS remnant depend on (i) the remnant mass ($M_R$) and (ii) the EOS of the NS, which dictates TOV mass ($M_{\rm TOV}$);
%%%%%%%%%%%%%%%%%
\begin{enumerate}
	\item if $M_R \geq 1.5 \ M_{\rm TOV}$, the system promptly collapses into a black hole.
	\item for $1.2 \ M_{\rm TOV} \leq M \leq 1.5 \ M_{\rm TOV}$, a hypermassive neutron star survives the collision but collapses to form a black hole on dynamical timescales.
	\item in the case, $M_{\rm TOV}<M_R \leq 1.2 \ M_{\rm TOV}$, a supermassive NS survived the collision and will collapse to form a black hole on secular timescales.
	\item if $M_R\leq M_{\rm TOV}$, a stable NS is formed after the merger.
\end{enumerate}
%%%%%%%%%%%%%%%
We refer the readers to see Fig. 1 of this Ref. \cite{Sarin_2021} for a clear picture. The post-merger remnant of DM admixed BNS merger depends on the percentage of DM inside the NS. In Ref. \cite{EllisPLB_2018}, it is reported that for a contribution of DM $\sim 5-10$\% of the NS mass, a different peak in the power spectral density of the GW signal is found. Also, a self-interacting DM of about $\leq10\%$ leads to the early formation of a black hole \cite{Pollack_2015}, which can be a subsequent seed for a supermassive black hole. Therefore, to understand the nature of the post-merger remnant of DM admixed BNS, one has to know (i) the exact nature of DM particles and (ii) the percentage of DM present inside the NS.
%%%%%%%%%%%%%%%%%%%%%
\section*{Conclusion}
%%%%%%%%%%%%%%%%%%%%%
In this chapter, we have calculated the DM admixed binary NS properties with RMF and E-RMF equation of states. The shape of the BNS is deformed due to the mutual interactions between them. The rate of deformations can be measured by various Love numbers and their corresponding deformabilities. Here, we calculated those Love numbers and deformabilities for both gravitoelectric and gravitomagnetic cases for DM admixed NS. It has been observed that the magnitude of different Love numbers increases with DM percentage for a fixed value of $l$. However, their corresponding tidal deformabilities decreased with increasing the value of DM percentage. 

In the late inspiral stage, the distance between binary stars decreases by the emission of GWs. We have calculated the properties of the BNS mainly in the late stage of inspiral with the post-Newtonian expansion for the equal mass binary cases. The inspiral properties, such as its phase, frequency, and post-Newtonian parameter, as well as various waveforms, are calculated. The properties of waveforms in the inspiral phase are compared with and without tidal interactions. We found that if the tidal interactions are switched off, the BNS sustain more time in their inspiral phases. Also, the polarization waveform $h_+$, $h_\times$, and the strain amplitude of $h_{22}$ modes have been calculated for different EOSs in the retarded time interval at a distance of 100 Mpc between the source and the observer.

We also calculated the last inspiral time of BNS. It is observed that the binary system favors a longer time for softer EOS (G3) than the stiffer one (NL3). The other quantities (frequencies and phase) are almost similar to all the considered sets in the calculations, but the inspiral times in the last orbit are found to be force-dependent. We get that with the addition of DM, the binary system becomes less deformed and sustains more time in its inspiral phases. The DM also affects the other properties compared to the ordinary NS, and its effects on the inspiral properties of the BNS were significant. Finally, we suggest that one has to take DM inside the compact objects while modeling the inspiral waveforms for the BNS systems. One can also use more accurate methods, such as numerical relativity techniques, to study the DM effects on the inspiral-merger-ringdown and post-merger properties.
%%%%%%%%%%%%%%%%%%%%%%%%%%%%% END %%%%%%%%%%%%%%%%%%%%%%%%%%%%%%%%%%%%%%%%%%%%%%%%
%\blankpage 
%%%%%%%%%%%%%%%%%%%%%%%%% Chapter-7 %%%%%%%%%%%%%%%%%%%%%%%%%%%%%%%
%%%%%%%%%%%%%%%%%%%%%%%%% CHAPTER - 7 %%%%%%%%%%%%%%%%%%%%%%%%%
\chapter{Summary and Conclusion}
\label{C7}
%%%%%%%%%%%%%%%%%%%%%%%%%%%%%%%%%%%%%%%%%%%%%%%%%%%%%%%%%%%%%%%
This thesis explored several areas of nuclear astrophysics, such as nuclear structures, NM, NSs, and gravitational waves. We briefly introduced the NS, its birth, and its evolution in the Universe. The detection mechanism and the internal structure of the NS are also discussed. To study the properties of NS, one has to solve the TOV equations, where the star EOS is used as an input. The RMF and E-RMF formalisms are chosen to get the NS equation of states. In the first part of the thesis, we validated the formalism by applying the model to finite nuclei and infinite nuclear matter. The last part of the introduction explained the methodology, problem statement, and motivation for the thesis. 

After a brief introduction, we tried to present the formalism of the relativistic mean-field models in Chapter-\ref{C2}. In the RMF theory, the nuclear interactions are carried out by exchanging different mesons. We used the effective field theory motivated RMF model because it successfully predicts the properties of finite nuclei and nuclear matter. In recent years, it has also well explained the NS properties such as $\sim 2 \, M_\odot$ constraint. Therefore, in the thesis, we have used RMF and E-RMF to calculate the properties of the NS. In the E-RMF formalism, we include the self and cross-couplings of the mesons such as $\sigma$, $\omega$, $\rho$, $\delta$ up to $4^{\rm th}$ order. Hence, the system's energy is the addition of nucleons and mesons. The finite nuclei properties, such as binding energy per nucleon, skin thickness, charge radius, density distributions, single particle energy levels, and two neutron separation energy for different successful forces, along with the recently reported BigApple, are calculated. It has been observed that almost all predicted properties of the finite nuclei are well consistent with the experimental data available for known magic and doubly magic spherical nuclei. The properties of nuclear matter have been calculated with various RMF parameter sets to check their consistency with the experimental/empirical data. Lastly, we extended the calculations to understand the EOS for the beta-equilibrated NS and also checked the speed of sound for consistency throughout the star for different parameter sets.

In Chapter-\ref{C3}, in addition to nucleons, we have added extra candidates known as DM to enumerate their effects on NM quantities. For that, we choose a simple DM model, where the DM particles interact with nucleons by exchanging standard model Higgs. The DM density is hypothesized by choosing one-third of the total baryon density. We have chosen Neutralino as a DM candidate, a weakly interacting massive particle with a mass of 200 GeV. The coupling constants between DM-nucleons and nucleons-Higgs are considered by following the earlier calculations \cite{Panotopoulos_2017} and constrained with the help of experimental data. Therefore, the energy density and pressure of the NM system are the combinations of nucleons, mesons, and DM. To estimate the effects of DM on NM properties such as binding energy, incompressibility, effective mass, symmetry energy, and its different coefficients with different fractions of DM and neutron-proton asymmetry, the RMF and E-RMF models are used on top of the DM EOS. The EOS becomes softer with the increasing percentage of DM. The energy density also increases with $k_f^{\rm DM}$ without adding much to the pressure. We observed that the DM has marginal effects on the NM properties, except the EOSs, the binding energy, and $Q_{\rm sym}$. This is because it has less effect on the system's pressure due to the contribution of the Higgs field in the order of $10^{-6}-10^{-8}$ MeV. One can also choose other DM models by varying their mass and percentage to explore the DM admixed NM properties.

The isolated static/rotating DM admixed NS properties such as mass, radius, tidal deformability, and moment of inertia have been calculated with the EOSs obtained from the RMF and E-RMF models in Chapter-\ref{C4}. It has been observed that DM has significant effects on both static and rotating NS. The mass and radius decrease by increasing the DM percentage. However, if we compare the static and rotating cases, the change is $\sim 20 \% $ for different macroscopic properties. We have also calculated the curvatures for both inside and outside the NS with DM content. The change of curvature with and without DM is found to be approximately 33\% for the canonical star. This value increases with the maximum mass of the NS. The surface curvature and the compactness are also estimated, and it is obtained that their magnitude increased with DM percentage. For a stable system, the binding energy is always negative (here, we used the notation negative for a stable system). However, with the addition of DM inside the NS, the binding energy moves towards positive values, making the NS unstable. From this, we concluded that a tiny amount of DM can accumulate inside the NS. A higher amount of DM warms the NS and accelerates the Urca process, which enhances the cooling of NS and causes instability. GW19814 event opened a debatable issue regarding the type of secondary component ``whether it is the lightest black hole BH or heaviest NS?". We have suggested that the secondary component might be a DM admixed NS if the underlying nuclear EOS is sufficiently stiff. 

In Chapter-\ref{C5}, we have calculated the $f$-mode oscillations of the DM admixed hyperon stars within the relativistic Cowling approximations. Inside the DM admixed hyperon star, they interact with DM by exchanging Higgs. Also, the coupling between DM-Higgs is assumed to be the same as nucleons. We have calculated the EOSs for various combinations of hyperons-mesons couplings with different group theory methods. Therefore, total energy density and pressure are the combinations of nucleons, hyperons, mesons, leptons, and DM. With those EOSs, we calculated the $f$-mode frequency of the hyperon star with DM content. The values of these frequencies for softer EOSs are larger (lower mass NSs) and vice-versa for stiffer EOS. The angular frequencies scaled by mass and radius are calculated as a function of compactness. It has been found that both quantities follow a linear relationship. We found the correlations between $f_{1.4} – \Lambda_{1.4}$ and $f_{\rm max} – \Lambda_{\rm max}$ are almost consistent with the earlier prediction \cite{Wen_2019}. Our results correspond to $f_{1.4}$ is in the range $1.78-2.22$ kHz for the considered EOSs. We hope the discovery of more BNS merger events will open up tight constraints on the $f$-mode frequencies in the future.

Chapter-\ref{C6} is dedicated to binary NS in the presence of DM. It is one of the most exciting aspects after the discovery of the gravitational waves from the merger of binary NSs. In this chapter, we have explored the DM admixed binary NS properties with RMF and E-RMF EOSs. Various Love numbers and corresponding tidal deformabilities are essential quantities to explain the degree of deformation for a BNS. Therefore, we have calculated those Love numbers and their corresponding deformabilities for both gravitoelectric and gravitomagnetic cases. It has been observed that the magnitude of different Love numbers increases with DM percentage for a fixed value of $l$. However, their corresponding tidal deformabilities decrease with increasing DM percentage. In the late inspiral stage, the BNS system rotates quite first by emitting a huge amount of GWs. We have calculated those late inspiral properties, such as the frequency of emitted GWs, waveforms, phase, etc., with the post-Newtonian expansion for the equal mass binary cases. The waveforms emitted in the inspiral phase are compared with and without tidal interactions. We found that if the tidal interactions are switched off, the BNS sustain more time in their inspiral phases. The polarization waveform $h_+$,  $h_\times$, and the strain amplitude of $h_{22}$ modes are calculated in the retarded time interval at a distance of 100 Mpc between the source and the observer. We observed that the binary system favors a longer time for softer EoS (G3) than the stiffer NL3. The other quantities (frequencies and phase) are almost similar for assumed EOSs. However, the inspiral times in the last orbit are found to be model dependent. For a DM admixed BNS, the binary system becomes less deformed and sustains more time in its inspiral phases. The DM also affects the other properties as compared to the ordinary NS, and the DM effects on the inspiral properties of the BNS were significant. As a result, we advise that when modeling the inspiral waveforms for the BNS systems, one can take DM inside the compact objects. Numerical relativity tools can be used for an accurate description to explore the DM effects on the inspiral-merger-ringdown and post-merger properties. 
%%%%%%%%%%%%%%%%%%%%%%%%%%%%%%%%
\subsection*{Future Prospective}
%%%%%%%%%%%%%%%%%%%%%%%%%%%%%%%%
%%%%%%%%%%%%%%%%%%%%%%%%%%%%%%%%%%%%%%%%%%%%%%%%%
\subsubsection*{1. Different types of DM model:-}
%%%%%%%%%%%%%%%%%%%%%%%%%%%%%%%%%%%%%%%%%%%%%%%%%
An effective Lagrangian density with different types of DM particles will be constructed. This may satisfy the relic abundance, and DM detection limit observed from various experimental data. In our present studies,  we are taking only fermionic DM; however, there is a possibility that the DM may be bosonic also \citep{Maity_2021, Rafiei_2022}. In that case, one has to change the framework to calculate the properties of static and rotating NS and compare with and without DM cases \cite{Grandclement_2014}. In addition to this, the study of symmetric/asymmetric self-annihilating DM is also interesting. 
%%%%%%%%%%%%%%%%%%%%%%%%%%%%%%%%%%%%%%%%%%%%%%%%%%%%%
\subsubsection*{2. Effects of DM on hyperon puzzle:-}
%%%%%%%%%%%%%%%%%%%%%%%%%%%%%%%%%%%%%%%%%%%%%%%%%%%%%
Hyperons may appear in the inner core of NSs at densities of about $2-3 \rho_0$. Although the presence of hyperons in NSs seems energetically unavoidable, their substantial softening of the EOS leads to a reduction of maximum mass, which is not compatible with observation. This is known as the ``Hyperon Puzzle". The DM particles can be accreted not only inside the NS but also hyperon stars. However, the exact coupling between hyperons with nucleons and mesons is still a debatable issue. To overcome these difficulties, several works have been dedicated to explaining both the hyperon puzzle and coupling strength between different species \cite{Weissenborn_2012, Oertel_2015, Vida_a_2016}. From the previous works \cite{Weissenborn_2012, Oertel_2015, Vida_a_2016}, we may extend to DM admixed hyperon star. We can put constraints on the hyperon-DM coupling strength from the different observational and experimental data. That may provide a narrow window for the existence of DM and can explain the scattering cross-section estimations from terrestrial DM detectors.
%%%%%%%%%%%%%%%%%%%%%%%%%%%%%%%%%%%%%%%%%%%%%%%%%%%%%%%%%%%%%%%%%
\subsubsection*{3. Binary NS merger and post-merger properties:-}
%%%%%%%%%%%%%%%%%%%%%%%%%%%%%%%%%%%%%%%%%%%%%%%%%%%%%%%%%%%%%%%%%
The numerical relativity simulation already opens the door to explore the second most enigmatic merger between two NSs. We propose to understand more details in this direction using the DM equation of states. So far, our analysis reveals that the DM admixed BNS sustains more time in the inspiral phase than the  BNS without DM admixture. We find it interesting to emphasize the merger and post-merger properties for DM admixed NS with the help of numerical relativity techniques. In future work, we propose to write our numerical relativity packages or Einstein toolkit thorns to enumerate the DM effects on the merger and post-merger systems.
%%%%%%%%%%%%%%%%%%%%%%%%%%%%%%%%%%%%%%%%%%%%%%%%%
\subsubsection*{4. Emitted gravitational waves:-}
%%%%%%%%%%%%%%%%%%%%%%%%%%%%%%%%%%%%%%%%%%%%%%%%%
The GWs emitted from the inspiral-merger-ringdown phases are the primary source for the GW detectors. The detected GWs properties are compared with different theoretical templates to estimate the parameters of the GW's origin, distance, and mass of the sources. Therefore, making some templates using the present DM admixed NS equation of states with different analytical techniques is a fascinating subject. Using such templates, one can estimate the GWs properties and the source details more accurately. We can use the observational data of LIGO/Virgo, which will provide direct constraints on our theoretical model and fix the amount of the accumulated DM inside the star.
%%%%%%%%%%%%%%%%%%%%%%%%%%%%%%%
\subsection*{Expected Outcomes}
%%%%%%%%%%%%%%%%%%%%%%%%%%%%%%%
It has been proposed that the merger of DM admixed binary NS provides an extra peak in the post-merger signal. From this post-merger data, we may constrain the percentage of DM inside the NS. Also, we can improve the various existing templates for the inspiral-merger-ringdown phases for BNS, which may better estimate the GWs source and its origin in the presence of DM.
%%%%%%%%%%%%%%%%%%%%%%%%%%%%%%%%%%%%%%%%%%%% END %%%%%%%%%%%%%%%%%%%%%%%%%%%%%%%%%%%%%%%%%%%
%\blankpage 
%%%%%%%%%%%%%%%%%%%%%%%%%%%%%%%%%%%%%%%%%%%%%%%%%%%%%%%%%%%%%%%%%%%
\addcontentsline{toc}{chapter}{REFERENCES}
\bibliographystyle{hcdas}
\bibliography{thesis}
\end{document}